\documentclass[a4paper,10pt,twoside]{report}

\usepackage[utf8]{inputenc}
\usepackage{graphicx}
\graphicspath{ {images/} }

\usepackage{latexsym}
\usepackage{amsmath}
\usepackage{amsfonts}
\usepackage{listings}
\usepackage{natbib}
\usepackage{bibentry}
\usepackage{anyfontsize} 
\usepackage{ogonek}
\usepackage{subfigure}
\usepackage{fancyhdr}
\usepackage{chngcntr}

\fancyheadoffset[]{0pt}

\fancypagestyle{plain}{%
\fancyhf{} 
\fancyfoot[C]{\bfseries \thepage} 

}

\usepackage[a4paper, left=2cm, right=2cm, top=3.7cm, bottom=2.5cm, headsep=1.2cm, headheight=16pt, bindingoffset=1cm]{geometry}

\usepackage{hyperref} 
\hypersetup{pdfborder={0 0 0 0}}
\providecommand\phantomsection{} 
\def\D{\mathrm{d}} 

\makeatletter
\renewcommand*{\cleardoublepage}{\clearpage\if@twoside \ifodd\c@page\else
\hbox{}%
\thispagestyle{empty}%
\newpage%
\if@twocolumn\hbox{}\newpage\fi\fi\fi}
\makeatother

\def\one{\mbox{1\hspace{-3.85pt}\fontsize{11}{14.4}\selectfont\textrm{1}}} 

\begin{document}

\pagenumbering{roman}

\title{
	\textbf{Analytic estimation of parameters of stochastic volatility diffusion models with exponential-affine characteristic function for currency option pricing}
	\\[1.65cm]
	\includegraphics[width=45mm]{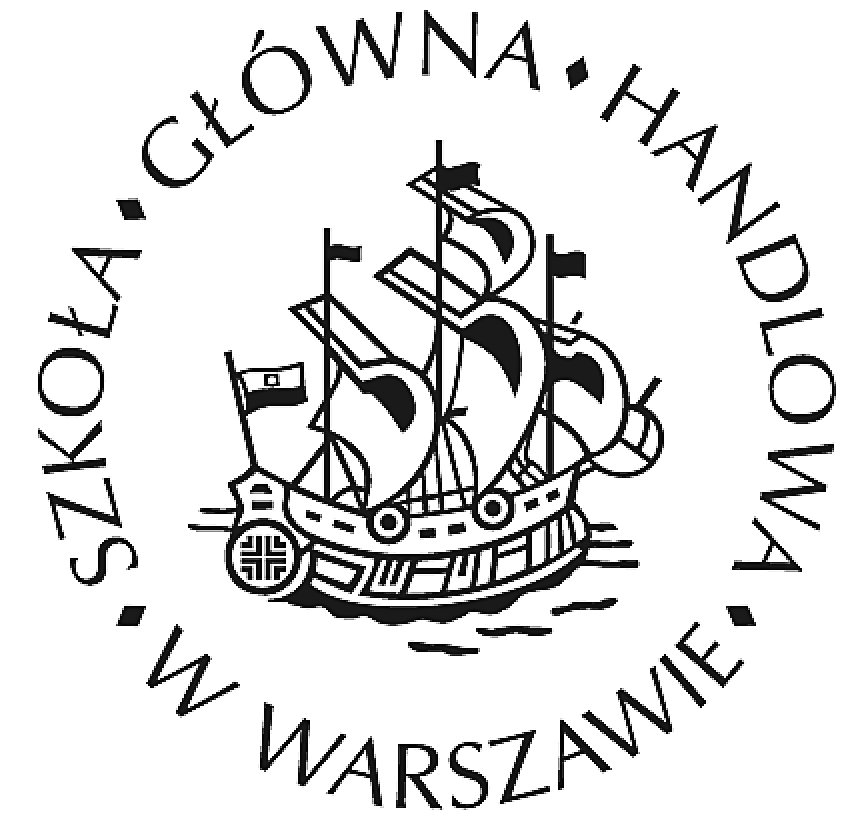}~	
	\\[0.55cm]
}
\author{
	\Large \textbf{Miko\l{}aj~\L{}ab\k{e}dzki} \normalsize\\
	\\[0.8cm]
	\textit{A dissertation submitted to the}\\
	\textit{Collegium of Management and Finance}\\
	\textit{Warsaw School of Economics}\\
	\textit{in partial fulfillment of the}\\
	\textit{requirements for the degree of}\\
	\textbf{Doctor of Philosophy}\\
	\\
	\textit{Supervisor: prof. Tomasz Kuszewski, PhD}\\
	\\
}
\date{{December 2017}}

\maketitle


\newpage\null\thispagestyle{empty}
\vspace{10cm}
\begin{center}
\Large Copyright \textcircled{c} Miko\l{}aj~\L{}ab\k{e}dzki, 2017 \\ ALL RIGHTS RESERVED \normalsize
\end{center}
\newpage

\chapter*{Abstract}
The dissertation principal research objective is a development and justification of a method for derivation of approximate formulas for estimation of two parameters in the stochastic volatility diffusion models with exponentially-affine characteristic function of price distribution and single-factor or two-factor variance. The new formulas should be more advantageous than other formulas in terms of accuracy of option pricing with regard to observed option prices and in terms of the accuracy and risk of model calibration using this formula to determine the starting point for the local minimization algorithm. Considered parameters correspond to the volatility of the stochastic factor in the instantaneous variance dynamics and the correlation of stochastic factors in the instantaneous variance dynamics and the price dynamics.\par

The dissertation consists of five chapters. In the first chapter there is short characteristic of currency option market, methods of option pricing and there is presented general form of stochastic volatility models. The second chapter contains derivation of replication strategy dynamics for the general form of considered models. The chapter contains also the  derivation of the solution for a new proposed model with a two-factor volatility, namely OUOU model. Third chapter contains statistical analyses of the distribution of implied volatilities and dynamics of implied volatility surface. This includes principal component analysis and common factor analysis applied to the two dimensions of the implied volatility surface dynamics. The fourth chapter concerns methods of calibration of stochastic volatility models, in particular the Heston model. Moreover, approximate formulas for estimation of the Heston model parameters are presented. This chapter includes also the development of a new method, namely, Implied Central Moments method, which enables derivation of approximate formulas for parameters of Heston and Sch\"obel-Zhu models. Besides, additional methods for using those formulas for models with two-factor variance, namely, the Bates two-factor variance model and the author's OUOU model, are presented. The fifth chapter contains empirical tests of methods from the fourth chapter, namely author's Implied Central Moment formulas and Durrleman formulas for the Heston model, in order to compare and determine the quality of parameters estimates with approximate formulas. The tests have been carried out on the EURUSD options market and have proved the advantage of the new method.\par

All research objectives in the dissertation have been realized. This gives an expansion of existing tools for derivatives pricing and their risk assessment in financial institutions. The added value is the more accurate approximate formula for estimation of parameters of stochastic volatility models, in particular models of Heston and Sch\"obel-Zhu, as well as models with two-factor variance. Fast method of parameters estimations can then be used as a starting point in a numerical calibration scheme. Additional scientific and practical value is the introduction of a new model, i.e. the OUOU model which is an extension of the Sch\"obel-Zhu model and has a semi-analytical formula for the valuation of european options. Such a~model was so far not considered nor studied in the literature.


\chapter*{Streszczenie (Polish Abstract)}

G\l{}\'{o}wnym celem badawczym tej rozprawy doktorskiej jest opracowanie i uzasadnienie metody wyprowadzania przybli\.{z}onych formu\l{} do estymacji dw\'{o}ch parametr\'{o}w w dyfuzyjnych modelach zmienno\'{s}ci stochastycznej z wyk\l{}adniczo-afiniczn\k{a} funkcj\k{a} charakterystyczn\k{a} rozk\l{}adu ceny i jednoczynnikow\k{a} lub dwuczynnikow\k{a} wariancj\k{a}.
Nowe formu\l{}y powinny by\'{c} bardziej korzystne ni\.{z} inne formu\l{}y w aspekcie precyzji wyceny opcji wzgl\k{e}dem obserwowanych cen opcji oraz w aspekcie precyzji i ryzyka kalibracji modelu, przy zastosowaniu tej metody do wyznaczenia punktu startowego dla algorytmu minimalizacji lokalnej.
Rozwa\.{z}ane parametry odpowiadaj\k{a} zmienno\'{s}ci czynnika stochastycznego w dynamice wariancji natychmiastowej i korelacji czynnik\'{o}w stochastycznych z dynamiki wariancji natychmiastowej i dynamiki ceny.

Praca sk\l{}ada si\k{e} z pi\k{e}ciu rozdzia\l{}\'{o}w.
W pierwszym rozdziale jest kr\'{o}tka charakterystyka rynku opcji walutowych, metody wyceny opcji oraz przedstawiona jest og\'{o}lna posta\'{c} modeli zmienno\'{s}ci stochastycznej.
Rozdzia\l{} drugi zawiera wyprowadzenie dynamiki strategii replikuj\k{a}cej dla og\'{o}lnej postaci rozwa\.{z}a-nych modeli. Rozdzia\l{} zawiera r\'{o}wnie\.{z} wyprowadzenie rozwi\k{a}zania dla nowego, proponowanego modelu o dwuczynnikowej zmienno\'{s}ci, a mianowicie modelu OUOU.
Trzeci rozdzia\l{} zawiera analizy statystyczne rozk\l{}adu implikowanych zmienno\'{s}ci i dynamiki implikowanej powierzchni zmienno\'{s}ci. Obejmuje to analiz\k{e} g\l{}\'{o}wnych sk\l{}adowych i analiz\k{e} czynnikow\k{a} zastosowan\k{a} do dw\'{o}ch wymiar\'{o}w dynamiki implikowanej powierzchni zmienno\'{s}ci.
Czwarty rozdzia\l{} dotyczy metod kalibracji modeli zmienno\'{s}ci stochastycznej, w szczeg\'{o}lno\'{s}ci modelu Hestona. Ponadto, przedstawiono przybli\.{z}one formu\l{}y do estymacji parame-tr\'{o}w modelu Hestona. Rozdzia\l{} ten obejmuje r\'{o}wnie\.{z} opracowanie nowej metody, czyli metody Impli-kowanych Moment\'{o}w Centralnych, kt\'{o}ra umo\.{z}liwia wyprowadzanie przybli\.{z}onych wzor\'{o}w dla parametr-\'{o}w modeli Hestona i Sch\"obla-Zhu. Ponadto, przedstawiono dodatkowe metody do wykorzystania tych formu\l{} dla modeli z dwuczynnikow\k{a} wariancj\k{a}, a mianowicie dwuczynnikowego modelu wariancji Batesa oraz autorskiego modelu OUOU.
Rozdzia\l{} pi\k{a}ty zawiera testy empiryczne metod z czwartego rozdzia\l{}u, a~mianowicie formu\l{} Implikowanych Moment\'{o}w Centralnych oraz formu\l{} Durrlemana dla modelu Hestona, w~celu por\'{o}wnania i okre\'{s}lenia jako\'{s}ci estymacji parametr\'{o}w przy pomocy przybli\.{z}onych wzor\'{o}w. Testy zosta\l{}y przeprowadzone na rynku opcji kursu EURUSD i pokaza\l{}y przewag\k{e} wynik\'{o}w nowej metody. 



Wszystkie cele badawcze w rozprawie zosta\l{}y zrealizowane. Daje to mo\.{z}liwo\'{s}\'{c} rozszerzenia istniej\k{a}cych narz\k{e}dzi do wyceny instrument\'{o}w pochodnych i oceny ich  ryzyka w instytucjach finansowych. Warto\'{s}ci\k{a} dodan\k{a} jest dok\l{}adniejsza przybli\.{z}ona formu\l{}a do estymacji parametr\'{o}w modeli zmienno\'{s}ci stochastycznej, w szczeg\'{o}lno\'{s}ci modeli Hestona i Sch\"obla-Zhu, a tak\.{z}e modeli z dwuczynnikow\k{a} war-iancj\k{a}. Szybka metoda estymacji parametr\'{o}w mo\.{z}e by\'{c} nast\k{e}pnie wykorzystana jako punkt pocz\k{a}tkowy w~kalibracji numerycznej. Dodatkow\k{a} warto\'{s}ci\k{a} naukow\k{a} i praktyczn\k{a} jest wprowadzenie nowego mode-lu, tj. modelu OUOU, kt\'{o}ry jest rozszerzeniem modelu Sch\"obla-Zhu i ma p\'{o}\l{}jawn\k{a} formu\l{}\k{e} do wyceny opcji europejskich. Takiego modelu do tej pory nie rozwa\.{z}ano ani nie badano w literaturze.

\newpage\null\thispagestyle{empty}
\vspace{10cm}
\begin{center}
	\Large \textit{To my parents} \normalsize
\end{center}
\newpage

\chapter*{Acknowledgements}
I would like to thank my supervisor professor Tomasz Kuszewski for the continuous support of my PhD study and related research, for his patience and motivation. His guidance helped me a lot.

Above all, I would like to thank my parents for all their care and love. Their support helped me a lot before and during my research.


\renewcommand{\thesection}{\thechapter.\arabic{section}}

\fancyhead{} 
\fancyhead[C]{\leftmark}
\fancyhead[RO,LE]{\thepage}
\renewcommand{\MakeUppercase}[1]{\textit{#1}}
\tableofcontents
\nobibliography* 
\clearpage
\phantomsection
\renewcommand{\MakeUppercase}[1]{\uppercase{#1}}

\fancyhead{} 
\fancyhead[RO,LE]{\thepage}
\fancyhead[LO]{\rightmark}
\fancyhead[RE]{\leftmark}

\addcontentsline{toc}{chapter}{List of tables}
\listoftables
\clearpage{\pagestyle{empty}\cleardoublepage}

\addcontentsline{toc}{chapter}{List of figures}
\listoffigures
\clearpage
\phantomsection

\addcontentsline{toc}{chapter}{List of abbreviations and chosen symbols}
\chapter*{List of abbreviations and chosen symbols}
\markright{\hfill \textit{List of abbreviations and chosen symbols} \hfill}{}

\begin{tabular}{l l}
\hline
\textbf{Abbreviation} & \textbf{Description} \\
\hline
10FLY & 10-delta Vega-weighted Butterfly\\
10RR & 10-delta Risk Reversal \\
25FLY & 25-delta Vega-weighted Butterfly\\
25RR & 25-delta Risk Reversal \\
AIC & Akkaike Information Criterion \\
APT & Arbitrage Price Theory \\
ATM & At The Money \\
AUROC & Area Under Receiving Operating Characteristic \\
BIC & Bayes-Schwarz Information Criterion\\
CIR & Cox-Ingersol-Ross \\
EWMA & Exponentialy Weigthed Moving Average \\
ITM & In The Money \\
MA & Moving Average  \\
MAE & Mean Absolute Error \\
MAPE & Mean Absolute Percentage Error \\
ICM & Implied Central Moments\\
MSE & Mean Square Error \\
MSPE & Mean Square Percentage Error \\
OTC & Over The Counter \\
OTM & Out Of The Money \\
OU & Ornstein-Uhlenbeck \\
PCA & Principal Component Analysis \\
RMSE & Root Mean Square Error \\
SABR & Stochastic Alpha Beta Rho \\
SMA & Simple Moving Average \\
VIX & Square root of implied variance calculated according to CBOE methodology\\
\hline
\end{tabular}

\begin{tabular}{l l}
\hline
\textbf{Symbol} & \textbf{Description} \\
\hline
$B_t^d$ & price of domestic bond in domestic currency in moment $t$\\
$B_t^f$ & price of foreign bond in foreign currency in moment $t$\\
$\mathbb{C}_t$ & curvature of volatility smile in moment $t$\\
$C_t$ & call option price in moment $t$ \\
$dW_t^S$ & Wiener process of underlying instrument prices equation \\
$dW_t^\nu$ & Wiener process of variance/volatility equation \\
$F(t,T)$ & forward price of maturity $T$ in moment $t$ \\
$K$ & option strike price \\
$\mathbb{M}_t$ & volatility smile term structure premium in moment $t$\\
$\mathbb{Q}^d$ & risk-neutral measure of domestic investors \\
$\mathbb{Q}^f$ & risk-neutral measure of foreign investors \\
$P_t$ & put option price in moment $t$ \\
$R_t$ & underlying instrument logarithmic return in moment $t$\\
$r_f$ & foreign logarithmic interest rates \\
$r_d$ & domestic logarithmic interest rates \\
$\mathbb{S}_t$ & slope of volatility smile in moment $t$\\
$S_t$ & spot price in moment $t$ \\
$V_t$ & option price in moment $t$ \\
$\beta$ & exponent of spot price in general form of stochastic volatility models \\ 
$\gamma_i$ & $i$-th eigenvector of a matrix \\
$\delta t$ & time interval in form of year fraction \\
$\mathbf{\epsilon}$ & random vector of errors \\
$\theta$ & long term mean of the variance/volatility process \\
$\kappa$ & speed of mean-reversion in the variance/volatility process \\
$\lambda$ & Box-Cox transformation parameter \\ 
$\mu$ & level of the drift of the underlying instrument price process \\
$\mu_n$ & $n$-th central moment \\
$\nu_t$ & variance/volatility in moment $t$ \\
$\rho$ & correlation between prices Wiener process and variance/volatility Wiener process \\
$\sigma_{K}(\tau)$ & implied volatility of options of maturity $\tau$ and strike price $K$\\
$\tau$ & time to maturity \\
$\upsilon$ & variance in auxiliary equation in Sch\"obel-Zhu model\\
$\phi$ & characteristic function of $\ln(S)$ \\
$\psi$ & characteristic function of dampened call option price\\
$\omega$ & volatility of the variance/volatility process \\
$\Delta_t$ & share of foreign bonds in replication portfolio in moment $t$\\
$\Delta$ & differentiation operator, if not defined before \\
$\Theta$ & vector containing all free parameters of stochastic volatility model \\
$\Xi_t$ & share of variance derivatives in replication portfolio in moment $t$\\
$\Pi_t$ & value of replication portfolio in moment $t$\\
$\Sigma_t(K,T,S_t)$ & implied volatility surface in moment $t$ \\
\hline
\end{tabular}
\clearpage{\pagestyle{empty}\cleardoublepage}

\makeatletter
\renewcommand{\thesection}{%
  \ifnum\c@chapter<1 \@arabic\c@section
  \else \thechapter.\@arabic\c@section
  \fi
}
\makeatother

\pagenumbering{arabic}

\counterwithin{figure}{chapter}

\renewcommand{\thesection}{\Roman{section}} 
\chapter*{Introduction and methodological basis of research}
\markboth{\hfill \textit{Introduction and methodological basis of research} \hfill}{}
\addcontentsline{toc}{chapter}{Introduction and methodological basis of research}

\section{Importance of the research area}

On the FX options market the majority of the turnover takes place in a decentralized direct trade form, therefore it is categorized as Over-The-Counter (OTC) market. The European vanilla options are quoted as implied volatility, rather than the option premium, which is the value of options. In the case of the exchange option, the implied volatility can also be calculated by searching for the value of the volatility parameter in the Black-Scholes (1973) formula, which substituted into the formula can return the traded option premium. Hence, the implied volatility is understood as a value associated with a given option premium with other parameters at fixed levels. In addition, on the currency options market for a given currency pair each option is characterized by two parameters: the time to maturity and the option delta, expressing moneyness of the options, namely the relationship between the forward exchange rate and strike rate of the option. The existence of two parameters differentiating the market makes sense for the analysis of implied volatilities quotations expressed as a two-dimensional surface for a given moment in time. \par

Modelling of implied volatility surface is of great importance in the correct valuation of options, i.e. finding such a price that reflects the cost of replication of this instrument. In the case of European vanilla options the problem of modelling of the implied volatility surface is reduced to problems of interpolation and extrapolation of existing quotations for options traded on the market. In the case of exotic options this problem is more complex than interpolation, because more important is a good model describing the dynamics of price and its volatility. Then, on the basis of such model, the valuation method can be applied. The result of non-constant price volatilities is the existence of non-flat implied volatility surfaces. Models, which are often used for pricing exotic options, are stochastic volatility models (e.g. the Heston model), which reflect the dynamics of variance and include two sources of risk, first in the dynamics of price and the second in the dynamics of variance. They allow to simulate both underlying instrument price path and variance path and thus to calculate the expected value of options with complicated payout profile, including options depending on the levels of variance in the future. \par

A major problem when using stochastic volatility models is their calibration. A calibration of the model is the process of searching for such model parameters, which will minimize the distance between the option prices from the model and the corresponding option prices from the market. Usually, European vanilla option prices are used for the calibration process. \par

The popular Heston model has five parameters, which have to be calibrated. In market practice, among important things is not only the accuracy of the model, but also the speed of the calibration, because during higher volatility times on the market the need for rapid recalibration of the model can be very important in the commercial activities of a bank, a broker or other institutional market participant.
The area of research regarding the choice of the model for option pricing is also important, because it concerns an accurate pricing of all observed European vanilla options on the market, i.e. the whole implied volatility surface. Almost perfect match between model and observed European vanilla option prices can be interpreted as good calibration of market dynamics model, which can be used then for pricing of exotic options. \par

\section{Justification for the choice of the topic}

The values of calibrated parameters in stochastic volatility models have a significant impact on the results of the valuation of currency options, including exotic options. Particular interest of researchers in this field concerns models with exponential-affine form of the characteristic function of the underlying instrument price. In such case, there is usually an analytic formula for the characteristic function. After calculating the definite integral of characteristic function and its further transformation we obtain the value of the option. Thanks to the analytic formula the option pricing and model calibration execution time is shorter than for models, which do not have an analytic solution. \par

In addition, part of the research in the field of option pricing with stochastic volatility models directly concerns deriving approximate formulas for estimation of model parameters. For most of the models a closed-form formula for a European vanilla option price either exist or can be analytically approximated. Such formula contains dependencies between market observed option prices and model parameters. Therefore, such function can be inverted to calculate the analytic formula for some of the model parameters. An approximation of a function can be done in several steps of this process. An approximate formula is understood as a theoretically justified closed-form formula, which itself has no unknown parameters. In this way a parameters are estimated without numerical methods, which are usually used in market practice for option pricing models. Although it is not guaranteed that such estimated parameters will be a global minimum of cost function, they can be used further as starting values in model calibration with numerical methods. Considered models are calibrated numerically by finding the minimum of an objective function, which has absolute or squared differences between observed option prices and model option prices.\par

It should be noted that in the area of option pricing models market data of the current moment are of main interest and not time series of historical data. Observed market prices of all options on the implied volatility surface at any given moment are basic data used for an estimation. Observed in a given moment options prices are used in this estimation. This is due to the use of risk neutral measure in the asset valuation.\par

When a good method of fast parameter estimation is available, then it is advantageous to apply it at the beginning of the model calibration process, as the starting value for the local minimization algorithm. Finding good starting point is significant issue, because objective function has a large number of local minima and due to the limited amount of time during market operations only local algorithms are used instead of global. \par

A good starting point for calibration is understood as a point that is as close to the global minimum as possible. On the other hand, this point should be estimated with an analytic formula and not as the result of the numeric minimization algorithm. Therefore, the estimation of model parameters with approximate formulas is directly applicable for this issue. \par

In addition, knowing initial estimates for part of the parameter vector, we can modify the calibration method by fixing some of parameters at constant level obtained from initial estimates and applying the local minimization algorithm only in relation to other model parameters. Moreover, for some algorithms, calibration results of some parameters are clearly unstable in time, i.e. the time series of calibrated parameter has an abnormally large value changes. The possibility to change the method to such one in which the unstable parameter of the model is always estimated with approximate formula, allows to verify the reasonableness and stability of the results of the full calibration. Such solution with fixing a parameter at a given level is reasonable when an estimation method of parameter is not using complex numerical algorithm itself. Lesser effect would be obtained from applying two separate calibrations on two disjoint subsets of parameters in two steps. A calibration of first subset of parameters, although is computationally less problematic than the full calibration, still takes more time than estimating with an analytic formula obtained with some approximation methods.\par

Another important aspect of determining the starting point with the use of estimates of model parameters from an approximate formula is a reduction of the susceptibility to diversity in calibration results using different algorithms and different cost functions. This diversity of found local minima should be as small as possible and if all found minima are global minimum, then it should be zero. Some researchers identified problem of such diversity as a calibration risk. \par

In summary, the added value from the application of an approximate formula is not only due to faster or better in terms of lower cost function values calibration of stochastic volatility model. The added value is also due to the reduced calibration risk by starting a calibration process in a point which is closer to the global minimum, thereby reducing the risk of finding local minima in the process. It is also due to reducing the quantity of parameters, which are calibrated by the algorithm.

Among sources of the calibration risk is the presence of a variety of local minima around a global minimum. In market practice the global minimum is rarely being discovered and algorithms are stopped at a satisfactory local minimum or after exceeding a pre-set maximum amount of iterations due to limited time which is available for the search. At the same time, when comparing local minima, it should be checked if all values of the calibrated parameters are at levels, which are possible in practice. Time series of calibrated parameters should also have a justifiable level of volatility, so differences between previous and current values of calibrated parameters should not be due to finding local minima in various points around global minimum.\par

\section{Research problems}

\textit{Principal research problem is:}\par

What should be the new method for derivation of approximate formulas for estimation of parameters corresponding to the volatility of the stochastic factor in the instantaneous variance dynamics and the correlation of stochastic factors in the instantaneous variance dynamics and the price dynamics in the stochastic volatility diffusion models with exponential-affine characteristic function of the price distribution and single-factor or two-factor stochastic variance, which would use information from all options from volatility surface and would be more advantageous than other formulas in terms of accuracy of option pricing with regard to observed European vanilla option prices and in terms of the accuracy and risk of model calibration using this formula to determine the starting point for the local minimization algorithm? \par

~\\
\textit{Additional research problem are:}

\begin{enumerate}
\item What one-factor stochastic volatility models have been studied for options pricing and what model form could generalise them?
\item What are dynamics of the replication strategy for the general form of stochastic volatility model and what is a closed-form formula for a characteristic function of a model which is using the Sch\"obel-Zhu model form but has two independent stochastic factors in spot price equation with two corresponding equations of the volatility dynamics?
\item What are the statistic factors in the dynamics of the currency options implied volatility surface in both of its dimensions: option delta and the term structure of the time to maturity ?
\item What is the behaviour of dynamics of implied volatility and implied variance of currency options in the time series models with the form corresponding to the functional form equivalent to variance equation in the Heston and the Sch\"obel-Zhu model?
\item What is the the accuracy of the forecasting of changes in implied volatility on the basis of variables constructed from the term structure of implied volatility and its parametrization in the Heston and the Sch\"obel-Zhu model, as well as the other information contained in the surface of the implied volatilities?
\end{enumerate}

\section{Dissertation objectives}


\textit{Principal objective is:}\par

Development and justification of a method for derivation of approximate formulas for estimation of parameters corresponding to the volatility of the stochastic factor in the instantaneous variance dynamics and the correlation of stochastic factors in the instantaneous variance dynamics and the price dynamics in the stochastic volatility diffusion models with exponential-affine characteristic function of the price distribution and single-factor or two-factor variance, which would use information from all options from volatility surface and would be more advantageous than other formulas in terms of accuracy of option pricing with regard to observed option prices and in terms of the accuracy and risk of model calibration using this formula to determine the starting point for the local minimization algorithm.\par

~\\
\textit{Additional objectives are:}

\begin{enumerate}
	\item The formulation of a general model which is a unification of most famous stochastic volatility diffusion models with a one-factor variance.
	\item The derivation of the equation for the replication strategy dynamics in the general one-factor variance model and the derivation of a closed-form formula for a characteristic function of a model which is using the Sch\"obel-Zhu model form but has two independent stochastic factors in spot price equation with two corresponding equations of the volatility dynamics.
	\item Identification of statistical factors in the dynamics of the currency options implied volatility surface in both of its dimensions: option delta and the term structure of the time to maturity,
	\item Study of the dynamics of implied volatility and implied variance of currency options in the time series models with the form corresponding to the functional form equivalent to variance equation in the Heston and the Sch\"obel-Zhu model,
	\item Study of the accuracy of the forecasting of changes in implied volatility on the basis of variables constructed from the term structure of implied volatility and its parametrization in the Heston and the Sch\"obel-Zhu model, as well as the other information contained in the surface of the implied volatilities.
\end{enumerate}

\section{Dissertation thesis}

\textit{Derived with Implied Central Moments method approximate formula for estimation of parameters corresponding to the volatility of the stochastic factor in the instantaneous variance dynamics and the correlation of stochastic factors in the instantaneous variance dynamics and the price dynamics in the Heston model is a preferable alternative to other approximate formulas for estimation of parameters in terms of accuracy of option pricing with regard to observed option prices and in terms of the accuracy and risk of model calibration using this method to determine the starting point for the local minimization algorithm.}

\section{Research hypotheses}

Tests of the following hypotheses will be carried out in dissertation : \\
\textbf{The first hypothesis} (H1) refers to the principal components of the implied volatility surface for the EURUSD options market. \\
\textit{H1: The first 3 principal components for the 5 element vector of the volatility smile explain less variance than the first 3 principal components for the 5 element vector of the volatility term structure.} \\
\textbf{The second hypothesis} (H2) refers to the fit of models with the functional form of the variance equation, which corresponds to the Heston or the Sch\"obel-Zhu model, to the dynamics of implied volatility and implied variance. \\
\textit{H2: Model with the functional form, which corresponds to the Sch\"obel-Zhu model, explains more variance of observed implied variance and implied volatility than the model with the functional form, which corresponds to the Heston model.} \\
\textbf{The third hypothesis} (H3) refers to the accuracy of the forecasting of changes in implied volatility on the basis of variables constructed from the term structure of implied volatility and its parametrization in the Heston and the Sch\"obel-Zhu model, as well as the other information contained in the surface of the implied volatilities. \\
\textit{H3: Volatility term structure parametrization from the Sch\"obel-Zhu model and differences between implied volatilities on volatility term structure are better variables for a logistic regression model of directions of changes in implied volatility in terms of higher AIC, BIC and AUROC, than variables from the Heston model variance term structure parametrization and differences between implied volatilities on volatility term structure.} \\
\textbf{The fourth hypothesis} (H4) concerns the quality of the fit of results of the Heston model with parameters estimated with approximate formulas to observed option prices, which imply the volatility surface. \\
\textit{H4: The Heston model parameters which were estimated by Implied Central Moments formulas are closer to the global minimum in terms of the size of the vega-weighted option price mean squared error than in the case of parameters which were estimated by the Durrleman (2004) formulas.} \\
\textbf{The fifth hypothesis} (H5) refers to the calibration risk of the Heston model variance term structure parameters in calibrations from different starting points. \\
\textit{H5: The Heston model variance term structure parameters which were calibrated using starting points from Implied Central Moments method estimation have a lower calibration risk measure than in the case of their calibration using starting points from Durrleman (2004) method estimation.}\\
\textbf{The sixth hypothesis} (H6) concerns the quality of the fit of option prices of the Bates two-factor variance model and the OUOU model, which were calibrated using two-stage calibration method and one-stage calibration with starting points from Implied Central Moments method with equal variance equation method, to observed option prices, which imply the volatility surface. \\
\textit{H6: Parameters in the Bates two-factor variance model and the OUOU model, which were calibrated using starting points from Implied Central Moments method with equal variance equation method and modified equal variance method, are closer to the global minimum in terms of the size of the vega-weighted option price mean squared error than in the case of the parameters in models which were calibrated using two-stage full calibration.} \\
\textbf{The seventh hypothesis} (H7) concerns the quality of the fit of results of the Bates two-factor variance model with Feller condition and the author's OUOU model to observed option prices, which imply the volatility surface, after the full calibration. \\
\textit{H7: Results from the author's OUOU model are closer to the global minimum in terms of the size of the vega-weighted option price mean squared error than in the case of the Bates two-factor variance model with Feller condition, when using the same calibration method with same method of obtaining starting points for both models.}\\

\section{Justification of presented objectives, thesis and hypotheses}

The first additional objective is an introduction to the existing literature, which concerns the valuation of options and stochastic volatility models with a one-factor variance. It is also the summary and the analysis of knowledge on these models and corresponding equations. On the other hand, the second additional objective is an enhancement of an existing literature, which concerns the valuation of the options in the stochastic volatility diffusion models with exponential-affine characteristic function of price distribution with both single and two-factor variance. This objective is also enhancing the existing knowledge, which has been summarized in the first additional objective. For both the first and second additional objectives, the implementation of these objectives is qualitative in its nature and there is no hypotheses relating directly to them. \par

The third additional objective complements the existing statistical research on the dynamics of the volatility surface. Most of the earlier studies were related to the equity indices option markets. The research hypothesis H1 is related to the third additional objective. Although, acceptance of this hypothesis is highly probably, in the beginning a replication of previous studies on the dissertation dataset is needed. The hypothesis refers to previous studies on the implied volatility surface, in which the entire volatility surface, as well as the volatility smile and the volatility term structure were tested separately, but there was no research which would relate the variance of the volatility smile factors to the variance of the volatility term structure factors. Such a study is essential for a more complete look at modelling the implied volatility surface as an output of stochastic volatility models. \par

The fourth and the fifth additional objective are in the same area as the third additional objective. Nevertheless, unlike in the third, the study related to next two additional objective contains the option pricing mathematics. However, the direct analysis of option pricing is absent in these objectives. The research hypothesis H2 is related to the fourth additional objective, while the research hypothesis H3 is related to the fifth additional objective. In addition, the hypotheses H2 and H3 are also related to the second additional objective, because their confirmation confirms the legitimacy of research on the development of the two-factor alternative to the Sch\"obel-Zhu model. \par

The principal objective of the dissertation has both a theoretical and an empirical character, because it presents a modification of the existing theory and also it contains tests of this theory. The research hypotheses H4, H5, H6 and H7 are related to the principal objective. These hypotheses refer directly to the thesis of the dissertation and their acceptance would also confirm the main thesis of the dissertation and solve the principal problem. The attempt to answer whether the hypotheses are true is implementation of the principal objective of the dissertation, regardless of the confirmation of the principal thesis. Hypothesis H4 refers to the quality of the model fit to the market data in terms of the size of the average error in the proposed algorithm. The hypothesis H5 refer to an essential element, namely the risk of calibration. Hypotheses H6 and H7 refer to the use of the principal objective of the dissertation in the author's enhancement of the theory of stochastic volatility models, which is planned as the second additional objective. \par

The studied problem may have no solution, namely, a new proposed method will not be more favourable than existing methods within defined criteria. However, even an attempt to solve the problem will have some value for further research on this problem in the future.

Regardless of the dissertation thesis, never done statistical analysis of the relationship between maturity and strike price dimensions of the dynamics of the implied volatility surface of currency options and statistical comparisons of the Heston and the Sch\"obel-Zhu model have added value itself. Furthermore, the theoretical correctness of the method, even in case of additional results, can provide its usefulness in other studies in this area. \par

On the other hand, confirmation of the dissertation thesis means that banks and other option portfolio managers will be able to use even more accurate approximate formula for estimation of parameters to determine the starting point in their preferable calibration method for the Heston or the Sch\"obel-Zhu model, as well as their two-factor variance alternatives, which have advantages over other solutions.\par

Additional scientific and practical value is the introduction of a new model, i.e. the model which uses the Sch\"obel-Zhu model form and has two equations of the dynamics of volatility. This model should have an analytic form of the characteristic function of  underlying instrument price distribution, so also a semi-analytic formula for the valuation of options, similarly as the Sch\"obel-Zhu model has. Such a model was not so far considered nor studied in the literature. Scheme of connections of the principal and additional objectives is shown in Figure \ref{figure:I.1}. \par



\renewcommand\thefigure{I.\arabic{figure}}   
\begin{figure}[h]
\centering
\includegraphics[width=10cm,height=10cm,keepaspectratio]{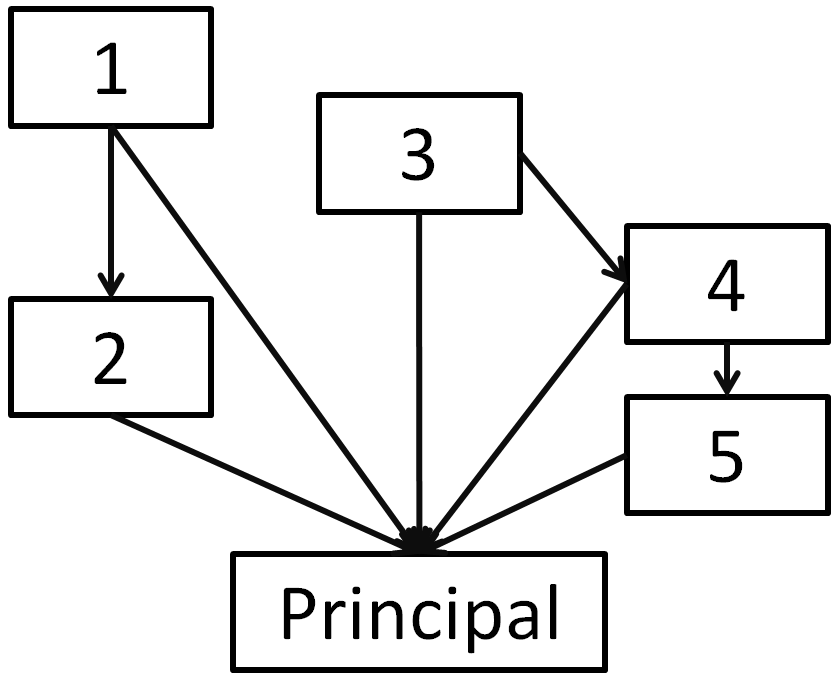}
\caption{Scheme of links between principal objective and additional objectives}
\label{figure:I.1}
\end{figure}
\renewcommand\thefigure{\arabic{chapter}.\arabic{figure}}

\section{Dissertation plan}

\textbf{Introduction and methodological basis of research} contains the definition and the discussion on the importance of the research area, a description of the methodological foundations of the research, a presentation of dissertation objectives, research problems, the principal thesis and dissertation research hypotheses. The results of the dissertation objectives are presented in the following five chapters.

\textbf{Chapter 1: Option market, option pricing and volatility models}, is an introduction to option pricing and modelling of volatility in the stochastic volatility context. It consists analysis of literature on the valuation of options and stochastic volatility modelling, including option pricing by the integral of the characteristic function. The implementation of the first additional objective is presented in this chapter.

\textbf{Chapter 2: Characteristic functions of price distribution in stochastic volatility models}, using the generalised stochastic volatility model from the first chapter, contains a derivation of the replication strategy dynamics equation and analytic formulas for the characteristic functions of the underlying instrument price distribution for the considered stochastic volatility diffusion models. Some of derivations are summary of knowledge related to this area, but the chapter also contains derivation of solution for a new model with two-factor volatility. The implementation of the second additional objective is presented in this chapter.

\textbf{Chapter 3: Statistical analysis of implied volatility surface dynamics}, starts from the introduction to the PCA method, then the results of research on the statistical properties of the distribution of implied volatilities and dynamics of implied volatility surface is summarized. This is followed by principal component analysis to the two dimensions of the implied volatility surface dynamics. The implementation of the third, the fourth and the fifth additional objective are presented in this chapter, including testing of research hypotheses H1, H2 and H3. \par


\textbf{Chapter 4: Estimation of stochastic volatility models parameters} concerns methods of calibration of stochastic volatility models, in particular the Heston model. Moreover, approximate formulas for estimation of the Heston model parameters  are presented. Parameters estimated with an approximate formula can also be used as starting points for calibration with numerical algorithm. This chapter includes also the development of a new method, namely, Implied Central Moments method. Then, derivation of approximate formulas for the Heston model and the Sch\"obel-Zhu model parameters is performed.
Besides, an additional method for using those formulas for models with two-factor variance, namely the Bates two-factor variance model and the OUOU model proposed in the second chapter, is presented. Development of the proposed method is an implementation of the theoretical part of the principal objective of the dissertation. \par

\textbf{Chapter 5: Empirical analysis of model calibration and parameters estimation methods}, contains empirical tests of methods from the fourth chapter, namely author's Implied Central Moment formulas and Durrleman formulas for the Heston model, in order to compare and determine the quality of parameters estimates with approximate formulas. The tests are carried out on the EURUSD options market. This chapter provides implementation of the empirical part of the principal objective of the dissertation, including tests of following hypotheses: H4, H5, H6 and H7. \par

In \textbf {Research conclusion} the author presents conclusions about developed models, proposed methods for estimation of model parameters and performed empirical research. \par

\section{Methodological approach}

The author used the following theoretical methods in the dissertation to develop a new model used for the valuation of options as well as to develop a new approximate formula for estimation of parameters: analysis of literature and mathematical modelling. The theory which served as the basis of the developed methods, belongs to the following branches of mathematics: mathematical statistics and stochastic differential equations. \par

In addition, with regard to the chapter which concerns the analysis of the dynamics of the implied volatility surface and to the chapter on testing presented methods, the author uses empirical methods, such as principal component analysis, factor analysis, testing auto-regression models, discriminant analysis and testing of the stochastic volatility models on historical data. \par

Statistical package R was used for data processing, statistical analysis and visualization. Option pricing methods and Nelder-Mead algorithm for minimization were implemented by the author in C ++.

\section{Assumptions and limitations}

Following limitations have been set for the research principal objective. \par

Firstly, based on the identification analysis from the fourth chapter and due to the fact that the most popular stochastic volatility model in exponential-affine class is the Heston model and all previous work on approximate formulas for estimation of parameters have considered and tested this particular model, the author assumed that basic version of developed method should be also presented from the Heston model perspective. Similarly, tests with alternative methods will also apply to the case of the Heston model.\par

Secondly, identified approximate formulas are related to some parameters of the Heston model, but not all parameters at once. Therefore, the author made the synthesis of these methods and set the objective of finding the appropriate values for two of the model parameters, $ \omega $ and $ \rho $, which corresponds to the volatility of variance and correlation of variance and returns. The other three are assumed to be already known, either from parameters estimates with approximate formula (e.g. by Guilliame and Schoutens (2010) method) or by the calibration of the term structure of variance. Identified approximate formulas for parameters $ \omega $ and $ \rho $ are based on Durrleman method (2004) and the Gauthier and Rivalle (2009) method. Because the latter method uses only three of available five points from the FX options volatility smile, so it is considered to be potentially more risky. Obtained parameters estimates for $ \omega $ and $ \rho $ may not reflect real degree of convexity in the volatility smile, because the 10-delta or 25-delta options are not taken into account in the method. This is particularly important for the listed options market, for which the volatility smile is formed by more than five points. Thus the Durrleman (2004) method is the only so far known method, which allows to determine parameters $ \omega $ and $ \rho $ basing on all of the information in the volatility smile. However, this method uses only information from the volatility smile of the maturity which is nearest to the valuation date. Therefore, it can omit theoretically different information present in other volatility smiles. Empirical comparison tests with regard to the Durrleman method are presented in the fifth chapter. \par

Thirdly, estimates computed from parameters of analogous statistical models which are estimated on time series of returns of the underlying instrument, e.g. GARCH class models, were not considered due to the aforementioned characteristic of stochastic volatility models. The option valuation is usually done using only the market data for the valuation moment and the literature confirms lower usefulness of time series models of historical returns to the determination of option prices and implied volatility forecasting in a situation in which implied volatility data are available.\par

Fourthly, because of the identified in the first chapter significant share of EURUSD pair in the currency market trading volume, empirical tests were carried out on market data for options on this currency pair. The period 2010-2015 has been chosen for empirical tests.
\clearpage{\pagestyle{empty}\cleardoublepage}

\renewcommand{\thesection}{\thechapter.\arabic{section}}

\chapter{Option market, option pricing and volatility models}

\section{Introduction}
This chapter is the outline of the issues relating to the OTC foreign exchange option market and the problem of option valuation. Valuation of options is discussed since the earliest work on the subject, which contains assumption of constant volatility. Later, local volatility and stochastic volatility models were added into consideration. The biggest part of the chapter is an introduction to the option pricing by replication strategy in risk neutral world, which includes also pricing by a numerical integration of the price characteristic function. The rest part is the summary of previous works on stochastic volatility.

\section{Differences between OTC FX option market and exchange}
\label{sec:1.2}

On the Over The Counter (OTC) market the trading takes place directly between pairs of financial institutions (e.g. banks), without the mediation of an exchange. On the FX options market, which is an OTC market, the options are quoted as implied volatility, unlike stock options, which are quoted on stock exchange in the form of option premium, which is the value of the option. In the case of the latter options, the implied volatility can also be calculated by searching for such volatility parameter in the Black-Scholes (1973) formula, which will give in return the option premium quoted on the stock exchange. Hence, the implied volatility is understood as a value associated with a given option premium with other parameters being fixed.\par

In the case of OTC market option prices are not quoted directly in the form of a premium but are presented as implied volatilities. Nevertheless, in this case the term "implied volatility" still has a same sense, despite the fact that the premium is not known and it is calculated on the basis of quoted implied volatility. The term implied volatility exists as the equivalent of the price of option and at the same time it still has some relation to the volatility in the statistical sense, for example historical volatility or realised volatility, which are square roots of returns variance over some period of time. Actually, the volatility in the physical sense and implied volatilities are interrelated. Implied volatility is market expectation of the realized volatility of the underlying asset observed in the future. Since the current and future value of realized volatility are related to some extent, then if on the market would be too much difference between the implied volatility and realized volatility, it would lead to the possibility of an arbitrage between options and dynamically hedged position in the underlying instrument.\par

\begin{table}[h!]
\centering
\small
\caption{Average daily turnover on FX options versus the rest of FX market and the world GDP}
\label{table:1.1}
\begin{tabular}{lccccccc}
  \hline
Year & 1995 & 1998 & 2001 & 2004 & 2007 & 2010 & 2013 \\ 
  \hline
FX options turnover (bn USD) & 41 & 87 & 60 & 119 & 212 & 207 & 337 \\
FX derivatives turnover (bn USD) & 696 & 959 & 853 & 1303 & 2319 & 2483 & 3299 \\
Total FX turnover (bn USD) & 1190 & 1527 & 1239 & 1934 & 3324 & 3971 & 5345 \\
Share in FX derivatives market & 5.9\% & 9.1\% & 7.0\% & 9.1\% & 21.1\% & 13.9\% & 16.5\% \\
Share in whole FX market & 3.4\% & 5.7\% & 4.8\% & 6.2\% & 6.4\% & 5.2\% & 6.3\% \\
World GDP (current tn USD) & 30.84 & 31.32 & 33.37 & 43.77 & 57.79 & 65.91 & 76.93 \\
Annualized FX options turnover / World GDP & 33.5\% & 70.0\% & 45.3\% & 68.5\% & 92.4\% & 79.1\% & 110.4\% \\
  \hline
\multicolumn{8}{l}{\footnotesize \textit{Source: Bank for International Settlements 2002, 2013, World Bank 2016}}
\end{tabular}
\end{table}

The size of the FX options market in terms of its turnover\footnote{~Since 1998 it contains also complex structured trades with non-linear components that are harder to compare to other.}, the fractions of its turnover to the turnover of all FX derivatives market and the whole FX market (spot, options, futures, forwards, swaps), the world GDP and the ratio of annualized\footnote{~Average daily turnover was annualized by multiplication by 252.} FX options turnover to the world GDP in the years 1995-2013 are presented in Table \ref{table:1.1}. The presented data shows that the share of FX options market in the whole currency market (i.e. spot and derivatives) started from 1998 to remain in the relative stability with the fluctuations in the range of around 5-6\% share of the total. Besides, the ratio of FX options turnover to the world GDP was rising for almost 20 years, decreasing only in two of all six triennial periods.\par

Share of FX options turnover in all FX derivatives turnover presented rather growth tendency, which indicates that hedging currency risk of future cash flows with options has become more popular. A peak is present in 2007, when the FX options market was over 20\% of all FX derivatives market. It should not be a surprise, because the year 2007 is considered to be the last year the prevailing bull market, just before the global financial crisis. In contrast, since 2010, we are observing a growing trend of FX options turnover and their share in all FX derivatives market is again getting bigger. It can be considered that with the growth of both markets the importance of proper valuation of currency options is increasing. \par

\begin{table}[h!]
\centering
\caption{Average daily net turnover in April of chosen years as fraction of total turnover}
\label{table:1.2}
\begin{tabular}{lcc|lcc}
  \hline
 & \multicolumn{2}{c}{Spot} & & Options & Spot\\ 
Currency pair & 2013 & 2010 & Currency & 2013 & 2013 \\
  \hline
EURUSD & 24.1 & 2.7 & USD & 43.5 & 41.3 \\
USDJPY & 18.3 & 14.3 & JPY & 22.7 & 15.0 \\
GBPUSD & 8.8 & 9.1 & EUR & 10.4 & 18.4 \\
AUDUSD & 6.8 & 6.3 & GBP & 4.3 & 5.5 \\
USDCAD & 3.7 & 4.6 & AUD & 4.0 & 4.8 \\
USDCHF & 3.4 & - & CNY & 2.5 & 0.8 \\
USDMXN & 2.4 & - & CHF & 2.1 & 2.1 \\
USDCNY & 2.1 & 0.8 & CAD & 1.8 & 2.3 \\
NZDUSD & 1.5 & - & BRL & 1.6 & 0.3 \\
USDRUB & 1.5 & - & MXN & 0.9 & 1.4 \\
USDHKD & 1.3 & 2.1 & KRW & 0.6 & 0.5 \\
USDSGD & 1.2 & - & NZD & 0.4 & 1.0 \\
USDTRY & 1.2 & - & RUB & 0.4 & 0.9 \\
USDPLN & 0.4 & - & SGD & 0.4 & 0.5 \\
EURJPY & 2.8 & 2.8 & TRY & 0.4 & 0.4 \\
EURGBP & 1.9 & 2.7 & INR & 0.4 & 0.4 \\
EURCHF & 1.3 & 1.8 & SEK & 0.3 & 0.7 \\
EURPLN & 0.3 & - & PLN & 0.1 & 0.3 \\
Other & 10.3 & 13.7 & Other & 3.0 & 3.6 \\
  \hline
\multicolumn{6}{l}{\footnotesize \textit{Source: Bank for International Settlements 2013}}
\end{tabular}
\end{table}

Table \ref{table:1.2} presents data on trade in the foreign exchange spot and options markets by individual currency pairs \footnote{~For all Dollar pairs, except those with EUR, GBP, AUD, NZD, Dollar is always at the first place, i.e. as the base currency. In the case of pairs with Euro it is always the base currency.} for the spot market and by currency for the spot and options markets. The data shows that in 2013 with a significant lead a large part (24.1 \%) of turnover took place on the EURUSD pair. Next pair is USDJPY, then with a significantly lower market share GBPUSD pair. The first two currency pairs make together 44.4 \% of the FX spot market. In the case of FX options market the share of the first two currencies is almost two-thirds of the total (66.2 \%), although considered data set contains all currency pairs related to these two currencies.

With information about spot price of the underlying instrument, strike price, time to maturity, interest rates and of course an implied volatility, OTC market participants using the Garman-Kohlhagen (1983) formula \footnote{~\bibentry{Garman1983}.} calculate the option premium. It is the market standard. For options on a given currency pair implied volatilities $\sigma_\delta(\tau)$ depend on two parameters: \par

\begin{enumerate}
\item $\delta$ - option price sensitivity to the spot price, denominated in percentage, which due to the standard used on OTC market for currency options, has a value of 10, 25, 50 for the call option and the 10, 25, 50 for the put options, the 50 delta for call and put options is assigned to the same quote of implied volatility, which is called ATM volatility, and it is due to the option call-put parity\footnote{~Call-put option parity implies the following relationship: $C_t - P_t = e^{r_d(T-t)} (F(t,T)-K)$.}, hence implied volatilities for options with 5 different levels of delta sensitivity can be calculated from the market traded options strategies; OTC option exercise price can be determined using delta sensitivity, current spot price of the underlying instrument, interest rates, time to maturity and implied volatility,
\item $\tau$ - option time to maturity in the form of a year fraction which for OTC traded options usually has one of the following values: 1/365 (ON), 7/365 (SW), 1/12 (1M), 2/12 (2M), 3/12 (3M), 6/12 (6M), 9/12 (9M), 1 (1Y), 2 (2Y), 3 (Y), 5 (Y), 7 (Y), 10 (Y). However, options with time to maturity, which equals less than 1 month or more than 2 years, have limited liquidity.
\end{enumerate}

It should be noted that implied volatilities are not quoted on the OTC market directly as a function of these two parameters. The implied volatility surface for the currency options is quoted in relation to certain types of option strategies. OTC market uses quotation of prices of the following types of option strategies: \par

\begin{itemize}
\item ATM Straddle (ATMF),
\item 25 - delta Risk Reversal (25RR),
\item 25 - delta Vega-weighted Butterfly (25FLY),
\item 10 - delta Risk Reversal (10RR),
\item 10 - delta Vega-weighted Butterfly (10FLY).
\end{itemize}

Based on the quotes of presented options strategies, we can determine the value of the so-called volatility smile for a given maturity. Volatility smile is a set of strike-volatility pairs. In OTC FX options case it will consist of values of implied volatility for 10-delta put, 25-delta put, 50-delta call/put, 25-delta call, 10-delta call. The market convention is quoting prices of options in relation to the level of the strike price expressed as delta. Implied volatility surface on OTC market is described by the nodal points of a volatility smile for each maturity (ON, SW, 1M, 2M, 3M, 6M, 1Y, 2Y). Used nodal points correspond to the following values of the delta parameter: \par

\begin{itemize}
\item 10 delta put (10P),
\item 25 delta put (25P),
\item 50 delta put/call (ATM),
\item 25 delta call (25C),
\item 10 delta call (10C).
\end{itemize}

Nodal points for implied volatility surface are determined using following formulas: \par

\begin{equation}\sigma_{10P} = \sigma_{ATM}+\sigma_{10FLY}-0.5\sigma_{10RR},\end{equation}
\begin{equation}\sigma_{25P} = \sigma_{ATM}+\sigma_{25FLY}-0.5\sigma_{25RR},\end{equation}
\begin{equation}\sigma_{50P} = \sigma_{ATM},\end{equation}
\begin{equation}\sigma_{25C} = \sigma_{ATM}+\sigma_{25FLY}+0.5\sigma_{25RR},\end{equation}
\begin{equation}\sigma_{10C} = \sigma_{ATM}+\sigma_{10FLY}+0.5\sigma_{10RR}.\end{equation}

In the case of Vega-weighted Butterfly quotations, the reversal of equations (1.1)-(1.5), i.e. $\sigma_{25FLY}=0.5(\sigma_{25C} + \sigma_{25C}) - \sigma_{ATM}$ (and analogous equation for 10 delta) is only an approximation of the implied volatility of the quoted strategy. Malz (1997)\footnote{~\bibentry{Malz1997}.}, Lipton (2002)\footnote{~\bibentry{Lipton2002}.} and Bakshi, Carr and Wu (2008)\footnote{~\bibentry{Bakshi2008}.} considered this approximation in their studies. The method of calculating the precise value can be found, among others, in Reiswich and Wystup (2012)\footnote{~\bibentry{Reiswich2012}.} work.

\begin{figure}[h]
\centering
\includegraphics[width=10cm,height=10cm,keepaspectratio]{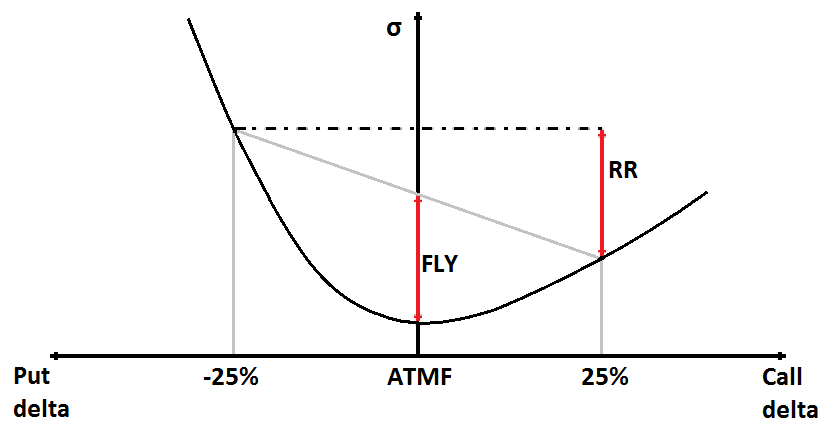}
\caption{Volatility smile}
\label{figure:1.1}
\end{figure}

The dependence of the discussed option strategies on implied volatilities of 25 delta put and 25 delta call options is shown on Figure \ref{figure:1.1} (analogous relationship is for 10 delta put and 10 delta call options). \par

Since $\sigma_\delta(\tau)$ is a function of two variables, a surface of implied volatility can be considered. The implied volatility surface of foreign exchange options changes its shape in such a way that the movement of the individual points of this plane is highly correlated. For this reason, the practical approach to the modelling of implied volatility surface movement is the use of principal component analysis and factor analysis with respect to all of the implied volatilities forming this surface. \par

\section{Black-Scholes and Garman-Kohlhagen pricing formulas}

\subsection{Historic view}

Valuation of options was scientifically examined for the first time by Bachelier (1900)\footnote{~\bibentry{Bachelier1900}.}, who assumed normal distribution of returns in his model. Since then, it took more than half a century before researchers returned to this theme. Another model, proposed by Boness'a in 1964\footnote{~\bibentry{Boness1964}.}, contained the assumption of log-normal distribution of the underlying instrument returns. But the real milestone was the work of the next decade. In 1973 Black and Scholes\footnote{~\bibentry{Black1973}.} Published their article on option pricing. In the same year Merton\footnote{~\bibentry{Merton1973}.} did the same thing independently.

The pioneering works of Black-Scholes and Merton have brought a lot to the world of finance. They have derived the option pricing model by constructing option replication strategy, which has the same payout as a valued option, regardless of market conditions. They began by defining the model of stochastic differential equations for the share price $ S $, assuming that it follows log-normal random walk process, and finished by obtaining the formula for the valuation of vanilla European options \footnote{~Vanilla option is the most popular type of option, the profile of its payment is $\mathrm{max}(S-K,0)$ for call options and $\mathrm{max}(K-S,0)$ for put options. European style option may be exercised only at its expiration date.} on this share. Later, the version of the model for futures contracts was developed by Black (1976)\footnote{~\bibentry{Black1976}.} and the version for the spot foreign exchange rate was developed by Garman and Kohlhagen (1983)\footnote{~\bibentry{Garman1983}.}.\par

Black-Scholes model (1973) is not free from drawbacks and limitations. It is built exclusively on the basis of the idea of delta hedging, therefore, it does not include other types of financial risk. Hedging aspect is very important because it is the main method to maintain a neutral position towards risk. The use of the model results in a potential risk for the valuation and market risk management. First of all, the Black-Scholes model (1973) assumes that the relative change in the price of the underlying is a drifting Wiener process of constant volatility.

The assumption that the future realized volatility on the market is a known constant or a deterministic function of time and the value of the underlying instrument is a simplification and leads to the omission of the risk of volatility changes and price spikes which are observed in the market data. In addition, in time series of returns an autocorrelation may be present, which is a violation of another assumption. Next dubious assumption is that in a replication strategy the hedging process is continuous and there are no transaction costs and taxes, which leads to a non-zero error in the effectiveness of option hedging in the real world on the market, because hedging is not performed in a continuous manner, but discreetly and with certain transaction costs.\par

Despite these theoretical limitations Black-Scholes (1973) formula is considered the standard market model, also the process of mark-to-market accounting on OTC markets depends on this model. In the case of currency options the standard for the mark-to-market accounting is the modification of the formula, namely the Garman-Kohlhagen formula. Therefore, a key element of the model is the choice of the appropriate implied volatility from implied volatility surface. The standard market practice, taking into account the imperfections of Black-Scholes model, was, and is a quotation of implied volatilities at different levels for different maturities and strike prices of an option. This is related to the presence of a non-flat implied volatility surface. Another market practice is the quoting of the bid and ask prices spread to take into account the margin for potential error in the model. Robustness of Black-Scholes model (1973) in this configuration has been verified theoretically by El Karoui et al. (1998)\footnote{~\bibentry{ElKaroui1998}.}.\par

Other option pricing models were developed since the Black-Scholes seminal work. Especially important is the discrete equivalent of the model, namely the model of Cox-Ross-Rubinstein (1979)\footnote{~\bibentry{Cox1979}.}. It assumes that changes of the underlying price are not continuous, but discrete, and can be described by a binomial or trinomial tree. This modification carried an improvement for hedging strategies, but it has not eliminated the problems of observing non-flat volatility surface, described in the work Derman and Kani (1996)\footnote{~\bibentry{Derman1996}.}.\par

Note also that the Black-Scholes approach to derive the formula included the construction of a partial differential equation with the appropriate boundary conditions and its analytic solution. Another possible approach is the martingale method, which was presented by Cox and Ross (1976)\footnote{~\bibentry{Cox1976}.} and Harrison and Kreps (1979)\footnote{~\bibentry{Harrison1979}.}. In this approach the value of the future cash flows is taken as the expected value of those cash flows in risk-neutral measure and then the expected value is calculated according to the theory of probability.\par

\subsection{Black-Scholes model assumptions}

Despite the existence of formulas for the option valuation before 1973, the Black-Scholes model (1973) is popular for its derivation and thus the justification of this model by the Black-Scholes differential equation, which describes the dynamics of changes in the value of the option. A version of this model for the FX market, i.e. Garman-Kohlhagen (1983) model is a simple modification of the first model by introduction of a spot exchange rate instead of a share price and a foreign interest rate in place of a dividend rate from share. \par

Assumptions of the Black-Scholes model (for the FX market) can be formulated briefly in following points:

\begin{enumerate}
\item Price of the underlying (i.e. spot exchange rate) $ S_t $ follows a log-normal process (geometric Brownian motion) with constant volatility $ \sigma $ and the expected value of $ \mu $, $ W_t $ in the following formula is 1-dimensional Wiener process (i.e. Brownian Motion):

\begin{equation} \D S_t = \mu S_t \D t + \sigma S_t \D W_t .\end{equation}

\item Domestic and foreign currencies have risk-free interest rates, $r_d$ and $r_f$, which are fixed for all maturities.
\item There are no transaction costs and no taxes.
\item There are no opportunities for a risk-free arbitrage.
\item Short selling is possible (i.e. the asset can be borrowed and sold with purpose of later buy-back).
\item Trading between buying (or selling) an option and the moment of its execution is carried out in a continuous manner (not discreet).

\end{enumerate}

The dynamics of the bonds value with considered risk-free rates is given by the relation $\D B_t^d = r_d B_t^d \D t$ (i.e. bond prices are neutral to risk).

\subsection{Garman-Kohlhagen version of the Black-Scholes equation}

Black and Scholes in their work start with the equation presenting the dynamics of the contingent claim price, $ V (S_t, t) $, which is known, and the price of the underlying, $ S_t $, which is also known. Assuming $ V (S_t, t) = V_t $ and using It\^{o} Lemma (1944)\footnote{~\bibentry{Ito1944}.} on this variable, they have found that

\begin{equation} \D V_t = \frac{\partial V}{\partial t} \D t + \frac{\partial V}{\partial S} \D S_t + \frac{1}{2} \frac{\partial^2 V}{\partial S^2} \D S_t^2 .\end{equation}

In the considered equation $\D S_t^2$ mean quadratic variance of $ S_t $. Then, using the first assumption (about the dynamics of the process $ S_t $), we can substitute $\D S_t^2 = \sigma^2 S_t^2 \D t$ yielding:

\begin{equation} \D V_t = \Bigg(\frac{\partial V}{\partial t} + \frac{1}{2} \sigma^2 S^2 \frac{\partial^2 V}{\partial S^2}\Bigg) \D t + \frac{\partial V}{\partial S} \D S_t .\end{equation}

Part of the equation that is contained in parentheses is deterministic, the only stochastic part is the factor $ \D S_t $. Next, the authors introduced the so-called replicating portfolio $ \Pi_t $ to eliminate the stochastic part of the equation. The portfolio consists of one purchased unit of conditional claim which is worth $ V_t $ and $\frac{\partial V}{\partial S}$ sold units of the underlying asset with price $ S_t $:

\begin{equation} \Pi_t = V_t - \frac{\partial V}{\partial S} S_t .\end{equation}

For FX version of this problem Garman and Kohlhagen (1983) define the replicating portfolio as:

\begin{equation} \Pi_t = V_t - \Delta_t S_t B_t^f ,\end{equation}

which can be interpreted as the purchase of one unit of conditional claim which is worth $ V_t $ and sale of $ \Delta_t $ units of foreign bonds, whose price in the domestic currency is $ S_t B_t^f $. Then, after differentiating $ \Pi_t $ by time it is obtained that:

\begin{equation} \D \Pi_t = \D V_t - \Delta_t \D (S_t B_t^f) = 
\D V_t - \Delta_t B_t^f \D S_t - \Delta_t S_t \D B_t^f = \D V_t  - \Delta_t B_t^f (r_d S_t \D t + \sigma S_t \D W_t) ,\end{equation}

where the last part may be obtained after reordering of the part with the element $ d(S_t B_t^f) $, then substituting the formula for $ \D S_t $ and then simplifying the equation.
Further, by inserting formulas for $ \D V_t $ and $ \D S_t $ and reordering the equation, it is obtained that:

\begin{equation} \D \Pi_t = \Bigg(\frac{\partial V}{\partial t} + \frac{1}{2} \sigma^2 S^2 \frac{\partial^2 V}{\partial S^2} - \Delta_t B_t^f r_d S_t + \frac{\partial V}{\partial S}(r_d-r_f)S_t \Bigg) \D t + \Bigg( \frac{\partial V}{\partial S} - \Delta_t B_t^f \Bigg) \sigma S_t \D W_t .\end{equation}

Only after these operations it can be seen that if $ \Delta_t $ has such value that dependence $\Delta_t B_t^f = \frac{\partial V}{\partial S}$ is fulfilled, then component with dependency from $ \D W_t $ will disappear in the final equation. The value of the portfolio $ \Pi_t $ in domestic currency according to the assumptions should be neutral towards risk, similar to the prices of bonds in domestic currency and meet the same dependence, $\D \Pi_t = r_d \Pi_t \D t$. In the latter equation the assumed definition (1.19) can be added for $ \Pi_t $ which results in:

\begin{equation} \D \Pi_t = r_d \Bigg(V_t - \frac{\partial V}{\partial S} S_t \Bigg) \D t .\end{equation}

Finally, we can compare two formulas for $ \D \Pi_t $, i.e. (1.12) and (1.13) to get after some simplification the Garman-Kohlhagen (1983)  equation:

\begin{equation} \frac{\partial V}{\partial t} + \frac{1}{2} \sigma^2 S^2 \frac{\partial^2 V}{\partial S^2} + (r_d-r_f) S \frac{\partial V}{\partial S} - r_d V = 0 .\end{equation}

An important element of this model is the lack of $ \mu $, which is the expected value of returns of the underlying instrument, according to the assumed definition of its dynamics. This means that the valuation (and replication) of option does not depend on the expected trends in the underlying instrument price dynamics.

\subsection{Risk neutrality of domestic investors}

Previously defined $ \D S_t $ is a differential of the following function $ S_t $:

\begin{equation} S_t = S_0 \exp\Bigg(\sigma W_t + \Bigg(\mu - \frac{1}{2}\sigma^2 \Bigg) t \Bigg) .\end{equation}

Risk-neutral investors see the domestic foreign bonds, $ B_t^f $, as a risky asset, worth $ B_t^f S_t $. To maintain their risk neutrality the ratio of the value of foreign bonds denominated in the domestic currency in relation to domestic bonds should be a martingale \footnote{~In the continuous version, it is assumed that the stochastic process $ Y_t $ is a martingale with respect to the stochastic process $ X_t $ if for any $ t $ occurs: $\mathbb{E} |Y_t| < \infty$ and $\mathbb{E}(Y_t|{X_\tau,\tau \leq s})=Y_s$ for any $ s \leq t $.}. Using the previous formula for the $ S_t $ and bond prices which depends on the interest rate, $ B_t^d = e^{r_t^d t} $, considered relationship is:

\begin{equation}\begin{split} Z_t = S_t B_t^f / B_t^d = S_0 \exp\Bigg(\sigma W_t + \Bigg(\mu - \frac{1}{2}\sigma^2 \Bigg) t \Bigg) \exp\big((r_f - r_d )t\big) \\
= S_0 \exp\Bigg(\sigma W_t - \frac{1}{2}\sigma^2 t \Bigg) \exp\big((\mu + r_f - r_d )t\big) .\end{split}\end{equation}

Because the component $(\sigma W_t - \frac{1}{2}\sigma^2 t)$ is already a martingale, then $ Z_t $ is a martingale when the equality: $ \mu = r_d - r_f $, is fulfilled for the drift $ \mu $. Then in the domestic risk-neutral measure $ \mathbb{Q}^d $, we can state that:

\begin{equation} S_t = S_0 \exp\Bigg(\sigma W_t^d + \Bigg(r_d - r_f - \frac{1}{2}\sigma^2 \Bigg) t \Bigg) \end{equation}

and

\begin{equation} \D S_t = (r_d - r_f)S_t \D t + \sigma S_t \D W_t^d ,\end{equation}

where $W_t^d$ is a drift adjusted Wiener process, and under considered assumptions the relationship between $ W_t $ and $ W_t^d $ must be:

\begin{equation} W_t^d = W_t + \frac{\mu - (r_d - r_f)}{\sigma}t .\end{equation}

In this way, introducing the $W_t^d$ we will receive the equivalent of the exchange rate dynamics in the measure $\mathbb{Q}^d$. This measure is a risk-neutral measure, and is equivalent to the real world measure, i.e. $\mathbb{P}$, because both measures are defined in the same sample space and impossible events under one of this measures remain impossible under another. The last entry leads to a Radon-Nikodym derivative \footnote{~\bibentry{Nikodym1930}.} at the time $ T $, which is a rate of change of density of measure $\mathbb{Q}^d$ in respect to changes of density of measure $ \mathbb{P} $, namely\footnote{~Radon-Nikodym derivative allows calculation of the conditional expected value in relation to equivalent measures.}:

\begin{equation} \frac{\D \mathbb{Q}^d}{\D \mathbb{P}} = \exp\Bigg(-\gamma^d W_T - \frac{1}{2}(\gamma^d)^2 T \Bigg),  \mbox{	where } \gamma^d = \frac{\mu - (r_d - r_f)}{\sigma} ,\end{equation}

in which right hand side of the equation can be obtained from the Girsanov theorem \footnote{~\bibentry{Girsanov1960}.}, while the factor $ \gamma^d $ is called the market price of risk, and $ T $ is some moment in the future, e.g. expiry of the option.

This formula is useful in the valuation of options, it makes possible to change the expected value in the real world measure, $ \mathbb{P} $, to expected value in the risk-neutral measure, $ \mathbb{Q}^d $, which enables further transformation of the equation.

\subsection{Risk neutrality of foreign investors}

The derivation that is analogous to the one from the previous section can be repeated for foreign investors. Foreign investors see domestic bonds $ B_T^d $ as an risky asset, which are measured in foreign currency by $ B_t^d / S_t $ or otherwise by $ B_t^d \hat{S_t} $, if the foreign currency is expressed in the domestic exchange rate units, i.e. as reciprocal of exchange rate, $ \hat{S_t} = 1 / S_t $. In this setting, the previous argument can be repeated for $ \hat{S_t} = \hat{S_0} \exp\bigg(-\sigma W_t + (\frac{1}{2}\sigma^2 - \mu)t\bigg) $ by construction of a relationship of domestic bonds prices denominated in foreign currency to foreign bonds prices:

\begin{equation}\begin{split} \hat{Z_t} = \hat{S_t} B_t^d / B_t^f = \hat{S_0} \exp\Bigg(-\sigma W_t + \Bigg(\frac{1}{2}\sigma^2 - \mu \Bigg) t \Bigg) \exp\big((r_f - r_d )t\big) \\
= \hat{S_0} \exp\Bigg(-\sigma W_t - \frac{1}{2}\sigma^2 t \Bigg) \exp\big((-\mu + r_d - r_f + \sigma^2)t\big) .\end{split}\end{equation}

Again, $\hat{Z_t}$ is a martingale when $\mu = r_d - r_f + \sigma^2$, so there is some difference in respect to the case of domestic investors. This leads to the dynamics of the exchange rate, $\hat{S_t}$, in risk-neutral measure, $\mathbb{Q}^f$, as:

\begin{equation} \hat{S_t} = \hat{S_0} \exp\Bigg(-\sigma W_t^f + \Bigg(\frac{1}{2}\sigma^2 - (r_d - r_f + \sigma^2)\Bigg)t\Bigg) .\end{equation}

However, in the same measure the price $ S_t $ can be described as:

\begin{equation}\begin{split} S_t = S_0 \exp\Bigg(\sigma W_t^f + \Bigg((r_d - r_f + \sigma^2) - \frac{1}{2}\sigma^2\Bigg)t\Bigg) \\
= S_t = S_0 \exp\Bigg(\sigma W_t^f + \Bigg(r_d - r_f + \frac{1}{2}\sigma^2\Bigg)t\Bigg) ,\end{split}\end{equation}

and

\begin{equation} \D S_t = (r_d - r_f + \sigma^2)S_t \D t + \sigma S_t \D W_t^f .\end{equation}.

In both presented equations change of drift is introduced as  $W_t^f = W_t + \frac{\mu - (r_d - r_f + \sigma^2)}{\sigma}t$, which leads to a Radon-Nikodym derivative in moment $ T $:

\begin{equation} \frac{\D \mathbb{Q}^f}{\D \mathbb{P}} = \exp\Bigg(-\gamma^f W_T - \frac{1}{2}(\gamma^f)^2 T \Bigg),  \mbox{	where } \gamma^f = \frac{\mu - (r_d - r_f + \sigma^2)}{\sigma} .\end{equation}

Additionally, the $W_t^f$ process can be shown as adjusted $W_t^d$ process, since both were previously presented as a transformation of $ W_t $, so the relationship between them should be:

\begin{equation} W_t^f = W_t^d + \frac{(r_d - r_f) - (r_d - r_f + \sigma^2)}{\sigma}t = W_t^d - \sigma t .\end{equation}

Knowing the above relationship, the Radon-Nikodym derivative of the change of the measure $\mathbb{Q}^d$ to the measure $\mathbb{Q}^f$ can be written as:

\begin{equation} \frac{\D \mathbb{Q}^f}{\D \mathbb{Q}^d} = \exp\Bigg(-\gamma^{fd} W_t - \frac{1}{2}(\gamma^{fd})^2 t \Bigg),  \mbox{	where } \gamma^{fd} = \frac{(r_d - r_f) - (r_d - r_f + \sigma^2)}{\sigma} = -\sigma ,\end{equation}

So in the end we have:

\begin{equation} \frac{\D \mathbb{Q}^f}{\D \mathbb{Q}^d} = \exp\Bigg(\sigma W_t - \frac{1}{2}\sigma^2 t \Bigg) .\end{equation}

\subsection{Garman-Kohlhagen formula for European options}

To derive a formula for the value of the option, firstly the function of its payout must be introduced. Consider a call option payout function for the moment $ T $ given by $ V_T = \mathrm{max}(S_T-K,0)=(S_T-K)^+ $. The price of such options at point $ t = 0, t < T $ can be transformed as follows, using the expected values in risk-neutral measure, i.e. $\mathbb{E}^{\mathbb{Q}^d}$ for domestic risk-neutral measure and $\mathbb{E}^{\mathbb{Q}^f}$ for foreign risk-neutral measure:

\begin{equation}\begin{split}
V_0 & = e^{-r_d T} \mathbb{E}^{\mathbb{Q}^d} [(S_T-K)^+]\\
& = e^{-r_d T} \mathbb{E}^{\mathbb{Q}^d} [(S_T-K)\one_{S_T \geq K}]\\
& = e^{-r_d T} \mathbb{E}^{\mathbb{Q}^d} [S_T\one_{S_T \geq K} - K\one_{S_T \geq K}]\\
& = e^{-r_d T} \mathbb{E}^{\mathbb{Q}^d} [S_T\one_{S_T \geq K}] - K e^{-r_d T} \mathbb{E}^{\mathbb{Q}^d} [\one_{S_T \geq K}]\\
& = e^{-r_d T} \mathbb{E}^{\mathbb{Q}^d} [S_0 \exp\bigg(\sigma W_t^d + \bigg(r_d - r_f - \frac{1}{2}\sigma^2 \bigg) T \bigg)\one_{S_T \geq K}] - K e^{-r_d T} \mathbb{E}^{\mathbb{Q}^d} [\one_{S_T \geq K}]\\
& = S_0 e^{-r_d T}e^{(r_d-f^f) T} \mathbb{E}^{\mathbb{Q}^d} [\exp\bigg(\sigma W_t^d - \bigg(-\frac{1}{2}\sigma^2 \bigg) T \bigg)\one_{S_T \geq K}] - K e^{-r_d T} \mathbb{E}^{\mathbb{Q}^d} [\one_{S_T \geq K}]\\
& = S_0 e^{-r_f T} \mathbb{E}^{\mathbb{Q}^d} \bigg[\frac{\D \mathbb{Q}^f}{\D \mathbb{Q}^d}\one_{S_T \geq K}\bigg] - K e^{-r_d T} \mathbb{E}^{\mathbb{Q}^d} [\one_{S_T \geq K}]\\
& = S_0 e^{-r_f T} \mathbb{E}^{\mathbb{Q}^f} [\one_{S_T \geq K}] - K e^{-r_d T} \mathbb{E}^{\mathbb{Q}^d} [\one_{S_T \geq K}]
.\end{split}\end{equation}

Similar derivation can be repeated for $ 0 <t <T $. In considered equations the dynamics of $ S_T $ process has been used, then the change of the measure, from $\mathbb{Q}^d$ to $\mathbb{Q}^f$, was applied in the expected value. At the end the expected values can be transformed to probabilities:

\begin{equation}
V_0 = S_0 e^{-r_f T} \mathbb{P}^{\mathbb{Q}^f} (S_T \geq K) - K e^{-r_d T} \mathbb{P}^{\mathbb{Q}^d} (S_T \geq K)
,\end{equation}

where $\mathbb{P}^{\mathbb{Q}^d}$ and $\mathbb{P}^{\mathbb{Q}^f}$ are respectively the probabilities under domestic and foreign risk-neutral measure. Then, considering the dynamics of $ S_T $ under measure $\mathbb{Q}^f$ (and $\mathbb{Q}^d$), we can calculate the probability of $\mathbb{P}^{\mathbb{Q}^f} (S_T \geq K)$ (and $\mathbb{P}^{\mathbb{Q}^d} (S_T \geq K)$), making the transformation of following inequalities:

\begin{equation}\begin{split}
\mathbb{P}^{\mathbb{Q}^f}(S_T \geq K) & = \mathbb{P}^{\mathbb{Q}^f}\bigg( S_0 \exp\bigg(\sigma W_T^f + \bigg(r_d - r_f + \frac{1}{2}\sigma^2 \bigg) T \bigg) \geq K \bigg) \\
& = \mathbb{P}^{\mathbb{Q}^f}\bigg( W_T^f \geq \frac{\ln(K/S_0)-(r_d-r_f+\sigma^2/2)T}{\sigma} \bigg)\\
& = \mathbb{P}\bigg( \sqrt{T}\xi \geq \frac{\ln(K/S_0)-(r_d-r_f+\sigma^2/2)T}{\sigma} \bigg)\\
& = \mathbb{P}\bigg( \xi \geq -\frac{\ln(S_0/K)+(r_d-r_f+\sigma^2/2)T}{\sigma\sqrt{T}} \bigg)\\
& = \mathbb{P}( \xi \geq -d_1 ) = \Phi(d_1)
,\end{split}\end{equation}

where $ \xi $ is a random variable with a standard normal distribution ($ W_T ^ f $ has a distribution of $ \mathcal{N}(0, T)$), so all can be reduced to the form of a normal distribution function $ \Phi(d) $ where $ d_1 $ is deterministic. Similar transformations can be made for $\mathbb{P}^{\mathbb{Q}^d} (S_T \geq K)$:

\begin{equation}\begin{split}
\mathbb{P}^{\mathbb{Q}^d}(S_T \geq K) & = \mathbb{P}^{\mathbb{Q}^d}\bigg( S_0 \exp\bigg(\sigma W_T^d + \bigg(r_d - r_f - \frac{1}{2}\sigma^2 \bigg) T \bigg) \geq K \bigg) \\
& = \mathbb{P}^{\mathbb{Q}^d}\bigg( W_T^d \geq \frac{\ln(K/S_0)-(r_d-r_f-\sigma^2/2)T}{\sigma} \bigg)\\
& = \mathbb{P}\bigg( \sqrt{T}\xi \geq \frac{\ln(K/S_0)-(r_d-r_f-\sigma^2/2)T}{\sigma} \bigg)\\
& = \mathbb{P}\bigg( \xi \geq -\frac{\ln(S_0/K)+(r_d-r_f-\sigma^2/2)T}{\sigma\sqrt{T}} \bigg)\\
& = \mathbb{P}( \xi \geq -d_2 ) = \Phi(d_2)
\end{split}\end{equation}

Substituting $\mathbb{P}^{\mathbb{Q}^d} (S_T \geq K)$ and $\mathbb{P}^{\mathbb{Q}^f} (S_T \geq K)$ to the formula (1.30) and making generalizations from the point in time 0 for the moment $ 0 \leq t < T $, i.e. substituting $ t $ for $ 0 $ and $ \tau = T-t $ for $ T $, gives the Garman-Kohlhagen (1983) formula:

\begin{equation}
V_t = S_t e^{-r_f \tau} \Phi\bigg(\frac{\ln(S_t/K)+(r_d-r_f+\sigma^2/2)\tau}{\sigma\sqrt{\tau}} \bigg) - K e^{-r_d \tau} \Phi\bigg(\frac{\ln(S_t/K)+(r_d-r_f-\sigma^2/2)\tau}{\sigma\sqrt{\tau}} \bigg)
.\end{equation}

Formula (1.33) is identical to the Black (1976)\footnote{~\bibentry{Black1976}.} model when in the latter the forward price will be substituted by the following formula: $ F = Se^{(r_d-r_f) \tau } $.

\section{Option pricing methods}

Beside using the Black-Scholes formula, vanilla European options can be priced by following methods:
\begin{enumerate}
\item by computation and transformation of the definite integrals of the characteristic function of the underlying instrument price distribution, if the form of the characteristic function is well known,
\item by the numerical solution, e.g. by finite difference method, of differential equations describing the dynamics of the option replication strategy, if the equations are known,
\item by N-fold \footnote{~The values of $ N $ in practice are in the range from $10^4$ to $10^6$.} simulation of the trajectory of the underlying instrument price until the option maturity $ T $, followed by computation of the option payout in each scenario, and then calculation of the average of all results (Monte Carlo method).
\end{enumerate}

The first method is used for models for which the exact form of the characteristic function is well known. This implies that the use of this method is limited to models with an exponential-affine characteristic function and some cases of non-affine class. If the explicit formula for the characteristic function does not exist, then this method can still be applied, if a numerical solution for the characteristic function parameters can be found.
Last two methods are more general and can be used as well for non-vanilla options with complex features. Usually, for low number of dimensions, the least computationally complex is the first method and the third method is the most complex.
The third method, although is the most computationally complex, provides a simple way of pricing of options with a more complex payout profile than in the case of vanilla European options. Main advantage of this method is that it can be used for path dependent options and that adding more than one option on the same underlying instrument to the valuation algorithm is not changing its complexity in big $\mathcal{O}$ notation. The second method, which computational complexity is usually intermediate between the first and the third method is used for the valuation of options mainly in the case of stochastic volatility models, which are outside the exponential-affine class. This method is not good when there are many options to be valued and neither good when the number of considered dimensions in equation is 4 or more. The advantage of the differential equation methods is that they can be applied to options with an early exercise features. Due to the area that is related to objective of this research, namely stochastic volatility diffusion models with exponential-affine characteristic function, starting from this chapter the first method will be examined in details and all models will be considered in terms of application of this method.

\section{Option pricing with characteristic function}

\subsection{Heston method}

Heston (1993) \footnote{~\bibentry{Heston1993}.} assumed that, given the dynamics of the price of the underlying instrument and the dynamics of their volatility, formula for the price of vanilla European call option should have a form similar to the Black-Scholes (1973) formula. For currency options, it can be written as:

\begin{equation}
V_t = S_t e^{-r_f (T-t)} P_1 (S,\nu,t) - K e^{-r_d (T-t)} P_2 (S,\nu,t)
,\end{equation}

where $P_1$ and $P_2$ are probabilities of some distributions. It is known that $ P_j (t = T) = 0 $, $ i = 1,2 $. However, unlike the Black-Scholes model (1973), probabilities $ P_1 $ and $ P_2 $ are generally different in considered model and cannot be reduced to the normal distribution CDF due to the stochastic volatility, which is leading to other distribution than the normal distribution, that is present in the assumptions of Black-Scholes (1973 model. However, these probabilities can be computed by integration, if the characteristic function of the distribution of logarithmic prices is known. Taking into consideration the differential equation of replication strategy dynamics it is able to find an analytic formula for the characteristic function of the underlying instrument price distribution, at least for some versions of this equation (e.g. in Heston model). For the Heston model, the characteristic function has an exponential-affine form: $\phi_j(u,x,\nu,\tau) = e^{i u x_0 + A_{j,\tau} + B_{j,\tau} \nu_0}$, $\tau=T-t$, $x=\ln(S)$.

Knowledge of the exact form of the characteristic function can be used in Fourier transform reversion theorem (Gil-Pelaez (1951)\footnote{~\bibentry{Gil-Pelaez1951}.}), which states that:

\begin{equation}
F(k) = \frac{1}{2} - \frac{1}{\pi} \int_0^\infty \mathrm{Re} \bigg( \frac{e^{-i u k} \phi(u)}{i u} \bigg) \,\D u
,\end{equation}

where $x=\ln(S)$ and $k=\ln(K)$. Because $\mathbb{P}(x>k) = 1 - F(k)$, it can be written that:

\begin{equation}
P_j = \frac{1}{2} + \frac{1}{\pi} \int_0^\infty \mathrm{Re} \bigg( \frac{e^{-i u k} \phi_{j,x,\nu,T}(u)}{i u} \bigg) \,\D u
.\end{equation}

Alternatively to the version with two different characteristic functions, a version of this method with one characteristic function can be derived, where $\phi_T=\phi_{2,x,\nu,T}$. Assuming that $P_2=\mathbb{P}^{\mathbb{Q}^d}(S_T \geq K)$ and $S_0 e^{(r_d-r_f)\tau} P_1 = \mathbb{E}^{\mathbb{Q}^d}[S_T\one_{S_T \geq K}]$, it can be written that:

\begin{equation}
P_2 = \mathbb{P}^{\mathbb{Q}^d}(S_T>K) = \frac{1}{2} + \frac{1}{\pi} \int_0^\infty \mathrm{Re} \bigg( \frac{e^{-i u k} \phi_{2,x,\nu,T}(u)}{i u} \bigg) \,\D u
\end{equation}

and after defining additional measure:

\begin{equation}
\D \mathbb{Q}^f = \frac{S_T}{\mathbb{E}^{\mathbb{Q}^d}[S_T]} \D \mathbb{Q}^d
,\end{equation}

the first distribution function can be formulated as:

\begin{equation}
\mathbb{E}^{\mathbb{Q}^d}[S_T\one_{S_T \geq K}] = \int_{{S_T>K}} S_T \D \mathbb{Q}^d = \mathbb{E}^{\mathbb{Q}^d}[S_T] \int_{{S_T>K}} \frac{S_T}{\mathbb{E}^{\mathbb{Q}^d}[S_T]} \D \mathbb{Q}^d = \mathbb{E}^{\mathbb{Q}^d}[S_T] \mathbb{P}^{\mathbb{Q}^f}(S_T>K)
.\end{equation}

Given that $\mathbb{E}^{\mathbb{Q}^d}[S_T] = {\phi_T(-i)}$, the probability under the measure $\mathbb{Q}^f$ can be expressed by the characteristic function as:

\begin{equation}
\int_\omega e^{iux} \D \mathbb{Q}^f = \int_\omega e^{iux} \frac{S_T}{\mathbb{E}^{\mathbb{Q}^d}[S_T]} \D \mathbb{Q}^d = \frac{\mathbb{E}^{\mathbb{Q}^d}[e^{i(u-i)x}]}{\mathbb{E}^{\mathbb{Q}^d}[S_T]} = \frac{\phi_T(u-i)}{\phi_T(-i)}
,\end{equation}

and since $\mathbb{E}^{\mathbb{Q}^d}[S_T]$ is equal to the forward price ($S_0 e^{(r_d-r_f)\tau}$), there is also:

\begin{equation}
P_1 = \mathbb{P}^{\mathbb{Q}^f}(S_T>K) = \frac{1}{2} + \frac{1}{\pi} \int_0^\infty \mathrm{Re} \bigg( \frac{e^{-i u k}}{i u}\frac{\phi_T(u-i)}{\phi_T(-i)}\bigg) \,\D u
.\end{equation}

\subsection{Carr-Madan method}

One of the problems of the Heston method is that the Fourier transform of the underlying instrument price distribution density may not exist. Carr and Madan (1999) \footnote{~\bibentry{Carr1999}.} show how to deal with this problem. They presented the call option price with time to maturity $\tau$ as modified price of a dumped call option $c_{\tau}(k)$, which they defined as:

\begin{equation}
c_{\tau}(k) = exp(\alpha k) C_{\tau}(k)
,\end{equation}

where $\alpha>0$ is a parameter\footnote{~Authors of this method tested parameter $\alpha$  values, which were greater than one, i.e. 1.1 and 1.5.} and $k=\ln(K)$. The function $c_{\tau}(s)$ should be integrable on all of its domain. The price of the call option $ C_{\tau} (K) $ with the time to maturity $ \tau $ can be calculated from numeric integral if risk neutral distribution of underlying instrument logarithmic prices $ q_{\tau}(s) $ is known. Knowledge of this distribution allows to calculate following integral:

\begin{equation}\begin{split}
C_{\tau}(K) & = e^{-r_d \tau} \mathbb{E}^{\mathbb{Q}^d} [(S_{t+\tau}-K)\one_{S_{t+\tau} \geq K}] = e^{-r_d \tau} \int_{k}^\infty (e^{s}-e^k)q_{\tau}(s)\,\D s
.\end{split}\end{equation}

Fourier transform of distribution density $q_{\tau}(s)$, characteristic function of variable $s$, is defined as:
\begin{equation}
\phi_{\tau}(u) = \int_{-\infty}^\infty e^{ius} q_{\tau}(s)\,\D s
.\end{equation}

Fourier transform of $c_{\tau}(s)$ is given by:

\begin{equation}
\psi_{\tau}(v) = \int_{-\infty}^\infty e^{ivk} c_{\tau}(k)\,\D k
,\end{equation}

and because the price $C_{\tau}$ is an even real function, hence relation between $\psi_{\tau}$ and $\phi_{\tau}$ is given by:

\begin{equation}
C_{\tau}(k) = \frac{e^{-\alpha k}}{2\pi} \int_{-\infty}^\infty e^{ivk} \psi_{\tau}(v)\,\D v = \frac{e^{-\alpha k}}{\pi} \int_{0}^\infty e^{ivk} \psi_{\tau}(v)\,\D v
.\end{equation}

On the other hand, the component $\psi_{\tau}$ after taking into consideration the complete formula for $c_{\tau}$ is equal to:

\begin{equation}\begin{split}
\psi_{\tau}(v) & = e^{-r_d \tau} \int_{-\infty}^\infty e^{ivk} e^{\alpha k} \int_{k}^\infty (e^{s}-e^k)q_{\tau}(s)\,\D s \,\D k \\
& = e^{-r_d \tau}\int_{-\infty}^\infty q_{\tau}(s) \int_{-\infty}^s e^{ivk} (e^{s+\alpha k}-e^{(1+\alpha)k})\,\D k \,\D s \\
& = e^{-r_d \tau} \int_{-\infty}^\infty q_{\tau}(s) \bigg( \frac{e^{(\alpha+iv+1)s}}{\alpha+iv} - \frac{e^{(\alpha+1+iv)s}}{\alpha+1+iv} \bigg) \,\D s \\
& = e^{-r_d \tau} \int_{-\infty}^\infty q_{\tau}(s) \frac{e^{(\alpha+iv+1)s}}{\alpha^2+\alpha-v^2+i(2\alpha+1)v} \,\D s \\
& = \frac{e^{-r_d \tau} \phi_{\tau}(v-(\alpha+1)i)}{\alpha^2+\alpha-v^2+i(2\alpha+1)v}
.\end{split}\end{equation}

Thus, ultimately the price of call options is:

\begin{equation}
C_{\tau}(k) = \frac{e^{-\alpha k}}{\pi} \int_{0}^\infty e^{ivk} \frac{e^{-r_d \tau} \phi_{\tau}(v-(\alpha+1)i)}{\alpha^2+\alpha-v^2+i(2\alpha+1)v} \,\D v
.\end{equation}

The presented derivation also applies to foreign currency options prices, after substitution of forward price, $ f = s + (r_d-r_f) \tau $, in place of the underlying instrument spot price. Prices of call option can be calculated from put-call parity. An important advantage of this method is better convergence to zero of integrated function as well as the fact that only one function is integrated.

\subsection{Attari method}

An alternative formula for the valuation of the call option by the characteristic function is presented in Attari (2004)\footnote{~\bibentry{Attari2004}}. The basis for the discussion was formulation of the final price of the underlying instrument as $ S_T = S_t e^{rt + \epsilon} $, where $ \epsilon $ is a stochastic component of  price process. Like Heston he considers the probability of $ \mathrm{P1} $ and $ \mathrm{P2} $ in the formula for the option price. However, by introduction of a variable $ l = \mathrm{ln}(Ke^{-(rd-rf)(T-t)}/S)$ he makes following changes:

\begin{equation}\begin{split}
C(K) & = e^{-r_d \tau} (\mathrm{E}^{\mathbb{Q}^d}[S_T \mathrm{1}_{S_T>K}] - K\mathrm{E}^{\mathbb{Q}^d}[\mathrm{1}_{S_T>K}] ) ,\\
& = S_t  e^{-r_f\tau} \mathrm{E}^{\mathbb{Q}^d}[e^x \mathrm{1}_{x>l}] - K e^{-r_d \tau} \mathrm{E}^{\mathbb{Q}^d}[\mathrm{1}_{x>l}] ,\\
& = S_t  e^{-r_f\tau} \mathrm{P}_1 - K e^{-r_d \tau} \mathrm{P}_2
.\end{split}\end{equation}

In addition, because for the density of $ x $ given by $ q (x) $ it is true that $e^x q(x)>0$ and $0 \leq \mathrm{P}_1 \leq 1$, so $e^x q(x) = p(x)$ can be treated as a function of the density and then:

\begin{equation}\begin{split}
\mathrm{P}_1 & = \mathrm{E}^{\mathbb{Q}^d}[e^x\mathrm{1}_{x>l}] = \int_l^\infty e^x q(x) \,\D x = \int_l^\infty p(x) \,\D x ,\\
\mathrm{P}_2 & = \mathrm{E}^{\mathbb{Q}^d}[\mathrm{1}_{x>l}] = \int_l^\infty q(x) \,\D x
.\end{split}\end{equation}

Moreover, after defining the characteristic function for the $ q (x) $ as $ \phi_2(u) $ and for $ p(x) $ as $ \phi_1(u) $, it is true that $ \phi_1(u) = \phi_2(u-i)$, because:

\begin{equation}
\phi_1(u) = \int_{-\infty}^\infty e^{iux} p(x) \,\D x = \int_{-\infty}^\infty e^{iux} e^x q(x) \,\D x = \int_{-\infty}^\infty e^{i(u-i)x} q(x) \,\D x \phi_2(u-i) ,\\
.\end{equation}

After that the equation for P1 can be written as follows:

\begin{equation}
\mathrm{P}_1 = \tfrac{1}{2\pi} \int_{-\infty}^\infty \phi_1(\nu) \bigg( \int_{l}^\infty e^{i\nu x} \,\D x \bigg) \,\D \nu = \tfrac{1}{2\pi} \int_{-\infty}^\infty \phi_2(\nu-i) \bigg( \int_{l}^\infty e^{i\nu x} \,\D x \bigg) \,\D \nu
,\end{equation}

which can be further transformed with substitution, $u = \nu - i$:

\begin{equation}
\mathrm{P}_1 = \tfrac{1}{2\pi} \int_{-\infty}^\infty \phi_2(u) \bigg( \int_{l}^\infty e^{i (u+i) x} \,\D x \bigg) \,\D u
.\end{equation}

Inner integral of considered equation can be analysed with following representation:

\begin{equation}
\mathrm{P}_1 = \tfrac{1}{2\pi} \int_{-\infty}^\infty \phi_2(u) \tfrac{e^{-i(u+i)l}}{i(u+i)} \,\D u - \tfrac{1}{2\pi} \lim_{R \to \infty} \int_{-\infty}^\infty \phi_2(u) \tfrac{e^{-i(u+i)R}}{i(u+i)} \,\D u = I_1 - I_2
.\end{equation}

The second integral has a pole in complex plane at $u=-i$, so the residue there is $\phi_2(-i)/i$. After application of Residue Theorem, the integral yields:

\begin{equation}
I_2 = \lim_{R \to \infty} \tfrac{1}{2\pi} \bigg( -2\pi i \tfrac{\phi_2(-i)}{i} \bigg) = -\phi_2(-i) = -\phi_1(0) = -1
.\end{equation}

Including the result in the main equation produces:

\begin{equation}
\mathrm{P}_1 = \tfrac{e^l}{2\pi} \int_{-\infty}^\infty \phi_2(u) \tfrac{e^{-iul}}{i(u+i)} \,\D u + 1
.\end{equation}

For $\mathrm{P}_2$ a form, which is analogous to the one from the Heston method, can be used:

\begin{equation}
\mathrm{P}_2 = \tfrac{1}{2} + \tfrac{1}{2\pi} \int_{-\infty}^\infty \phi_2(u) \tfrac{e^{-iul}}{iu} \,\D u
.\end{equation}

Then, substituting both equations to the general equation of a call option price, we can transform the equation for the option price so that it will include only one integral, which has a positive effect on the speed of calculation:

\begin{equation}\begin{split}
C(K) & = S_t e^{-r_f\tau} \bigg( 1 + \tfrac{e^l}{2\pi} \int_{-\infty}^\infty \phi_2(u) \tfrac{e^{-iul}}{i(u+i)} \,\D u \bigg) - K e^{-r_d\tau} \bigg( \tfrac{1}{2} + \tfrac{1}{2\pi} \int_{-\infty}^\infty \phi_2(u) \tfrac{e^{-iul}}{iu} \,\D u \bigg) ,\\
& = S_t e^{-r_f\tau} - K e^{-r_d\tau} \bigg( \tfrac{1}{2} + \tfrac{1}{\pi} \int_{0}^\infty \mathrm{Re} \bigg( \phi_2(u) e^{-iul} \tfrac{1-i/u}{u^2+1} \bigg) \,\D u \bigg)
,\end{split}\end{equation}

where the last part is finally obtained after substitution of $l = K e^{-(r_d-r_f)\tau}/S_t$. Attari method in comparison to Carr-Madan method has a great advantage of being free from additional unknown parameters and still having fast converging function being integrated, due to the denominator $u^2+1$.

\section{Local volatility models}

The existence of the market implied volatilities, which form non-flat volatility surface in terms of both the exercise price and the time to maturity, was the reason for the search for new models describing the dynamics of instantaneous variance which is based on current quotations of implied volatilities. The phenomenon of non-flat volatility in the dimension of strike price is called the volatility smile. Market quotations of implied volatilities, as has already been shown (in Section \ref{sec:1.2}), usually form a parabolic shape in the strike price dimension and an exponential curve in time to maturity dimension. The reason for the existence of such a relationship is that market participants price options that are far in the money or far out of the money as being more risky than at the money options, whose strike price is close to the forward price of the underlying instrument. \par

Before the local volatility approach models did not take into account the volatility smile property of the market. However, those models allow to use different volatilities forming the smile as an input data in the valuation. Volatility surface was therefore exogenous feature, which is shaped only by the market. One of problems that remained open is how to parametrize volatility surfaces in such a way that can be calculated from implied volatilities market data for any exercise price $ K $ and the time to maturity $ T $. \par

Parametrisation of the strike price dimension, $ K $, (smile volatility) has been studied before the second dimension. Among others this problem is handled by Malz (1997)\footnote{~\bibentry{Malz1997}.} and by the Vanna-Volga interpolation in Castagna and Mercurio (2007)\footnote{~\bibentry{Castagna2007}.}. Malz presents a parabolic parametrisation of volatility smile, while Vanna-Volga represents the adjustment to the valuation of a hedging strategy due to the risks associated with changing volatility during the life of the option, those risks are partially reflected by Vanna and Volga sensitivities. Those sensitivities are derivatives of Vega\footnote{~Vega is a derivative of option price with respect to implied volatility.} with respect to the price of the underlying and to the implied volatility.\par

The work Derman and Kani (1994)\footnote{~\bibentry{Derman1994}.} and work of Dupire (1994)\footnote{~\bibentry{Dupire1994}.} present the model of local volatility surface. In the former of these works it is a model with the assumption of discrete price movements of the instrument, but in the latter movements are continuous. Deterministic local volatility models are considered to be inconsistent with the dynamics of the implied volatility surface of stock indices options. An interesting feature of the local volatility models is that the only source of volatility of implied volatilities are the underlying instrument price changes. This leads to a complete market in case of a market where the only option replicating strategy is the strategy that is based on underlying instrument trading. Local volatility model of Dupire can be described as: \par

\begin{equation}\begin{split}
\D S & = \mu (t) S \D t + \sigma (K,t,S_0) S \D W ,\\
\sigma (K,t,S_0) & = \sqrt{\frac{\frac{\partial C}{\partial t}-\mu (t)[C(K,t,S_0)-K\frac{\partial C}{\partial K}]}{\frac{K^2}{2}\frac{\partial^2 C}{\partial K^2}}}
,\end{split}\end{equation}

where:
\begin{itemize}
\item $C(K,t,S_0)$ is the price of the option with maturity $t$, strike price $K$ and a starting price of underlying $ S_0 $,
\item $\mu (t)$ is the risk-neutral drift.
\end{itemize}

The formula of local volatility of Dupire is using the numerical values of two derivatives of the option price, one with respect to strike price and the other with respect to the time to maturity. That feature corresponds to an assumption that in both dimensions of volatility surface the function representing the price of the option is continuous. This means in practice that this assumption will be true only when an infinite number of quoted options with respect to both dimension would be available. As the number of quoted option prices is finite, to apply the Dupire formula discreetly quoted option prices should be interpolated in both dimensions. This is an additional obstacle.\par

\section{Diffusion models of stochastic volatility}

A milestone in the valuation of options and also a milestone in modelling the volatility surface and its dynamics was the introduction of stochastic volatility models, which in the case of models with one-factor stochastic variance or volatility has the following general form:

\begin{equation}\begin{split}
& \D S_t = \mu S_t \D t + \nu_t^r S_t^\beta \D W_t ^S ,\\
& \D \nu_t = \kappa \nu_t^q (\theta - h(\nu_t)) \D t + \omega \nu_t^p \D W_t^\nu ,\\
& \langle \D W_t^S , \D W_t^\nu \rangle = \rho \D t , \quad r \in \{\tfrac{1}{2},1\} ,\\
& h(\nu) = \left\{
  \begin{array}{lr}
    \frac{1}{\lambda}(\nu^\lambda-1) & : \lambda > 0\\
    log(\nu) & : \lambda = 0
  \end{array}
\right.
.\end{split}\end{equation}

The parameters in the model can be interpreted as follows:
\begin{itemize}
\item $ \mu $ is the level of the drift of the underlying instrument price process, for most models for the FX market it is equal to the differential in interest rates, $ r_d-r_f $,
\item $\kappa$ is the speed of mean-reversion in the variance process,
\item $\theta$ is the long term mean of the variance process,
\item $\omega$ is the volatility of the variance process,
\item $\rho$ is the correlation between Wiener process of underlying instrument prices and Wiener process of the variance,
\item $\nu_0$ is the initial level of variance, i.e. at the time $t=0$,
\item $ \beta $ is the exponent of underlying instrument prices,
\item $ r $ is the exponent and is equal to 0.5 or 1, it determines whether $ \nu $ is interpreted as a variance or volatility,
\item $ p $ is the exponent of the variance that is multiplying the volatility of variance process,
\item $ q $ is the exponent of the variance that is multiplying the mean of the variance process,
\item $ \lambda $ is a parameter of Box-Cox transformations applied to the variance that is the mean-reverting element of the variance process, for the value of this parameter equal to 0, this function is a natural logarithm, and for a value of 1 is a linear function.
\end{itemize}

The formulation, could be complemented by a more specific form for a case with $ \lambda = 1 $, which occurs in several models:

\begin{equation}\begin{split}
& \D S_t = \mu S_t \D t + \nu_t^r S_t^\beta \D W_t ^S ,\\
& \D \nu_t = \kappa \nu_t^q (\hat{\theta} - \nu_t) \D t + \omega \nu_t^p \D W_t^\nu ,\\
& \langle \D W_t^S , \D W_t^\nu \rangle = \rho \D t, \quad r \in \{\tfrac{1}{2},1\}
,\end{split}\end{equation}

where $\hat{\theta} = \theta+1$.

Table \ref{table:1.3} shows the parametrization of considered general form in relation to the most important stochastic volatility models.

\begin{table}[h!]
\centering
\caption{Parametrization of general form of stochastic volatility models}
\label{table:1.3}
\begin{tabular}{lccccccccc}
  \hline
Model & $p$ & $q$ & $r$ & $\lambda$ & $\rho$ & $\kappa$ & $\mu$ & $\beta$ & $\hat{\beta}$ \\
  \hline
Hull-White (1987), \\ Melino-Turnbull (1990) & 0.5 & 0 & 0.5 & 1 & 0 & $\mathbb{R}_+$ & $(r_d-r_f)$ & 1 & - \\
Scott (1987), Wiggins (1987) & 0.5 & 0 & 0.5 & 0 & 0 & $\mathbb{R}_+$ & $(r_d-r_f)$ & 1 & - \\
Heston (1993) & 0.5 & 0 & 0.5 & 1 & $[-1;1]$ & $\mathbb{R}_+$ & $(r_d-r_f)$ & 1 & - \\
Stein-Stein (1991) & 0 & 0 & 1 & 1 & 0 & $\mathbb{R}_+$ & $(r_d-r_f)$ & 1 & - \\
Sch\"obel-Zhu (1999) & 0 & 0 & 1 & 1 & $[-1;1]$ & $\mathbb{R}_+$ & $(r_d-r_f)$ & 1 & - \\
GARCH diffusion (1990) & 1 & 0 & 0.5 & 1 & $[-1;1]$ & $\mathbb{R}_+$ & $(r_d-r_f)$ & 1 & - \\
Lewis (2000) & 1.5 & 1 & 0.5 & 1 & $[-1;1]$ & $\mathbb{R}_+$ & $(r_d-r_f)$ & 1 & - \\
Hagan et al. (2002) & 1 & 0 & 1 & 1 & $[-1;1]$ & 0 & 0 & $\hat{\beta}+\frac{(r_d-r_f)(\hat{\beta}-1)}{\ln S}$ & $\mathbb{R}$ \\
  \hline
\end{tabular}
\end{table}

Stochastic volatility models were presented at approximately the same time in the works of: Hull and White (1987)\footnote{~\bibentry{Hull1987}.}, Scott (1987)\footnote{~\bibentry{Scott1987}.}, Wiggins (1987)\footnote{~\bibentry{Wiggins1987}.}, Stein and Stein (1991)\footnote{~\bibentry{Stein1991}.}. For currency options stochastic volatility model was then examined by Melino and Turnbull (1990, 1991)\footnote{~\bibentry{Melino1990}.} \footnote{~\bibentry{Melino1991}.}, with results suggesting the model is correct for the FX market. However, none of these works did not present an explicit formula for the valuation of options with stochastic volatility, instead they depended on calculation of the numerical solution of two-dimensional partial differential equations. Given this background, important improvement was the Heston (1993)\footnote{~\bibentry{Heston1993}.} model. This model is obtained by substitution of parameters $r=p=1/2$, $q=0$, $\beta=1$ and $\lambda=1$ in the considered general form of models with one-factor stochastic variance. For currency options, this model was studied, among others, by Weron and Wystup (2005)\footnote{~\bibentry{Weron2005}.}.

Considered general form of model contains all of the previously mentioned models, and in almost all of them is $ \lambda = 1 $. The exception are Wiggins and Scott models, which contain the dynamics of the log-normal volatility and therefore they have $ \lambda = 0 $. In addition, for each stochastic variance model we can change the parameter $ p = 1/2 $ for $ p = 1 $ and using It\^{o} lemma transform the equation of variance dynamics to the equation of the volatility dynamics.

Special cases of the Heston model, as well as some of its extensions, are called the exponential-affine stochastic volatility models, due to the fact that the characteristic function of the underlying instrument price distribution is in exponential-affine form. In addition, in cases like the Heston model, an analytic solution for the parameters of this function exists. When $r=p=1/2$, $q=0$, $\beta=1$ and $\lambda=1$ are not satisfied then stochastic volatility model does not belong to the exponential-affine class and only in a few cases there is an analytic solution for parameters of the characteristic function used for the valuation of options.

In the assumptions of stochastic volatility models, in addition to the equation for the differential of the underlying instrument price, there is also the equation for the differential of variance or volatility of the underlying instrument price. Each of the equations has its own random factor in the form of the Wiener process, with the additional assumption that the two processes are correlated and that correlation is constant. After the calibration of the parameters, all of these models for different strike prices and times to maturity output option prices, which are related to the implied volatility surface that is non flat in both of its dimensions.

An important advantage of the Heston model is the existence of semi-closed formula\footnote {~The analytic formula exists for characteristic function and its parameters. The numerical integration of this function is needed to price an option, hence the formula for the option price is called semi-closed.} to price options with this model. This advantage was one of the reasons for the popularity of the model. Although the formula requires the use of the numerical integration, the existence of a formula for the valuation process is an important reduction of complexity. In addition, starting from the calibrated Heston model, options can be priced by an application of the finite-difference method to partial differential equation from the model or using a Monte Carlo simulation with variance reduction techniques. The Heston model has 5 parameters which are calibrated in a way which ensures that the model valuation of all quoted options is as close to their market prices as it is possible. \par

The Heston model and its expansions are particularly useful for pricing exotic options by Monte Carlo simulations. The more complex exotic options, which depend on future levels of volatility, can also be priced by an application of the Heston model, because the volatility term structure is also present in the model. A notable example of an extension of the Heston model, especially in the case of stock prices and credit instruments, is the jump-diffusion model of Bates (1996) \footnote{~\bibentry{Bates1996}.}. In comparison to the Heston model, the new model has added the jump process to the underlying instrument prices dynamics. Its author increased the number of parameters in the Heston model to eight. The construction of the model with use of the jump-diffusion process for the underlying instrument prices dynamics results in the presence of additional random discrete price movements. As a result, options replication with delta hedging strategy can not be used effectively, because it involves continuous time trading of variable amount of the underlying asset and the risk-free Treasury bills .\par

The same equation as in the case of variance in the Heston model, but used for a volatility (i.e. difference is in the factor $ p = 1 $ in the general form), provides the model of Stein and Stein (1991) and its extension with non-zero correlation, i.e. the Sch\"obel-Zhu (1999)\footnote{~\bibentry{SchoebelZhu1999}.} model. Nevertheless, from the beginning the latter model was presented with the corresponding characteristic function and analytic solution for its parameters, is not in the exponential-affine class with respect to the variance of the underlying instrument. However, Lord and Kahl (2006)\footnote{~\bibentry{LordKahl2006}.} have showed a new derivation of analytic solution for characteristic function and its parameters for this model. The characteristic function in their solution has the exponential-affine form  if the model is presented with an additional variance process equation, which is actually strictly related to the volatility process equation. This feature allows to present the characteristic function as an exponential-affine function of the vector [$\ln(S)$,$\nu$,$\nu^2$]. With these properties the model can be classified into exponential-affine class of stochastic volatility models considered by Duffie et al. (2003)\footnote{~\bibentry{Duffie2003}.}. Despite the additional component in the characteristic function it is exponential-affine and the additional component is endogenous in the model.

Two important models outside of the exponential-affine class are the GARCH diffusion model introduced by Nelson (1990)\footnote{~\bibentry{Nelson1990}.} and the 3/2 model by Lewis (2000)\footnote{~\bibentry{Lewis2000}.}. The GARCH\footnote{~GARCH stands for Generalized Autoregressive Conditional Heteroskedasticity (model).} diffusion model is a continuous limit for many models from discrete GARCH class, including GJR-GARCH model, for which limits of its parameters can be represented by parameters of the GARCH diffusion model. For the 3/2 model $ r = 1/2 $ and $ \beta = 1 $ just as in the Heston model, but $ p = 3/2 $ and $ q = 1 $, which in the Heston model were different, i.e. $ p = 1 $ and $ q = 0 $. For the GARCH diffusion model an analytic solution for the parameters of characteristic function has not been found, but for the 3/2 model its form of characteristic function and analytic solution for its parameters were presented in Lewis (2001). Since in both of these models used processes are not the CIR (Cox-Ingersoll-Ross, Cox et al. (1985)\footnote{~\bibentry{Cox1985}}) process as in the Heston model case, there is no condition that guarantees positivity of variance, which was present in the latter model.

\subsubsection{SABR model}

Another milestone in stochastic volatility models was SABR model from Hagan et al. (2002)\footnote{~\bibentry{Hagan2002}.}. SABR is an abbreviation, which stands for \textit{Stochastic Alpha Beta Rho}. On the one hand, in comparison to the Heston and the Sch\"obel-Zhu model, the model has a simplification of the volatility process and partially simplification of the underlying prices process ($ \mu = 0 $ and $ \kappa = 0 $). On the other hand, due to the $ \beta $ taking any value, and $ p = r = 1 $, the model does not belong to the exponential-affine class. In addition, to have an original formulation of the SABR model from considered general form, following substitution should be performed: $\beta = \hat{\beta}+\frac{(r_d-r_f)(\hat{\beta}-1)}{\ln S}$. This is due to the fact that general form relates to spot prices and the original form of SABR concerns forward prices.

The model has only 4 parameters and as a result of having less parameters is the simplification of the volatility dynamics equation. The SABR model volatility is not mean-reverting, which makes impossible to directly model the term structures of volatility. However, in market practice, for every volatility smile of the time to maturity $ T $ a separate SABR model is calibrated (i.e. there are 4 unique values of its parameters for any nodal point in time). This is a significant limitation with respect to the Heston model. Nevertheless, an important advantage is the existence of an analytic formula for implied volatility as a function of the strike price, which accelerates both option pricing and the calibration of the model in relation to the Heston model. Moreover, this formula is free of arbitrage. For this reason, SABR model is often used for interpolation and extrapolation of the volatility smile (for a fixed time to maturity) from several points of market data. Considered in the literature values of the parameter $\beta$ were usually in [0,1] range.\par

SABR model is a model of stochastic volatility primarily used for the market of interest rate derivatives. Using it for the stock and currency options markets is also possible. SABR model is one of the market standards for arbitrage-free interpolation and extrapolation of swaption implied volatilities. \par


SABR model for the forward price (or interest rates or the exchange rate), $ F $, time to maturity, $ \tau = T-t $, is given by a system of stochastic differential equations:

\begin{equation}\begin{split}
& dF_t = \hat{\alpha} F_t ^ \beta \D W_t^F ,\\
& \D \hat{\alpha_t} = \nu \hat{\alpha} \D W_t^{\hat{\alpha}} ,\\
& \langle \D W_t^F , \D W_t^{\hat{\alpha}} \rangle = \rho \D t
.\end{split}\end{equation}

$F$ and $\hat{\alpha}$ are martingales with initial conditions:

\begin{equation}\begin{split}
& F(0) = f ,\\
& \hat{\alpha}(0) = \alpha
.\end{split}\end{equation}

The parameters in the model can be interpreted as follows:
\begin{itemize}
\item $ \alpha $ is the initial volatility,
\item $ \nu $ is the volatility of the volatility process,
\item $ \beta $ is the exponent of forward prices,
\item $ \rho $ is a correlation between Wiener process of the underlying instrument prices and Wiener process of volatility.
\end{itemize}

A case in which $ \beta = 0 $ is equivalent to a normal model, and when $ \beta = 1 $ then it is equivalent with log-normal model, and when $ \beta = \tfrac{1}{2} $ is equivalent with a simplified CIR model. Mentioned analytic solution for implied volatility is as follows:

\begin{equation}\begin{split}
\sigma(K,f) = & \frac{\alpha}{(f K)^{(1-\beta)/2} [1 + \frac{(1-\beta)^2}{24} \mathrm{log}^2 \frac{f}{K} + \frac{(1-\beta)^4}{1920} \mathrm{log}^4 \frac{f}{K}]} \cdot \bigg( \frac{z}{x(z)} \bigg) \cdot \\
& \bigg[ 1 + \bigg(\frac{(1-\beta)^2}{24} \frac{\alpha^2}{(fK)^{1-\beta}} + \frac{1}{4} \frac{\rho \beta \nu \alpha}{(fK)^{(1-\beta)/2}} + \frac{2-3\rho^2}{24} \nu^2 \bigg) T \bigg]
,\end{split}\end{equation}

where

\begin{equation}\begin{split}
& z = \frac{\nu}{\alpha} (fK)^{(1-\beta)/2} \mathrm{log} \frac{f}{K} ,\\
& x(z) = \mathrm{log} \bigg( \frac{\sqrt{1-2 \rho z + z^2}+z-\rho}{1-\rho} \bigg)
.\end{split}\end{equation}

It can be seen that this formula can be simplified when $ K = f $, which is close to the ATM strike. SABR model is comparable with the Heston model, when the latter is calibrated only on a set of options with the same maturity. After calibration of all parameters, $\alpha,\beta,\rho$ and $\nu$, implied volatility depends only on the strike price, $ K $, the forward price, $ f $, and the time to maturity $T$. 

\subsubsection{Heston-Nandi model}

The model that is not in previously considered continuous time models classification, but has unique and important features is the model of Heston-Nandi (1997)\footnote{~\bibentry{Heston1997}.}. It is a model with discrete time, which has a special relationship with the Heston model and discrete GARCH models.

Starting from a base in the form of discrete GARCH econometric models, which were initiated by Engle (1982)\footnote{~\bibentry{Engle1982}.} and Bollerslev (1986)\footnote{~\bibentry{Bollerslev1986}.}, a concept of pricing option in this class of time series models was developed. The first model of this type was proposed by Duan (1995)\footnote{~\bibentry{Duan1995}.}, which represented an approximation of option price in a variant of GARCH model. Another article on this topic, this time showing the valuation of options with an analytic formula for the characteristic function and its parameters, while still using a discrete GARCH class model, was presented by Heston and Nandi (1997). Their model is of the following form:

\begin{equation}\begin{split}
& R_t = \Delta ln S_t = r + \lambda h_t + \sqrt{h_t} z_t ,\\
& h_t = \xi + \beta h_{t-1} + \alpha (z_{t-1} - \gamma \sqrt{h_{t-1}})^2
,\end{split}\end{equation}

where: $\alpha$, $\beta$, $\gamma$, $\lambda$ and $\xi$ are the parameters of the system of equations, $ R_t $ is a logarithmic rate of return of the underlying instrument, $ r $ is part of the drift level of the returns process, $ h_t $ is the variance of returns, $ z_t $ is a random variable with a standard normal distribution and is the only random factor in the model. In his work Heston and Nandi showed that degenerate version of Heston model can also be obtained as the limit of the particular type of discrete GARCH processes. Heston-Nandi model can be called a discrete equivalent of the Heston model. On the other hand, if the time interval in the time series tends to 0 the results of Heston-Nandi model should be similar to the results of Heston model with parameters $\theta$ and $\kappa$ expressed by the parameters of the Heston-Nandi model in the following manner:

\begin{equation}\begin{split}
\theta & = \frac{1}{\delta t}\bigg(1-\beta-\alpha \gamma^2\bigg) ,\\
\kappa & = \frac{\xi}{\delta t (1-\beta-\alpha \gamma^2)}
,\end{split}\end{equation}

where $ \delta t $ is the interval of time expressed as a fraction of the year. $ \delta t = 1/252 $, when the model is estimated on on a daily basis. As shown by Christoffersen (2014)\footnote{~\bibentry{Christoffersen2014}.}, in the case of other parameters their limits are dependent on the variance of $ h_{t + 1} $ and covariance of $ h_{t + 1} $ with returns at $ t + 1 $ moment:

\begin{equation}\begin{split}
\mathrm{Cov_t}(R_{t+1},h_{t+1}) & = -2\alpha \gamma h_t ,\\
\mathrm{Var}_t(h_{t+1}) & = 2 \alpha^2 (1+2\gamma^2 h_t)
.\end{split}\end{equation}



Similar relationships have been reported previously for the GARCH diffusion (continuous time) stochastic model and GJR-GARCH discrete model which is asymmetric and was introduced by Glosten, Jaganatan and Runkle (1993)\footnote{~\bibentry{GJR1993}.}. Nelson (1990)\footnote{~\bibentry{Nelson1990}.} showed (using procedures of moment fitting) that the GJR-GARCH model has a continuous time limit in form of GARCH diffusion model. In this case, the GARCH diffusion parameters can all be expressed using GJR-GARCH model parameters when the time interval in time series tends to 0. This problem has been also studied by Drost and Nijman (1993)\footnote{~\bibentry{Drost1993}.} in more general context. 



\section{Conclusions}


In the first chapter an introduction to the currency option market and its features has been made. Then, Black-Scholes formulas for pricing European vanilla options were derived. Besides, following methods of option valuation were summarized, namely, partial differential equations, integrals of the characteristic function of the underlying instrument price distribution and the Monte Carlo simulations. Option valuation methods containing the integral of the characteristic function of the underlying instrument price distribution have been discussed in more details.

Moreover, as shown in the first part of this chapter (Section \ref{sec:1.2}), currency pair with the highest turnover and the importance in the foreign exchange market is the EURUSD pair. This justifies empirical tests on models for pricing options on this pair. This is similar to the approach commonly used for the stock market, for which stock indices options pricing models are examined primarily for the S\&P500 index.

\clearpage{\pagestyle{empty}\cleardoublepage}

\chapter{Characteristic functions of price distribution in stochastic volatility models}

\section{Introduction}

In the previous chapter the option valuation methods containing the integral of the characteristic function of the underlying instrument price distribution have been thoroughly discussed. This method is particularly important, since many of stochastic volatility models, including many models from the exponential-affine class, have an analytic solution for the characteristic function.

In the first part of this chapter, the equation of replication strategy dynamics for the author's proposed general form of diffusion stochastic volatility models is derived. This equation is used as a basis for a comparison of different models of stochastic volatility in terms of possibilities for further transformation of the equation. Next, its special cases for the Heston and the Sch\"obel-Zhu model are analysed and summaries of analytic formulas for the characteristic functions of the underlying instrument price distribution for those models are presented.

Then, maintaining the generality of considerations that allows a better understanding of the analytic formulas derivation problem the author is extending the considered general form of stochastic volatility diffusion models to models with N stochastic factors in price process and N stochastic factors in variance or volatility process. The considerations start from the derivation of the general form of the replication strategy dynamics equation for those models.Author is also introducing the new model, namely the OUOU model, for which the derivation of analytic formula for the characteristic function is then presented. At the end are also summarized existing models from different class, which have two stochastic factors in variance or volatility process and one stochastic factor in price process.
 
\section{Replication strategy dynamics equation}

Let the following general model be a subject of considerations:

\begin{equation}\begin{split}
& \D S_t = \mu S_t \D t + \nu_t^r S_t^\beta \D W_t ^S ,\\
& \D \nu_t = \kappa \nu^q (\theta - h(\nu_t)) \D t + \omega \nu_t^p \D W_t^\nu ,\\
& \langle \D W_t^S , \D W_t^\nu \rangle = \rho \D t , \quad r \in \{\tfrac{1}{2},1\} ,\\
& h(\nu) = \left\{
  \begin{array}{lr}
    \frac{1}{\lambda}(\nu^\lambda-1) & : \lambda > 0\\
    log(\nu) & : \lambda = 0
  \end{array}
\right.
.\end{split}\end{equation}

Option replication strategy in the stochastic volatility model with one-factor variance consists of one asset more than the Black-Scholes option replication strategy. This is a consequence of the introduction of a second Wiener process to the model, i.e. $ W_t^\nu $, in the  considered general form of stochastic volatility models, while in the Black-Scholes model there is only one Wiener process, $ W_t^S $. The portfolio in the strategy consists of one option that is worth $ V_t (S_t \nu_t) $, a certain amount of the underlying asset, that is priced $ S_t $, and a certain amount of a second derivative, that is worth $ U_t (S_t \nu_t) $, to hedge the risk of the variance process. The value of the portfolio and changes of its value, due to the assumption of a self-financing, are as follows:

\begin{equation} \Pi_t = V_t(S_t,\nu_t) - \Delta_t S_t B_t^f - \Xi_t U_t(S_t,\nu_t) ,\end{equation}
\begin{equation} \D \Pi_t = \D V_t(S_t,\nu_t) - \Delta_t \D (S_t B_t^f) - \Xi_t \D U_t(S_t,\nu_t) ,\end{equation}

where $ \Delta_t $ and $ \Xi_t $ are certain parameters corresponding to proportions in the portfolio.

Next, to simplify calculations, we can introduce new variable, $Y_t = V_t - \Xi_t U_t$, then the equation for the dynamics is:

\begin{equation} \D \Pi_t = \D Y_t(S_t,\nu_t) - \Delta_t \D (S_t B_t^f) .\end{equation}

Starting from the considered formula for $\Pi_t$ we can follow the same procedure as in the case of the original Black-Scholes equation, along with the assumption that the risk-neutral drift equals $ \mu = r_d-r_f $, to obtain a new formula after expanding $ \Delta_t \D (S_t B_t^f) $ and reducing it:

\begin{equation} \D \Pi_t = \D Y_t(S_t,\nu_t) - \Delta_t B_t^f (r_d S_t \D t + \nu_t^r S_t^\beta \D W_t^S) .\end{equation}

Then for the instrument $Y_t(S_t,\nu_t)$ (with simplified notation), just as in the approach of the Black-Scholes, It\^{o} Lemma can be applied to expand the component $\D Y_t(S_t,\nu_t)$ and then inserted it into the $ S_t $ and $ \nu_t $ dynamics equations in risk-neutral measure, which results in following formula in a simplified notation $(S=S_t,\nu=\nu_t,Y=Y_t,V=V_t,U=U_t,\Delta=\Delta_t,\Xi=\Xi_t)$:

\begin{equation} \D Y = \frac{\partial Y}{\partial t} \D t + \frac{\partial Y}{\partial S} \D S + \frac{\partial Y}{\partial \nu} \D \nu + \frac{1}{2} \frac{\partial^2 Y}{\partial S^2}(\D S)^2 + \frac{\partial^2 Y}{\partial S \partial \nu}(\D S \D \nu) + \frac{1}{2}\frac{\partial^2 Y}{\partial \nu^2}(\D \nu)^2 .\end{equation}

Quadratic variances and covariances included in the equation can be expanded to:

\begin{equation}\begin{split}
(\D S)^2 & = \D \langle S\rangle  = \nu^{2r} S^{2\beta} \D \langle W^S\rangle  = \nu^{2r} S^{2\beta} \D t ,\\
(\D S \D \nu) & = \D \langle S,\nu\rangle  = \nu^{r+p} S^\beta \omega \D \langle W^S,W^\nu\rangle  = \nu^{r+p} S^\beta \omega \rho \D t ,\\
(\D \nu)^2 & = \D \langle \nu\rangle  = \omega^2 \nu^{2p} \D \langle W^\nu\rangle  = \omega^2 \nu^{2p} \D t
.\end{split}\end{equation}

On the other hand, since quadratic variances and quadratic covariances with the time factor (i.e. $\D \langle t\rangle $ $\D \langle t,W^\nu\rangle $, $\D \langle t,W^S\rangle $) are always equal to 0, these components can be omitted. Therefore we can simplify equation as follows by inserting formulas of $\D S$ and $\D \nu$ from their definition:

\begin{equation}\begin{split}
\D Y = & \frac{\partial Y}{\partial t} \D t + \frac{\partial Y}{\partial S} \D S + \frac{\partial Y}{\partial \nu} \D \nu + \frac{1}{2} \nu^{2r} S^{2\beta} \frac{\partial^2 Y}{\partial S^2} \D t + \nu^{r+p} S^\beta \omega \rho \frac{\partial^2 Y}{\partial S \partial \nu} \D t + \frac{1}{2} \omega^2 \nu^{2p} \frac{\partial^2 Y}{\partial \nu^2} \D t \\
= & \bigg(\frac{\partial Y}{\partial t} + \frac{1}{2} \nu^{2r} S^{2\beta} \frac{\partial^2 Y}{\partial S^2}+ \nu^{r+p} S^\beta \omega \rho \frac{\partial^2 Y}{\partial S \partial \nu} + \frac{1}{2} \omega^2 \nu^{2p} \frac{\partial^2 Y}{\partial \nu^2} \bigg) \D t + \frac{\partial Y}{\partial S} \D S + \frac{\partial Y}{\partial \nu} \D \nu \\
= & \bigg(\frac{\partial Y}{\partial t} + \frac{1}{2} \nu^{2r} S^{2\beta} \frac{\partial^2 Y}{\partial S^2} + \nu^{r+p} S^\beta \omega \rho \frac{\partial^2 Y}{\partial S \partial \nu} + \frac{1}{2} \omega^2 \nu^{2p} \frac{\partial^2 Y}{\partial \nu^2} + (r_d-r_f) S \frac{\partial Y}{\partial S} + \kappa\nu^q(\theta-h(\nu)) \frac{\partial Y}{\partial \nu} \bigg) \D t \\
& + \nu^r S^\beta \frac{\partial Y}{\partial S} \D W^S + \omega \nu^p \frac{\partial Y}{\partial \nu} \D W^\nu
,\end{split}\end{equation}

further after inserting $ \D Y $ to the general equation, it can be written that:

\begin{equation}\begin{split}
\D \Pi = & \bigg(\frac{\partial Y}{\partial t} + \frac{1}{2} \nu^{2r} S^{2\beta} \frac{\partial^2 Y}{\partial S^2} + \nu^{r+p} S^\beta \omega \rho \frac{\partial^2 Y}{\partial S \partial \nu} +  \frac{1}{2} \omega^2 \nu^{2p} \frac{\partial^2 Y}{\partial \nu^2} + (r_d-r_f) S \frac{\partial Y}{\partial S} + \kappa\nu^q(\theta-h(\nu)) \frac{\partial Y}{\partial \nu} \bigg) \D t \\
& + \nu^r S^\beta \frac{\partial Y}{\partial S} \D W^S + \omega \nu^p \frac{\partial Y}{\partial \nu} \D W^\nu - \Delta B^f (r_d S \D t + \nu^r S^\beta \D W^S) \\
= & \bigg(\frac{\partial Y}{\partial t} + \frac{1}{2} \nu^{2r} S^{2\beta} \frac{\partial^2 Y}{\partial S^2}+ \nu^{r+p} S^\beta \omega \rho \frac{\partial^2 Y}{\partial S \partial \nu} +  \frac{1}{2} \omega^2 \nu^{2p} \frac{\partial^2 Y}{\partial \nu^2} + (r_d-r_f) S \frac{\partial Y}{\partial S} + \kappa\nu^q(\theta-h(\nu)) \frac{\partial Y}{\partial \nu} - \Delta B^f r_d S \bigg) \D t \\
& + \bigg(\frac{\partial Y}{\partial S} - \Delta B^f \bigg) \nu^r S^\beta \D W^S + \omega \nu^p \frac{\partial Y}{\partial \nu} \D W^\nu 
.\end{split}\end{equation}

Then, the parameter $\Xi$ should be set to the value, which makes $\frac{\partial Y}{\partial \nu} = 0$ and offsets the impact of $\D W^\nu$ on the value of the portfolio.

\begin{equation} \frac{\partial Y}{\partial \nu} = \frac{\partial (V - \Xi U)}{\partial \nu} 
= \frac{\partial V}{\partial \nu} - \Xi \frac{\partial U}{\partial \nu} = 0 .\end{equation}

The same should be done for the parameter $ \Delta $ to make $(\frac{\partial Y}{\partial S} - \Delta B^f ) = 0$ and offset the impact of $\D W^S$ on the value of the portfolio. Requested impact will be achieved by following levels of the parameters:

\begin{equation} \Xi = \frac{\partial V}{\partial \nu} \bigg/ \frac{\partial U}{\partial \nu} ,\end{equation}
\begin{equation} \Delta = \frac{\partial Y}{\partial S} \bigg/ B^f = (V - \Xi U) / B^f = V / B^f - \frac{\partial V}{\partial \nu} \bigg/ \frac{\partial U}{\partial \nu} U / B^f ,\end{equation}

and taking into account those values of the parameters, the main equation simplifies to:

\begin{equation}
\D \Pi = \bigg(\frac{\partial Y}{\partial t} + \frac{1}{2} \nu^{2r} S^{2\beta} \frac{\partial^2 Y}{\partial S^2} + \nu^{r+p} S^\beta \omega \rho \frac{\partial^2 Y}{\partial S \partial \nu} + \frac{1}{2} \omega^2 \nu^{2p} \frac{\partial^2 Y}{\partial \nu^2} + (r_d-r_f) S \frac{\partial Y}{\partial S} + \kappa\nu^q(\theta-h(\nu)) \frac{\partial Y}{\partial \nu} - \Delta S B^f r_d \bigg) \D t 
.\end{equation}

On the other hand, the $ \Pi $ portfolio is without risk, so its value must grow only by domestic risk-free rate $ r_d $, i.e. $\D \Pi = r_d \Pi \D t = r_d (Y - \Delta S B^f) \D t$. As a result, after reduction of both sides by $ \D t $ there is:

\begin{equation}\begin{split}
& \frac{\partial Y}{\partial t} + \frac{1}{2} \nu^{2r} S^{2\beta} \frac{\partial^2 Y}{\partial S^2} + \nu^{r+p} S^\beta \omega \rho \frac{\partial^2 Y}{\partial S \partial \nu} + \frac{1}{2} \omega^2 \nu^{2p} \frac{\partial^2 Y}{\partial \nu^2} + (r_d-r_f) S \frac{\partial Y}{\partial S} \\
& + \kappa\nu^q(\theta-h(\nu)) \frac{\partial Y}{\partial \nu} - \Delta S B^f r_d = r_d (Y - \Delta S B^f)
.\end{split}\end{equation}

Both sides of equation have same parts with the factor $ \Delta $ and therefore it can be omitted, leading to the equation:

\begin{equation}
\frac{\partial Y}{\partial t} + \frac{1}{2} \nu^{2r} S^{2\beta} \frac{\partial^2 Y}{\partial S^2} + \nu^{r+p} S^\beta \omega \rho \frac{\partial^2 Y}{\partial S \partial \nu} + \frac{1}{2} \omega^2 \nu^{2p} \frac{\partial^2 Y}{\partial \nu^2} + (r_d-r_f) S \frac{\partial Y}{\partial S} + \kappa\nu^q(\theta-h(\nu)) \frac{\partial Y}{\partial \nu} - r_d Y = 0\\
.\end{equation}

The left hand side of this equation can be written as some differentiation operator $\mathcal{L}(Y)$, then using the accepted definition of $ Y $ ($Y = V - \Xi U = V - \frac{\partial V}{\partial \nu} / \frac{\partial U}{\partial \nu} U$) we can rewrite the equation in a form which includes instruments $ V $ and $ U $:

\begin{equation}
\mathcal{L}(Y) = \mathcal{L}(V) - \Xi \mathcal{L}(U) = 0
,\end{equation}

and after reordering:

\begin{equation}
\mathcal{L}(V) \bigg/ \frac{\partial V}{\partial \nu} = \mathcal{L}(U) \bigg/ \frac{\partial U}{\partial \nu}
.\end{equation}

This equation must hold independently for all instruments $ V $ and $ U $, therefore it is assumed that the right hand side is equal to a some function $\eta(S,\nu,t)$, which can be interpreted as the price of volatility risk. Heston assumed that this function is proportional to $ \nu $ and in the form of $\eta(S,\nu,t) = \eta \nu$, where $ \eta $ is a constant. Therefore, after substitution of $ \mathcal{L}(V) $ on the left hand side we obtain:

\begin{equation}\begin{split}
& \bigg(\frac{\partial V}{\partial t} + \frac{1}{2} \nu^{2r} S^{2\beta} \frac{\partial^2 V}{\partial S^2} + \nu^{r+p} S^\beta \omega \rho \frac{\partial^2 V}{\partial S \partial \nu} + \frac{1}{2} \omega^2 \nu^{2p} \frac{\partial^2 V}{\partial \nu^2} + (r_d-r_f) S \frac{\partial V}{\partial S} \\
& + \kappa\nu^q(\theta-h(\nu)) \frac{\partial V}{\partial \nu} - r_d V \bigg) \bigg/ \frac{\partial V}{\partial \nu} = \eta(S,\nu,t)
.\end{split}\end{equation}

After reordering we finally obtain the replication strategy dynamics equation, which is equivalent to the original equation of Heston (1993) with following parametrisation: $r=p=\tfrac{1}{2}$, $q=0$, $\beta=1$ and $\lambda=1$, namely:

\begin{equation}\begin{split}
& \frac{\partial V}{\partial t} + \frac{1}{2} \nu^{2r} S^{2\beta} \frac{\partial^2 V}{\partial S^2} + \nu^{r+p} S^\beta \omega \rho \frac{\partial^2 V}{\partial S \partial \nu} + \frac{1}{2} \omega^2 \nu^{2p} \frac{\partial^2 V}{\partial \nu^2} + (r_d-r_f) S \frac{\partial V}{\partial S} \\
& + [\kappa\nu^q(\theta-h(\nu))-\eta(S,\nu,t)] \frac{\partial V}{\partial \nu} - r_d V = 0
,\end{split}\end{equation}

where the component, $\kappa\nu^q(\theta-h(\nu)) - \eta$, in the considered equation is called the risk-neutral drift rate. In addition, it is worth noting that for models in which $ \beta = 1 $ (e.g. the Heston model) differential equation of replication strategy dynamics simplifies after the substitution of the logarithms of the underlying instrument prices in place of the underlying instrument prices, i.e.  $x=\ln(S)$ or $S=e^x$, because:

\begin{equation}\begin{split}
\frac{\partial V}{\partial S} & = \frac{\partial V}{\partial x} \frac{\partial x}{\partial S} = \frac{1}{S} \frac{\partial V}{\partial x} =
e^{-x} \frac{\partial V}{\partial x} ,\\
\frac{\partial^2 V}{\partial S^2} & = \frac{\partial}{\partial x} \bigg( \frac{\partial V}{\partial S} \bigg) \frac{\partial x}{\partial S} = 
\frac{\partial}{\partial x} \bigg( e^{-x} \frac{\partial V}{\partial x} \bigg) \frac{\partial x}{\partial S} = 
\bigg( e^{-x} \frac{\partial^2 V}{\partial x^2} - e^{-x} \frac{\partial V}{\partial x} \bigg) \frac{\partial x}{\partial S} \\
& = \bigg( \frac{1}{S} \frac{\partial^2 V}{\partial x^2} - \frac{1}{S} \frac{\partial V}{\partial x} \bigg) \frac{1}{S} = 
\frac{1}{S^2} \frac{\partial^2 V}{\partial x^2} - \frac{1}{S^2} \frac{\partial V}{\partial x} ,\\
\frac{\partial^2 V}{\partial S \partial \nu} & = \frac{\partial}{\partial x} \bigg( \frac{\partial V}{\partial \nu} \bigg) \frac{\partial x}{\partial S} = \frac{1}{S} \frac{\partial^2 V}{\partial x \partial \nu}
,\end{split}\end{equation}

which after substitution in the main equation lead to:

\begin{equation}\begin{split}
\label{eq:repstratdyn}
& \frac{\partial V}{\partial t}
+ \frac{1}{2} \nu^{2r} \frac{\partial^2 V}{\partial x^2}
+ \nu^{r+p} \omega \rho \frac{\partial^2 V}{\partial x \partial \nu} 
+ \frac{1}{2} \omega^2 \nu^{2p} \frac{\partial^2 V}{\partial \nu^2}
+ \bigg(r_d-r_f - \frac{1}{2} \nu^{2r} \bigg) \frac{\partial V}{\partial x} \\
& + [\kappa\nu^q(\theta-h(\nu))-\eta(S,\nu,t)] \frac{\partial V}{\partial \nu}
- r_d V = 0
.\end{split}\end{equation}

In the equation, reduction of the element associated with the underlying instrument price helps in further search of the characteristic function solution.

\section{General form of the characteristic function}
\label{sec:2.3}

In general, the dynamics of the underlying instrument price and the dynamics of their volatility given by general equation from the previous sections, the price of vanilla call option should have a formula similar to the Black-Scholes formula (1973), which for the currency options can be written as:

\begin{equation}
V_t = S_t e^{-r_f (T-t)} P_1 (S,\nu,t) - K e^{-r_d (T-t)} P_2 (S,\nu,t)
\end{equation}

or using logarithms of the underlying instrument price:

\begin{equation}
V_t = e^{x_t-r_f (T-t)} P_1 (x,\nu,t) - K e^{-r_d (T-t)} P_2 (x,\nu,t)
.\end{equation}

However, unlike the Black-Scholes model (1973), the probability of $ P_1 $ and $ P_2 $ are generally different under stochastic volatility assumption and cannot be reduced to the normal distribution CDF. This is leading to other distribution than the normal distribution, present in the assumptions of Black-Scholes (1973 model.

Knowledge of the explicit formula for the function $V_t$ allows computing of the partial derivatives of this function, which are present in the differential equation of replication strategy dynamics (\ref{eq:repstratdyn}):

\begin{equation}\begin{split}
\frac{\partial V}{\partial x} & = e^{x_t-r_f (T-t)} P_1 + e^{x_t-r_f (T-t)} \frac{\partial P_1}{\partial x} - e^{-r_d(T-t)} K \frac{\partial P_2}{\partial x} ,\\
\frac{\partial^2 V}{\partial x^2} & = e^{x_t-r_f (T-t)} P_1 + 2 e^{x_t-r_f (T-t)} \frac{\partial P_1}{\partial x} + e^{x_t-r_f (T-t)} \frac{\partial^2 P_1}{\partial x^2}  - e^{-r_d(T-t)} K \frac{\partial^2 P_2}{\partial x^2} ,\\
\frac{\partial^2 V}{\partial x \partial \nu} & = e^{x_t-r_f (T-t)} \frac{\partial P_1}{\partial \nu} + e^{x_t-r_f (T-t)} \frac{\partial^2 P_1}{\partial x \partial \nu} - e^{-r_d(T-t)} K \frac{\partial^2 P_2}{\partial x \partial \nu} ,\\
\frac{\partial V}{\partial \nu} & = e^{x_t-r_f (T-t)} \frac{\partial P_1}{\partial \nu} - e^{-r_d(T-t)} K \frac{\partial P_2}{\partial \nu} ,\\
\frac{\partial^2 V}{\partial \nu^2} & = e^{x_t-r_f (T-t)} \frac{\partial^2 P_1}{\partial \nu^2}  - e^{-r_d(T-t)} K \frac{\partial^2 P_2}{\partial \nu^2} ,\\
\frac{\partial V}{\partial t} & = r_f e^{x_t-r_f (T-t)} P_1 + e^{x_t-r_f (T-t)} \frac{\partial P_1}{\partial t} - r_d e^{-r_d(T-t)} K P_2 - e^{-r_d(T-t)} K \frac{\partial P_2}{\partial t}
.\end{split}\end{equation}

Then these derivatives can be substituted in the equation along with the formula for $ V(P_1, P_2)$ in place of the $r_d V$ component. Next, after rearrangement of equation in a way that left hand side resembles again a formula similar to the Black-Scholes (1973) and the right hand side is equal to 0, following order is obtained:

\begin{equation}
e^{x_t-r_f (T-t)} f(P_1( S,\nu,t)) - K e^{-r_d (T-t)} f(P_2 (S,\nu,t)) = 0
,\end{equation}

where functions $f(P_1)$ and $f(P_2)$ are polynomial operators of differentiation with respect to the cumulative distribution functions $ P_1 $ and $ P_2 $, and in non-simplified form, while maintaining the simplified notation, i.e. $\nu = \nu_t$, $x = x_t$, are equal to:

\begin{equation}\begin{split}
\label{eq:fP1fP2}
f(P_1) & = \frac{1}{2} \nu^{2r} \bigg(P_1 + 2 \frac{\partial P_1}{\partial x} + \frac{\partial^2 P_1}{\partial x^2} \bigg) + \rho \omega \nu^{r+p} \bigg(\frac{\partial P_1}{\partial \nu} + \frac{\partial^2 P_1}{\partial x \partial \nu}\bigg) + \frac{1}{2} \omega^2 \nu^{2p} \frac{\partial^2 P_1}{\partial \nu^2} \\
& + \bigg(r_d - r_f + \frac{1}{2} \nu^{2r} \bigg) \bigg(P_1 + \frac{\partial P_1}{\partial x} \bigg) + [\kappa\nu^q(\theta-h(\nu))-\eta_{S,\nu,t} + \rho \omega \nu^{r+p}] \frac{\partial P_1}{\partial \nu} - r_d P_1 + \bigg( r_f P_1 + \frac{\partial P_1}{\partial t} \bigg) ,\\
f(P_2) & = \frac{1}{2} \nu^{2r} \bigg(\frac{\partial^2 P_2}{\partial x^2} \bigg) + \rho \omega \nu^{r+p} \bigg( \frac{\partial^2 P_2}{\partial x \partial \nu} \bigg) + \frac{1}{2} \omega^2 \nu^{2p} \frac{\partial^2 P_2}{\partial \nu^2} \\
& + \bigg(r_d - r_f - \frac{1}{2} \nu^{2r} \bigg) \bigg(\frac{\partial P_2}{\partial x} \bigg) + [\kappa\nu^q(\theta-h(\nu))-\eta_{S,\nu,t}] \frac{\partial P_2}{\partial \nu} - r_d P_2 + \bigg(r_d P_2 + \frac{\partial P_2}{\partial t} \bigg)
.\end{split}\end{equation}

Components $r_d P_1$, $r_f P_1$ and $r_d P_2$ are present in the equations twice: once with a positive sign and once with a negative, so that they are not present after the simplification. The solution of equation (\ref{eq:fP1fP2}) is $f(P_1) = f(P_2) = 0 $, so after the simplification and reduction to the form of a single equation with parameters $ a_j $ and $ c_j $, which are dependent on other parameters of the model, there is:

\begin{equation}\begin{split}
f(P_j) = \frac{\partial P_j}{\partial t} + \frac{1}{2} \nu^{2r} \frac{\partial^2 P_j}{\partial x^2} + \rho \omega \nu^{r+p} \frac{\partial^2 P_j}{\partial x \partial \nu} + & \frac{1}{2} \omega^2 \nu^{2p} \frac{\partial^2 P_j}{\partial \nu^2} + \bigg(r_d - r_f + a_j \nu^{2r} \bigg) \frac{\partial P_j}{\partial x} \\
& + [\kappa\nu^q(\theta-h(\nu))-\eta_{S,\nu,t} + c_j \nu^{r+p}] \frac{\partial P_j}{\partial \nu} = 0
,\end{split}\end{equation}
$$\mathrm{where:} \quad j=1,2, \quad a_1=\frac{1}{2}, \quad a_2=-\frac{1}{2}, \quad c_1 = \rho \omega, \quad c_2 = 0.$$

In the case when $q=0$, $\lambda=1$ and $\eta_{S,\nu,t}=\eta \nu$ (and same as before $ \beta = 1 $) previous formula simplifies to:

\begin{equation}\begin{split}
f(P_j) = \frac{\partial P_j}{\partial t} + \frac{1}{2} \nu^{2r} \frac{\partial^2 P_j}{\partial x^2} + \rho \omega \nu^{r+p} \frac{\partial^2 P_j}{\partial x \partial \nu} + & \frac{1}{2} \omega^2 \nu^{2p} \frac{\partial^2 P_j}{\partial \nu^2} + \bigg(r_d - r_f + a_j \nu^{2r} \bigg) \frac{\partial P_j}{\partial x} \\
& + [\kappa \hat{\theta} - (\kappa + \eta) \nu + c_j \nu^{r+p}] \frac{\partial P_j}{\partial \nu} = 0
,\end{split}\end{equation}
$$\mathrm{where:} \quad j=1,2, \quad a_1=\frac{1}{2}, \quad a_2=-\frac{1}{2}, \quad c_1 = \rho \omega, \quad c_2 = 0 .$$

A further simplification would be to introduce a requirement for $ r + p = 1 $, then parameter $c_j$ could be converted to another one, $ b_j $, to get the generalized equation from the Heston paper:

\begin{equation}\begin{split}
\label{eq:HestonPj}
&\frac{\partial P_j}{\partial t} + \frac{1}{2} \nu^{2r} \frac{\partial^2 P_j}{\partial x^2} + \rho \omega \nu \frac{\partial^2 P_j}{\partial x \partial \nu} + \frac{1}{2} \omega^2 \nu^{2-2r} \frac{\partial^2 P_j}{\partial \nu^2} + \bigg(r_d - r_f + a_j \nu^{2r} \bigg) \frac{\partial P_j}{\partial x} + [\kappa \hat{\theta} - b_j \nu] \frac{\partial P_j}{\partial \nu} = 0 ,\\
&\mathrm{where:} \quad j=1,2, \quad a_1=\frac{1}{2}, \quad a_2=-\frac{1}{2}, \quad b_1 = \kappa + \eta - \omega \rho, \quad b_2 = \kappa + \eta
.\end{split}\end{equation}

Heston (1993) has showed, using Feynman-Kac theorem \footnote{~\bibentry{Kac1949}.}, that characteristic functions of a random variable $x=\ln(S)$ ($\phi_j(u,x,\nu,\tau)$, $j=1,2$) must satisfy the same partial differential equations, which are satisfied by cumulative distribution functions of this variable ($ P_j $) with the boundary condition in the form of:

\begin{equation} \phi_j(x,\nu,0,u) = e^{i u x_0} .\end{equation}

Therefore, to calculate the cumulative distribution functions $ P_1 $ and $ P_2 $ following steps should be performed. First, some assumption about the form of the characteristic function should be made (e.g. exponential-affine $\phi_j(u,x,\nu,\tau) = e^{i u x_0 + \alpha_{j,\tau} + \mathbf{\beta}_{j,\tau}^T \mathbf{z}}$, where $\alpha_{j,\tau}$ are scalars, $\mathbf{\beta}_{j,\tau}$ and $\mathbf{z}$ are vectors and holds $\alpha(\tau=0)=0$ and $\mathbf{\beta}(\tau=0)=\mathbf{0}$). Then, all of partial derivatives of $ \phi_j $ have to be computed and put into the equation (\ref{eq:HestonPj}) in place of the partial derivatives of $ \mathrm{P}_j $ probabilities, then finally the resulting system of equations have to be solved to find considered elements of characteristic function ($\alpha_{j,\tau}$ and $\mathbf{\beta}_{j,\tau}$), which satisfy this system of equations.

In the case of the Heston model the expected form of the characteristic function and the boundary conditions are:

\begin{equation}\begin{split}
& \phi_j(u,x,\nu,\tau) = e^{i u x_0 + A_{j,\tau} + B_{j,\tau} \nu_0} ,\\
& A(\tau=0)=B(\tau=0)=0
.\end{split}\end{equation}

In case of models with $ r = 1 $, which are models with the second equation representing the dynamics of volatility and not the variance (i.e. $ \nu $ is the volatility in this equation, and $ \nu^2 $ is the variance), considered assumptions should be extended to obtain the analytic solution. Using the relationship between volatility and variance and the It\^{o} lemma, additional auxiliary equation of the dynamics of the variance can be introduced to the model of stochastic volatility. This way the logarithm of the characteristic function can be presented as an affine with respect to the vector [$\ln(S)$,$\nu$,$\nu^2$]. Therefore such a model is also in the class of affine models. Then, the projected formula for characteristic function and the boundary conditions are:

\begin{equation}\begin{split}
& \phi_j(u,x,\nu,\tau) = e^{i u x_0 + A_{j,\tau} + B_{j,\tau} \nu_0 + C_{j,\tau} \nu^2_0  } ,\\
& A(\tau=0)=B(\tau=0)=C(\tau=0)=0
.\end{split}\end{equation}

Next, using the simplified notation ($\phi = \phi_j$, $A = A_{j,\tau}$, $B = B_{j,\tau}$, $C = C_{j,\tau}$), partial derivatives of $\phi_j$, which occur in the main equation are as follows:

\begin{equation}\begin{split}
&\frac{\partial \phi}{\partial x} = i u \phi, \quad \frac{\partial^2 \phi}{\partial x^2} = - u^2 \phi, \quad \frac{\partial \phi}{\partial \nu} = (B + 2 \nu C) \phi ,\\
&\frac{\partial^2 \phi}{\partial x \nu} =  (B + 2 \nu C) i u \phi, \quad \frac{\partial \phi}{\partial t} = \bigg(\frac{\partial A}{\partial t} + \nu \frac{\partial B}{\partial t} + \nu^2 \frac{\partial C}{\partial t} \bigg) \phi ,\\
& \frac{\partial^2 \phi}{\partial \nu^2} = B(B + 2 \nu C) \phi + 2 C (\phi + \nu \phi (B + 2 \nu C)) = \phi (B^2 + 4 C B \nu + 2 C + 4 C^2 \nu^2)
.\end{split}\end{equation}

After insertion of the partial derivatives of characteristic function in place of the partial derivatives of functions $\mathrm{P}_j$ in equation (\ref{eq:HestonPj}) (i.e., without the introduction of restriction $ r + p = 1 $) there is:

\begin{equation}\begin{split}
& \bigg( \frac{\partial A}{\partial t} + \nu \frac{\partial B}{\partial t} + \nu^2 \frac{\partial C}{\partial t} \bigg) \phi - \frac{1}{2} \nu^{2r} u^2 \phi + \bigg(r_d - r_f + a_j \nu^{2r} \bigg) i u \phi + \rho \omega \nu^{r+p} i u (B + 2 \nu C) \phi \\
& + \frac{1}{2} \omega^2 \nu^{2p} (B^2 + 4 C B \nu + 2 C + 4 C^2 \nu^2) \phi + [\kappa \hat{\theta} - (\kappa + \eta) \nu + c_j \nu^{r+p}] (B + 2 \nu C) \phi = 0
,\end{split}\end{equation}

and after reordering and dividing by $ \phi $:

\begin{equation}\begin{split}
\label{eq:genDynEq}
& \frac{\partial A}{\partial t} + (r_d - r_f) i u + B \kappa \hat{\theta} + \nu \bigg( \frac{\partial B}{\partial t} - B (\kappa+\eta) + 2C \kappa \hat{\theta} \bigg) + \nu^2 \bigg( \frac{\partial C}{\partial t} - 2C (\kappa+\eta) \bigg) \\
& + \nu^{2r} \bigg( a_j i u - \frac{1}{2} u^2 \bigg) + \nu^{2p} \bigg( \frac{1}{2} \omega^2 B^2 + \omega^2 C \bigg) + \nu^{2p+1} \bigg( \frac{1}{2} \omega^2 4 C B \bigg) + \nu^{2p+2} \bigg( \frac{1}{2} \omega^2 4 C^2 \bigg) \\
& + \nu^{r+p} \bigg( \rho \omega i u B + B c_j \bigg) + \nu^{r+p+1} \bigg( \rho \omega i u 2 C + 2C c_j \bigg) = 0
.\end{split}\end{equation}

Taking into consideration that $ \nu $ may not be equal to 0,  considered equation can be represented as a system of equations created on the basis of the factors in parentheses standing by successive powers of $ \nu $. It can be seen from this equation, that without further restrictions on the possible values of the parameter $ p $ and $ r $, the solution of the equation imposes restrictions with respect to other model parameters, or the elements $ B $ and $ C $ of the characteristic function. One possible limitation is the aforementioned condition $ r + p = 1 $, which is present in the Heston and the Sch\"obel-Zhu model. After including it there is:

\begin{equation}\begin{split}
& \frac{\partial A}{\partial t} + (r_d - r_f) i u + B \kappa \hat{\theta} + \nu \bigg( \frac{\partial B}{\partial t} - B b_j + 2C \kappa \hat{\theta} + \rho \omega i u B \bigg) \\
& + \nu^2 \bigg( \frac{\partial C}{\partial t} - 2C b_j + \rho \omega i u 2 C \bigg) \\
& + \nu^{2-2p} \bigg( a_j i u - \frac{1}{2} u^2 \bigg) + \nu^{2p} \bigg( \frac{1}{2} \omega^2 B^2 + \omega^2 C \bigg) + \nu^{2p+1} \bigg( \frac{1}{2} \omega^2 4 C B \bigg) + \nu^{2p+2} \bigg( \frac{1}{2} \omega^2 4 C^2 \bigg) = 0
.\end{split}\end{equation}

In such parametrisation there is a new parameter, $b_j=(\kappa+\eta)-c_j$. Solutions of this equation are presented in the following sections.

\section{Heston model}
\subsection{Model assumptions}

In the Heston model the price of the underlying $ S_t $ follows log-normal stochastic process as in the Black-Scholes model. In contrast to the latter, the variance, $ \nu $, evolves with CIR (Cox-Ingersoll-Ross, Cox et al. (1985)\footnote{~\bibentry {Cox1985}.}) process, which was originally described in the literature by Feller (1951) \footnote{~\bibentry{Feller1951}.}. The advantage of this process is the fact that when $ \nu_0> 0 $ and so called Feller condition is fulfilled, then the variance $\nu_t$ remain strictly positive with probability 1 for every $ t $. The Heston model is described by following equations:

\begin{equation}\begin{split}
& \D S_t = (r_d-r_f) S_t \D t + \sqrt{\nu_t} S_t \D W_t ^S ,\\
& \D \nu_t = \kappa (\theta - \nu_t) \D t + \omega \sqrt{\nu_t} \D W_t^\nu ,\\
& \langle \D W_t^S , \D W_t^\nu \rangle = \rho \D t
.\end{split}\end{equation}

The parameters in the model have the following interpretation:
\begin{itemize}
\item $(r_d-r_f)$ is the level of the drift of the underlying instrument price process,
\item $\kappa$ is the speed of mean-reversion in the variance process,
\item $\theta$ is the long term mean of the variance process,
\item $\omega$ is the volatility of the variance process,
\item $\rho$ is the correlation between Wiener process of underlying instrument and Wiener process of the variance,
\item $\nu_0$ is the initial level of variance, i.e. at the time $t=0$.
\end{itemize}

In addition, as stated in introduction, an important feature of the Heston model is the Feller condition\footnote{~Ibid.}. When it is satisfied then the stochastic process of variance is strictly positive. This condition defines the behaviour of the equation in the boundary condition $ \nu = 0 $. In short, its meaning can be summarized by the fact that when it is not satisfied then paths of the variance evolution in the Monte Carlo simulations can go below zero. Feller condition is described by following inequality:

\begin{equation} \frac{2\kappa \theta}{\omega^2} > 1 .\end{equation}

There is no closed formula for $\nu_t$ for the Heston model, but integrated variance equation can be presented in the following way:

\begin{equation}
\nu_T = \theta + (\nu_t - \theta) e^{-\kappa (T-t)} + \omega \int_t^T e^{-\kappa(T-s)} \sqrt{\nu_s} \D \, W_s^{\nu}
\end{equation}

Feller showed\footnote{~Ibid.} that the transformed variable which follows the considered process has non-central chi-square distribution, namely:

\begin{equation}
2c \nu_T \sim \chi \bigg( \frac{4 \kappa \theta}{\omega^2}, 2c\nu_0 e^{-\kappa T}\bigg)
,\end{equation}

where $c=\tfrac{2\kappa}{\omega^2}(1-e^{-\kappa T})^{-1}$. This implies:

\begin{equation}
\mathbb{E}_t [\nu_T] = \theta + (\nu_t - \theta) e^{-\kappa (T-t)}
\end{equation}

and

\begin{equation}\begin{split}
\mathrm{Var}_t(\nu_T) & = \omega^2 \mathbb{E}_t \Big( \int_t^T \nu_s \D \,s \Big) + \kappa^2 \mathrm{Var}_t \Big( \int_t^T \nu_s \D \,s \Big) \\
& = \omega^2 \Big( \tfrac{\nu_t}{\kappa}(e^{-\kappa (T-t)}-e^{-2 \kappa (T-t)}) + \tfrac{\theta}{2 \kappa}(1-e^{-\kappa (T-t)})^2 \Big)
.\end{split}\end{equation}

\subsection{Analytic formula for characteristic function}


The general equation (\ref{eq:genDynEq}) from the previous section simplifies to the affine form with respect to $ \nu $ due to fulfilled restrictions: $ p = \tfrac{1}{2} $ and $ C = 0 $, because in the Heston model $ r = p = \tfrac{1}{2} $ and $ \nu $ is a variance so $ C = 0 $:

\begin{equation}\begin{split}
& \frac{\partial A}{\partial t} + (r_d - r_f) i u + B \kappa \hat{\theta}
+ \nu \bigg( \frac{\partial B}{\partial t} - \frac{1}{2} u^2 + \rho \omega i u B + \frac{1}{2} \omega^2 B^2 + a_j i u - b_j B \bigg) = 0
.\end{split}\end{equation}

Given that $ \nu $ may not be equal to 0, and after substitution of partial derivatives $\frac{\partial f}{\partial t} = - \frac{\partial f}{\partial \tau}$, which arises from relation $ \tau = T-t $, the equation can also be reformulated as following system of two equations:

\begin{equation}\begin{split}
\label{eq:hestonEqSystem}
& \frac{\partial A}{\partial \tau} = (r_d - r_f) i u + B \kappa \hat{\theta} ,\\
& \frac{\partial B}{\partial \tau} = a_j i u - \tfrac{1}{2} u^2 + (\rho \omega i u  - b_j ) B + \tfrac{1}{2} \omega^2 B^2
.\end{split}\end{equation}

The considered system of equations has the analytic solution for $A$ and $B$.\footnote{~The solution can be treated as special case of the Sch\"obel-Zhu system solution.} This solution implies the following characteristic function:

\begin{equation}\begin{split}
& \phi_H(u,x,\nu,\tau) = e^{i u x_0 + A + B \nu_0} ,\\
& A(\tau,u) = (r_d - r_f) i u \tau + \frac{\kappa \theta}{\omega^2} \bigg( (\beta_j + d_j) \tau - 2 \ln \frac{1-g_j e^{d_j \tau}}{1-g_j} \bigg) ,\\
& B(\tau,u) = \bigg(\frac{\beta_j + d_j}{\omega^2}\bigg) \bigg(\frac{1-e^{d_j \tau}}{1-g_j e^{d_j \tau}}\bigg)
,\end{split}\end{equation}

where:

\begin{equation}\begin{split}
& \beta_j = b_j - \rho \omega i u, \quad d_j = \sqrt{\beta_j^2 - \omega^2 (2 a_j i u - u^2)}, \quad g_j = \frac{\beta_j + d_j}{\beta_j - d_j} ,\\
& j=1,2, \quad a_1=\frac{1}{2}, \quad a_2=-\frac{1}{2}, \quad b_1 = \kappa + \eta - \omega \rho, \quad b_2 = \kappa + \eta
.\end{split}\end{equation}

This solution is an original solution from the Heston paper, part of this solution can also be presented in different way, in form derived by Duffie, Pan and Singleton (2000)\footnote{~\bibentry{Duffie2000}.} and Schoutens, Simons, Tistaert (2004)\footnote{~\bibentry{Schoutens2004}.} and precisely studied by Albrecher et al. (2007)\footnote{~\bibentry{Albrecher2007}.}:

\begin{equation}\begin{split}
& A(\tau,u) = (r_d - r_f) i u \tau + \frac{\kappa \theta}{\omega^2} \bigg( (\beta_j - d_j) \tau - 2 \ln \frac{1-G_j e^{-d_j \tau}}{1-G_j} \bigg) ,\\
& B(\tau,u) = \bigg(\frac{\beta_j - d_j}{\omega^2}\bigg) \bigg(\frac{1-e^{-d_j \tau}}{1-G_j e^{-d_j \tau}}\bigg)
.\end{split}\end{equation}

To obtain this result there has been made substitution of $G(u)= 1/g(u) $. For this formula, although it is algebraically equivalent to the previous one, there are no discontinuities in the element $A(\tau,u)$ caused by the operation of the logarithm in the complex plane.

\subsection{Bates jump-diffusion model}

An important extension of the Heston model, especially in the case of stock prices and credit instruments, was the Bates (1996)\footnote{~\bibentry{Bates1996}.} jump-diffusion model, which introduces jumps to the price process in the Heston model. Author increased the number of parameters in the Heston model to eight formulating the model of the form:

\begin{equation}\begin{split}
& \D S_t = (r_d-r_f) S_t \D t + \sqrt{\nu_t} S_t \D W_t ^S -\lambda \hat{k} S_t \D t + S_t k\D q ,\\
& \D \nu_t = \kappa (\theta - \nu_t) \D t + \omega \sqrt{\nu_t} \D W_t^\nu ,\\
& \langle \D W_t^S , \D W_t^\nu \rangle = \rho \D t ,\\
& \mathbb{P}(\D q=1) = \lambda \D t, \quad \ln(1+k) \sim \mathcal{N}(\ln(1+\hat{k}) - \tfrac{1}{2} \delta^2, \delta^2)
,\end{split}\end{equation}

where the parameters that are also present in the Heston model have the same interpretation, while $ \lambda $ is an annual frequency of jumps, $ k $ is a random percentage jump which is occurring only when a jump event occurs, $\hat{k}$ is a parameter influencing the mean of jump, $ \delta $ is the volatility of jump, and $ q $ is the Poisson counter with intensity $ \lambda $.

The assumption about jumps in the underlying instrument prices process is close to the reality of the stock market and the credit market, where significant jumps in the price of the instrument can occur after the publication of important data, e.g. financial report of some company. But this is not so often in the case of the FX market and the interest rate market, where due to the size of the market and a large number of depending factors, prices adjust in a more gradual way. Commodity market is in this qualitative aspect positioned somewhere between the previous two cases. The most important consequence of the introduction of jumps is in terms of option replication, i.e. the construction of the hedging strategy.

When discontinuous (jump) evolution of prices is present then option replication with delta hedging strategy cannot be used effectively, because continuous time trade in the underlying asset and the risk-free treasury bills is not possible. The adoption of such model results in the incompleteness of the market, since replication of options in such market is only possible when there is a continuum of options with different strike prices, and in fact it never is. This is one of the main reasons for which other concepts of stochastic volatility models are developed, which, despite greater overall complexity do not take into account the Poisson process generating spikes in the price of the underlying.

In addition, in this model, the negative skewness may be both due to a negative correlation between the underlying returns process and volatility process ($\rho<0$), as well as due to jumps with negative mean ($\hat{k}<0$). However, the influence of both parameters is different for options with different maturities, because jumps mainly affect options with short maturities, while the stochastic volatility more affects options with long maturities.

It should be added that the introduction of this last element to the Heston model only slightly complicates the formula for the characteristic function of the underlying instrument prices distribution. Namely, the characteristic function in Bates model is equal to the characteristic function of the Heston model multiplied by a factor corresponding to jumps: \par

\begin{equation}
\phi_B(u,x,\nu,\tau,\Theta,\lambda,\delta,\hat{k}) = \phi_H(u,x,\nu,\tau,\Theta) \cdot \exp(\lambda \tau (1+\hat{k})^{a_j+1/2} [(1+\hat{k})^u e^{\delta^2 (a_j u + u^2/2)}-1]-\lambda \hat{k} u \tau)
,\end{equation}

where $\Theta$ is vector of the Heston model parameters, $(\kappa,\theta,\omega,\rho)$.

\section{Sch\"obel-Zhu model}

\subsection{Model assumptions}

In the Sch\"obel-Zhu (1999)\footnote{~\bibentry{SchoebelZhu1999}.} model the price of the underlying $ S_t $ follows log-normal stochastic process as in the Black-Scholes model. In contrast to the Heston model, not only $ \nu $ is a volatility and not a variance, but also it evolves with an Ohrstein-Uhlenbeck process\footnote{~\bibentry{OU1930}.}, which is equivalent to the AR(1) process in a continuous time. OU process in this class of models was considered first by Wiggins (1987), with the difference that it was applied to the variable $ \ln (\nu_t) $. The Sch\"obel-Zhu model can be considered as a modification of the Stein and Stein (1991) model with a non-zero correlation. Unlike the Heston model, there is no requirement ensuring that the value of the volatility, $ \nu_t $, will remain strictly positive with probability 1 for every $ t $. The Sch\"obel-Zhu model is described by following equations:

\begin{equation}\begin{split}
& \D S_t = (r_d-r_f) S_t \D t + \nu_t S_t \D W_t ^S ,\\
& \D \nu_t = \kappa (\theta - \nu_t) \D t + \omega \D W_t^\nu ,\\
& \langle \D W_t^S , \D W_t^\nu \rangle = \rho \D t
.\end{split}\end{equation}

The parameters in the model have the following interpretation:
\begin{itemize}
\item $(r_d-r_f)$ is the level of the drift of the underlying instrument price process,
\item $\kappa$ is the speed of mean-reversion in the volatility process,
\item $\theta$ is the long term mean of the volatility process,
\item $\omega$ is the volatility of the volatility process,
\item $\rho$ is the correlation between Wiener process of underlying instrument and Wiener process of the volatility,
\item $\nu_0$ is the initial level of volatility, i.e. at the time $t=0$.
\end{itemize}

Additionally, an auxiliary equation can be introduced. This equation describes evolution of the variance, $\upsilon=\nu^2$:

\begin{equation}
\D \upsilon = 2\nu_t \D \nu_t + \tfrac{1}{2} \omega^2 = (- 2 \kappa \upsilon_t + 2 \kappa \theta \nu_t + \omega^2) \D t + 2 \omega \nu_t \D W_t^\nu
.\end{equation}

It can be seen that the variance equation of the Heston model reduces to the case of the auxiliary variance equation in the Sch\"obel-Zhu model with $ \theta = 0 $, when the following substitution for the Heston model will be performed:

\begin{equation}
\label{eq:heston2shz}
\kappa_h = 2 \kappa \quad \omega_h = 2 \omega, \quad \theta_h = \tfrac{\omega^2}{2 \kappa}
.\end{equation}

Similar conclusion can be formulated regarding substitution to the other side, namely, that the constrained Heston model is a special case of the Sch\"obel-Zhu model. In general, there are some possible implied volatility surfaces, which can be generated by both models, but there are also unique surfaces, which can be attributed only to one of those models. \par

It should be noted that, unlike in the Heston model, the negative value of the volatility can be achieved regardless of model parameters, but also this feature has different effect. Firstly, there is no square root operation being performed on negative value, and secondly, even if the volatility reaches a negative value then  $ S_t $ process will still be a Brownian motion, because $\hat{\D W_t^S}=\mathrm{sgn}(\nu(t))\D W_t^S$ is also a Brownian motion. One important aspect of the transition of volatility to the side of negative values is its impact on the immediate correlation of both processes in the model, because:

\begin{equation}
\mathrm{Corr}(\D \ln(S_t),\D \nu_t) = \frac{\rho \nu_t \tau}{\sqrt{\nu_t^2 \tau^2}} = \rho \mathrm{sgn}(\nu_t)
.\end{equation}

The solution to this problem may be the substitution of $ | \nu_t | $ in place of $ \nu_t $. The former variable can be calculated using Tanaka Lemma \footnote{~\bibentry{Prokaj2013}.}.

\subsection{Analytic formula for the characteristic function}

The general equation (\ref{eq:genDynEq}) is simplified to the affine form with respect to the vector [$\ln(S)$,$\nu$,$\nu^2$] due to the additional condition $ p = 0 $ (because in the Sch\"obel-Zhu model $ p = 0 $ and $ r = 1 $) that is being fulfilled:

\begin{equation}\begin{split}
& \frac{\partial A}{\partial t} + (r_d - r_f) i u + B \kappa \hat{\theta} + \frac{1}{2} \omega^2 B^2 + \omega^2 C + \nu \bigg( \frac{\partial B}{\partial t} - B b_j + 2C \kappa \hat{\theta} + \rho \omega i u B + \frac{1}{2} \omega^2 4 C B \bigg) \\
& + \nu^2 \bigg( \frac{\partial C}{\partial t} - 2C b_j + \rho \omega i u 2 C + a_j i u - \frac{1}{2} u^2 + \frac{1}{2} \omega^2 4 C^2 \bigg)  = 0
.\end{split}\end{equation}

Given that $ \nu $ may not be equal to 0, and substitution of partial derivatives $\frac{\partial f}{\partial t} = - \frac{\partial f}{\partial \tau}$, which arises from relation $ \tau = T-t $, the equation can also be reformulated as following system of three equations:

\begin{equation}\begin{split}
\label{eq:shzEq}
& \frac{\partial A}{\partial \tau} = (r_d - r_f) i u + B \kappa \hat{\theta} + \tfrac{1}{2} \omega^2 B^2 + \omega^2 C ,\\
& \frac{\partial B}{\partial \tau} = - B b_j + B \rho \omega i u + 2 B C \omega^2 + 2 C \kappa \hat{\theta} ,\\
& \frac{\partial C}{\partial \tau} = - 2C b_j + 2 \rho \omega i u C + a_j i u - \tfrac{1}{2} u^2 + 2 \omega^2 C^2
.\end{split}\end{equation}


The considered system of differential equations has the analytic solution for the $ A $, $ B $ and $ C $.\footnote{~It is worth noting that the equation of $C$ is the Heston equation of $B$ with different parameters and the equation of $A$ contains part of the Heston equation of $A$.} Therefore, the characteristic function of price distribution in the Sch\"obel-Zhu model is:

\begin{equation}\begin{split}
\label{eq:SZchf}
& \phi_{SZ}(u,x,\nu,\tau) = e^{i u x_0 + A + B \nu_0 + C \nu^2_0  } ,\\
& A(\tau,u) = \hat{A}(\tau,u) + (r_d - r_f) i u \tau + \frac{1}{4}(\beta_j - d_j) \tau - \frac{1}{2} \mathrm{ln} \bigg(\frac{G_j e^{-d_j\tau}-1}{G_j-1}\bigg) ,\\
& \hat{A}(\tau,u) = \frac{\kappa^2 \hat{\theta}^2(\beta_j-d_j)}{d_j^2 \omega^2} \bigg( \tau \frac{\beta_j+d_j}{2} + \frac{4\beta_j e^{-d_j \tau/2}-(2\beta_j-d_j)e^{-d_j \tau}-2\beta_j-d_j}{d_j(1-G_j e^{-d_j \tau})}\bigg) ,\\
& B(\tau,u) = \kappa \hat{\theta} \frac{(\beta_j-d_j)(1-e^{-d_j\tau/2})^2}{d_j\omega^2(1-G_j e^{-d_j\tau})} ,\\
& C(\tau,u) = \bigg(\frac{\beta_j - d_j}{4 \omega^2}\bigg) \bigg(\frac{1-e^{-d_j\tau}}{1-G_j e^{-d_j\tau}}\bigg) ,\\
& \beta_j = 2(b_j - i \omega \rho u), \quad d_j = \sqrt{\beta_j^2-4\omega^2(2a_jiu-u^2)}, \quad G_j = \frac{\beta_j-d_j}{\beta_j+d_j} ,\\
& j=1,2, \quad a_1=\frac{1}{2}, \quad a_2=-\frac{1}{2}, \quad b_1 = \kappa + \eta - \omega \rho, \quad b_2 = \kappa + \eta
.\end{split}\end{equation}

The formula for $\hat{A}$ that is presented here has different order and amount of operations than the analogous formula from Lord and Kahl (2006) \footnote{~\bibentry{LordKahl2006}.}, however both formulas are algebraically equivalent. In their paper it is equal to:

\begin{equation}
\hat{A}(\tau,u) = \frac{(\beta_j-d_j)\kappa^2 \theta^2}{2 d_j^3 \omega^2}
\Bigg( \beta(d_j\tau-4)+d_j(d_j\tau-2) + \bigg( \frac{d_j^2-2\beta^2}{\beta+d_j}e^{-d_j\tau/2}+2\beta \bigg) \frac{4e^{-d_j\tau/2}}{1-G_j e^{-d_j\tau}} \Bigg)
.\end{equation}

The formula for $\hat{A}$ in (\ref{eq:SZchf}), has a lower computational complexity than the formula from Lord and Kahl (2006). Hence, the formula (\ref{eq:SZchf}) will be used as a standard for this model in further considerations.

\section{Models with N-factor price and N-factor variance or volatility}

\subsection{General structure and analysis}

Presented in the previous sections, the author's general form of stochastic volatility models can be extended to the following form:

\begin{equation}\begin{split}
& \D S_t = (r_d-r_f) S_t \D t + \sum^N_k \nu_{k,t}^{r_k} S_t^\beta \D W_{k,t}^S ,\\
& \D \nu_{k,t} = \kappa_k \nu_{k,t}^{q_k} (\hat{\theta_k} - h_k(\nu_{k,t})) \D t + \omega_k \nu_{k,t}^{p_k} \D W_{k,t}^\nu ,\\
& \langle \D W_{k,t}^S , \D W_{k,t}^\nu \rangle = \rho_k \D t, \quad r_k \in \{\tfrac{1}{2},1\} ,\\
& \langle \D W_{k,t} , \D W_{j,t} \rangle = 0, \quad k \neq j ,\\
& h_k(\nu) = \left\{
  \begin{array}{lr}
    \frac{1}{\lambda_k}(\nu^\lambda_k-1) & : \lambda_k > 0\\
    log(\nu) & : \lambda_k = 0
  \end{array}
\right.
,\end{split}\end{equation}

where in place of one shock of the underlying instrument process is the sum of $N$ independent shocks with the stochastic volatility or stochastic variance. Each variance process is correlated only with the one Wiener process, that is multiplying that variance in underlying instrument price process. Due to the latter property, considered general form does not include all stochastic volatility models with multi-factor variance and price known in the literature. On the other hand, discussed feature enables derivation of an explicit formula for the characteristic function of the underlying instrument price distribution, same as in the case of models with a one-factor variance or volatility.

Using It\^{o} lemma we can write the dynamics of $ V $ instrument as:

\begin{equation}
\D V = \frac{\partial V}{\partial t} \D t + \frac{\partial V}{\partial S} \D S + \frac{1}{2} \frac{\partial^2 V}{\partial S^2}(\D S)^2 + \sum^N_k \bigg( \frac{\partial V}{\partial \nu_k} \D \nu_k + \frac{\partial^2 V}{\partial S \partial \nu_k}(\D S \D \nu_k) + \frac{1}{2}\frac{\partial^2 V}{\partial \nu_k^2}(\D \nu_k)^2 \bigg)
,\end{equation}

and after performing transformations analogous to the previously described we obtain a following result:

\begin{equation}\begin{split}
& \frac{\partial V}{\partial t} - r_d V
+ \frac{1}{2} \frac{\partial^2 V}{\partial x^2} \sum^N_k \nu_k^{2r_k}
+ \bigg(r_d-r_f - \frac{1}{2} \sum^N_k \nu_k^{2r_k} \bigg) \frac{\partial V}{\partial x} \\ 
& + \sum^N_k \bigg(
\nu_k^{r_k+p_k} \omega_k \rho_k \frac{\partial^2 V}{\partial x \partial \nu_k} 
+ \frac{1}{2} \omega_k^2 \nu_k^{2p_k} \frac{\partial^2 V}{\partial \nu_k^2}
+ [\kappa_k \nu_k^{q_k}(\theta_k-h_k(\nu_k))-\eta_k(S,\nu_k,t)] \frac{\partial V}{\partial \nu_k} \bigg) = 0
.\end{split}\end{equation}

Analysing the considered dynamics of the instrument, it should be noted that components, which depend on the $ \nu $, have been multiplied in comparison with the equation from the one-dimensional variance case. 

Due to the independence of $ \nu_k $ and considered form of replication strategy dynamics equation, characteristic function of the underlying instrument price in the studied class of models, i.e. with $ \lambda = 1 $, $ q = 0 $ and $ r_i = r_j $ for $ i \neq j $ will have the form:

\begin{equation}\begin{split}
& \phi_j(u,x,\nu,\tau) = e^{i u x + \sum_k^N \Big( A_k + B_k \nu_{k,0} + C_k \nu_{k,0}^2 \Big)} ,\\
& A_k(\tau=0)=B_k(\tau=0)=C_k(\tau=0)=0
.\end{split}\end{equation}

This form of the characteristic function, the previously presented replication strategy dynamics equation with independent sums and solutions for differential equations with characteristic function elements for models with one-factor variance imply solutions in considered models with multi-factor variance. These solutions will be in the similar form to the solution for analogous models with one-factor variance.

\subsection{OUOU model}

Developed by the author the OUOU model is a model similar to the Bates two-factor variance model, but the OUOU model, similarly to the Sch\"obel-Zhu model, contains evolution of volatility processes and not variance processes directly. Both processes of volatility are expressed through an OU (Ohrstein-Uhlenbeck) process. The OUOU model is described by following equations:

\begin{equation}\begin{split}
& \D S_t = (r_d-r_f) S_t \D t + \nu_{1,t} S_t \D W_{1,t}^S + \nu_{2,t} S_t \D W_{2,t}^S ,\\
& \D \nu_{1,t} = \kappa_1 (\theta_1 - \nu_{1,t}) \D t + \omega_1 \D W_{1,t}^\nu ,\\
& \D \nu_{2,t} = \kappa_2 (\theta_2 - \nu_{2,t}) \D t + \omega_2 \D W_{2,t}^\nu ,\\
& \langle \D W_{1,t}^S , \D W_{2,t}^S \rangle = \langle \D W_{1,t}^\nu , \D W_{2,t}^\nu \rangle = 0 ,\\
& \langle \D W_{1,t}^S , \D W_{1,t}^\nu \rangle = \rho_1 \D t, \quad \langle \D W_{2,t}^S , \D W_{2,t}^\nu \rangle = \rho_2 \D t
.\end{split}\end{equation}

This model has four Wiener processes with same assumptions regarding relationships between them as in Bates (2000)\footnote{~\bibentry{Bates2000}.} two-factor variance model. The parameters in the model can be analogically interpreted as in the Sch\"obel-Zhu model. However, it should be noted, that in this model the effect of each of the parameters on the properties of the total variance is difficult to systematize.\par

The form of the characteristic function and the previously presented dynamics equation with independent sums, as well as previous research on the Sch\"obel-Zhu model are useful in derivation of the closed form formula for the characteristic function in case of the two-factor volatility OUOU model. Following steps lead to obtain the semi-closed formula for European vanilla option prices in the OUOU model.\par

In further considerations following restrictions for the parameters were adopted: $ r + p = 1 $, $ q = 0 $ and $ \lambda = 1 $, which come from considerations of Section \ref{sec:2.3} and the aim of obtaining an explicit formula for exponential-affine characteristic function. Transformations are done for $N$ volatility equations to maintain higher level of generality.

Using author's proposition for dynamics of replication strategy in multi-factor variance models (2.60), similarly as for one-factor variance models, partial derivatives of $ V $ in (2.60), can be computed starting from the expected formula for the value of the option, i.e. $V_t = e^{x_t-r_f (T-t)} P_1 (x,\nu_1, ... ,\nu_k,t)$ $-K e^{-r_d (T-t)} P_2 (x,\nu_1, ... , \nu_k,t)$. Most of these derivatives are structurally the same as for the case of one-factor variance, while some of them are only distinguished by the index $ k $ present in $ \nu_k $. This will result in being $N$-times more of following partial derivatives: $\frac{\partial V}{\partial \nu}$, $\frac{\partial^2 V}{\partial \nu^2}$ and $\frac{\partial^2 V}{\partial x \partial \nu}$. As a result, similarly as previously, following form will be obtained after reordering of the equation:

\begin{equation}
e^{x_t-r_f (T-t)} f(P_1(S,\nu_1,...,\nu_k,t)) - K e^{-r_d (T-t)} f(P_2 (S,\nu_1,...,\nu_k,t)) = 0
,\end{equation}

so a potential solution may result from $ f(P_1) = f(P_2) = 0 $, except that the polynomial operator of differentiation with respect to the cumulative distribution functions $ P_1 $ and $ P_2 $ is in this case equal to following formula, with the simplified notation, i.e. $ \nu_k = \nu_{k, t} $:

\begin{equation}\begin{split}
f(P_j) = & \frac{\partial P_j}{\partial t} + \frac{1}{2} \frac{\partial^2 P_j}{\partial x^2} \sum^N_k \nu_k^{2r_k}
+ \sum^N_k \rho_k \omega_k \nu_k^{r_k+p_k} \frac{\partial^2 P_j}{\partial x \partial \nu_k} 
+ \frac{1}{2} \sum^N_k \omega_k^2 \nu_k^{2p_k} \frac{\partial^2 P_j}{\partial \nu_k^2} \\
& + \bigg(r_d - r_f + a_j \sum^N_k \nu_k^{2r_k} \bigg) \frac{\partial P_j}{\partial x} + \sum^N_k [\kappa_k\nu_k^{q_k}(\theta_k-h_k(\nu_k))-\eta_k(S,\nu_k,t) + c_{j,k} \nu_k^{r_k+p_k}] \frac{\partial P_j}{\partial \nu_k} = 0
,\end{split}\end{equation}

$$\mathrm{where:} \quad j=1,2, \quad a_1=\tfrac{1}{2}, \quad a_2=-\tfrac{1}{2}, \quad c_{1,k} = \rho_k \omega_k, \quad c_{2,k} = 0, \quad k=1,...,N.$$

and under the limitations of $q_k=0$, $r_k+p_k=1$, $\lambda_k=1$, $\eta_k(S,\nu_k,t)=\eta_k \nu_k$ and same as before $ \beta = 1 $, which are present in the OUOU model there is:

\begin{equation}\begin{split}
\label{eq:fPj}
f(P_j) = \frac{\partial P_j}{\partial t}
+ \frac{1}{2} \frac{\partial^2 P_j}{\partial x^2} \sum^N_k \nu_k^{2-2p_k} 
& + \sum^N_k \rho_k \omega_k \nu_k \frac{\partial^2 P_j}{\partial x \partial \nu_k} + \frac{1}{2} \sum^N_k \omega_k^2 \nu_k^{2p_k} \frac{\partial^2 P_j}{\partial \nu_k^2} \\
& + \bigg(r_d - r_f + a_j \sum^N_k \nu_k^{2-2p_k} \bigg) \frac{\partial P_j}{\partial x} + \sum^N_k [\kappa_k \hat{\theta_k} - b_{j,k} \nu_k] \frac{\partial P_j}{\partial \nu_k} = 0
,\end{split}\end{equation}

where: $j=1,2, \quad a_1=\tfrac{1}{2}, \quad a_2=-\tfrac{1}{2}, \quad b_{1,k} = \kappa_k + \eta_k - \omega_k \rho_k, \quad b_{2,k} = \kappa_k + \eta_k, \quad k=1,...,N.$

On the other hand, due to the independence of $ \nu_k $ following general form of the characteristic function can be expected:

\begin{equation}\begin{split}
& \phi_j(u,x,\nu,\tau) = e^{i u x + \sum_k^N \Big( A_{k,j,\tau} + B_{k,j,\tau} \nu_{k,0} + C_{k,j,\tau} \nu_{k,0}^2 \Big)} ,\\
& A_k(\tau=0)=B_k(\tau=0)=C_k(\tau=0)=0
.\end{split}\end{equation}

Then, using the simplified notation ($\phi = \phi_j$, $A_k = A_{k,j,\tau}$, $B_k = B_{k,j,\tau}$, $C_k = C_{k,j,\tau}$),  derivatives of $ \phi_j $, that are present in the main equation, are as follows:

\begin{equation}\begin{split}
&\frac{\partial \phi}{\partial x} = i u \phi, \quad \frac{\partial^2 \phi}{\partial x^2} = - u^2 \phi, \quad  \frac{\partial \phi}{\partial t} = \phi \sum_k^N \bigg(\frac{\partial A_k}{\partial t} + \nu_k \frac{\partial B_k}{\partial t} + \nu_k^2 \frac{\partial C_k}{\partial t} \bigg),\\
&\frac{\partial^2 \phi}{\partial x \nu_k} =  (B_k + 2 \nu_k C_k) i u \phi, \quad \frac{\partial \phi}{\partial \nu_k} = (B_k + 2 \nu_k C_k) \phi ,\\
& \frac{\partial^2 \phi}{\partial \nu_k^2} = B_k(B_k + 2 \nu_k C_k) \phi + 2 C_k (\phi + \nu_k \phi (B_k + 2 \nu_k C_k)) = \phi (B_k^2 + 4 C_k B_k \nu_k + 2 C_k + 4 C_k^2 \nu_k^2)
.\end{split}\end{equation}

Substituting these partial derivatives in place of the partial derivatives of $ P_j $ in the formula (\ref{eq:fPj}), then ordering and dividing both sides by $ \phi $ lead to:

\begin{equation}\begin{split}
& \sum^N_k \Bigg(\tfrac{1}{N}(r_d - r_f) i u + \frac{\partial A_k}{\partial t} + B_k \kappa_k \hat{\theta_k} + \nu_k \bigg( \frac{\partial B_k}{\partial t} - B_k b_{j,k} + 2C_k \kappa_k \hat{\theta_k} + \rho_k \omega_k i u B_k \bigg) \\
& + \nu_k^2 \bigg( \frac{\partial C_k}{\partial t} - 2C_k b_{j,k} + \rho_k \omega_k i u 2 C_k \bigg) + \nu_k^{2-2p_k} \bigg( a_j i u - \frac{1}{2} u^2 \bigg) \\
& + \nu_k^{2p_k} \bigg( \frac{1}{2} \omega_k^2 B_k^2 + 2 C_k \bigg) + \nu_k^{2p_k+1} \bigg( \frac{1}{2} \omega_k^2 4 C_k B_k \bigg) + \nu_k^{2p_k+2} \bigg( \frac{1}{2} \omega_k^2 4 C_k^2 \bigg) \Bigg) = 0
.\end{split}\end{equation}

Due to the previously discussed feature of the considered general form of models with multi-factor variance we will consider solutions which arise from zeroing of all components of the sum. This in case of $ p_k = $ 1, and $ r_k = 0 $ and given that $ \nu_k $ cannot be 0 and after substitution of partial derivatives $\frac{\partial f}{\partial t} = - \frac{\partial f}{\partial \tau}$, which arises from the dependence $ \tau = T-t $, this implies $ N $ independent systems of three equations:

\begin{equation}\begin{split}
& \frac{\partial A_k}{\partial \tau} = \tfrac{1}{N}(r_d - r_f) i u + B_k \kappa_k \hat{\theta_k} + \tfrac{1}{2} \omega_k^2 B_k^2 + 2 C_k ,\\
& \frac{\partial B_k}{\partial \tau} = - B_k b_{j,k} + B_k \rho_k \omega_k i u + 2 B_k C_k \omega_k^2 + 2C_k \kappa_k \hat{\theta_k} ,\\
& \frac{\partial C_k}{\partial \tau} = - 2C_k b_{j,k} + 2 \rho_k \omega_k i u C_k + a_j i u - \tfrac{1}{2} u^2 + 2 \omega_k^2 C_k^2
.\end{split}\end{equation}

with boundary conditions $A_k(\tau=0)=B_k(\tau=0)=C_k(\tau=0)=0$. Considered system of differential equations is same as the one shown in the formula (\ref{eq:shzEq}) when the indexation $ k $ is absent and $N=1$. This system has an analytic solution for $A_k$, $B_k$ and $C_k$ and it is derived in Appendix A. Therefore, the characteristic function for $ N = 2 $ has the form of:

\begin{equation}\begin{split}
& \phi_{OUOU}(u,x,\nu_{1},\nu_{2},\tau) = e^{i u x_0 + A_1 + A_2 + B_1 \nu_{1,0} + B_2 \nu_{2,0} + C_1 \nu^2_{1,0} + C_2 \nu^2_{2,0} } ,\\
& A_k(\tau,u) = \frac{1}{2} (r_d - r_f) i u \tau + \frac{1}{4}(\beta_{j,k} - d_{j,k}) \tau - \frac{1}{2} \ln \bigg(\frac{G_{j,k}e^{-d_{j,k}\tau}-1}{G_{j,k}-1}\bigg) + \hat{A_k}(\tau,u),\\
& \hat{A}_k(\tau,u) = \frac{\kappa_j^2 \hat{\theta_j}^2(\beta_{j,k}-d_{j,k})}{d_{j,k}^2 \omega_j^2} \bigg( \tau \frac{\beta_{j,k}+d_{j,k}}{2} + \frac{4\beta_{j,k} e^{-d_{j,k} \tau/2}-(2\beta_{j,k}-d_{j,k})e^{-d_{j,k} \tau}-2\beta_{j,k}-d_{j,k}}{d_{j,k}(1-G_{j,k} e^{-d_{j,k} \tau})}\bigg) ,\\
& B_k(\tau,u) = \kappa_k \hat{\theta_k} \frac{(\beta_{j,k}-d_{j,k})(1-e^{-d_{j,k}\tau/2})^2}{d_{j,k}\omega_k^2(1-G_{j,k} e^{-d_{j,k}\tau})} ,\\
& C_k(\tau,u) = \bigg(\frac{\beta_{j,k} - d_{j,k}}{4 \omega_k^2}\bigg) \bigg(\frac{1-e^{-d_{j,k} \tau}}{1-G_{j,k} e^{-d_{j,k} \tau}}\bigg) ,\\
& \beta_{j,k} = 2(b_{j,k} - i \omega_k \rho_k u), \quad d_{j,k} = \sqrt{\beta_{j,k}^2 - 4 \omega_k^2 (2a_j iu-u^2)}, \quad G_{j,k} = \frac{\beta_{j,k}-d_{j,k}}{\beta_{j,k}+d_{j,k}} ,\\
& k=1,2, \quad j=1,2, \quad a_1=\frac{1}{2}, \quad a_2=-\frac{1}{2}, \quad b_1 = \kappa_k + \eta_k - \omega_k \rho_k, \quad b_2 = \kappa_k + \eta_k
.\end{split}\end{equation}

\subsection{Bates two-factor variance model}

The first described and studied model of two-factor variance was a Heston model equivalent with two-factor variance, hence it is called sometimes the double Heston model. This model was proposed by Bates (2000)\footnote{~\bibentry{Bates2000}.}, then it was studied by Christoffersen et al. (2009)\footnote{~\bibentry{Christoffersen2009}.} and by Gauthier and Possma{\"i} (2011)\footnote{~\bibentry{Gauthier2011}.}.

It is a model that is expanding the Heston model by an additional equation and an additional two Wiener processes. It is therefore in the form of three equations with four Wiener processes: two of them are independent processes affecting the price of the underlying instrument, and another two are in the equations that are describing the dynamics of variances of both Wiener processes of underlying instrument. For this model, similarly as in the case of the Heston model, analytic formula for the characteristic function exists, and each of the stochastic variance processes has its own Feller condition, which must be met in order for the variances $ \nu_i (t) $ remain strictly positive with a probability of 1 for each $ t $. Bates two-factor variance model in a particular version in which the process of the underlying instrument price is a diffusion without jumps is described by the following equations:

\begin{equation}\begin{split}
& \D S_t = (r_d-r_f) S_t \D t + \sqrt{\nu_{1,t}} S_t \D W_{1,t}^S + \sqrt{\nu_{2,t}} S_t \D W_{2,t}^S ,\\
& \D \nu_{1,t} = \kappa_1 (\theta_1 - \nu^1_t) \D t + \omega_1 \sqrt{\nu_{1,t}} \D W_{1,t}^\nu ,\\
& \D \nu_{2,t} = \kappa_2 (\theta_2 - \nu^2_t) \D t + \omega_2 \sqrt{\nu_{2,t}} \D W_{2,t}^\nu ,\\
& \langle \D W_{1,t}^S , \D W_{2,t}^S \rangle = \langle \D W_{1,t}^\nu , \D W_{2,t}^\nu \rangle = 0 ,\\
& \langle \D W_{1,t}^S , \D W_{1,t}^\nu \rangle = \rho_1 \D t, \quad \langle \D W_{2,t}^S , \D W_{2,t}^\nu \rangle = \rho_2 \D t
.\end{split}\end{equation}

The parameters in the model can be analogically interpreted as in the Heston model. However, it should be noted, that in this model the effect of each of the parameters on the properties of the total variance is difficult to systematize.\par

Considered form of the model has an analytic solution for the characteristic function of underlying instrument prices. This function is of the form:

\begin{equation}\begin{split}
& \phi_j(u,x,\nu_{1},\nu_{2},\tau) = e^{i u x + A_1 + A_2 + B_1 \nu_{1,0} + B_2 \nu_{2,0}} ,\\
& A_k(\tau,u) = \frac{1}{2} (r_d - r_f) i u \tau + \frac{\kappa_k \theta_k}{\omega_k^2} \bigg( (\beta_{j,k} + d_{j,k}) \tau - 2 \ln \frac{1-g_{j,k} e^{d_{j,k} \tau}}{1-g_{j,k}} \bigg) ,\\
& B_k(\tau,u) = \bigg(\frac{\beta_{j,k} + d_{j,k}}{\omega_k^2}\bigg) \bigg(\frac{1-e^{d_{j,k} \tau}}{1-g_{j,k} e^{d_{j,k} \tau}}\bigg)
,\end{split}\end{equation}

where

\begin{equation}\begin{split}
& \beta_{k,j} = b_{k,j} - \rho_k \omega_k i u, \quad d_{k,j} = \sqrt{\beta_{k,j}^2 - \omega_k^2 (2 a_j i u - u^2)}, \quad g_{k,j} = \frac{\beta_{k,j} + d_{k,j}}{\beta_{k,j} - d_{k,j}} ,\\
& k=1,2, \quad j=1,2, \quad a_1=\frac{1}{2}, \quad a_2=-\frac{1}{2}, \quad b_1 = \kappa_k + \eta_k - \omega_k \rho_k, \quad b_2 = \kappa_k + \eta_k
.\end{split}\end{equation}

The formula for the characteristic function can be derived as a special case of part of the proof presented for the OUOU model, starting from the formula (\ref{eq:fPj}) with $ p_k = 0.5 $ and elements $ C_k = 0 $ in the assumptions about the form of the characteristic function. Then, for given value of the index $ k $ the implied system of two differential equations is same as the Riccati system of the Heston model, so knowing the solution of the latter a formula for the characteristic function for the Bates two-factor variance model is obtained. Additionally, same as in the case of the Heston model, a substitution of $ G_j = 1 / g_j $ can be applied to introduce components $ e ^ {- d_j \tau} $ in the formula for the same reason as in the formula of Albrecher et al. (2007).

\section{Models with one-factor price and two-factor variance or volatility}

\subsection{Duffie-Pai-Singleton model}

Apart from Bates model from previous section the other model with two-factor stochastic variance is Duffie-Pai-Singleton (2000)\footnote{~\bibentry{Duffie2000}.} model, but in this case there is only one stochastic factor in price equation. This model has the stochastic mean in mean-return variance equation. Only the process of the equation of the underlying instrument and the process of the variance equation have non-zero correlations. The model is described by the following equations:

\begin{equation}\begin{split}
& \D S_t = (r_d-r_f) S_t \D t + \sqrt{\nu_t} S_t \D W_t ^S ,\\
& \D \nu_t = \kappa_1 (\bar{\nu}_t - \nu_t) \D t + \omega_1 \sqrt{\nu_t} \D W_t^\nu ,\\
& \D \bar{\nu}_t = \kappa_2 (\theta - \bar{\nu}_t) \D t + \omega_2 \sqrt{\bar{\nu}_t} \D W_t^{\bar{\nu}} ,\\
& \langle \D W_t^S , \D W_t^{\bar{\nu}} \rangle = \langle \D W_t^\nu, \D W_t^{\bar{\nu}} \rangle = 0 ,\\
& \langle \D W_t^S , \D W_t^\nu \rangle = \rho \D t
.\end{split}\end{equation}

This model belongs to the affine class, but in contrast to the double Heston model of Bates (2000) an analytic solution for the parameters of its characteristic function does not exist. Parameters of the characteristic function can be estimated only by numerical algorithm.

\subsection{Cheng-Scaillet model}

So far, there was little research on the stochastic volatility models with two-factor volatility, also analytic formulas for the characteristic function was not derived in any of these models, because this was not possible. One of these models was proposed by Cheng and Scaillet (2002)\footnote{~\bibentry{Cheng2002}.}. It is a model with following form:

\begin{equation}\begin{split}
& \D S_t = (r_d-r_f) S_t \D t + \sqrt{2}(X_{1,t} + X_{2,t}) S_t \D W_t ^S ,\\
& \D X_{1,t} = \kappa_1 (\theta_1 - X_{1,t}) \D t + \omega_1 \D W_{1,t}^X ,\\
& \D X_{2,t} = - \kappa_2 X_{2,t} \D t + \omega_2 \D W_{2,t}^X ,\\
& \langle \D W_t^S , \D W_{1,t}^X \rangle = \rho_1 \D t ,\\
& \langle \D W_t^S , \D W_{2,t}^X \rangle = \rho_2 \D t ,\\
& \langle \D W_{1,t}^X , \D W_{2,t}^X \rangle = 0
.\end{split}\end{equation}

With such a system of equations it should be noted that the dynamics of the underlying instrument can also be presented as:

\begin{equation}
\D S_t = (r_d-r_f) S_t \D t + \nu_t S_t \D W_t^S
.\end{equation}

Processes $X_{k,t}$ are similar in the interpretation to the volatility, but, as opposed to the OUOU model, considered model has three Wiener processes. In addition, Cheng-Scaillet model belongs to the class of linear-quadratic models, which contains a class of affine models. Gaspar (2004)\footnote{~\bibentry{Gaspar2004}.} and Cheng and Scaillet (2002)\footnote{~\bibentry{Cheng2002}.} put this classification into their works.

Unfortunately, for this model there is no analytic solution for the characteristic function. Because the model is directly related to the dynamics of volatility in the form of a sum of two Ohrstein-Uhlenbeck\footnote{~\bibentry{OU1930}.} processes, there is no additional requirements, such as the Feller condition. This is similar to the Sch\"obel-Zhu model case.

\section{Conclusions}

The general form of the replication strategy dynamics equation with use of characteristic function elements for different stochastic volatility models was presented in the second chapter. It enables to compare different models of stochastic volatility in terms of the possibility of finding an analytic formula for the characteristic function.

Moreover, detailed assumptions of each model were presented and discussed, as well as the differences and similarities between the considered models. Apart from models previously well studied in the literature, a new stochastic volatility model was proposed. It was named OUOU, because it contains two volatility processes of the Ornstein-Uhlenbeck form. Then, the analytic formula for the characteristic function for the OUOU model was derived. The new OUOU model, as well as very close to it Bates two-factor variance model, should be tested empirically in the aspect of their fit to the FX options implied volatility surface. 
\clearpage{\pagestyle{empty}\cleardoublepage}

\chapter{Statistical analysis of implied volatility surface dynamics}

\section{Introduction}

In previous chapters topics related to option valuation have been discussed. Main context of the discussion was pricing with a replication strategy. Equally important as model theoretical assumptions and pricing methods is an analysis of underlying instrument price dynamics, especially dynamics of whole implied volatility surface (i.e. analysis in two dimensions, option maturity and option strike price).

In this chapter principal component analysis has been performed for first differences time series of implied volatility surface. This analysis helps with answering to the question: \textit{how many factors are needed to explain sufficient fraction of whole volatility surface dynamics?}

It may be also interpreted as search for the proper number of parameters in stochastic volatility models, as they can be used as kind of semi-parametrisation of implied volatility surface. Additionally, there is performed analysis of time series models of implied variance with aim of finding which model form used for stochastic volatility dynamics, which were discussed in previous chapters, has better grounding in empiric outcomes of analogous time series model.

\section{Review of PCA outcomes in literature}

Separate part of analysis of implied volatility surface are statistical analyses of its shape, its shape dynamics and its distribution. In many articles implied volatility surface was analysed by statistical methods such as principal component analysis (PCA) without adding financial mathematics theory including option pricing methods. There is noticeable relation between analysis of factors which explain implied volatility surface dynamics and parameters of stochastic volatility models (e.g. Hull-White, Heston, Bates), which can be interpreted as semi-parametrisation of this surface.\par

Early applications of PCA to option market were analyses of implied volatility smile observed in stock option markets as in Derman and Kamal (1997)\footnote{~\bibentry{Derman1997}}, Skiadopoulos et al. (1999)\footnote{~\bibentry{Skiadopoulos1999}} and Fengler (2003)\footnote{~\bibentry{Fengler2003}}. The latter of presented articles contains application of PCA to whole implied volatility surface. Alexander (2001)\footnote{~\bibentry{Alexander2001}} has shown that in the case of exchange traded options PCA is more effective in application to implied volatility spread over ATM volatility, than in direct application to implied volatilities. The choose of spread instead of levels is not only empirically justified by higher variance explained by PCA, but also according to Derman's rigid volatility model correlation between daily changes of implied volatility spreads over ATM volatility (for given spread of strike price) should be much higher than in between daily changes of implied volatility levels for fixed strike price.\par

There are some articles, in which PCA has been applied to implied volatility surface or to its dynamic, i.e. to time series of first differences instead of time series of levels. However, in many articles implied volatilities have been tested only in one dimension, which means that only volatility smile (strike price dimension) for fixed option maturity or only volatility term structure (option maturity dimension) for ATM strike price were tested.\par

In the mainstream of the research on the implied volatility surface and its dynamics, the surface is usually estimated on daily basis using parametric or non-parametric methods. In the parametric approach, which began Shimko (1993)\footnote{~\bibentry{Shimko1993}} volatility is modelled by polynomial functions in dimension of moneyness and option maturity. Non-parametric methods used in the study of this subject are kernel regression (Fengler (2005)\footnote{~\bibentry{Fengler2005}}) and the techniques of grouping (Pena et al. (1999)\footnote{~\bibentry{Pena1999}}). After applying these methods dynamics of volatility is then analysed by principal component analysis (PCA) or related methods. Among other methods worth mentioning is Value-at-Risk (VaR) applied to the parameters of polynomial smile regressions presented in Goncalves and Guidolin (2006)\footnote{~\bibentry{Goncalves2006}}.\par

Numerous studies show that a small number of two to four factors explain much of the daily variation in the shape of the implied volatility surface. These factors are associated with the following four transformations: parallel shifts of surface (1), changes of its slope (2), changes of its curvature (3) and changes in the term structure of volatility (4) (Skiadopoulos et al. (1999), Fonseca (2002)\footnote{~\bibentry{Fonseca2002}}, Hafner (2004)\footnote{~\bibentry{Hafner2004}}, Fengler et al. (2003), Fengler et al. (2007)\footnote{~\bibentry{Fengler2007}}, Daglish (2007)\footnote{~\bibentry{Daglish2007}}, Gatheral (2010)\footnote{~\bibentry{Gatheral2010}}, Badshah (2010)\footnote{~\bibentry{Badshah2010}}). Studies also show that the factor that is associated with parallel shifts has a strong negative correlation to the returns of the underlying index. A summary of the most important research using PCA and concerning implied volatility surface has been shown in Table \ref{table:3.1}.\par

\begin{table}[h!]
\centering
\caption{Summary of the most important research using PCA and concerning implied volatility surface}
\label{table:3.1}
\begin{tabular}[c]{| c | c | c | p{1.1cm} | p{1.2cm} |}
  \hline
Source & Market, frequency, period & PC1-3 interpretation & PC1 var. & PC1-3 var. \\ 
  \hline
Derman1997 & S\&P500, weekly, '94-'97& Level, Term Struct., Skewness & 81.6\% & 90.7\% \\
Derman1997 & Nikkei, daily, '94-'97 & Level, Term Struct., Skewness & 85.6\% & 95.9\% \\
Skiadopoulos1999 & S\&P500, daily, '92-'95 & Level, Term Struct., Skewness & 38.2\% & 70.1\% \\
Cont2002 & S\&P500, daily, '00-'01 & Level, Skewness, Curvature & 94\% & 97.8\% \\
Cont2002 & FTSE100, daily, '99-'01 & Level, Skewness, Curvature & 96\% & 98.8\% \\
Daglish2007 & S\&P500, mies., '98-'02 & Level, Term Struct., Skewness & 92.6\% & 99.3\% \\
Badshah2010 & FTSE100, daily, '04-'07 & Level, Term Struct., Curvature & 56.4\% & 79.0\% \\
Gatheral2010 & S\&P500, daily, '01-'09 & Level, Term Struct., Skewness & 95.3\% & 98.2\% \\
\hline 
\multicolumn{5}{l}{\footnotesize{Except Skiadopoulos (1999) and Badshah (2010), which presented the results of PCA for some months in the analysed}}\\
\multicolumn{5}{l}{\footnotesize{sample so average values of these results was calculated, all results are outcomes of one PCA for whole period.}}\\
\end{tabular}
\end{table}

Table \ref{table:3.1} shows that the level is always identified with the first principal component, which explains the biggest fraction of variance. The second factor is often identified as the term structure of volatility and much less as one of the two characteristics of the volatility smile (skewness or curvature).\par

Most of the studies concerns option on stock market index. Currency option markets have been covered much less. In Krylova (2009)\footnote{~\bibentry{Krylova2009}} PCA and Common Factor Analysis have been applied to term structures of OTC currency options implied volatilities. Analysed pairs were GBPUSD, USDCAD, USDCHF, USDJPY for daily data in 2001-2004 period. First two principal components explained 57\% of variance in volatility term structures. Volatility smiles were not in scope of the analysis.\par

\section{Principal Components Analysis}

Principal Component Analysis (PCA) is a statistical method that uses an orthogonal transformation to transform a set of observations with potentially correlated variables into a set of linearly uncorrelated variables that are called principal components. The amount of principal components is less than or equal to the number of original variables. Applied transformation is defined in a such way that the first principal component has the greatest variance. Each next principal component will have the greatest possible variance while maintaining a boundary condition, which is the orthogonality to the preceding principal components. The vectors resulting from the use of this method are uncorrelated orthogonal basis. The principal components are orthogonal, because they are eigenvectors of the covariance matrix, which is symmetrical. An important feature of the PCA is the sensitivity of the results to the relative scaling of the original variables.\par

For random vector $\mathbf{X}$ principal components transformation can be interpreted as rotation and recentering of this vector, because this transformation is defined as:

\begin{equation} \mathbf{Y} = \mathbf{\Gamma}'(\mathbf{X}-\mathbf{\mu}) \end{equation}

where: $\mathbf{\mu}$ is vector of means, $\mathbf{\Gamma}$ is orthogonal matrix that is created from decomposition of matrix $\mathbf{\Sigma_X}$ (being $\mathbf{X}$ vector covariance matrix) to the eigenvalues matrix $\mathbf{\Lambda}$ and eigenvector matrices $\mathbf{\Gamma}$ (i.e. $\mathbf{\Sigma_X = \Gamma' \Lambda \Gamma}$). 

The matrix $\mathbf{Y}$ columns are named principal components of $\mathbf{X}$ and can be presented as:

\begin{equation} \mathbf{Y}_j = \mathbf{\gamma_j}'(\mathbf{X}-\mathbf{\mu}) \end{equation}

where: $\mathbf{\gamma_j}$ is an eigenvector of the covariance matrix $\mathbf{\Sigma_X}$ corresponding to j-th eigenvalue.

Moreover, according to assumptions, the result of covariance matrix computation for $ \mathbf{Y} $ is a diagonal matrix of eigenvalues. Thus, vectors of the matrix $ \mathbf{Y} $ are uncorrelated with each other:

\begin{equation} \mathrm{cov}(\mathbf{Y}) = \mathbf{\Gamma' \Sigma_X \Gamma} = \mathbf{\Gamma' \Gamma \Lambda \Gamma' \Gamma} = \mathbf{\Lambda} \end{equation}

Research in following part of the section concerns the application of PCA to dynamics of implied volatility surface of currency options in three aspects. Firstly, volatility smiles for different option maturities will be examined (option moneyness dimension), secondly term structure of volatility (option maturity dimension) and finally analysis will concern whole volatility surface in two dimensions.\par
The analysis will help not only to answer how many parameters should be in a model describing the dynamics of the volatility surface to get sufficient quality fit to the market data, but also to find what differences are between principal components in the two dimensions of the volatility surface.\par


Implied volatilities of EURUSD currency pair (25 time series in total) of the period from 22-07-2010 to 31-08-2015 in daily frequency were used for analyses. Selected time series are quotations of implied volatilities for options with an assigned tenor $\tau$ (1M, 3M, 6M, 1Y and 2Y) and the parameter $\delta$ (10P, 25P, ATM, 25C or 10C). These series were obtained by making the transformation according to the previously described formulas to quoted option strategies (ATM, 25RR, 10RR, 25FLY and 10FLY) on the underlying currency. All data were downloaded from Reuters platform. Before transformation, each time series element was obtained by averaging bid and offer prices. During selected 5-year period the market experienced different states, from the strengthening of the prices of the Euro (EUR) against the American Dollar (USD), through the collapse of the market sentiment in 2011 when the price of the Dollar significantly strengthened (due to the debt crisis in Greece), followed by a period of improving confidence in markets and the recovery of the Euro value, until late 2015, when the EURUSD rate was on record low levels.\par

Time series of first differences in the daily quotations of implied volatility were calculated in order to study the dynamics of volatility surface. Summary statistics for time series of first differences are shown in Table \ref{table:3.2}. \par

It can be observed that volatility of short term volatility is higher. Besides, the distribution is leptokurtic, which suggests that volatility of volatility is probably either also stochastic or it is deterministic and volatility changes come from non-normal distribution, e.g. mixed normal distribution.

\begin{table}[h!]
\centering
\caption{Summary statistics for daily changes in implied volatility levels}
\label{table:3.2}
\begin{tabular}{rrrrr}
  \hline
 & Mean & S. D. & Skewness & Kurtosis \\ 
  \hline
EUR10P1M & -0.0016 & 0.523 & 0.551 & 6.783 \\ 
  EUR25P1M & -0.0010 & 0.471 & 0.660 & 7.242 \\ 
  EURATM1M & -0.0004 & 0.424 & 0.566 & 6.961 \\ 
  EUR25C1M & -0.0001 & 0.395 & 0.564 & 7.088 \\ 
  EUR10C1M & 0.0001 & 0.395 & 0.498 & 7.104 \\ 
  EUR10P2M & -0.0022 & 0.430 & 0.578 & 7.317 \\ 
  EUR25P2M & -0.0014 & 0.380 & 0.717 & 8.067 \\ 
  EURATM2M & -0.0008 & 0.334 & 0.652 & 7.768 \\ 
  EUR25C2M & -0.0004 & 0.313 & 0.690 & 7.945 \\ 
  EUR10C2M & -0.0003 & 0.309 & 0.554 & 7.744 \\ 
  EUR10P3M & -0.0029 & 0.386 & 0.735 & 8.327 \\ 
  EUR25P3M & -0.0020 & 0.337 & 0.922 & 9.068 \\ 
  EURATM3M & -0.0013 & 0.295 & 0.853 & 8.910 \\ 
  EUR25C3M & -0.0010 & 0.274 & 0.917 & 9.459 \\ 
  EUR10C3M & -0.0009 & 0.280 & 0.500 & 8.163 \\ 
  EUR10P6M & -0.0036 & 0.324 & 0.615 & 8.360 \\ 
  EUR25P6M & -0.0025 & 0.277 & 0.879 & 9.072 \\ 
  EURATM6M & -0.0019 & 0.237 & 0.831 & 9.528 \\ 
  EUR25C6M & -0.0018 & 0.223 & 0.907 & 9.914 \\ 
  EUR10C6M & -0.0018 & 0.233 & 0.372 & 9.892 \\ 
  EUR10P1Y & -0.0044 & 0.295 & 0.222 & 7.143 \\ 
  EUR25P1Y & -0.0032 & 0.247 & 0.609 & 7.504 \\ 
  EURATM1Y & -0.0024 & 0.194 & 0.447 & 7.826 \\ 
  EUR25C1Y & -0.0023 & 0.188 & 0.460 & 7.313 \\ 
  EUR10C1Y & -0.0025 & 0.200 & -0.080 & 7.053 \\ 
  EUR10P2Y & -0.0040 & 0.278 & 0.147 & 8.099 \\ 
  EUR25P2Y & -0.0030 & 0.231 & 0.505 & 6.861 \\ 
  EURATM2Y & -0.0022 & 0.172 & 0.450 & 7.404 \\ 
  EUR25C2Y & -0.0022 & 0.170 & 0.440 & 6.980 \\ 
  EUR10C2Y & -0.0023 & 0.171 & 0.325 & 7.013 \\ 
   \hline
\end{tabular}

\end{table}

Then principal components analysis algorithm was used to the following sets of first differences time series:

\begin{itemize}
\item all 25 time series (whole surface),
\item 5 time series: ATM1M, ATM3M, ATM6M, ATM1Y, ATM2Y (ATM volatility term structure),
\item 5 time series: 10P1M, 25P1M, ATM1M, 25C1M, 10C1M (1M tenor volatility smile),
\item 5 time series: 10P3M, 25P3M, ATM3M, 25C3M, 10C3M (3M tenor volatility smile),
\item 5 time series: 10P6M, 25P6M, ATM6M, 25C6M, 10C6M (6M tenor volatility smile),
\item 5 time series: 10P1Y, 25P1Y, ATM1Y, 25C1Y, 10C1Y (1Y tenor volatility smile),
\item 5 time series: 10P2Y, 25P2Y, ATM2Y, 25C2Y, 10C2Y (2Y tenor volatility smile).
\end{itemize}

The explanatory power of the principal components resulting from the use of PCA (on the covariance matrix) to these 3 sets are presented in Tables \ref{table:3.3} and \ref{table:3.4}.

\begin{table}[h!]
\centering
\caption{Explanatory power of the principal components for dynamics of whole volatility surface}
\label{table:3.3} 
\begin{tabular}{rrrrrrrrr}
  \hline
 & PC1 & PC2 & PC3 & PC4 & PC5 & PC6 & PC7 & PC8 \\ 
  \hline
Standard deviation & 1.376 & 0.429 & 0.271 & 0.209 & 0.182 & 0.150 & 0.129 & 0.107 \\ 
  Proportion of Variance & 0.819 & 0.080 & 0.032 & 0.019 & 0.014 & 0.010 & 0.007 & 0.005 \\ 
  Cumulative Proportion & 0.819 & 0.899 & 0.930 & 0.949 & 0.964 & 0.973 & 0.981 & 0.986 \\ 
   \hline
\end{tabular}

\end{table}

\begin{table}[h!]
\centering
\caption{Explanatory power of the principal components for dynamic of ATM volatility term structure and volatility smiles (1M-2Y tenors)}
\label{table:3.4}
\begin{tabular}{rrrrrr}
  \hline
 & PC1 & PC2 & PC3 & PC4 & PC5 \\ 
  \hline
  \multicolumn{6}{c}{ATM volatility term structure} \\  
  \hline
Standard deviation & 0.590 & 0.172 & 0.085 & 0.056 & 0.044 \\ 
  Proportion of Variance & 0.892 & 0.076 & 0.019 & 0.008 & 0.005 \\ 
  Cumulative Proportion & 0.892 & 0.968 & 0.987 & 0.995 & 1.000 \\ 
   \hline
 \multicolumn{6}{c}{1M volatility smile} \\  
  \hline
Standard deviation & 0.977 & 0.158 & 0.078 & 0.031 & 0.022 \\ 
  Proportion of Variance & 0.967 & 0.025 & 0.006 & 0.001 & 0.000 \\ 
  Cumulative Proportion & 0.967 & 0.992 & 0.999 & 1.000 & 1.000 \\
   \hline   
\multicolumn{6}{c}{3M volatility smile} \\   
 \hline
Standard deviation & 0.690 & 0.135 & 0.083 & 0.034 & 0.025 \\ 
  Proportion of Variance & 0.947 & 0.036 & 0.014 & 0.002 & 0.001 \\ 
  Cumulative Proportion & 0.947 & 0.983 & 0.997 & 0.999 & 1.000 \\ 
   \hline   
\multicolumn{6}{c}{6M volatility smile} \\     
  \hline
Standard deviation & 0.560 & 0.132 & 0.087 & 0.041 & 0.034 \\ 
  Proportion of Variance & 0.919 & 0.051 & 0.022 & 0.005 & 0.003 \\ 
  Cumulative Proportion & 0.919 & 0.969 & 0.992 & 0.997 & 1.000 \\ 
   \hline   
   \multicolumn{6}{c}{1Y volatility smile} \\  
  \hline
Standard deviation & 0.470 & 0.155 & 0.105 & 0.054 & 0.038 \\ 
  Proportion of Variance & 0.849 & 0.092 & 0.042 & 0.011 & 0.006 \\ 
  Cumulative Proportion & 0.849 & 0.941 & 0.983 & 0.994 & 1.000 \\ 
   \hline   
   \multicolumn{6}{c}{2Y volatility smile} \\  
  \hline
Standard deviation & 0.422 & 0.164 & 0.099 & 0.042 & 0.038 \\ 
  Proportion of Variance & 0.817 & 0.123 & 0.045 & 0.008 & 0.007 \\ 
  Cumulative Proportion & 0.817 & 0.941 & 0.985 & 0.993 & 1.000 \\ 
   \hline
   \multicolumn{6}{c}{All volatility smiles (1M-2Y) - average outcomes} \\  
\hline
Standard deviation & 0.624 & 0.149 & 0.090 & 0.040 & 0.031 \\ 
  Proportion of Variance & 0.900 & 0.066 & 0.026 & 0.006 & 0.003 \\ 
  Cumulative Proportion & 0.900 & 0.965 & 0.991 & 0.997 & 1.000 \\ 
   \hline
   \end{tabular}
\end{table}

The results for the whole volatility surface are similar to previous works on the subject. The first 3 principal components explain more than 90\% of dynamic, and the first 4 about 95\%. It is worth to analyse the impact of next principal components due to popularity of stochastic volatility models, which can be considered as the semi-parametrization of the volatility surface, e.g Heston (1993) and Bates (1997) models. The former one has 5 free parameters and 1 stochastic factor and the latter 8 parameters and 2 stochastic factors. Comparing this with the principal components, the number of factors in the model translate into a 81.9\%, for one factor, up to 98.6\% of the total variance for 8 factors. However, such comparison should be made with some special care. According to PCA assumption, first N principal components explain the greatest possible proportion of total variance, while variables that parametrize surface by stochastic volatility model can explain less variation, because they are the result of the minimization algorithm on the objective function being mean square error between observed surface and model implied surface. However, in this case PCA helps answering how many parameters, which can be called risk factors, should be in the model. Given the above results, the conclusion is that some 4 parameters model for the volatility surface would be only slightly worse to more complicated 5 parameters model.\par

The results of separate analyses of the dynamics of volatility smiles and ATM volatility term structure provide additional information that was not present in previous studies on the dynamics of the implied volatility surface of currency options. For the analysis all volatility smiles that are present in data (1M, 3M, 6M, 1Y, 2Y) have been selected to determine differences in their dynamics. As a term structure of volatility the one for ATM volatilities has been selected due to their location in the middle of the volatility smile and the highest market liquidity for these options. From the statistical point of view, to make the analysis comparable, all the sets, ie, both the term structure and smiles, must have the same number of variables, i.e. 5, which is the result of limited amount of options forming a volatility smile in OTC option markets. In the term structure dimension there is more traded options than 5 so the analysis was limited to 5 of them.

In addition to analysing separate results, the results of PCA have been averaged for the volatility smiles. Averaging the 5 input variables was worse option, because the averaging of input variables causes smoothing of time series and reduce their volatility. Levels of explained variance by the first three principal components are comparable and similar to previous studies on the subject. On the other hand cumulative proportions of  variance are greater for an average of volatility smiles principal components than for the term structure of volatility for the first factor and the sum of the first three factors (90\% vs 89.2\%, 99.1\% vs 98.7\%). By comparing separated volatility smiles and the term structure, it can be observed that the volatility smiles up to the 6M tenor always have more variance that is explained by the PC1-PC3 than term structure of the ATM volatility. An important finding of the study is therefore that it is harder to explain variance of the implied volatility term structure than the volatility smile with the same number of factors from PCA.\par

In addition, the distribution of principal components loadings from the set of all 25 time series are presented in Table \ref{table:3.5}. Due to the fact that the whole surface was in scope of analysis the outcomes are presented in a 5x5 matrix. Interpreting these matrices as 3-dimensional graph can help in answering what movements in these two dimensions correspond to particular principal components.\par
\begin{table}[h!]
\centering
\caption{Principal components loadings for dynamic of whole volatility surface}
\label{table:3.5} 
\begin{tabular}{rrrrrr}
  \hline
 & 10P & 25P & ATM & 25C & 10C \\ 
  \hline   
   \multicolumn{6}{c}{PC1} \\  
  \hline
1M & 0.360 & 0.328 & 0.292 & 0.265 & 0.255 \\ 
  3M & 0.264 & 0.234 & 0.205 & 0.187 & 0.178 \\ 
  6M & 0.213 & 0.185 & 0.161 & 0.147 & 0.141 \\ 
  1Y & 0.171 & 0.150 & 0.128 & 0.117 & 0.111 \\ 
  2Y & 0.141 & 0.127 & 0.109 & 0.101 & 0.097 \\ 
   \hline   
   \multicolumn{6}{c}{PC2} \\  
  \hline
1M & -0.290 & -0.265 & -0.293 & -0.312 & -0.356 \\ 
  3M & 0.135 & 0.111 & 0.065 & 0.053 & 0.011 \\ 
  6M & 0.247 & 0.189 & 0.115 & 0.107 & 0.079 \\ 
  1Y & 0.312 & 0.227 & 0.120 & 0.116 & 0.081 \\ 
  2Y & 0.320 & 0.233 & 0.115 & 0.116 & 0.068 \\ 
  \hline   
   \multicolumn{6}{c}{PC3} \\  
  \hline
1M & -0.224 & -0.215 & -0.126 & -0.056 & 0.086 \\ 
  3M & 0.022 & 0.054 & 0.175 & 0.231 & 0.409 \\ 
  6M & -0.115 & -0.025 & 0.145 & 0.181 & 0.350 \\ 
  1Y & -0.264 & -0.142 & 0.090 & 0.122 & 0.302 \\ 
  2Y & -0.372 & -0.223 & 0.044 & 0.048 & 0.189 \\ 
  \hline   
   \multicolumn{6}{c}{PC4} \\  
  \hline 
1M & 0.221 & 0.012 & -0.093 & -0.235 & -0.268 \\ 
  3M & 0.497 & 0.266 & 0.207 & 0.077 & 0.105 \\ 
  6M & 0.205 & 0.003 & 0.024 & -0.101 & -0.068 \\ 
  1Y & -0.041 & -0.208 & -0.123 & -0.234 & -0.207 \\ 
  2Y & -0.114 & -0.291 & -0.141 & -0.262 & -0.192 \\
  \hline 
\end{tabular}
\end{table}

The first component has the values of all loadings above 0 and all are at a comparable level, although the differences between the loadings in the upper left and lower right corner of the matrix are larger. This component can be assigned to parallel shifts, although volatilities of long tenor far OTM Call options move slower. The second component has in almost every column a growing series of loadings, beginning below 0 and ending above 0. This component can be attributed to changes in the slope of the volatility term structure.

In third component matrix all row series are monotonically growing, which can be attributed to changes in the slope of volatility smile. Besides, column series are non-monotonic and having a maximum in the middle, which links them with changes in curvature of the volatility term structure. The fourth component matrix is similar to the third.

Column series form the same shapes as in third factor matrix with more visible peaks, which relates them to changes in curvature of the volatility term structure. Row series have negative slope like in third component matrix. The difference is that all row series are no longer monotonic with similarity to the shape of the letter "w", the strength of this observation is increasing for longer tenors. In summary, fourth component can be considered as an additional factor explaining the change in curvature of the volatility surface.

These results can be considered quite similar to the results of Derman (1997)\footnote{~\bibentry {Derman1997}} and Daglish (2007)\footnote{~\bibentry{Daglish2007}}, whose interpretations of the first three factors are: level changes, changes in the slope of the term structure and skewness changes. The similarity is less visible in the third factor, because the second dimension of the surface has the characteristic of a change in curvature of the term structure. This difference may be a result of studying the volatility of options for different type of underlying asset, in this study they are for the currency options, which usually have symmetrical volatility smile, while in Derman and Daglish articles options on stock index were examined. For the stock index option market, the smile usually has only a little curvature and is tilted. \par

\section{Factor Analysis}

In addition to using PCA to investigate statistical factors in the dynamics of implied volatility surface, a similar method such as the Common Factor Analysis can be used. This method except for one work (Krylova (2009)\footnote{~\bibentry{Krylova2009}.}) was not used in the literature to study the dynamics of implied volatility surface. However, it was used in other areas of quantitative finance such as Arbitrage Pricing Theory (APT). By using factor analysis in addition to the PCA other aspects can be further investigated, for example a specific variance in each of the studied time series. In summary, Common Factor Analysis is a method to study the structure of internal links in the multidimensional dataset\footnote{~\bibentry{Lawley1962}.}. All of the observed variables are translated by the model as a linear combination of N unobservable factors, which are common to all variables, and one non-observable specific factor for each variable.

Moreover, Common Factor Analysis has an additional assumption, which is the lack of correlation between common factors and specific factors. The number of common factors, N, is a given value from the starting point of the method. There are several algorithms for finding such a set of N common factors that explain the greatest amount of variance in the set of all observed variables. The correlation between variables is explained by the existence of common factors, which co-creates them.

For random vector $\mathbf{X}$, common factor analysis equals to estimation of $\mathbf{a}$, $\mathbf{B}$ and $\mathbf{F}$ in equation:

\begin{equation} \mathbf{X} = \mathbf{a} + \mathbf{B} \mathbf{F} + \mathbf{\epsilon} \end{equation}

where: $\mathbf{F}$ is random vector of common factors, matrix $\mathbf{B}$ is matrix of common factor loadings, $\mathbf{a}$ is vector of constants and $\mathbf{\epsilon}$ is random vector of errors.

Moreover, following boundary conditions need to be fulfilled:
\begin{itemize}
\item random vector of common factors $\mathbf{F}$ has to have positive-definite covariance matrix,
\item random vector of errors $\mathbf{\epsilon}$ has to have elements, which are uncorrelated with each other and have mean equal to 0,
\item covariance of $\mathbf{F}$ and $\mathbf{\epsilon}$ has to be equal to 0, i.e. $\mathrm{cov}(\mathbf{F},\mathbf{\epsilon}) = E((\mathbf{F}-E(\mathbf{F}))\mathbf{\epsilon}') = 0$.
\end{itemize}

Fulfilling of mentioned assumptions results in following equation:

\begin{equation} \Sigma_X = \mathrm{cov}(\mathbf{X}) = \mathbf{B} \Omega_F \mathbf{B}' + \Upsilon_{\epsilon} \end{equation}

where: $\Omega_F$ is covariance matrix of $\mathbf{F}$ and $\mathbf{\epsilon}$ is covariance matrix of $\Upsilon_{\epsilon}$.

In the case of applying the Common Factor Analysis to implied volatility time series, that are forming implied volatility surface, in addition to the PCA, the most valuable information from this application, is not information about the common factors, but the answer to the question: \textit{which parts of the implied volatility surface have dynamics, which is characterized by the highest specific variance?}.

This information allows the separation, in a statistical sense, of parts for which the results of most models in relation to the observed market data will be associated with mean-squared error that is greater than for other points on the surface. \par

As in previous analysis with use of PCA, Common Factor Analysis has been applied to the same data set, which consists of 25 time series representing the whole volatility surface. Common Factor Analysis has been calculated with maximum likelihood algorithm with varimax rotation. The explanatory power of the common factors resulting from the application of factor analysis to whole volatility surface is presented in Table \ref{table:3.6}.

\begin{table}[h!]
\centering
\caption{Explanatory power of the common factors for dynamics of whole volatility surface}
\label{table:3.6} 
\begin{tabular}{rrrrr}
  \hline
 & F1 & F2 & F3 & F4 \\ 
  \hline
SS loadings & 8.219 & 6.690 & 5.682 & 2.001 \\ 
  Proportion Var & 0.329 & 0.268 & 0.227 & 0.080 \\ 
  Cumulative Var & 0.329 & 0.596 & 0.824 & 0.904 \\ 
   \hline
\end{tabular}

\end{table}

The analysis has been started with a number of factors equal to 4, which is satisfactory, taking into account the facts from the literature \footnote{~During calculation of factor analysis with a number of factors greater than 4 (i.e. 5 and 6) the algorithm was not able to find a solution}. Based on the results of the factor analysis with 4 factors, using the Kaiser criterion it can be stated that all factors are important, because they have eigenvalues greater than 1. Also analysing the declines in the eigenvalues between following factors, i.e. scree test, a similar conclusion can be formulated. As expected, factor analysis represents a smaller share of explained variance than PCA for the same number of principal components (90\% vs 94.9\%).

\begin{table}[h!]
\centering
\caption{Specific factors for dynamic of whole implied volatility surface}
\label{table:3.7} 
\begin{tabular}{rrrrrr}
  \hline
 & 1M & 3M & 6M & 1Y & 2Y \\ 
  \hline
10P & 0.045 & 0.055 & 0.122 & 0.234 & 0.353 \\ 
  25P & 0.007 & 0.009 & 0.048 & 0.083 & 0.164 \\ 
  ATM & 0.005 & 0.005 & 0.058 & 0.056 & 0.115 \\ 
  25C & 0.015 & 0.025 & 0.076 & 0.088 & 0.161 \\ 
  10C & 0.034 & 0.067 & 0.150 & 0.198 & 0.237 \\ 
   \hline
\end{tabular}

\end{table}

Values of specific factors for implied volatility surface, presented in table 3.7, allow to formulate following conclusions:
\begin{enumerate}
\item Values of specific factors increase in time to maturity dimension. This can be interpreted as a possible difficulty of 4 factor models in reflecting the movements of the surface at its end that are corresponding to long tenors.
\item Specific factors are greater at the ends of the volatility smile, so 4 factor models also have significant difficulty in reflecting changes in the curvature of the volatility smile.
\end{enumerate}

\section{Autoregression models analysis}

An important element in the analysis of the dynamics of implied volatility, is to examine the problem: \textit{how the functional form of the model that is used for the valuation of the options is justified by the results of studies on corresponding econometric models?}.

In this section functional forms that are corresponding to the Heston and Sch\"obel-Zhu models are tested. For these models both volatility and variance equations are tested. The equations are summarized in Table \ref{table:3.8}.

\begin{table}[h!]
\centering
\caption{Functional form of variance (volatility) equation for Heston and Sch\"obel-Zhu models}
\label{table:3.8} 
\begin{tabular}{rrr}
  \hline
 & Variance form & Volatility form \\ 
  \hline
  Heston & $d\sigma^2_t = \kappa (\theta - \sigma^2_t) dt + \omega \sigma_t dZ_t$ & $d\sigma_t = ([\tfrac{1}{2}\kappa \theta - \tfrac{1}{8} \omega^2] \tfrac{1}{\sigma_t} - \tfrac{1}{2} \kappa \sigma_t) dt + \tfrac{1}{2} \omega dZ_t$ \\ 
  Sch\"obel-Zhu & $d\sigma^2_t = (\omega^2 + 2 \kappa \theta \sigma - 2 \kappa \sigma^2_t) dt + 2 \omega \sigma_t dZ_t$ & $d\sigma_t = \kappa (\theta - \sigma_t) dt + \omega dZ_t$ \\ 
   \hline
\end{tabular}
\end{table}

In order to examine forms of the variance (volatility) function of these two models with data observed in the market, implied variance index has been constructed according to VIX index methodology applied to OTC options (the method is described in details in Chapter 4) and its square root on the basis of quotations of 3M currency options for the EURUSD currency pair for a period from 22-07-2010 to 31-08-2015. The data source is the same as previously. Both series were transformed into series of first differences of variance or volatility and were tested by following models:

\begin{itemize}
\item Model 1: $\Delta \sigma_t^2 = \beta_0 + \beta_1 \sigma_{t-1}^2 $, for variance in the Heston model
\item Model 2: $\Delta \sigma_t^2 = \beta_0 + \beta_1 \sigma_{t-1}^2 + \beta_2 \sigma_{t-1}$, for variance in the Sch\"obel-Zhu model
\item Model 3: $\Delta \sigma_t = \beta_1 \sigma_{t-1} + \beta_2 \tfrac{1}{\sigma_{t-1}}$, for volatility in the Heston model 
\item Model 4: $\Delta \sigma_t = \beta_0 + \beta_1 \sigma_{t-1} $, for volatility in the Sch\"obel-Zhu model
\end{itemize}

These models are discrete equivalent of considered equations of variance and volatility from the stochastic volatility models in continuous time. The process that is an equivalent to Ornstein-Uhlenbeck for discrete time is AR(1) process, which can also be formulated as a regression model for the time series of the first differences of the base series. Although the volatility equation in the Heston model and the variance equation in the Sch\"obel-Zhu model have some specific coefficient for model residuals, but in order to investigate the variance that is explained by given model and the significance of its parameters testing models 1-4 is sufficient. Estimation results for these models are shown in Table \ref{table:3.9}.

\begin{table}[h!]
\centering
\caption{Estimation results for models with alternative functional forms}
\label{table:3.9} 
\begin{tabular}{l c c c c}
\\[-1.8ex]\hline 
\hline \\[-1.8ex] 
 & \multicolumn{4}{c}{\textit{Dependent variable:}} \\ 
\cline{2-5} 
\\[-1.8ex] & \multicolumn{2}{c}{$\Delta \sigma^2_t$} & \multicolumn{2}{c}{$\Delta \sigma_t$} \\ 
\\[-1.8ex] & \multicolumn{1}{c}{(1)} & \multicolumn{1}{c}{(2)} & \multicolumn{1}{c}{(3)} & \multicolumn{1}{c}{(4)}\\ 
\hline \\[-1.8ex] 
 (Intercept) & $0.0002^{**}$  & $-0.0009^{*}$   &               & $0.0011^{**}$  \\
            & $(0.0001)$     & $(0.0003)$      &               & $(0.0004)$     \\
$\sigma^2_{t-1}$          & $-0.0153^{**}$ & $-0.1084^{***}$ &               &                \\
            & $(0.0047)$     & $(0.0299)$      &               &                \\
$\sigma_{t-1}$          &                & $0.0205^{**}$   & $-0.0047^{*}$ & $-0.0111^{**}$ \\
            &                & $(0.0065)$      & $(0.0020)$    & $(0.0040)$     \\
$\tfrac{1}{\sigma_{t-1}}$       &                &                 & $0.0000^{*}$  &                \\
            &                &                 & $(0.0000)$    &                \\
\hline
R$^2$       & 0.0078         & 0.0151          & 0.0042        & 0.0056         \\
Adj. R$^2$  & 0.0070         & 0.0136          & 0.0027        & 0.0049         \\
Num. obs.   & 1332           & 1332            & 1332          & 1332           \\
RMSE        & 0.0011         & 0.0011          & 0.0044        & 0.0044         \\
F Statistic & 10.4043 & 10.2069 & 2.8174 & 7.5332 \\
\hline
\hline \\[-1.8ex] 
\multicolumn{5}{l}{\scriptsize{\textit{Note:} $^{***}p<0.001$, $^{**}p<0.01$, $^*p<0.05$}}  \\ 
\end{tabular} 
\end{table} 

Basing on the results from Table \ref{table:3.9} it can be concluded that the functional forms of Heston and Sch\"obel-Zhu models give comparable results, although a comparison of adjusted R$^2$ enables stating that the model Sch\"obel-Zhu is slightly better in terms of explained variance and this fact is true for both cases, i.e. for variance equation and for volatility equation.

\section{Discriminant analysis}

In the remaining part of this chapter, the term structure of implied volatility (time to maturity dimension of implied volatility surface for ATM strike) and the problem of prediction of direction of changes in implied volatility from information contained in the forward volatility, i.e. market expectations regarding the future value of implied volatility, are examined in more detail. The term structure of implied volatilities, similarly as in the case of the interest rates term structure, should represent market participants expectations of future levels of implied volatilities. Such relationship has been investigated in the literature, e.g in Jablecki et al (2014)\footnote{~\bibentry{Jablecki2014}}, in which a model for forecasting future levels of implied volatility using implied volatility term structure has been built.\par

The following test was constructed for this purpose. It is intended to verify that the data contained in the term structure of volatility enable to effectively predict direction of implied volatility changes.

Because basing on the observation of market behaviour it can be said that accurate predictions of changes in the market are difficult, only direction of changes in the level of implied volatility will be examined, i.e. the model for binary variable describing whether an increase or decrease in implied volatility has occurred will be tested.

For the same daily data as before (the EURUSD options implied volatility surface for 2010-2015), changes in the implied volatility of the ATM options for 1M tenor have been computed. Considered changes were between day $t$ and nearest following business day 3 months after $t$. Then  the time series of changes were transformed into the time series of binary variable, for which 1 stands for a positive change and 0 for a negative change. As the basic variables, i.e. the implied volatilities from the term structure of implied volatility are highly correlated, it was decided to use the transformed variables. Three sets of variables were used as independent variables:\par

\begin{itemize}
\item simple transformation of implied volatility term structure: level (EURATM1M), slope (EURATM2Y - EURATM1M), curvature (EURATM2Y + EURATM1M - 2 $\times$ EURATM6M) and 1M 25-delta Butter Fly strategy market quote, which partially depends on short term market expectations towards future volatility of implied volatility,
\item 3 factor parametrization of implied volatility term structure in the form of parameters of variance dynamics equation from Heston model,
\item 3 factor parametrization of implied volatility term structure in the form of parameters of volatility dynamics equation from Sch\"obel-Zhu model.
\end{itemize}

In the case of the Heston (1993)\footnote{~\bibentry{Heston1993}} stochastic volatility model, the instantaneous variance of the underlying instrument is:

\begin{equation} \nu(t) = \theta + (\nu_0-\theta)e^{-\kappa t} \end{equation}

In the considered equation instantaneous variance $ \nu $ tends to the long term mean level, $ \theta $, at a speed of mean-reversion, $ \kappa $, whereas at the beginning the initial level of variance is equal to $ \nu_0 $. On the other hand, in order to obtain a level of  implied variance for any tenor, it is necessary to compute the integral of instantaneous variance equation and to divide the result by the time to maturity $\tau$:

\begin{equation} \sigma(\tau)^2 = \frac{\int_0^\tau \nu(t)\,\D t}{\tau} = \theta + (\nu_0-\theta)\frac{1-e^{-\kappa \tau}}{\kappa \tau} \end{equation}

The last equation is assumed parametrization of the variance term structure, and the parameters: $\theta$, $\nu_0$ and $\kappa$ can be found by numerical minimization, e.g. with Nelder-Mead algorithm, of the function of mean absolute error between the observed variance term structure consisting of 5 points (2M, 3M, 6M, 1Y and 2Y) based on the squared quotations of the ATM option volatility, and the results of parametric equation. All points on the term structure have the same weight in the applied sum of errors.

In this way, parameters $\nu_0$, $\theta$ and $\kappa$ have been estimated. To simplify the notation the parameter $ \xi $ has been introduced in place of the component $\frac{1-e^{-\kappa}}{\kappa}$.

Similarly, instantaneous volatility of the instrument for the Sch\"obel-Zhu model is equal to:

\begin{equation} \sigma(t) = \theta + (\nu_0-\theta)e^{-\kappa t} \end{equation}

Similarly, the implied volatility, which is calculated as the integral of instantaneous variance equation divided by the time to maturity $\tau$, is equal to:

\begin{equation} \sigma(\tau) = \frac{\int_0^\tau \sigma(t)\,\D t}{\tau} = \theta  + (\nu_0-\theta)\frac{1-e^{-\kappa \tau}}{\kappa \tau} \end{equation}

The $exp(x)$ transformation is used for all parameters before calculating cost function to guarantee positivity in estimations of numerical algorithm. In the Heston model case, as a starting point for estimation through numerical algorithm, squared 1M ATM option volatility has been chosen for $\nu_0$, squared 2Y ATM volatility for $\theta$ and 2 for $\kappa$. For the Sch\"obel-Zhu case, untransformed volatilities instead of squared volatilities are used and $1.1$ is used for $\kappa$. Then, a logarithm is calculated for all elements of starting point vector due to the $exp(x)$ transformation used always before the cost function calculation.

Using Neldear-Mead algorithm and considered equations all parameters of Heston and Sch\"obel-Zhu term structures have been estimated. For discernibility of data both parametrization have been marked with superscripts $H$ and $S$, which leads to following two sets: \{$\theta_H$, $\nu_{0,H}$, $\kappa_H$\} and \{$\theta_S$, $\nu_{0,S}$, $\kappa_S$\}.

In addition, parameters $\nu_{0,S}$ and $\theta_S$ due to the fact that they represent initial and mean level of volatility were squared to match variance order of magnitude, which is used in the Heston model.

To investigate the probability of belonging of observation to the set of increases or decreases, the linear discriminant analysis (LDA) or the logistic regression model can be used, but in the case of data from financial markets the latter is the preferred model because of fewer methodological limitations \footnote{~LDA requires a normal distribution of explanatory variables, which is rarely seen in the data from financial markets, where distributions are usually leptokurtic; LDA also assumes equal level of covariances of these variables}. After selecting the type of model, i.e. logistic regression model, parameters in the following seven models have been estimated:

\begin{itemize}
\item Model 1, with the form of: \begin{equation} \mathbb{P}(Y=1) = \frac{1}{1+e^{-(\beta_0+\beta_1\mathrm{level}+\beta_2\mathrm{slope}+\beta_3\mathrm{curvature})}}\end{equation}
\item Model 2, with the form of: \begin{equation} \mathbb{P}(Y=1) = \frac{1}{1+e^{-(\beta_0+\beta_1\mathrm{level}+\beta_2\mathrm{slope}+\beta_3\mathrm{curvature}+\beta4\mathrm{BF1M})}}\end{equation}
\item Model 3, with the form of: \begin{equation} \mathbb{P}(Y=1) = \frac{1}{1+e^{-(\beta_0+\beta_1\nu_{0,H}+\beta_2\theta_H+\beta_3\xi_H)}} ,\end{equation}
\item Model 4, with the form of: \begin{equation} \mathbb{P}(Y=1) = \frac{1}{1+e^{-(\beta_0+\beta_1\nu_{0,S}^2+\beta_2\theta_S^2+\beta_3\xi_S)}}\end{equation}
\item Model 5, with the form of: \begin{equation} \mathbb{P}(Y=1) = \frac{1}{1+e^{-(\beta_0+\beta_1\nu_{0,H}+\beta_2\theta_H+\beta_3\xi_H+\beta_4\nu_{0,S}^2+\beta_5\theta_S^2+\beta_6\xi_S)}} ,\end{equation}
\item Model 6, with the form of: \begin{equation} \mathbb{P}(Y=1) = \frac{1}{1+e^{-(\beta_0+\beta_1\mathrm{level}+\beta_2\mathrm{slope}+\beta_3\mathrm{curvature}+\beta_4\mathrm{BF1M}+\beta_5\nu_{0,H}+\beta_6\theta_H+\beta_7\xi_H)}} .\end{equation}
\item Model 7, with the form of: \begin{equation} \mathbb{P}(Y=1) = \frac{1}{1+e^{-(\beta_0+\beta_1\mathrm{level}+\beta_2\mathrm{slope}+\beta_3\mathrm{curvature}+\beta_4\mathrm{BF1M}+\beta_5\nu_{0,S}^2+\beta_6\theta_S^2+\beta_7\xi_S)}} .\end{equation}
\end{itemize}


The first form of the model consists of information about the term structure, which can be achieved with simple transformations. The second form of the model includes variables from the first model, but is enhanced with information from the one-month 10-delta Butter Fly strategy, which contains information about the curvature of volatility smile and is linked to the volatility of volatility, which could potentially have an impact on forecasting probabilities of directions of volatility changes. The third and fourth form of the model have parameters from Heston and Sch\"obel-Zhu models used as variables. The fifth form consists of variables, which are a sum of two sets of parameters, from Heston and Sch\"obel-Zhu models. Finally, the sixth and seventh forms are combination of variables from the second form and the third or fourth form of the model in order to examine which of these sets of information are better in terms of model accuracy. The results of the estimation of parameters in all models are shown in Table \ref{table:3.10}.

\begin{table}[h!]
\begin{center}
\caption{Models explaining the probability of increases and decreases of implied volatility for the EURUSD 1M option in 1 quarter horizon}
\label{table:3.10} 
\begin{tabular}{l c c c c c c c}
\hline
                 & Model 1 & Model 2 & Model 3 & Model 4 & Model 5 & Model 6 & Model 7 \\
\hline
(Intercept)      & $6.0^{***}$    & $6.3^{***}$    & $2.8^{***}$   & $2.1^{***}$  & $3.2^{***}$  & $4.4^{***}$    & $3.0^{**}$     \\
                 & $(0.6)$        & $(0.6)$        & $(0.3)$       & $(0.3)$      & $(0.3)$      & $(0.8)$        & $(1.0)$        \\
level            & $-55.0^{***}$  & $-71.6^{***}$  &               &              &              & $-20.5$        & $9.1$          \\
                 & $(4.8)$        & $(5.9)$        &               &              &              & $(12.2)$       & $(16.1)$       \\
slope            & $-78.4^{***}$  & $-84.8^{***}$  &               &              &              & $-49.6^{***}$  & $-80.9^{***}$  \\
                 & $(9.5)$        & $(9.8)$        &               &              &              & $(14.8)$       & $(19.0)$       \\
curvature        & $-180.0^{***}$ & $-181.9^{***}$ &               &              &              & $-150.4^{***}$ & $-203.9^{***}$ \\
                 & $(15.3)$       & $(15.5)$       &               &              &              & $(17.1)$       & $(18.4)$       \\
BF1M             &                & $357.6^{***}$  &               &              &              & $458.4^{***}$  & $569.3^{***}$  \\
                 &                & $(67.2)$       &               &              &              & $(75.1)$       & $(79.0)$       \\
$\nu_{0,H}$ &                &                & $-63.5^{***}$ &              & $-25.7$      & $-140.8^{**}$  &                \\
                 &                &                & $(14.8)$      &              & $(26.3)$     & $(42.9)$       &                \\
$\theta_H$ &                &                & $-85.0^{***}$ &              & $-43.1$      & $-152.7^{***}$ &                \\
                 &                &                & $(17.4)$      &              & $(41.3)$     & $(41.3)$       &                \\
$\xi_H$           &                &                & $-3.2^{***}$  &              & $-2.7^{***}$ & $-1.5^{***}$   &                \\
                 &                &                & $(0.3)$       &              & $(0.4)$      & $(0.4)$        &                \\
$\nu_{0,S}^2$   &                &                &               & $-41.2^{**}$ & $-45.5$      &                & $-410.7^{***}$ \\
                 &                &                &               & $(15.1)$     & $(32.5)$     &                & $(60.9)$       \\
$\theta_S^2$   &                &                &               & $-49.6^{**}$ & $-35.0$      &                & $-82.0$        \\
                 &                &                &               & $(18.4)$     & $(41.1)$     &                & $(54.0)$       \\
$\xi_S$        &                &                &               & $-2.6^{***}$ & $-1.2^{**}$  &                & $-0.7$         \\
                 &                &                &               & $(0.3)$      & $(0.4)$      &                & $(0.4)$        \\
\hline
AIC              & 1560.4         & 1533.4         & 1610.1        & 1641.9       & 1597.9       & 1479.5         & 1437.5         \\
BIC              & 1581.0         & 1559.1         & 1630.7        & 1662.4       & 1633.9       & 1520.7         & 1478.7         \\
Log Likelihood   & -776.2         & -761.7         & -801.0        & -816.9       & -791.9       & -731.7         & -710.8         \\
Deviance         & 1552.4         & 1523.4         & 1602.1        & 1633.9       & 1583.9       & 1463.5         & 1421.5         \\
Num. obs.        & 1270           & 1270           & 1270          & 1270         & 1270         & 1270           & 1270           \\
AUROC & 0.71 & 0.73 & 0.70 & 0.67 & 0.71 & 0.76 & 0.78 \\
\hline
\multicolumn{8}{l}{\scriptsize{\textit{Note:} $^{***}p<0.001$, $^{**}p<0.01$, $^*p<0.05$}}
\end{tabular}
\end{center}
\end{table}

\begin{figure}[h!]
\centering
\includegraphics[width=150mm]{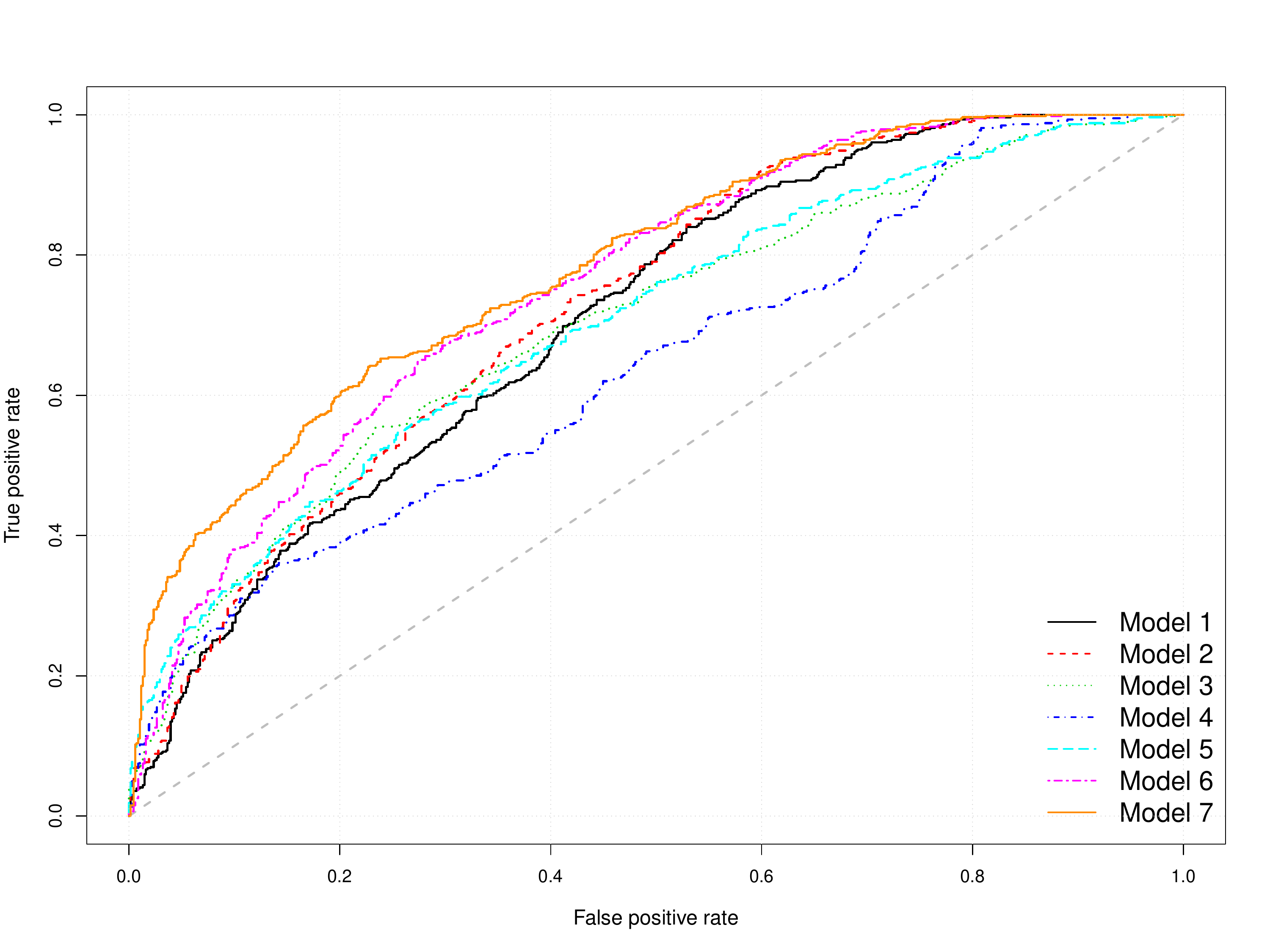}
\caption{ROC curves for Models 1-7}
\label{figure:3.1} 
\end{figure}

The estimation results indicate that all of the variables from four basic models (Model 1-4) are significant at 99 \% level of confidence. Based on the results of seven models the following conclusions can be formulated:

\begin{enumerate}
\item Model 1 and Model 2 are a better models in terms of Akkaike and Bayes-Schwarz information criteria and the Area Under the ROC (AUROC)\footnote{~ROC (\textit{Receiver Operating Characteristic}) shows the relationship between the proportion of correctly classified observations of type 1 (\textit{True Positive}) and proportion of incorrectly classified observations of type 1 (\textit{False Positive}).} than Model 3 or Model 4. It can be considered as an important element in the criticism of parametrization of Heston and Sch\"obel-Zhu models. 
\item Model 2 has a similar form to Model 1, but is additionally enhanced with a variable corresponding to the volatility of volatility and has this variable significant at a high confidence level and also has even better AIC/BIC and AUROC than Model 1. This means that the level of volatility affects the likelihood of direction of volatility changes.
\item All variables are significant at a confidence level of 99 \% in Model 3 and Model 4. However, Model 3 is slightly better in terms of the AIC / BIC criteria and AUROC. In the case of the merged sets of variables from these two models (Model 5), the resulting model is even better (AIC / BIC, AUROC), but only two variables (one from each parametrization) and intercept are significant at a confidence level of at least 95 \%.
\item Model 6 and Model 7, which have most number of variables, are the best two models of the proposed seven in terms of information criteria and area under the ROC curve, which is shown in Figure \ref{figure:3.1}. Although, in case of Model 6, which uses the Heston parametrization, all variables are significant at a high confidence level of 99 \%, but the other statistics, AIC / BIC and AUROC, are worse than for Model 7. Only one variable from the Sch\"obel-Zhu parametrization in Model 7 is significant at a confidence level of at least 95 \%, but the model has the highest accuracy of all seven models in terms of information criteria and AUROC. Therefore it can be concluded that the volatility term structure parameters of the Sch\"obel-Zhu model contain additional significant information, which is useful in formulating expectations for changes in the level of implied volatility in the future.
\end{enumerate}

\section{Conclusions}

The study was designed to test theories regarding the dynamics of the implied volatility surfaces with respect to EURUSD currency options implied volatilities. Most of the previous work on the dynamics of the implied volatility surface was related to the implied volatility of options on stock index, usually the S\&P500. The study consisted of an application of principal component analysis, common factor analysis, auto-regression analysis and discriminant analysis, i.e. logistic regression model. \par

The main components of the implied volatility surface dynamics that were obtained from PCA method were then interpreted . The results were similar in comparison to analogous work on the implied volatility of options on stock indices. However, there are differences that may result from studying currency options volatilities, which have usually symmetrical volatility smile with higher curvature, while other articles were examining options on stock index, for which the smile usually has a small curvature and is skewed. \par

Another important finding of the study is the fact that it is harder to explain the dynamics of the implied volatility term structure than the dynamics of the volatility smile with the same number of factors of PCA. This is the confirmation of the first research hypothesis. \par

The conclusion of the factor analysis applied to the same data set is the existence of difficulties in models with four factors in fitting of the volatility surface dynamics at its end that is corresponding to the longest tenors. Secondly, specific factors are greater at the ends of the volatility smile, so models with four factors also have some problems in fitting of the volatility smile curvature dynamics. \par

The conclusion from the analysis of auto-regressive models is that the functional form of the Sch\"obel-Zhu model is more flexible than the form of the Heston model in terms of fitting to the historical data, because an autoregressive model which is analogous to the Sch\"obel-Zhu model explains bigger fraction of the variance in the implied volatility time series. This confirms the second research hypothesis. \par

The conclusion from the discriminant analysis is that the parameters of the term structure from Sch\"obel-Zhu model contain additional and important information, which are useful in formulating expectations for changes in the level of implied volatility in the future. Moreover, those information are more useful in forecasting direction of changes in volatility than information from the Heston model variance term structure parametrization. This confirms the third research hypothesis. \par

On the basis of this study it can be generally stated that the stochastic volatility models, especially the five-parameter Sch\"obel-Zhu model, are justified in terms of the statistical properties of the implied volatility surface dynamics, although there are problems in the fitting of the dynamics of implied volatilities of options for distant maturities. \par

As a result of this study, a comparison of models that are extensions of Heston and Sch\"obel-Zhu models with two- or multi-dimensional equation of stochastic volatility is a proposition for further research. Additionally, the following hypothesis can be formulated: for the EURUSD currency options the extension of Sch\"obel-Zhu model will also have a better fit to the data market than analogous two-dimensional Heston model extension \par

\clearpage{\pagestyle{empty}\cleardoublepage}

\chapter{Estimation of stochastic volatility models parameters}

\section{Introduction}

In previous chapters we have dealt with stochastic volatility models, semi-analytic formulas for the valuation of options by these models and empirical properties of FX options market, i.e. market quoted implied volatilities.

In these chapters, parameters in stochastic volatility models were considered as given, but in the fourth chapter, the main attention is paid to estimating the value of these parameters using the observed market prices of options. This process is called model calibration and is made on the basis of numerical minimization of the cost function. Firstly, an outline of several algorithms for finding the minimum is presented. Then, methods for derivation of approximate formulas of Heston stochastic volatility model parameters are summarized. The approximate formula for estimation of model parameters can also be used to determine a starting point of minimization algorithm. The main part of the chapter is the presentation of the author's new method of derivation of approximate formulas for estimation of stochastic volatility models parameters. The method is based on implied central moments.

\section{Calibration of stochastic volatility models}

\subsection{Overview of current used methods}

Stochastic volatility models are usually calibrated using a numerical algorithm to minimize the cost function. Mean-squared error between option prices observed in the market in relation to those obtained from the model can be used as the cost function. Other possible cost functions are shown in Table \ref{table:4.1}. However, in market practice instead of differences in options prices, differences between options prices divided by options Vega \footnote{~Vega is a sensitivity of option price to changes in implied volatility, $\frac{\partial C}{\partial \sigma}$.} are preferred. This avoids the situation where errors from cheap far OTM options have little effect on the value of the cost function.

Another possible form of the cost function is mean-square error between the implied volatilities from the model and those that are observed on the market. However, this change in approach increases computational complexity of minimization algorithm, because in considered problem a direct output of stochastic volatility model is the option price, which then has to be converted to implied volatility using numerical method applied to the Black-Scholes formula.

\begin{table}[h!]
\centering
\caption{Summary of the most commonly used cost functions}
\label{table:4.1}
\begin{tabular}[c]{l c c}
  \hline
Name & Abbreviation & Formula \\
  \hline
Mean Absolute Error & MAE & $\sum^N_{i=1} |v_i^{\mathrm{model}}-v_i^{\mathrm{market}}|$ \\
Mean Square Error & MSE & $\sum^N_{i=1} (v_i^{\mathrm{model}}-v_i^{\mathrm{market}})^2$ \\
Mean Absolute Percentage Error & MAPE & $\sum^N_{i=1} \bigg|\frac{v_i^{\mathrm{model}}-v_i^{\mathrm{market}}}{v_i^{\mathrm{market}}}\bigg|$ \\
Mean Square Percentage Error & MSPE & $\sum^N_{i=1} \bigg(\frac{v_i^{\mathrm{model}}-v_i^{\mathrm{market}}}{v_i^{\mathrm{market}}}\bigg)^2$ \\
  \hline
  \multicolumn{3}{l}{where: $ v $ is the price of a call option, ($ C $), the Vega-weighted price of the call option ($ C / \nu $)} \\ 
  \multicolumn{3}{l}{or the implied volatility, $N$ is a number of the options that are observed on the market and have been} \\ 
  \multicolumn{3}{l}{selected for calibration}
\end{tabular}
\end{table}

After defining the cost function, $ F $, we must examine its arguments, which are being changed in the process of finding the minimum of the function. In cases of the Heston and the Sch\"obel-Zhu model we minimize the cost function over five parameters, $\nu_0, \theta, \omega, \rho, \kappa$:

\begin{equation}\begin{split}
\Theta^* & = \operatorname{arg\,min}_{\Theta} F(\Theta) ,\\
\Theta & = \{\nu_0, \theta, \omega, \kappa, \rho \}
.\end{split}\end{equation}

When there is no further special treatment then additional restrictions have to be applied to the minimization algorithm, because most of the parameters have to be in a given range. However, this can be omitted if a transformation of the original parameters, $\nu_0, \theta, \omega, \kappa, \rho$ to variables $ x_1 $, $ x_2 $, $ x_3 $, $ x_4 $ and $ x_5 $ is introduced to map the space $ \mathbb{R}^5 $ to $\mathbb{R}_+^4\times(-1,1)$. Inverse functions of these transformations are:

\begin{equation}\begin{split}
\nu_0 & = \mathrm{exp}(x_1) ,\\
\theta & = \mathrm{exp}(x_2) ,\\
\omega & = \mathrm{exp}(x_3) ,\\
\kappa & = \mathrm{exp}(x_4) ,\\
\rho & = \mathrm{tanh}(x_5) 
.\end{split}\end{equation}

The same approach and analogous substitution is also used in two-factor and non-affine models as SABR model\footnote{~The SABR model has an additional parameter $ \beta $, which is in market practice often determined empirically, i.e. on the basis of econometric analysis on time-series of underlying instrument prices. This results in a need of estimation of only three parameters, $\alpha, \rho, \nu$, instead of four.}. Moreover, in the case of the Heston model typically an additional non-linear boundary condition has to be added to the problem of seeking the cost function minimum. This condition guarantees positivity of variance process and has been formulated by Feller in the form of following inequality:

\begin{equation} 2 \kappa \theta - \omega^2 > 0  .\end{equation}

In the case of the currency market, the Feller condition is rarely fulfilled for the volatility surfaces that are observed in practice (Clark (2011)\footnote{~\bibentry{Clark2011}.}). Hence, in the case of this market one should pay particular attention to the non-fulfillment of this condition and correct treatment of variance absorption on boundary condition $ \nu = 0 $.

Choosing of right minimization algorithm is a very important part of the calibration of stochastic volatility model. On the one hand, the minimization problem is given in the form of highly non-linear non-convex cost function\footnote{~\bibentry{Gilli2010}.}, with some additional boundary conditions (e.g. the Feller condition). Due to non-convexity, there are numerous local minima and maxima, so it may make sense to apply a global minimization algorithm. On the other hand, such algorithms applied to the function with such degree of complexity may have a long execution time, which is undesirable in market practice. For example, Gilli and Schumann (2010\footnote{~Ibidem},2011\footnote{~\bibentry{Gilli2012}.}) propose a hybrid algorithm in the form of a combination of the Differential Evolution\footnote{~\bibentry{StornPrice1997}.} algorithm or Particle Swarm\footnote{~\bibentry{Kennedy1995}.} algorithm with Nelder-Mead\footnote{~\bibentry{NelderMead1965}.} local minimization algorithm. Another work showing hybrid (global-local) algorithm is the work of Kilin (2007)\footnote{~\bibentry{Kilin2007}.} who combined Differential Evolution algorithm with Levenberg-Marquardt\footnote{~\bibentry{Marquardt1963}.} local minimization algorithm.

Bin (2007)\footnote{~\bibentry{Bin2007}.} studied the methods of calibration of the Heston model and the Bates jump-diffusion model and presented the HSAS (Hybrid Stochastic Approximation Search) algorithm, which is a multi-level structured global minimization algorithm, which includes stochastic approximation for the robustness and deterministic search for better performance. He also studied using each of five Heston model parameters as a deterministic function of option time to maturity ($ \tau $). The results of applying this method of calibration indicated that the correlation and volatility of variance in the model are the most unstable parameters, while the other parameters were relatively stable. The algorithm was particularly resistant and allowed minimization of functions containing up to ten parameters.

However, despite the fact that the usage of such global algorithm increases the probability of finding the global minimum, its computational complexity is still high. Therefore, in other papers on the stochastic volatility models and in market practice local minimization algorithms are mostly applied. In this group, the following algorithms are the most commonly used: Nelder-Mead, Powell\footnote{~\bibentry{Powell1964}.}, Levenberg-Marquardt\footnote{~\bibentry{Marquardt1963}.} and BFGS (Broyden\footnote{~\bibentry{Broyden1970}.}, Fletcher\footnote{~\bibentry{Fletcher1970}.}, Goldfarb\footnote{~\bibentry{Goldfarb1970}.} and Shanno\footnote{~\bibentry{Shano1970}.}), which is an approximation of Newton back-propagation algorithm. All of them are well suited for non-linear least squares problem, however, with the exception of the first two algorithms they require the differentiability of the cost function, because they use derivatives of the cost function.

However, it should be noted that for some combinations of parameters, e.g. those related to the specific shape of the volatility surface, and the option strike price discontinuities can occur in the characteristic function, which affects the possibility of its integration and computation of the option price. Because of this, the application of algorithms which require calculation of derivatives of the cost function is sometimes problematic. Due to these problems, in the following considerations the Nelder-Mead algorithm will be used as the minimization algorithm. The adopted setup of the algorithm as well as its short summary is presented in the following section.

\par~\par \noindent\textbf{Nelder-Mead algorithm} \par~\par

For a given function, $ f (x) $, where $x \in \mathbb{R}^n$ and test points $\{x_1, ... , x_{n+1}\} $ Nelder-Mead algorithm can be described in the following steps, assuming standard values for the parameters of the algorithm, i.e. $\alpha=1,\gamma=2,\rho=1/2,\sigma=1/2$, and accuracy of $\epsilon_1$ and $\epsilon_2$ and the maximum number of iterations $ N $:

\begin{enumerate}
	\item \textbf{Ordering}:
	\begin{itemize}
		\item Find the new numbering of points $x_i$ to get the following order: $f(x_1) \leq f(x_2) \leq ... \leq f(x_{n+1})$
		\item If $|f(x_{n+1})-f(x_1)|<\epsilon_1$ and $\mathrm{vol}(\{x_1, ... , x_{n+1}\})<\epsilon_2$ (where $\mathrm{vol}$ is volume of simplex) then stop the algorithm
		\item If the existing number of repeats of step 1 is greater than $N$ then stop the algorithm
	\end{itemize}
	\item \textbf{Centering}: for all points except $x_{n+1}$ calculate the centroid (i.e. centre of gravity) $x_o = \sum^n_{i=1} x_i / n$
	\item \textbf{Reflection}:
	\begin{itemize}
		\item Calculate the reflected point $x_r = x_o + \alpha (x_o-x_{n+1})(\alpha>0)$
		\item If $f(x_1) \leq f(x_r) \leq f(x_n)$, then $x_{n+1}:=x_r$, calculate the new simplex and go to step 1.
		\item Else, go to step 4.
	\end{itemize}
	\item \textbf{Expansion}:
	\begin{itemize}
		\item If $f(x_r) \leq f(x_1)$, then calculate expanded point $x_e = x_o + \gamma (x_r-x_o)(\gamma>0)$
		\begin{itemize}
			\item If $f(x_e) \leq f(x_r)$, then $x_{n+1}:=x_e$, 
			\item Else, $x_{n+1}:=x_r$
			\item Calculate the new simplex and go to step 1.
		\end{itemize}	
		\item Else, go to step 5.
	\end{itemize}
	\item \textbf{Contraction}:
	\begin{itemize}
		\item Calculate the contraction point $x_c = x_o + \rho (x_{n+1}-x_o)(0<\rho \leq 0.5)$
		\item If $f(x_c) \leq f(x_{n+1})$, then $x_{n+1}:=x_c$, calculate the new simplex and go to step 1.
		\item Else, go to step 6.
	\end{itemize}
	\item \textbf{Reduction}:
	\begin{itemize}
		\item All but the best points should be replaced by substituting: $x_i:=x_1 + \sigma (x_i-x_1)$ for each $i \in \{ 2, ... , n+1 \}$
		\item Go to step 1.
	\end{itemize}
\end{enumerate}

The starting point, $x_s$, is used as an algorithm input. On base of $x_s$ the test points $\{x_1, ... , x_{n+1}\}$ are created by the following rules:
\begin{itemize}
	\item $x_{n+1} = x_s$,
	\item $x_i = x_1 + h_i e_i$ for each $i \in \{ 1, ... , n \}$.
\end{itemize}
where: $e_i$ are the unit vectors for $ i $-th dimension of the $\mathbb{R}^n$ space (i.e. $ e_2 = \{0,1,0,...,0\}$), while $h_i$ are given by\footnote{~\bibentry{Gao2012}.}:

\begin{equation}\begin{split}
\left\{
\begin{array}{lr}
h_i = 0.05 & : x_0^T e_i \neq 0 \\
h_i = 0.00025 & : x_0^T e_i = 0
\end{array}
\right.
.\end{split}\end{equation}

\subsection{Model calibration risk}

The process of model calibration should be seen also in the context of model risk, i.e. the risks associated with the inadequacy of the model and its parameters in relation to reality. One of the first authors, who dealt with this issue was Cont (2006)\footnote{~\bibentry{Cont2006}.}. In addition to the basic model risk, i.e. the risk that the functional form of the model does not represent reality, the calibration process, which is carried out by numerical minimization of the cost function with many local minima, brings additional risk of inadequacy of parameters in the model. The calibration process and the adequacy of the calibration that is performed on artificially generated data have been described in Schoutens, Simons and Tistaert (2004)\footnote{~\bibentry{Schoutens2004}.}. The results of valuation of exotic options obtained by them using different models have significant differences. Such differences have appeared even when the Heston model output and implied volatilities observed in the market were perfectly matching. The conclusion from this work was that the prices of exotic options significantly depend on the choice of the model and the quality of the model calibration.

The problem of the risk of model calibration was also examined by Guilliame and Schoutens (2012)\footnote{~\bibentry{Guilliame2012}.}. They identified that the calibration method alone, i.e. choice of the form of the cost function, changes the calibration results. This implies uncertainty in values of calibrated parameters. Based on these conclusions they have proposed calibration of only three of the parameters in the Heston model. The other two parameters ($ \nu_0 $ and $ \theta $) have been estimated by them on the basis of option prices and historical data. The impact of the use of different cost function on the uncertainty in the model parameters (calibration risk) was also studied by Detlefsen and H\"ardle (2007)\footnote{~\bibentry{Detlefsen2007}.}.

Gilli and Schumann (2010\footnote{~\bibentry{Gilli2010}.},2011\footnote{~\bibentry{Gilli2012}.}) have found that even when there is no model risk and no calibration risk, perfect calibration, i.e. the global minimum, still may not be found. Due to the nonlinearity and non-convexity of the cost function there is a non-zero probability of finding a local minimum.

\section{Approximate formulas for estimation of stochastic volatility models parameters}
\label{sec:4.3}

Works on the analytic estimation of stochastic volatility models parameters with use of approximate formulas are a separate stream of research. The majority of those works concerns the calibration of only a part of model parameters in the numerical minimization algorithm, while the remaining parameters are estimated with some approximate analytic solution to the system of equations, which is based on current or historical prices of options and the underlying instrument.

\subsection{Guilliame and Shoutens method}

Guilliame and Shoutens (2010)\footnote{~\bibentry{Guilliame2010}.} proposed computational simplification in the Heston model calibration by introduction of a heuristic applied to two of the model parameters. While the initial value of the variance was already easily determined in some calibration approaches as the squared implied volatility of closest maturing option, then Guilliame and Shoutens go further by setting that parameter to the level of squared VIX, which is non-parametric implied volatility index. The long-term level of variance is estimated with the simple moving average of squared VIX time series.\par

The option market, which was studied empirically in their work, was the market of S\&P500 index options. Therefore, the VIX index, which is quoted by CBOE exchange, was given as directly observable variable. In the case of other option markets, instead of VIX index this would be square-root of implied variance that can be calculated with the VIX index methodology.\par

Later, Guilliame and Shoutens (2012)\footnote{~\bibentry{Guilliame2012}.} presented a potential risk of an inadequate Heston model calibration on the example of differences in the valuation of exotic options, namely barrier, lookback and cliquet options, as well as differences in the profit and loss of the delta hedging strategy. In their article, as in the previous one, the authors studied options on S\&P500 index. In addition, they presented improved scheme for the Heston model calibration with a new formula for the long-term level of the variance. They proposed to use new methods to estimate it. The first one is by the exponential moving average of the squared VIX time series and the second is by the non-parametric formula that is based on options on the VIX index. The article concludes that it is better to assume a margin of safety in the valuation of exotic options with the Heston model. \par

In summary, both works introduced following estimates of the Heston model parameters:

\begin{equation}\begin{split}
& \nu_0 = \bigg( \frac{\mathrm{VIX}(t_0)}{100} \bigg)^2 ,\\
& \theta = \frac{1}{T} \int_{t_0-T}^{t_0} \bigg( \frac{\mathrm{VIX}(t)}{100} \bigg)^2 \,\D t = \mathrm{MA}_{t_0-T \leq t \leq T} \bigg( \frac{\mathrm{VIX}(t_0)}{100} \bigg)^2
.\end{split}\end{equation}

where: VIX is an index of implied volatility, MA is a historical moving average (simple or exponentially weighted), while $ T $ is a year fraction, which is used in the historical moving average.
Simple and exponentially weighted moving averages are defined as follows for daily data:

\begin{equation}\begin{split}
& \mathrm{SMA}(x,t_0) = \frac{1}{252T} \sum_{i=0}^{252T-1} x_{t_0-i} ,\\
& \mathrm{EWMA}(x,t_0) = \tfrac{1}{252T} x_{t_0} + (1-\tfrac{1}{252T}) EWMA(x,t_0-1)
.\end{split}\end{equation}

In addition, when data on long-term VIX option prices are present, Guillaume and Shoutens (2012) proposed to use the following call-put options arbitrage:

\begin{equation}
P(K,T)-C(K,T) = \exp(-(r_d-r_f)T)(K-\mathrm{VIX}_T)
,\end{equation}

to formulate the analytic estimation that is based on particular solution, $K=\mathrm{VIX}_T$. Then, using the option with the longest maturity, $ T $, and the strike price, $ K^{\mathrm{ATM}} $, which is the average of call and put strikes, we can alternatively present the parameter $ \theta $ as follows:

\begin{equation}
\theta = \bigg( \frac{K^{\mathrm{ATM}}}{100} \bigg)^2 = \bigg( \frac{\mathrm{LVIX}(t_0)}{100} \bigg)^2
.\end{equation}

In both of their works, the time efficiency of considered calibration processes has not been studied. However, in both papers authors have observed that in general using two parameters that are estimated with the approximate formula method and calibrating remaining three parameters instead of all five parameters reduces the overall accuracy in terms of errors between market option prices and model option prices. However, it also reduces the risk of calibration in the sense of variation of the parameters in the set of calibration results for different forms of the cost function.

\subsection{Durrleman method}

Durrleman (2004)\footnote{~\bibentry{Durrleman2004}.} and Durrleman and El Karoui (2008)\footnote{~\bibentry{Durrleman2008}.} present an approximation of the implied volatility smile function, which can be reversed to calculate the analytic formula for some of the Heston model parameters. The approximate formula for estimation of model parameters was not the main purpose of their work, but its implication is the possibility to accelerate the calibration of the Heston model. This feature was not further explored. In conclusion, the authors present the following parameterization of the implied volatility by some given characteristics, $\mathbb{S}_t$, $\mathbb{C}_t$, $\mathbb{M}_t$, which describe the implied volatility surface, $\Sigma_t(K,T,S_t)$, that is observed in the market:

\begin{equation}\begin{split}
& \sigma(T,K) = \sigma^{ATM}_t + (T-t)\mathbb{M}_t + \bigg( \frac{K}{S_t}-1 \bigg) \mathbb{S}_t + \bigg( \frac{K}{S_t}-1 \bigg)^2 \frac{\mathbb{C}_t}{2} ,\\
& \sigma^{ATM}_t = \lim_{T \to t} \Sigma_t(K,T,S_t) ,\\
& \mathbb{S}_t = \lim_{T \to t} S_t \frac{\partial \Sigma_t(K,T,S_t)}{\partial K} ,\\
& \mathbb{C}_t = \lim_{T \to t} S_t^2 \frac{\partial^2 \Sigma_t(K,T,S_t)}{\partial K^2} ,\\
& \mathbb{M}_t = \lim_{T \to t} \frac{\partial \Sigma_t(K,T,S_t)}{\partial T} 
,\end{split}\end{equation}

where: the characteristics $\mathbb{S}_t$, $\mathbb{C}_t$ and $\mathbb{M}_t$ have following interpretation: the slope of the volatility smile, the curvature of the volatility smile, the term structure premium of the volatility smile. Then, authors show the formulas for the aforementioned characteristics of the volatility smile with regard to the Heston model parameters:

\begin{equation}\begin{split}
& \mathbb{S}_t = \frac{\omega\rho}{4\sqrt{\nu_t}} ,\\
& \mathbb{C}_t = \frac{\omega^2}{24\sqrt{\nu_t}^3}(2-5\rho^2) - \frac{\omega\rho}{4\sqrt{\nu_t}} ,\\
& \mathbb{M}_t = \frac{\kappa}{4\sqrt{\nu_t}} (\theta - \nu_t) - \frac{\omega^2}{48\sqrt{\nu_t}^3} (2-\rho^2/2) + \frac{\omega\rho\sqrt{\nu_t}}{8}
.\end{split}\end{equation}

Although, they do not show the reversion of the analysis, reversed formulas can be obtained after a series of transformations. Then, we can relate the Heston model parameters in terms of the volatility smile characteristics:

\begin{equation}\begin{split}
& \omega = 2 \sqrt{\nu_0} (3 \sqrt{\nu_0} \mathbb{C}_t +3 \sqrt{\nu_0} \mathbb{S}_t +10 \mathbb{S}_t^2)^{\frac{1}{2}} ,\\
& \rho = 2 \mathbb{S}_t (3 \sqrt{\nu_0} \mathbb{C}_t +3 \sqrt{\nu_0} \mathbb{S}_t +10 \mathbb{S}_t^2)^{-\frac{1}{2}} ,\\
& \kappa = \frac{1}{2(\theta-\nu_t)}(8 \mathbb{M}_t \sqrt{\nu_t} + \frac{\omega^2}{6\nu_t}(2-\rho^2/2) + \omega\rho\nu_t)
.\end{split}\end{equation}

Implied parameters $ \rho $ and $ \omega $ do not depend on the maturity, $ \tau $, but they have to be computed using values of $\mathbb{S}_t$ and $\mathbb{C}_t$, which were calculated on the volatility smile for options with the shortest time to maturity.

Next, on the basis of observed implied volatilities of traded options it can be assumed that the term premium, $\mathbb{M}_t = \tfrac{1}{\tau_{2}-\tau_{1}}(\sigma^{\mathrm{ATM}}(\tau_{2})-\sigma^{\mathrm{ATM}}(\tau_{1}))$, where $\tau_{1}$ and $\tau_{2}$ are respectively the smallest time to maturity of traded options and next longer time to maturity. On the other hand, smile sensitivities, $\mathbb{S}_t$ and $\mathbb{C}_t$, can be calculated for options with sufficiently short time to maturity by solving the problem of linear regression with the following regressand, $y$, and the regressors, $x$ and $x^2$:

\begin{equation}\begin{split}
& y =  \mathbb{S}_t x  + \tfrac{1}{2} \mathbb{C}_t x^2 ,\\
& y = \sigma(t,T,K)-\sigma^{ATM}(t,T) ,\\
& x = \bigg( \frac{K}{S_t}-1 \bigg)
.\end{split}\end{equation}

Then, knowledge of the numeric values of the characteristics $\mathbb{S}_t$, $\mathbb{C}_t$ and $\mathbb{M}_t$ enables estimation of the Heston model parameters with the approximate formula.

\subsection{Gauthier and Rivaille method}

Gauthier and Rivaille (2009)\footnote{~\bibentry{Gauthier2009}.} presented the method that is called by them "smart parameters". According to authors, this method can be used for estimation only or as a starting point for further local minimization algorithm. The study examined the robustness of the Levenberg-Marquardt algorithm during calibration of the Heston and the SABR models. The results indicated the increase in robustness after the application of smart parameters in comparison to the version without them. Calibration time has been reduced by more than 90\% compared to global minimization algorithms such as the Differential Evolution.

Gauthier and Rivaille method is based on the approximation of option prices in the Heston model by the Taylor expansion in the parameter $\omega$, which corresponds to a volatility of variance, as shown in Benhamou et al. (2010)\footnote{~\bibentry{Benhamou2010}.}. The derivation is made for a call option price, but an analogous derivation can be done for a put option price.
For logarithmic forward prices, which are defined as $x = \ln (S e^{(r_d-r_f)\tau}) $, and the Black-Scholes price of put option, $ P_{BS} $, the authors approximate the put option price from the Heston model as: 

\begin{equation}\begin{split}
P_{Hes} = A + B \omega^2 + C \rho \omega + D \rho^2 \omega^2
,\end{split}\end{equation}

where:

\begin{equation}\begin{split}
& A = P_{BS}(x,w_{\tau}), \quad B = (\nu_0 r_0 + \theta r_1)\frac{\partial^2 P_{BS}}{\partial w_{\tau}^2}, \quad C = (\nu_0 p_0 + \theta p_1)\frac{\partial^2 P_{BS}}{\partial x \partial w_{\tau}} ,\\
& D = \bigg[ (\nu_0 q_1 + \theta q_1)\frac{\partial^3 P_{BS}}{\partial x^2 \partial w_{\tau}} + \frac{1}{2}(\nu_0 p_0 + \theta p_1)\frac{\partial^4 P_{BS}}{\partial x^2 \partial w_{\tau}^2} \bigg] ,\\
& w_{\tau} = \theta + (\nu_0-\theta)\bigg( \frac{1-e^{\kappa \tau}}{\kappa} \bigg)
.\end{split}\end{equation}

In the considered equation $w_{\tau}$ represent the Black-Scholes variance, i.e. $w_{\tau}=\sigma_{BS}^2 \tau$. The remaining parameters are:

\begin{equation}\begin{split}
& r_0 = \tfrac{1}{4} \kappa^{-3} (- 4 e^{-\kappa\tau}\kappa\tau + 2 - 2 e^{-2\kappa\tau}) ,\\
& r_1 = \tfrac{1}{4} \kappa^{-3} (4  e^{-\kappa\tau}(\kappa\tau+1) + (2\kappa\tau-5)+ e^{-2\kappa\tau}) ,\\
& p_0 = \kappa^{-2} (-e^{-\kappa\tau}\kappa\tau+1-e^{-\kappa\tau}) ,\\
& p_1 = \kappa^{-2} (e^{-\kappa\tau}\kappa\tau+(\kappa\tau-2)+2e^{-\kappa\tau}) ,\\
& q_0 = \tfrac{1}{2} \kappa^{-3} (-e^{-\kappa\tau}\kappa\tau(\kappa\tau+2)+2-2e^{-\kappa\tau}) ,\\
& q_1 = \tfrac{1}{2} \kappa^{-3} (2(\kappa\tau-3)+e^{-\kappa\tau}\kappa\tau(\kappa\tau+4)+6e^{-\kappa\tau})
.\end{split}\end{equation}

The whole idea of smart parameters is constructed on the fact that using prices, $ P(K_1) $ and $ P(K_2) $, of two traded put options of the same maturity $ \tau $ and different strike prices, $ K_1 $ and $ K_2 $, the formula for the approximated Heston model option price can be used twice, i.e. once for each option, resulting in following system of equations:

\begin{equation}\begin{split}
& P(K_1) = A(K_1) + B(K_1) \sigma^2 + C(K_1) \rho \sigma + D(K_1) \rho^2 \sigma^2 ,\\
& P(K_2) = A(K_2) + B(K_2) \sigma^2 + C(K_2) \rho \sigma + D(K_2) \rho^2 \sigma^2
.\end{split}\end{equation}

As a result a system of equations with two unknowns, $ \rho $ and $ \omega $, is obtained. This system can be solved analytically or by numerical minimization algorithm, assuming some fixed values for other parameters of the Heston model, i.e. $ \kappa $, $ \theta $ and $ \nu_0 $. The set of solutions of these equations has four elements, of which only one is valid in the context of the Heston model, i.e. $|\rho| \leq 1$ and $\omega > 0$.


The analytic solution of this system is as follows\footnote{~The authors of the method did not provide an analytic solution, the author of this work presents his results.}:

\begin{equation}\begin{split}
\omega & = \sqrt{\tfrac{g^2}{2 D_1 f^2}(-B_1-C_1 \tfrac{f}{g} - 2 D_1 d \tfrac{f}{g} + \sqrt{(B_1+C_1 \tfrac{f}{g} + 2 D_1 d \tfrac{f}{g})^2 - 4 D_1 \tfrac{f^2}{g^2} (A_1 + C_1 \tfrac{d}{g} + D_1 \tfrac{d^2}{g^2})})} ,\\
\rho & = \tfrac{d}{g} \tfrac{1}{\omega} + \tfrac{f}{g}\omega
,\end{split}\end{equation}

where: $d = \tfrac{D_1 A_2}{D_2}-A_1$, $f = \tfrac{D_1 B_2}{D_2}-B_1$, $g = C_1 - \tfrac{D_1 C_2}{D_2}$, $A_i = A(K_i)$, $B_i = B(K_i)$, $C_i = C(K_i)$, $D_i = D(K_i)$, $i=1,2$.

\section{Implied Central Moments Method}

For stochastic volatility models with an explicit formula for the characteristic function, e.g. the Heston model, basing on the theorem of the characteristic function and moments we can calculate $ n $-th central moment of underlying instrument price process, if the moment exists and is finite. However, the formulas of those moments for the characteristic function of the degree of complexity such as in the Heston model, have to be obtained in a very tedious way. Hence, the Implied Central Moments approximation method is based on relatively more simple formulas for other moments namely: the variance of stochastic variance and correlation of stochastic variance and log-returns of underlying. Both of these moments have to be computed in two forms. First, as a theoretical value for given stochastic volatility model. Second, as an approximated value of the function that is computed from market implied values of central moments of underlying returns.

\subsection{Power payout portfolios}

Power payout portfolios are replication of expected values of consecutive powers (1st, 2nd, 3rd and 4th) of underlying returns. The replication is done with use of option portfolios. They are equivalents of market implied values of standard moments of returns. Same as in the case of the portfolio that is replicating the realized variance, such portfolios can be replicated by the delta hedging strategy for portfolios of at-the-money and out-of-the-money options. Carr and Madan (2002)\footnote{~\bibentry{Carr2002}.} showed that every continuous payout may be reduced to a sum of option payouts. This is sufficient proof that the power returns can be replicated by the portfolio of options. In addition, for the portfolio that is replicating realized variance it is necessary to also use the Ito lemma.\par

Bakshi, Kapadia and Madan (2000)\footnote{~\bibentry{Bakshi2003}.} derived formulas for expected values of contracts with power payout using integrals of traded option prices and assuming log-returns, $R=\ln(S_{t+1}/S_t)$:

\begin{equation}\begin{split}
\label{eq:powerPayout}
V(t,\tau) = \mathbb{E}[e^{-r_d\tau}R(t,\tau)^2] =  & \int_{S(t)}^\infty \frac{2(1-\ln[\tfrac{K}{S(t)}])}{K^2} C(t,\tau,K) \,\D K + \int_0^{S(t)} \frac{2(1+\ln[\tfrac{S(t)}{K}])}{K^2} P(t,\tau,K) \,\D K ,\\
W(t,\tau) = \mathbb{E}[e^{-r_d\tau}R(t,\tau)^3] = & \int_{S(t)}^\infty \frac{6\ln[\tfrac{K}{S(t)}]-3(\ln[\tfrac{K}{S(t)}])^2}{K^2} C(t,\tau,K) \,\D K \\ & - \int_0^{S(t)} \frac{6\ln[\tfrac{S(t)}{K}]+3(\ln[\tfrac{S(t)}{K}])^2}{K^2} P(t,\tau,K) \,\D K ,\\
X(t,\tau) = \mathbb{E}[e^{-r_d\tau}R(t,\tau)^4] = & \int_{S(t)}^\infty \frac{12(\ln[\tfrac{K}{S(t)}])^2-4(\ln[\tfrac{K}{S(t)}])^3}{K^2} C(t,\tau,K) \,\D K \\ & + \int_0^{S(t)} \frac{12(\ln[\tfrac{S(t)}{K}])^2+4(\ln[\tfrac{S(t)}{K}])^3}{K^2} P(t,\tau,K) \,\D K
.\end{split}\end{equation}

In addition, after equating the expected return definition with the forward price, to the Taylor expansion of the function $ \exp(R) $, we can compute $\mathbb{E}[R(t,\tau)]$:

\begin{equation}
\mathbb{E}[R(t,\tau)] = \mu(t,\tau) = e^{(r_d-r_f)\tau} - 1 - \tfrac{1}{2}e^{-r_d\tau}V(t,\tau) - \tfrac{1}{6}e^{-r_d\tau}W(t,\tau) - \tfrac{1}{24}e^{-r_d\tau}X(t,\tau)
.\end{equation}

The formulas (\ref{eq:powerPayout}) for the value of the contract with power payout can be written also in the discrete version. Although this approximation will have a discretization bias, Dennis and Mayhew (2002)\footnote{~\bibentry{Dennis2002}.} showed that such biases are usually small, even with a small set of put and call options. Therefore, the functions of expected value of non-discounted power contracts, which are sometimes also called portfolios of power payout, can be approximated in the following way with use of integration method that is similar to the trapezoidal rule:

\begin{equation}\begin{split}
P_1 & = \mathbb{E}[R(t,\tau)] = e^{(r_d-r_f)\tau} - 1 - \tfrac{1}{2} P_2 - \tfrac{1}{6} P_3 - \tfrac{1}{24} P_4 + \epsilon_1(\tfrac{F_0}{K_0}) ,\\
P_2 & = \mathbb{E}[R(t,\tau)^2] = e^{r_d\tau} \bigg(\sum_i \frac{2}{K_i^2} \bigg(1-\ln \bigg(\frac{K_i}{F_0} \bigg) \bigg) Q(K_i) \Delta K_i \bigg) + \epsilon_2(\tfrac{F_0}{K_0}) ,\\
P_3 & = \mathbb{E}[R(t,\tau)^3] = e^{r_d\tau} \bigg(\sum_i \frac{3}{K_i^2} \bigg(2\ln \bigg(\frac{K_i}{F_0} \bigg) - \ln^2 \bigg(\frac{K_i}{F_0} \bigg) \bigg) Q(K_i) \Delta K_i \bigg) + \epsilon_3(\tfrac{F_0}{K_0}) ,\\
P_4 & = \mathbb{E}[R(t,\tau)^4] = e^{r_d\tau} \bigg(\sum_i \frac{4}{K_i^2} \bigg(3\ln^2 \bigg(\frac{K_i}{F_0} \bigg) - \ln^3 \bigg(\frac{K_i}{F_0} \bigg) \bigg) Q(K_i) \Delta K_i \bigg) + \epsilon_4(\tfrac{F_0}{K_0})
,\end{split}\end{equation}

where:

\begin{itemize}
\item $ F_0 $ is the forward price of the underlying asset and is calculated from option prices,
\item $ K_0 $ is the first strike price below $ F_0 $,
\item $ K_i $ is the strike price of $ i $-th out-of-the-money option and concerns a call option for $ K_i>K_0 $ and a put option for $ K_i < K_0 $,
\item $ \Delta K_i $ is half of the difference between strike prices which are above and below the $ K_i $ price, i.e. $ \Delta K_i = 0.5 (K_{i + 1} -K_{i-1}) $, in the case of the highest (lowest) strike price $ \Delta K_i $ is a simple absolute difference between this price and the first price below (above) it,
\item $ Q (K_i) $ is the mid price of the available bid-ask spread for options with an exercise price of $ K_i $,
\item $ r_d $ is the domestic risk-free short interest rate,
\item $ r_f $ is the foreign risk-free short interest rate,
\item $ \tau $ it is the time to maturity of the option and is expressed as a fraction of the year,
\item $\epsilon_j(\tfrac{F_0}{K_0})$ is a factor for correction of the difference between the $ K_0 $ and $ F_0 $.\footnote{~CBOE Stock Exchange in its SKEW Index methodology uses the following correction factors: $\epsilon_1 = - \bigg(1 + \ln\bigg( \frac{F_0}{K_0} \bigg) - \frac{F_0}{K_0} \bigg)$, $\epsilon_2 = 2 \ln\bigg( \frac{K_0}{F_0} \bigg) \bigg(\frac{F_0}{K_0} - 1\bigg) + \frac{1}{2} \ln^2\bigg( \frac{K_0}{F_0} \bigg)$, $\epsilon_3 = 3 \ln^2\bigg( \frac{K_0}{F_0} \bigg) \bigg(\frac{1}{3} \ln\bigg( \frac{K_0}{F_0} \bigg) - 1 + \frac{F_0}{K_0} \bigg)$.}
\end{itemize}

Similarly defined functions of power payout portfolios are part of the CBOE exchange methodology that is used to calculate the SKEW index\footnote{~\bibentry{CBOE2010}.} \footnote{~The difference between the discretization of the version presented by Bakshi et al (2003) and the methodology of the CBOE is the value of $ P_1 $, which in CBOE version is equal to $P_1 = e^{r_d\tau} \bigg(- \sum_i \frac{1}{K_i^2} Q(K_i) \Delta K_i \bigg) + \epsilon_1(\tfrac{F_0}{K_0})$.}. It is worth noting that these functions can be generalized also for the foreign exchange OTC options, in this case, there are always five strike prices, $ K_i $, two for OTM call options, two for OTM put options and one for call (or sell) ATM option (i.e. $K_0=K_{\mathrm{ATMF}}$). The latter option has different value of $ Q(K_i) $ that is equal to the arithmetic average of call and put options for the $ K $ price. $ F_0 $ is the price of foreign exchange forward transaction, and $ K_0 $ is equal to the strike price of the ATM option. It is worth noting that $ F_0 \neq K_0 $ due to the fact that $ K_0 $ is equal to a forward price corrected with the level of the underlying price volatility to make the option delta equal to 0.5, but it is safe to assume that $ F_0 $ is close enough to $ K_0 $ in the case of the currency options, so we can skip adding the correction factors, because $\epsilon_j(\tfrac{F_0}{K_0}) \approx 0$.

\subsection{Implied central moments}

Using the knowledge related to the valuation of power payout portfolios, which allows computation of the ordinary moments of returns basing on option prices, we can also compute Implied Central Moments, $\mathbb{E}[(R-\mu)^2]$, $\mathbb{E}[(R-\mu)^3]$, $\mathbb{E}[(R-\mu)^4]$, as well as implied skewness and kurtosis.

\begin{equation}\begin{split}
\mu_2 = \mathbb{E}[(R-\mu)^2] & = \mathbb{E}(R^2)-2\mu \mathbb{E}(R) + \mu^2 ,\\
\mu_3 = \mathbb{E}[(R-\mu)^2] & = \mathbb{E}(R^3)-3\mu \mathbb{E}(R^2) + 3\mu^2 \mathbb{E}(R) - \mu^3 ,\\
\mu_4 = \mathbb{E}[(R-\mu)^2] & = \mathbb{E}(R^4) - 4 \mu \mathbb{E}(R^3) + 6 \mu^2 \mathbb{E}(R^2) - 4 \mathbb{E}(R) \mu^3 + \mu^4
,\end{split}\end{equation}

where: $\mu_2$, $\mu_3$ and $\mu_4$ are respectively second, third and fourth central moment. Then, using the fact that $\mu = \mathbb{E}(R)$ and previously defined expected values of power payout portfolios it can be written that:

\begin{equation}\begin{split}
\mathbb{E}[(R-\mu)^2] & = \mathbb{E}(R^2)-\mathbb{E}^2(R) = P_2 - P_1^2 , \\
\mathbb{E}[(R-\mu)^3] & = \mathbb{E}(R^3)-3 \mathbb{E}(R) \mathbb{E}(R^2) + 2 \mathbb{E}(R)^3 = P_3 - 3 P_1 P_2 + 2 P_1^3 ,\\
\mathbb{E}[(R-\mu)^4] & = \mathbb{E}(R^4) - 4 \mathbb{E}(R) \mathbb{E}(R^3) + 6 \mathbb{E}(R)^2 \mathbb{E}(R^2) - 3 \mathbb{E}(R)^4 = P_4 - 4 P_1 P_3 + 6 P_1^2 P_2 - 3 P_1^4
.\end{split}\end{equation}

To compute normalized central moments we have to divided them by appropriate power of standard deviation. By definition, standard deviation of returns is defined as:
\begin{equation}
\sigma = \sqrt{\mathbb{E}[(R-\mu)^2]} = \sqrt{P_2 - P_1^2}
.\end{equation}

Standard deviation calculated in this particular way is not normalized to the form of the annual volatility. To make that happen we must normalize the functions of $ P_1 $ and $ P_2 $ multiplying each of them by $ \frac{1}{T} $. However, the CBOE calculate its index of implied variance for options on the S\&P500 using a simpler, standardized by time to maturity formula\footnote {~\bibentry{CBOE2014}.}.

Using the implied standard deviation, $\sigma$, skewness and kurtosis of returns are equal to:
\begin{equation}\begin{split}
\mathbb{S} & = \frac{\mu_3}{\sigma^3} = \frac{P_3 - 3 P_1 P_2 + 2 P_1^3}{(P_2 - P_1^2)^{3/2}} ,\\
\mathbb{C} & = \frac{\mu_4}{\sigma^4} = \frac{P_4 - 4 P_1 P_3 + 6 P_1^2 P_2 - 3 P_1^4}{(P_2 - P_1^2)^{2}}
.\end{split}\end{equation}

For the options market and prices of options with given maturity date, we can calculate the implied skewness and the implied kurtosis for this market. In this way, in the case of the skewness index CBOE calculates skew for options on the S\&P500 by making the transformation of considered skewness to the form of $\mathrm{SKEW} = 100 - 10\mathbb{S}$. Currently CBOE is not maintaining any index related to the kurtosis.

In addition to the use of implied skewness by the CBOE stock exchange, Guilliame and Schoutens (2013)\footnote{~\bibentry{Guillaume2013}.} present the concept of calculating of implied skewness and kurtosis in order to use them for model calibration by matching them to the theoretical moments of the following distributions: Variance Gamma, Normal Reverse Gaussian and Meixner. All mentioned distributions can be used as the distribution of the Levy jump processes. However, in their paper stochastic volatility diffusion models were not investigated, in particular there was no connection to the Heston and the Sch\"obel-Zhu model, nor the derivation of analytic formulas for approximation of individual model parameters.

\subsection{Calibration of variance term structure}
\label{sec:4.4.3}

In various papers on the implied variance, as well as the VIX index \footnote{~\bibentry{CBOE2014}.}, annualized integrated instantaneous variance from a stochastic volatility model is identified with the implied variance. This approach is used to calibrate the Heston model parameters $ \nu_0 $, $ \theta $, and $ \kappa $ by formulating error function that contains squared differences between the integrated instantaneous variance and the implied variance. However, it should be noted that the variance of the stochastic volatility model, such as the Heston model, still contains additional components depending on the parameters $\rho $ and $ \omega $:

\begin{equation}
\mathbb{E}[(R-\mathbb{E}[R])^2] = \int_t^T \mathbb{E}(\upsilon_u) \D u - \rho \omega \int_t^T \tfrac{1-e^{-\kappa(T-u)}}{\kappa} \mathbb{E}(\upsilon_u) \D u + \tfrac{\omega^2}{4} \int_t^T \tfrac{(1-e^{-\kappa(T-u)})^2}{\kappa^2} \mathbb{E}(\upsilon_u) \D u
.\end{equation}

Nevertheless, considered components are usually small for small $T$, which is noted by Zhao, Zhang and Chang (2013)\footnote{~\bibentry{Zhao2013}.}. At the same time, the omission of that small variables simplifies the form of the cost function for the calibration of the term structure of variance. On the other hand, a compromise between computational complexity and amount of information used can be obtained by using historical estimates of $ \rho $ and $ \omega $ parameters. This is especially important for longer times to maturity that are present on the currency option market. Hence, in the considered Implied Central Moments method parameters: $ \nu_0 $, $ \theta $, and $ \kappa $ for the Heston model are calibrated to the adjusted annualized total variance term structure, which takes into account the components of variance that is dependent on the $ \rho $ and $ \omega $ and is created as follows:

\begin{enumerate}
	\item Approximation of annualized implied total variance term structure for each maturity by the methodology which is an equivalent to the VIX index methodology:
	
	\begin{equation}
	\mathrm{V}_{\tau}^2 = \frac{2}{\tau} \sum_i \frac{\Delta K_i}{K_i^2} e^{r_d \tau} Q(K_i) - \frac{1}{\tau} \bigg( \frac{F_0}{K_0} - 1 \bigg)^2
	,\end{equation}
	where:
	\begin{itemize}
		
		\item $ F_0 $ is the forward price of the underlying asset and is calculated from option prices,
		\item $ K_0 $ is the first strike price below $ F_0 $,
		\item $ K_i $ is the strike price of $ i $-th out-of-the-money option and concerns a call option for $ K_i>K_0 $ and a put option for $ K_i < K_0 $,
		\item $ \Delta K_i $ is half of the difference between the strike prices which are above and below the $ K_i $ price, i.e. $ \Delta K_i = 0.5 (K_{i + 1} -K_{i-1}) $, in the case of the highest (lowest) strike price $ \Delta K_i $ is a simple absolute difference between this price and the first price below (above) it,
		\item $ Q (K_i) $ is the mid price of the available bid-ask spread for options with an exercise price of $ K_i $,
		\item $ r_d $ is the domestic risk-free short interest rate,
		\item $ \tau $ it is the time to maturity of the option expressed as a fraction of the year,
	\end{itemize}
	
	\item Computation of the historical correlation between the implied variance and the logarithms of the price and historical volatility of implied variance using the exponential moving average method (with decay parameter $\lambda=1-\tfrac{1}{63}$) on the implied variance time series, which is constructed from option with the closest maturity (1M),
	\item Approximation of $ \nu_0 $ by the value of the implied variance for the closest maturity (1M),
	\item Computation of the adjusted term structure of the annualized implied total variance, which is derived from the implied variance term structure by subtraction of a vector which is the limit of previously considered formula for $ \kappa \to 0 $:
	\begin{equation}
	\widetilde{\mathrm{V}}_{\tau}^2 = \mathrm{V}_{\tau}^2 (1 - \tfrac{1}{2} \rho_H \omega_H \tau + \tfrac{\omega_H^2}{12} \tau^2)
	,\end{equation}
	\item The numerical calibration of parameters $\nu_0$, $\theta$, $\kappa$ using equality $\widetilde{\mathrm{V}}_{\tau}^2 = \tfrac{1}{\tau} \mathbb{E}[\upsilon_{\tau}]$, where $\mathbb{E}[\upsilon_{\tau}]$ is theoretical expected value of total variance in the Heston model.
\end{enumerate}

In the case of the Sch\"obel-Zhu model and other similar models, which have the equation of volatility dynamics instead of variance dynamics, there are two possible procedures of calibration of parameters $\nu_0$, $\theta$ and $\kappa$. 

The first one is the calibration of the variance term structure to the implied variance. Variance term structure in the Sch\"obel-Zhu is given by much more complicated equation, which is derived in Appendix B, than the standard deviation (volatility) term structure equation. The annualized total integrated variance is given by the following equation:

\begin{equation}
\mathbb{E}[\tfrac{1}{\tau}\upsilon_T] = \frac{e^{-2 \kappa \tau} (e^{2 \kappa \tau} (\theta^2 (2 \kappa \tau-3)+2 \theta \nu_0+ \tfrac{\omega^2}{2\kappa} (2 \kappa \tau-1)+\nu_0^2)+4 \theta (\theta-\nu_0) e^{\kappa \tau}-(\theta-\nu_0)^2+ \tfrac{\omega^2}{2\kappa} )}{2 \kappa \tau}
.\end{equation}

The parameter $\omega$ has to be estimated before the procedure, e.g. it can be obtained from its historical estimator from the implied variance time series. The minimized function is more complicated in the first approach, but there is less assumptions than in the second approach.

The second possible approach is a calibration of the term structure of the standard deviation to square root of implied variance multiplied by a convexity adjustment. It should be noted that the latter term is not synonymous with implied volatility observed in the market. Starting from the corrected implied variance, same as for Heston model case, the expected volatility is obtained from the following equation:

\begin{equation}\begin{split}
\mathbb{E}\bigg[\sqrt{\tfrac{1}{\tau}\bar{\upsilon}}\bigg] & = \sqrt{\mathbb{E}[\tfrac{1}{\tau}\bar{\upsilon}]}\bigg(1-\tfrac{\sigma_{\bar{\upsilon}/\tau}^2}{8\mathbb{E}[\tfrac{\bar{\upsilon}}{\tau}]^2}\bigg) ,\\
\sigma_{\bar{\upsilon}/\tau}^2 & = \tfrac{\omega^2}{2 \tau^2 \kappa^3} (2 (\nu_0-\theta) (1-2 \kappa \tau e^{-\kappa \tau} - e^{-2 \kappa \tau}) + \theta (4 e^{-\kappa \tau}-3+2 \kappa \tau- e^{-2 \kappa \tau}))
,\end{split}\end{equation}

where the second order Taylor expansion of the function $\sqrt{x}$ has been used to obtain the main part of the formula. Then, the formula for the expected value of variance of variance in the Heston model from Broadie (2008)\footnote{~\bibentry{Broadie2008}.} was applied\footnote{~The formula can be derived by calculating the second moment of the Laplace transform of the total variance.}. All parameters in considered equation refers to the Heston model. The parameter $\omega$ is known from its historical estimator. For the parameters $\nu_0$, $\theta$ and $\kappa$ their values from Heston term structure calibration can be used.

\subsection{Application of Implied Central Moments to other parameters}

The beginning of the considerations is the stochastic volatility model of the following form:

\begin{equation}\begin{split}
& \D S_t = \mu S_t \D \, t + \nu_t^p S_t \D \, W_t ^S ,\\
& \D \nu_t = \kappa a(\nu_t) \D \, t + \omega b(\nu_t) \D \, W_t^\nu ,\\
& \langle \D \, W_t^S , \D \, W_t^\nu \rangle = \rho \D \, t
,\end{split}\end{equation}




where $p$ is a parameter which can be equal to 1 or $\tfrac{1}{2}$. Next, using the It\^{o}'s lemma $ \D S_t $ can be converted into:

\begin{equation}
\D \ln S = (\mu-\tfrac{1}{2} \nu_t^{2p}) \D \, t + \nu_t^p \D \, W_t^S
.\end{equation}

On this basis, we can compute integrated form of $\ln S$, $\nu_t$ and $\nu_t^2$.  Computation of the latter element makes sense only when $p=1$, e.g. in case of the Sch\"obel-Zhu model\footnote{~For the Sch\"obel-Zhu model $g_1(\nu_t) = 2 \nu_t $.}:

\begin{equation}\begin{split}
& \ln S_T = \ln S_t + \mu (T-t) - \tfrac{1}{2} \int_t^T \nu_u^{2p} \D \, u + \int_t^T \nu_u^p \D \, W_u ^S ,\\
& \nu_T = \nu_t + \kappa \int_t^T  f(\nu_u) \D \, u + \omega \int_t^T  g_{1/2}(\nu_u) \D \, W_u^\nu ,\\
& \nu_T^2 = \nu_t^2 + \kappa \int_t^T h(\nu_u) \D \, u + \omega \int_t^T  g_1(\nu_u) \D \, W_u^\nu
.\end{split}\end{equation}

Thus, the rates of return, $ R_T $, are equal to:

\begin{equation}
R_T = \ln(S_T/S_t) = \mu (T-t) - \tfrac{1}{2} \int_t^T \nu_u^{2p} \D \,u + \int_t^T \nu_u^p \D \,W_u^S = \mu \tau -\tfrac{1}{2} \bar{\upsilon}_{t,T} + X_T
,\end{equation}

where: $\tau=T-t$, $\bar{\upsilon}_{t,T} = \int_t^T \nu_u^{2p} \D \,u$ and $X_T = \int_t^T \nu_u^p \D \,W_u^S$.

In the case of $\bar{\upsilon}_{t,T}$ process\footnote{~Variable $\upsilon_t$ was introduced to maintain the consistency. It is always equal to instantaneous variance, regardless of allowed $p$ parameter value, $p=1$ or $p=\tfrac{1}{2}$. In the same manner, $\bar{\upsilon}_{t,T}$ is integrated instantaneous variance function.} we can decompose it to stochastic and deterministic components, namely:

\begin{equation}
\bar{\upsilon}_{t,T} = \int_t^T \mathbb{E}[\nu_u^{2p}] \D \,u +  \int_t^T (\nu_u^{2p}-\mathbb{E}[\nu_u^{2p}]) \D \,u = \mathbb{E}[\bar{\upsilon}_{t,T}] + Y_T
,\end{equation}

where: $Y_T = \int_t^T \omega \int_t^u  g_p(\nu_s) \D \, W_s^\nu \D \,u$ and $\mathbb{E}[\bar{\upsilon}_{t,T}] = \int_t^T (\nu_t + \kappa \int_t^u  f(\nu_s) \D \, s) \D \,u $.

\subsubsection{Approximated implied values of moments}

With rates of return that are defined same as previously, centralized returns, $ \hat{R}_T $, can be expressed as the sum of two martingales\footnote{~Such decomposition of returns was also used by Zhang et al. (2016) for calculation of implied skewness in the Heston model.}:

\begin{equation}
\hat{R}_T = R_T-\mathbb{E}[R_T] = -\tfrac{1}{2} Y_T + X_T
.\end{equation}

Derivation of approximate formula for parameters $\omega$ and $\rho$ parameters is possible through equating theoretical values of two moments: $\mathbb{E}[Y_T^2]$ and $\mathbb{E}[X_T Y_T]$, with their approximated implied values, which can be calculated using $\mathbb{E}[\hat{R}_T^n]$ and $\mathbb{E}[\bar{\upsilon}_{t,T}]$.

Then, we can observe that:

\begin{equation}
\hat{R}_T^2 = (X_T -\tfrac{1}{2} Y_T)^2 = X_T^2 - X_T Y_T + \tfrac{1}{4} Y_T^2
,\end{equation}

and making another substitution that is enabled by the assumption about decomposition of centralized rates of return we obtain:

\begin{equation}
\hat{R}_T^2 = X_T^2 - (\hat{R}_T+\tfrac{1}{2} Y_T) Y_T + \tfrac{1}{4} Y_T^2 = X_T^2 - \hat{R}_T Y_T - \tfrac{1}{4} Y_T^2
.\end{equation}

Moreover, using the It\^{o} integral, the component $X_T^2$ can be further transformed into:

\begin{equation}\begin{split}
X_T^2 = & \int_t^T \D\,(X_u^2) = \int_t^T (\D\,X_u)^2 + 2 \int_t^T X_u \D\,X_u \\
= & \int_t^T \nu_u^{2p} \D\,u + 2 \int_t^T X_u \D\,X_u = \mathbb{E}[\bar{\upsilon}_{t,T}] + Y_T + 2 \int_t^T X_u \D\,X_u
.\end{split}\end{equation}

which in turn gives the following equation:

\begin{equation}
\hat{R}_T^2 = \mathbb{E}[\bar{\upsilon}_{t,T}] + Y_T + 2 \int_t^T X_u \D\,X_u - \hat{R}_T Y_T - \tfrac{1}{4} Y_T^2
,\end{equation}

which can be represented as a quadratic equation of $Y_T$:

\begin{equation}
\tfrac{1}{4} Y_T^2 + Y_T(\hat{R}_T - 1) + \hat{R}_T^2 - \mathbb{E}[\bar{\upsilon}_{t,T}] - 2 Z_T = 0
,\end{equation}

where $Z_T=\int X_u \D\,X_u$. Considered equation has two solutions, of which the smaller one is appropriate and has the form:

\begin{equation}
Y_T = 2 - 2 \hat{R}_T - 2 \sqrt{1 + \mathbb{E}[\bar{\upsilon}_{t,T}] - 2 \hat{R}_T + 2 Z_T} = 2(1 - \hat{R}_T - f(\tilde{R}_T))
.\end{equation}

where: $f(\tilde{R}_T)=\sqrt{1+\mathbb{E}[\bar{\upsilon}_{t,T}]-2 \tilde{R}_T}$, $\tilde{R}_T=\hat{R}_T - Z_T$ and $\mathbb{E}[\tilde{R}_T]=0$.

Taylor expansion for $f(\tilde{R}_T=0)$ is equal to:

\begin{equation}\begin{split}
f(\tilde{R}_T) = & A - \tilde{R}_TA^{-1} - \tfrac{1}{2}\tilde{R}_T^2 A^{-3} + \mathcal{O}(\tilde{R}_T^3) \\
= & A^{-3} (A^4 - A^2 \hat{R}_T - \hat{R}_T^2 + A^2 Z_T + 2 \hat{R}_T Z_T - Z_T^2) + \mathcal{O}((\hat{R}_T-Z_T)^3)
,\end{split}\end{equation}

where $A=\sqrt{1+\mathbb{E}[\bar{\upsilon}_{t,T}]}$. 

Therefore, sought expected values: $\mathbb{E}[Y_T^2]$ and $\mathbb{E}[X_T Y_T]$, are equal to:

\begin{equation}\begin{split}
\mathbb{E}[Y_T^2] & = \mathbb{E}[4(1 - \hat{R}_T - A^{-3} (A^4 - A^2 \hat{R}_T - \hat{R}_T^2 + A^2 Z_T + 2 \hat{R}_T Z_T - Z_T^2) + \mathcal{O}((\hat{R}_T-Z_T)^3))^2] \\
\mathbb{E}[X_T Y_T] & = \mathbb{E}[\hat{R}_T Y_T] + \tfrac{1}{2}\mathbb{E}[Y_T^2], \\
\mathbb{E}[\hat{R}_T Y_T] & = \mathbb{E}[2\hat{R}_T(1 - \hat{R}_T - A^{-3} (A^4 - A^2 \hat{R}_T - \hat{R}_T^2 + A^2 Z_T + 2 \hat{R}_T Z_T - Z_T^2) + \mathcal{O}((\hat{R}_T-Z_T)^3))]
.\end{split}\end{equation}

First of all it is known that $\mathbb{E}[Z_T] = 0$ and $\mathbb{E}[\hat{R}_T] = 0$. Secondly, because values of $\mathbb{E}[Z_T^n]$, $n \geq 2$, and $\mathbb{E}[Z_T^n R_T^k]$, $n,k \geq 1$ are relatively small, they can be approximated by 0. On the other hand, those approximated elements are dependent from $\rho$ and $\omega$ itself and their analytic formulas depend on the value of $p$ and the form of $g_p$.

Therefore, approximated expected values are equal to:

\begin{equation}\begin{split}
\mathbb{E}[Y_T^2] \approx & A^{-6} (4 A^8-8 A^7+4 A^6+ (4 A^6-8 A^5+4 A^3) \mathbb{E}[\hat{R}_T^2]+(4 A^2-4 A^3) \mathbb{E}[\hat{R}_T^3]+\mathbb{E}[\hat{R}_T^4]) ,\\
\mathbb{E}[X_T Y_T] \approx & A^{-6} (2 A^8-4 A^7+2 A^6+ (2 A^3-2 A^5) \mathbb{E}[\hat{R}_T^2]+(2 A^2-A^3) \mathbb{E}[\hat{R}_T^3]+\tfrac{1}{2} \mathbb{E}[\hat{R}_T^4])
,\end{split}\end{equation}

where $A=\sqrt{1+\mathbb{E}[\bar{\upsilon}_{t,T}]}$. In addition, instead of the assumed definition of $ A $, we can use its approximation, i.e. $A=\sqrt{1+\mathbb{E}[\hat{R}_T^2]}$, which uses the relationship that has been previously used for term structure calibration. After such change in the formula, it will depend only on implied moments and it will have no elements which depend on the stochastic volatility model parameters. The same approximation can also be justified by the similarity to the estimation of realized volatility by the average of squared rates of return, which was investigated empirically by Barndorff (2002)\footnote{~\bibentry{Barndorff2002}.} and McAleer (2008)\footnote{~\bibentry{McAleer2008}.}.

\subsubsection{Approximated theoretical values of moments}

On the other hand, we can calculate the theoretical expected values, $\mathbb{E}[Y_T^2]$ and $\mathbb{E}[X_T Y_T]$, for a given model of stochastic volatility and equate both versions of the expected values. Then, we can calculate implied correlation and volatility of the variance as a function of the expected values of the power payout portfolios. Derivations for the Heston model ($ p = \tfrac{1}{2} $) and the Sch\"obel-Zhu model are presented in the following section.

On the beginning, we need to go back to the definitions of $Y_T$ and $X_T$:

\begin{equation}\begin{split}
& Y_T = \int_t^T \omega \int_t^u  g_p(\nu_s) \D \, W_s^\nu \D \,u \\
& X_T = \int_t^T \nu_u^p \D \,W_u^S 
.\end{split}\end{equation}

More precisely, for the Heston model $ Y_T $ and $ X_T $ are equal to:

\begin{equation}\begin{split}
& Y_T = \int_t^T \omega \int_t^u e^{-\kappa(u-s)} \sqrt{\nu_s} \D \, W_s^\nu \D \,u = \omega \int_t^T \int_s^T e^{-\kappa(u-s)} \D \,u \sqrt{\nu_s} \D \, W_s^\nu = \omega \int_t^T \frac{1-e^{-\kappa(T-s)}}{\kappa} \sqrt{\nu_s} \D \, W_s^\nu \\
& X_T = \int_t^T \sqrt{\nu_u} \D \,W_u^S 
,\end{split}\end{equation}

At this step $ Y_T $ is not martingale. Zhang et al. (2016)\footnote{~\bibentry{Zhang2016}.} have showed how this problem can be tackled by introducing a new process, $ Y_u^* $, for which it holds $ Y_T^* = Y_T $ and which is defined as:

\begin{equation}
Y_u^* = \omega \int_t^u \frac{1-e^{-\kappa(T-s)}}{\kappa} \sqrt{\nu_s} \D \, W_s^\nu
.\end{equation}

The process  $ Y_u^* $ is a martingale, because the function which is weighing the volatility in the integral does not depend on $ u $. Therefore moments $\mathbb{E}[Y_T^2]$ and $\mathbb{E}[X_T Y_T]$ are equal:

\begin{equation}\begin{split}
\mathbb{E}[X_T Y_T] & = \mathbb{E}[X_T Y_T^*] = \mathbb{E} \int_t^T \D (X_s Y_s^*) = \mathbb{E} \int_t^T Y_s^* \D X_s + X_s \D Y_s^* + \D X_s \D Y_s^* \\
& = \mathbb{E} \int_t^T \D X_s \D Y_s^* = \rho \omega \int_t^T \frac{1-e^{-\kappa(T-s)}}{\kappa} \mathbb{E}[\nu_s] \D s = \rho \omega I_1, \\
\mathbb{E}[Y_T^2] & = \mathbb{E}[Y_T^{*2}] = \mathbb{E} \int_t^T \D (Y_s^{*2}) = \mathbb{E} \int_t^T 2 Y_s^* \D Y_s^* + (\D Y_s^*)^2 \\
& = \mathbb{E} \int_t^T (\D Y_s^*)^2 = \omega^2 \int_t^T \frac{(1-e^{-\kappa(T-s)})^2}{\kappa^2} \mathbb{E}[\nu_s] \D s = \omega^2 I_2
.\end{split}\end{equation}

where $\mathbb{E}[\nu_s] = \mathbb{E}[\upsilon_s] = \theta + (\nu_t - \theta)e^{-\kappa(s-t)}$. Whereas the integrals $ I_1 $ and $ I_2 $ are equal to:

\begin{equation}\begin{split}
I_1 & = \int_t^T \frac{1-e^{-\kappa(T-s)}}{\kappa} \mathbb{E}[\nu_s] \D s = \kappa^{-2} (e^{-\kappa \tau}(2\theta-\nu_t-\kappa \tau(\nu_t-\theta))+(\nu_t+\theta(\kappa \tau-2))), \\
& \approx \mathbb{E}[\hat{R}^2_{T}]\frac{\theta \kappa^2 \tau^2 e^{\kappa \tau} - 2 \theta \kappa \tau e^{\kappa \tau} + 2 \theta e^{\kappa \tau} - 2 \theta + 2 \kappa \tau \nu_t e^{\kappa \tau} - 2 \nu_t e^{\kappa \tau} + 2 \nu_t}{2 \kappa (-\theta e^{\kappa \tau} + \theta \kappa \tau e^{\kappa \tau} + \theta + \nu_t e^{\kappa \tau} - \nu_t)} \xrightarrow[\kappa \to 0]{} \tfrac{\tau}{2} \mathbb{E}[\hat{R}^2_{T}] ,\\
I_2 & = \int_t^T \frac{(1-e^{-\kappa(T-s)})^2}{\kappa^2} \mathbb{E}[\nu_s] \D s = \tfrac{(\theta (2 \kappa \tau - 5) + 2 \nu_t) + 4 e^{-\kappa \tau} (\theta (\kappa \tau + 1) - \kappa \nu_t \tau) + (\theta - 2 \nu_t) e^{-2 \kappa \tau}}{2 \kappa^3} \\
& \approx \mathbb{E}[\hat{R}^2_{T}] \frac{e^{-2 \kappa \tau} (\theta - 2 \nu_t) + 4 e^{-\kappa \tau} (\theta \kappa \tau + \theta - \nu_t \kappa \tau) + \theta (2 \kappa \tau - 5) + 2 \nu_t}{2 \kappa^2 ((1 - e^{-\kappa \tau}) (\nu_t - \theta) + \theta \kappa \tau)} \xrightarrow[\kappa \to 0]{} \tfrac{\tau^2}{3} \mathbb{E}[\hat{R}^2_{T}]
,\end{split}\end{equation}

where $\tau=T-t$ and the following approximation has been used: $\mathbb{E}[\hat{R}^2_{T}] \approx \mathbb{E}[\bar{\upsilon}_{t,T}] = \theta \tau + (\nu_t-\theta)\tfrac{1-e^{-\kappa \tau}}{\kappa}$. The considered approximation introduces a dependency of the second implied central moment, while the factor multiplying it can be estimated in the limit, $ \kappa \to 0 $.

Same as with implied values of moments, the final approximation of the formula is thus made to be independent from other parameters of the model, in particular the parameter $ \kappa $, which is highly variable between subsequent calibrations. The introduction of the expected value of $\mathbb{E}[\hat{R}^2_{T}]$ preserves part of the information about the variance term structure, despite using the approximation with the limit, $ \kappa \to 0 $.

Such action despite additional approximations, reduces the calibration risk, because it immunizes the method for the wrong calibration of parameter $ \kappa $, and to a lesser extent, $ \theta $ and $ \nu_0 $, which are independently calibrated to the term structure of variance.

On the other hand, for the Sch\"obel-Zhu model because of $g_1 = 2 g_{1/2} $ the moments $\mathbb{E}[Y_T^2]$ and $\mathbb{E}[X_T Y_T]$ are equal to:

\begin{equation}\begin{split}
\mathbb{E}[X_T Y_T] & = 2 \rho \omega \int_t^T \frac{1-e^{-\kappa(T-s)}}{\kappa} \mathbb{E}[\nu_s^2] \D s ,\\
\mathbb{E}[Y_T^2] & = 4 \omega^2 \int_t^T \frac{(1-e^{-\kappa(T-s)})^2}{\kappa^2} \mathbb{E}[\nu_s^2] \D s
.\end{split}\end{equation}

However, the expected value of the instantaneous variance in the Sch\"obel-Zhu model\footnote{~The derivation of formula (\ref{eq:ExpInstVarSZ}) for $\mathbb{E}[\upsilon_t]$ in the Sch\"obel-Zhu model is presented in Appendix B.} is not only more complex than its form for the Heston model, but is also dependent of $ \omega^2 $:

\begin{equation}\begin{split}
\label{eq:ExpInstVarSZ}
\mathbb{E}[\nu_s^2] & = \mathbb{E}[\upsilon_s] = \tfrac{\omega^2}{2\kappa} + e^{-2 \kappa s}(\nu_0^2-\tfrac{\omega^2}{2\kappa}) + \theta e^{-2 \kappa s} (e^{\kappa s} - 1) (\theta (e^{\kappa s} - 1) + 2 \nu_0)
.\end{split}\end{equation}

Integral of the function that consists of the expected value of the variance will also depend on the $ \omega^2 $ component. Undesirable dependence on $ \kappa $ is in this case more difficult to avoid due to the infinite limit of the component $\tfrac{\omega^2}{\kappa}$ at $ \kappa \to 0$. Alternatively, we can consider the basic equation of the dynamics of the standard deviation with $ \kappa = 0 $, which implies that alongside $ \D t $ component will be only $ \omega^2 $ in the variance equation, so the instantaneous variance diverges to infinity for infinitely long maturities ($\mathbb{E}[\nu_{\tau}^2] = \nu_0^2 + \omega^2 \tau $). For this case the moments $\mathbb{E}[Y_T^2]$ and $\mathbb{E}[X_T Y_T]$ are equal:

\begin{equation}\begin{split}
\mathbb{E}[X_T Y_T] & = 2 \omega \rho (\omega^2 \tfrac{\tau^2}{2} + \nu_0^2 \tau) \approx 2 \omega \rho \mathbb{E}[\hat{R}_T^2] \\
\mathbb{E}[Y_T^2] & = 4 \omega^2 (\omega^2 \tfrac{\tau^2}{2} + \nu_0^2 \tau) \approx 4 \omega^2 \mathbb{E}[\hat{R}_T^2] 
,\end{split}\end{equation}

from which the parameter $ \omega^2 $ can be calculated as a solution to quadratic equation and then the parameter $ \rho $ can be calculated after substitution of the numerical value of $ \omega $:

\begin{equation}\begin{split}
\omega^2 & = -\tfrac{\nu_0^2}{\tau} + \tfrac{1}{\tau}\sqrt{\nu_0^4+ \tfrac{1}{2}\mathbb{E}[Y_T^2]} \\
\omega \rho & = \frac{\mathbb{E}[X_T Y_T]}{\omega^2 \tau^2 + 2 \nu_0^2 \tau}
.\end{split}\end{equation}

As a simplified alternative, we can use approximations that were computed for the Heston model. Namely, following relations could be used: $\rho_{SZ} \approx \rho_{H}$ and $\omega_{SZ} \approx \tfrac{1}{2} \omega_{H}$.

\subsubsection{Formulas for parameters}

To sum up the calculations from previous parts of the section, we can provide the following formulas for the Heston model:

\begin{equation}\begin{split}
\omega_{\tau}^2 & \approx \frac{A^{-6} (4 A^8-8 A^7+4 A^6+ (4 A^6-8 A^5+4 A^3) \mathbb{E}[\hat{R}_T^2]+(4 A^2-4 A^3) \mathbb{E}[\hat{R}_T^3]+\mathbb{E}[\hat{R}_T^4])}{\tfrac{1}{3} \mathbb{E}[\hat{R}_T^2] \tau^2}, \\
\rho_{\tau} \omega_{\tau} & \approx \frac{A^{-6} (2 A^8-4 A^7+2 A^6+ (2 A^3-2 A^5) \mathbb{E}[\hat{R}_T^2]+(2 A^2-A^3) \mathbb{E}[\hat{R}_T^3]+\tfrac{1}{2} \mathbb{E}[\hat{R}_T^4])}{\tfrac{1}{2} \mathbb{E}[\hat{R}_T^2] \tau}
\end{split}\end{equation}

and the following for the Sch\"obel-Zhu model:

\begin{equation}\begin{split}
\omega_{\tau}^2 & \approx -\tfrac{\nu_0^2}{\tau} + \tfrac{1}{\tau}\sqrt{\nu_0^4+ \tfrac{1}{2}\mathbb{E}[Y_T^2]} \\
\omega_{\tau} \rho_{\tau} & \approx \frac{\mathbb{E}[X_T Y_T]}{\omega^2 \tau^2 + 2 \nu_0^2 \tau} ,\\
\mathbb{E}[Y_T^2] & = A^{-6} (4 A^8-8 A^7+4 A^6+ (4 A^6-8 A^5+4 A^3) \mathbb{E}[\hat{R}_T^2]+(4 A^2-4 A^3) \mathbb{E}[\hat{R}_T^3]+\mathbb{E}[\hat{R}_T^4]) ,\\
\mathbb{E}[X_T Y_T] & = A^{-6} (2 A^8-4 A^7+2 A^6+ (2 A^3-2 A^5) \mathbb{E}[\hat{R}_T^2]+(2 A^2-A^3) \mathbb{E}[\hat{R}_T^3]+\tfrac{1}{2} \mathbb{E}[\hat{R}_T^4])
,\end{split}\end{equation}

where $A=\sqrt{1+\mathbb{E}[\hat{R}_T^2]}$ and $\mathbb{E}[\hat{R}_T^n]$ are implied central moments of returns. Because, like in some other methods, parameters in the considered formulas depend on the option time to maturity, $ \tau $, so a vector of approximations of these parameters is obtained after using the formula for options from the entire volatility surface. The length of the vector is equal to the amount of times to maturity, $ \tau $, that are observed on the market. In order to calculate the parameters $ \rho $ and $ \omega $ it is necessary to know the average value of $ \rho \omega $ and $ \omega^2 $. In this case, the median may be more appropriate than the average, due to the bigger approximation error at the long end of the variance term structure.

\section{Calibration of two-factor variance models using Implied Central Moments formulas}

\subsection{Properties of two-factor variance models}

We can also find a way to use the Implied Central Moments formulas both in the case of Bates two-factor variance model (2000) and author's OUOU model, which are models with ten parameters. To do this we need an additional method of converting the approximate formulas for $\omega$ and $\rho$ to some parameters of the two-factor variance model, because formulas for parameters of unobservable variance equations cannot be directly derived by the Implied Central Moments method. Furthermore, by introducing such new method, it will be possible to use it, irrespective of the estimation method used for $\omega$ and $\rho$.

The search for the necessary transformation should begin by taking into account the following relationships of two-factor models. One property is the summation of the variances of both equations into a value which describes the total variance of the underlying process:

\begin{equation}
\upsilon_t = \upsilon_{1,t} + \upsilon_{2,t}
.\end{equation}

In addition, it is known that the variance of the variance and the correlation between it and the rates of return of the underlying instrument prices is also stochastic, unlike in the case of stochastic volatility models with a one-dimensional variance process. The variance of the sum of variance processes, $\mathrm{Var}(\D \upsilon_t)=\upsilon_t \omega_t^2$, can be expressed by the equation that is a weighted average of the two local variances of the variances and with simplified notation ($\upsilon_k = \upsilon_{k,t}$, $\omega_k = \omega_{k,t}$, $\rho_k = \rho_{k,t}$) it has a following form:

\begin{equation}
\omega_t^2 = \frac{\upsilon_1}{\upsilon_1+\upsilon_2} \omega_1^2 + \frac{\upsilon_2}{\upsilon_1+\upsilon_2} \omega_2^2
.\end{equation}

For correlation, $\mathrm{Corr}(\D S_t/S_t,\D \upsilon_t) = \upsilon_t \rho_t \omega_t $, product of the correlation and the variance of the variance can be similarly expressed as a weighted average of two products of the local volatilities of variances and correlations. This will result in the following formula for the parameter $\rho_t$:

\begin{equation}
\rho_t = \frac{1}{\omega_t} \bigg( \frac{\upsilon_1}{\upsilon_1+\upsilon_2} \omega_1 \rho_1 + \frac{\upsilon_2}{\upsilon_1+\upsilon_2} \omega_2 \rho_2 \bigg) = \frac{\upsilon_1 \omega_1 \rho_1 + \upsilon_2 \omega_2 \rho_2}{\sqrt{\upsilon_1 \omega_1^2 + \upsilon_2 \omega_2^2} \sqrt{\upsilon_1+\upsilon_2}}
.\end{equation}

In this case, if $\rho_{1,t}$ and $\rho_{2,t}$ have different signs, then the local correlation of the whole process will be able to be greater or less than zero, depending on the evolution of the processes $ \upsilon_{1,t} $ and $ \upsilon_{2,t} $ and ratio between them. Therefore, after introducing the variance processes ratio, $\upsilon_{R,t} = \upsilon_{1,t} / \upsilon_t$, we can write:

\begin{equation}
\rho_t = \frac{\upsilon_R (\omega_1 \rho_1 - \omega_2 \rho_2) + \omega_2 \rho_2}{\sqrt{\upsilon_R (\omega_1^2 - \omega_2^2) + \omega_2^2}}
.\end{equation}

This means that the correlation is independent of the total variance. However, it depends on the relationship between the first variance process and the total variance.

Although, considered relationship also hold for the OUOU model, they have different applications for this model. In the OUOU model the ratio $\upsilon_R$ is additionally dependent on $\omega_1$ and $\omega_2$, i.e. 

\begin{equation}
\upsilon_R = \tfrac{\upsilon(\omega_1^2,\theta_1,\nu_1,\kappa_1,\tau)}{\upsilon(\omega_1^2,\theta_1,\nu_1,\kappa_1,\tau)+\upsilon(\omega_2^2,\theta_2,\nu_2,\kappa_2,\tau)}
.\end{equation}

This implies that the symmetrical OUOU model\footnote{~A symmetrical two-factor variance model is a model which has same values of respective parameters in both variance equations.} is not the equivalent to the analogous Sch\"obel-Zhu model, which is true for the relation between the symmetrical Bates model and the Heston model. In such model due to assumptions about parameters there is $\upsilon_R = \tfrac{1}{2}$, so $\omega^2 = \tfrac{\omega_1^2+\omega_2^2}{2} $, but then $\upsilon(\omega^2,\theta,\nu,\kappa,\tau) = 2\upsilon(\omega^2,\alpha\theta,\alpha\nu,\kappa,\tau)$, where $\alpha$ is a function of $\omega$, $\theta$, $\nu$, $\kappa$ and $\tau$, so there is no equality in case of $\alpha=\tfrac{1}{\sqrt{2}}$, which would be analogous to Bates model. The solution for $\alpha$ is

\begin{equation}
\alpha = \tfrac{1}{2} (1-\tfrac{\omega^2 (1- e^{-2 \kappa \tau})}{2\kappa(e^{-2 \kappa \tau} \nu^2 + \theta e^{-2 \kappa \tau} (e^{-\kappa \tau}-1) (\theta (e^{-\kappa \tau}-1)+2 \nu))})
.\end{equation}

Moreover, some combinations of total variance of variance and total variance term structure values, which are obtainable in the Sch\"obel-Zhu models will be not obtainable in the symmetrical OUOU model.

\subsection{Two stage calibration method for two-factor models}

A good starting point for calibration algorithm is needed, because calibration of the model with 10 parameters is a difficult task. The calibration of the two-factor model can be done after a calibration of the symmetrical version of the model, which in the case of the unconditioned Bates model is its one-factor version, namely, the Heston model with parameters $\theta$ and $\nu_0$ that both were divided by 2. In the symmetrical version of the model following constraints are present: $\nu_{0,1}=\nu_{0,2}$, $\theta_1=\theta_2$, $\kappa_1=\kappa_2$, $\rho_1=\rho_2$, $\omega_1=\omega_2$. Therefore, the calibration of the symmetrical version of the model is a problem with 5 parameters calibration. Then, using calibrated parameters from the symmetrical model a calibration of 10 parameters in the unconstrained model can be done. Such estimation of starting points for calibration of Bates two-factor variance model is proposed by Christoffersen et al. (2009)\footnote{~\bibentry{Christoffersen2009}.}.
In the case of the Bates model with the Feller condition, the first calibration is understood as a calibration of the Heston model with the Feller condition. Then, dividing parameters $\nu_0$ and $\theta$ by 2 to get parameters $\nu_{0,1}$, $\nu_{0,2}$, $\theta_1$, $\theta_2$, and finally truncating $\omega_i$ with following transformation: $\widetilde{\omega}_i = \min(\omega_i, \sqrt{1.99 \theta_i \kappa_i})$.

\subsection{Approximate formulas for some of two-factor models parameters}

Following sections present possible approaches for estimation of starting points for all ten parameters in further numerical calibration of the unconditioned and conditioned Bates model and the OUOU model. All these methods need estimates of the parameters $\omega$ and $\rho$ from the Heston or the Sch\"obel-Zhu model. The second method is based on the first and has slight modifications due to the previously considered constraints in the OUOU model. 

\subsubsection{Equal Variance Parametrisations method for the unconditioned Bates model}

Equal variance parametrisation method is based on the same idea as the two stage calibration method. This is a naive method with two assumptions. The first assumption is about equality of variance equation parametrisations, i.e. they have to have same parameters, so $\upsilon_R = \tfrac{1}{2}$. The second assumption is about following symmetries for the rest of parameters: $ \omega_1 = \omega_2 $ and $ \rho_1 = \rho_2 $. This method is specified separately, becuase it is using external estimates of parameters $\omega$ and $\rho$, which are estimates from one-factor version of the model and then variance term structure is divided by 2 and then $\nu_0$, $\theta$ and $\kappa$ are calibrated in the standard way. Alternatively if the Heston model variance term structure was previousy calibrated then parameters $\nu_0$, $\theta$ can be taken from that calibration and dividend by 2, the parameter $\kappa$ is unchanged. Parameters of one-factor version of the model can be either numerically calibrated or estimated with approximate formulas.

\subsubsection{Modified Equal Variance Parametrisations method for the Sch\"obel-Zhu model}

In the case of the OUOU model parameters obtaining parameters of the symmetrical version of the two-factor model using one-factor model parameters is more difficult. The simplest possibility is to use estimates of $\nu_0$, $\theta$ from the Sch\"obel-Zhu variance term structure and divide them by $\sqrt{2}$. Then take the unchanged parameter $\kappa$ from that term structure and an estimate of the parameter $\rho$. Finally, for the parameter $\omega$ small change has to be done in order to maintain summation of variances to the total variance in one-factor model. Unfortunately, such operation is also underestimating the total variance of variance in comparison to the one-factor model, but some compromises have to been done. In summary, in the symmetrical OUOU model there is a trade-off between the underestimation of the total variance of variance and the overestimation of the total variance.

To handle with the considered problem following modification to the Equal Variance Parametrisations method are proposed. With presented in previous section relationships and after adding a few additional assumptions we can find approximate formulas for $\omega_1$ and $\omega_2$, which are using $\rho$ and $\omega$ as arguments.

Additional assumptions for some of the parameters are following:
\begin{itemize}
\item equations for $\upsilon_{1,t}$ and $\upsilon_{2,t}$ are identical, i.e. $\upsilon_{R,t} = 0.5$ for all times to maturity and $\nu_{1,0}=\nu_{2,0}$, 
\item $\rho_1$ and $\rho_2$ correlations have opposite signs and extreme values, $\rho_1$=1, $\rho_2=-1$.
\end{itemize}

While the first assumption is usually modified during calibration by minimization  algorithm, then usually after the calibration the second assumption is changed only slightly. This observation is confirmed in the literature. For example, Carr and Wu (2007) \footnote{~\bibentry{Carr2007}.} developed and studied a model, which, like the Bates two-factor variance model, includes two Wiener processes for rates of return of the underlying instrument and two Wiener processes for two rates of activity, which have an interpretation similar to the variance and their role is to control the strength of the Levy components in returns equations.

In their estimation of the model that was made for currency options, the correlations between the rates of return and rates of activity of Wiener processes have opposite signs and the absolute value of around 1. In practice, for the calibration process a value of $1-\epsilon$, e.g. 0.99, should be chosen instead of 1, when in the calibration process a transformation of $\rho$ variable, that is transforming $(-1;1)$ to $(-\infty,\infty)$, is used.

On the other hand, even setting $\rho_1$ and $\rho_2$ at constant levels equal to 1 and -1, the correlation of the total variance process and the process of the underlying instrument price in Bates two-factor variance model is still equal to the weighted average of $\rho_1$ and $\rho_2$ correlations, and all points on $(-1,1)$ range are still available and depend on the final fit of the parameter $\upsilon_{R,t}$.

First of all, the parameters $\rho_H $ and $\omega_H $ should be calculated from Implied Central Moments formulas for the Heston model. Then using relationships $\omega_{S} = \tfrac{1}{2} \omega_H $ and $\rho=\rho_H$, the Sch\"obel-Zhu model parameters are obtained. The initial value of the parameter $\omega$ for considerations is chosen to be smaller than its one-factor equivalent model due to the problems mentioned at the beginning of the section, i.e. $\omega = \tfrac{1}{\sqrt{2}} \omega_{S} $.

Then, starting from those $\rho$ and $\omega$, we can write following the equations:

\begin{equation}\begin{split}
& \omega^2 = \upsilon_R \omega_1^2 + (1-\upsilon_R) \omega_2^2 ,\\
& \rho \omega = \upsilon_R \rho_1 \omega_1 + (1-\upsilon_R) \rho_2 \omega_2
.\end{split}\end{equation}

Next, using the considered assumptions we get the following solution:

\begin{equation}\begin{split}
& \omega^2 = 0.5 \omega_1^2 + 0.5 \omega_2^2 ,\\
& \rho \omega = 0.5 \omega_1 - 0.5 \omega_2
,\end{split}\end{equation}

which is a system of equations with one quadratic equation and has only two unknown variables so they are equal to:

\begin{equation}\begin{split}
& \omega_1 = \omega \Big(\sqrt{1 - \rho^2}+\rho \Big) ,\\
& \omega_2 = \omega \Big(\sqrt{1 - \rho^2}-\rho \Big)
.\end{split}\end{equation}

Both $\omega_1$ and $\omega_2$ will be positive when $|\rho|<\tfrac{1}{\sqrt{2}} \approx 0.707$, which is often met in usual market conditions.
The parameter $\omega_1$ ($\omega_2$) will be negative for $\rho < -\tfrac{1}{\sqrt{2}}$ ($\rho > \tfrac{1}{\sqrt{2}}$). In such cases, some minimum value, which is greater than zero, e.g. $ 0.001 $, should be used instead of negative $\omega_1$ ($\omega_2$). 

The other parameters should be obtained from the symmetrical version of the two-factor model, which in the case of the OUOU model has a little different properties than its one-factor equivalent model. Namely, parameters $\nu_0$, $\theta$ and $\kappa$ should be obtained from calibration of the variance term structure to the half of the implied variance. Moreover, in this calibration the initial value of the parameter $\omega$ should be used, i.e. $\omega = \tfrac{1}{\sqrt{2}} \omega_{S} $. This simplification is allowing to maintain $\upsilon_R=\tfrac{1}{2}$ and is not affecting the total variance. Moreover, even after updating of the sum of variance functions with transformed parameters $\omega_1$ and $\omega_2$, the following equality holds:

\begin{equation}
\mathbb{E}[\upsilon_{S}(\omega_{S})] = \mathbb{E}[\upsilon_{1}(\omega)] + \mathbb{E}[\upsilon_{2}(\omega)] = \mathbb{E}[\upsilon_{1}(\omega_1)] + \mathbb{E}[\upsilon_{2}(\omega_2)],\\
,\end{equation}

because the instantaneous variance is equal to:

\begin{equation}
\mathbb{E}[\upsilon_t] = \tfrac{\omega^2}{2\kappa}(1 - e^{-2 \kappa t}) + e^{-2 \kappa t}\nu_0^2 + \theta e^{-2 \kappa t} (e^{\kappa t} - 1) (\theta (e^{\kappa t} - 1) + 2 \nu_0) = a(t) \omega^2 + b(t)
\end{equation}

and $ \omega_1^2 + \omega_2^2 = \omega^2 \Big( \Big(\sqrt{1 - \rho^2}+\rho \Big)^2 + \Big(\sqrt{1 - \rho^2}-\rho \Big)^2 \Big) = 2 \omega^2 = \omega^2_{S} $.

On the other hand, ratios $\upsilon_R$ after the updating with new values of transformed parameters $\omega_1$ and $\omega_2$ and setting $m=\Big(\sqrt{1 - \rho^2}+\rho \Big)$ will be equal to:

\begin{equation}\begin{split}
\upsilon_R & = \frac{\mathbb{E}[\upsilon_1]}{\mathbb{E}[\upsilon_1]+\mathbb{E}[\upsilon_2]} = \frac{a(t) \omega_1^2 + b(t)}{2a(t) \omega^2 + 2b(t)} = \frac{a(t) \omega^2 m^2 + b(t)}{2a(t) \omega^2 + 2b(t)} = \frac{a(t) \omega^2 + b(t) + a(t) \omega^2 (m^2-1)}{2(a(t) \omega^2 + b(t))} \\
& = \frac{1}{2} + (m^2-1)\frac{a(t) \omega^2}{2(a(t) \omega^2 + b(t))}
,\end{split}\end{equation}

where the last part has following limits:

\begin{equation}\begin{split}
& \lim_{t \to 0} \tfrac{a(t) \omega^2}{2(a(t) \omega^2 + b(t))} = 0 ,\\
& \lim_{t \to \infty} \tfrac{a(t) \omega^2}{2(a(t) \omega^2 + b(t))} = \tfrac{1}{2}
\end{split}\end{equation}

and for $t=2$ and exemplary values of parameters: $\omega=0.15$, $\nu_0=0.06$, $\theta=0.001$ and $\kappa=1.1$, it is equal to $\tfrac{1}{2+\log(-36.843)}$. The local extreme in 0 is also obtained when $t=\tfrac{\log(h)-\log(h-v)}{\kappa}$. So,

\begin{equation}\begin{split}
& \lim_{t \to 0} \upsilon_R(t) = \tfrac{1}{2} ,\\
& \lim_{t \to \infty} \upsilon_R(t) = \tfrac{1}{2} + \rho\sqrt{1-\rho^2}
\end{split}\end{equation}

and $\upsilon_R$ is in $[\tfrac{1}{2},\tfrac{1}{2}+\rho\sqrt{1-\rho^2}]$ for positive $\rho$ and in $[\tfrac{1}{2}+\rho\sqrt{1-\rho^2},\tfrac{1}{2}]$ for negative $\rho$. So, for a very short time the effective $\omega^2$ is

\begin{equation}
\omega_{e}^2 = \omega^2 = \tfrac{1}{2}\omega_{S}^2
,\end{equation}

which is the same as when using Equal Variance Parametrisations approach for the symmetrical OUOU model with $\omega_1=\omega_2=\tfrac{1}{\sqrt{2}}\omega_{S}$, but for long times

\begin{equation}
\omega_{e}^2 = \omega^2 \Big( \tfrac{1}{2} \Big(\sqrt{1 - \rho^2}+\rho \Big)^4 + \tfrac{1}{2}\Big(\sqrt{1 - \rho^2}-\rho \Big)^4 \Big) = \omega^2 (-4\rho^4 + 4 \rho^2 + 1) = \omega_{S}^2 (-2\rho_{S}^4 + 2 \rho_{S}^2 + \tfrac{1}{2})
,\end{equation}

where the multiplier of $\omega_{S}^2$ is an even function and its minimum equal to $\tfrac{1}{2}$ is obtained for $|\rho|=1$ and $\rho=0$ and its maximum equal to 1 is obtained for $|\rho|=\tfrac{1}{\sqrt{2}}$.
For example, for long times and $\rho=-0.4$, the ratio is $\upsilon_R=0.1334$ and the multiplier of $\omega_{S}^2$ is equal to 0.77, which is only slight underestimation of the Sch\"obel-Zhu parameter $\omega$ equivalent by 12\% for long times in comparison to 31\% in base method. In extreme case, when $|\rho|=1$ then underestimation for long times is same as in the base method. Averaging the error along the variance term structure curve makes the effective $\omega$ much more closer to $\omega_{S}$ than in the symmetrical OUOU model with $\omega$ fixed to $\tfrac{1}{\sqrt{2}}\omega_{S}$.

Similar conclusions can be made for the Sch\"obel-Zhu parameter $\rho$ equivalent, for which for a very short time the effective product of $\omega$ and $\rho$ is:

\begin{equation}
\omega_{e} \rho_{e} = \omega \rho = \tfrac{1}{\sqrt{2}}\omega_{S} \rho_{S}
,\end{equation}

which is the same as when using the standard equal variance parametrisation approach for the symmetrical OUOU model with $\omega_1=\omega_2=\tfrac{1}{\sqrt{2}}\omega_{S}$ and for long times

\begin{equation}
\omega_{e} \rho_{e} = \omega \Big( \tfrac{1}{2} \Big(\sqrt{1 - \rho^2}+\rho \Big)^3 - \tfrac{1}{2} \Big(\sqrt{1 - \rho^2}-\rho \Big)^3 \Big) = \omega \rho (3-2 \rho^2) = \omega_{S} \rho_{S} \tfrac{3-2\rho_{S}^2}{\sqrt{2}}
,\end{equation}

where the multiplier of $\omega_{S} \rho_{S}$ is an even function, its maximum equal to $\tfrac{3}{\sqrt{2}}$ is obtained for $\rho=0$ and its minimum equal to $\tfrac{1}{\sqrt{2}}$ is obtained for $|\rho|=1$. The value 1 is obtained for $|\rho|= \sqrt{\tfrac{3}{2} - \tfrac{1}{\sqrt{2}}} \approx 0.89$.

In this way, the initial values for all ten parameters are known. In addition, same as in the case of a one-factor variance models a restricted calibration can be performed, i.e. with some parameters being fixed, in this case the $\rho_1=1$ and $\rho_2=-1$, which results in 8 parameters being calibrated.

\subsubsection{Modified Equal Variance Parametrisations method for the Bates model with the Feller condition}

Instead of using the standard equal variance parametrisation approach for the Bates model with the Feller condition and then truncating parameters $\omega_1=\omega_2$ to the level, which is implied by the condition, an approach similar to the one, that was proposed for the OUOU model, can be used. Namely, parameters $\rho_1$ and $\rho_2$ can be fixed at levels 0.99 and -0.99 and $\omega_1$ and $\omega_2$ can be transformed in the same way, ie:

\begin{equation}\begin{split}
& \omega_1 = \omega \Big(\sqrt{1 - \rho^2}+\rho \Big) ,\\
& \omega_2 = \omega \Big(\sqrt{1 - \rho^2}-\rho \Big)
\end{split}\end{equation}

using estimates of $\omega$ and $\rho$ from the Heston model. Then, in the same way as in the standard equal variance parametrisation method, parameters $\kappa $, $\nu_0$ and $\theta$ are taken form the Heston model variance term structure and the last two of them are divided by 2.

Using the modified equal variance parametrisation method for the Bates model, when a condition is present in the calibration, can grant following advantages:
\begin{itemize}
\item Starting levels of $\rho_1$ and $\rho_2$, which are near their boundaries, can help avoid local minima, because then $\omega_i$ $\rho_i$ products can still reach high levels despite of $\omega_i$ being limited.
\item Truncation of transformed $\omega_1$ and $\omega_2$ in the case of markets, for which the Feller condition is not satisfied, will consist of one bigger truncation and one smaller truncation. The latter can be even close to 0, which is giving inefficiency only for one variance equation instead of smaller inefficiency for both variance equations.
\end{itemize}

\section{Conclusions}

Presented in the fourth chapter parameters estimation methods allow not only to shorten the time of model calibration and thus the time of option pricing, which is important in environment of rapidly changing implied volatilities, but also to improve the accuracy of a local minimization algorithm by using more appropriate starting points.

The author's approximate formulas for estimation of parameters is using all of observed option prices, which contain fundamental characteristics of the option market. This is done in a similar way as in case of CBOE Stock Exchange VIX and SKEW indices. In this respect, although it is similar to the formulas from Durrleman (2004) method, it is different from the latter method by the fact that it does not use parabolic approximation of volatility smile and linear regression. Also, estimated parameters are dependent on more than only two options as in the method of Gauthier and Rivaille.

The Implied Central Moments method can be used for different models. Derivations of approximate formulas for the Heston and the Sch\"obel-Zhu model parameters have been presented in the fourth chapter. Besides, an additional method for making use of parameters which were estimated with Implied Central Moments formulas has been introduced for models with two-factor variance. The new method of estimation should be further verified empirically in order to confirm its effectiveness in comparison with the analogous methods described in the literature.
\clearpage{\pagestyle{empty}\cleardoublepage}

\chapter{Empirical analysis of model calibration and parameters estimation methods}

\section{Introduction}

In previous chapters stochastic volatility models and option pricing with these models were considered. Next, number of statistical factors in implied volatility surface has been analysed. Then, methods of parameters estimation for these models were presented.

This chapter contains a study of two problems. In the first place, there is a test of the quality of the fit of the Heston model to market data using methods of parameters analytic estimation, which were presented in the fourth chapter. In the latter chapter the author presented also his own approach to the problem. The studied methods for estimation of $\rho$ and $\omega$ are:

\begin{itemize}
\item Durrleman (2004) method,
\item Implied Central Moments method,
\item calibration with Nelder-Mead method,
\item exponentially weighted moving averages used on currency VIX index.
\end{itemize}

Then, in a separate part of the study, there is a test of the quality of the fit to the market data for stochastic volatility models with two-factor variance or volatility, which were presented in the first and the second chapters.

In this way, for the first time the new OUOU model has been empirically verified. The Bates two-factor variance model (2000) has been chosen as a competitive solution against which the OUOU model was compared to. Both models have all processes in the form of diffusion only, there are no jumps in their processes. In both parts of the study the following setup of the pricing model was used: the Albrecher et al. (2007)\footnote{~\bibentry{Albrecher2007}.} form of the Heston model characteristic function, the Attari (2004)\footnote{~\bibentry{Attari2004}.} pricing method with integration over $w=\ln(u)$, instead of $u$, and trapezoidal quadrature for numerical integration on points from interval [-17,5] with 0.4 spacing between points. The change of integral in the pricing function is as follows:

\begin{equation}\begin{split}
C(K) & = S_t e^{-r_f\tau} - K e^{-r_d\tau} \bigg( \tfrac{1}{2} + \tfrac{1}{\pi} \int_{0}^\infty \mathrm{Re} \bigg( \phi_2(u) e^{-iul} \tfrac{1-i/u}{u^2+1} \bigg) \,\D u \bigg)\\
& = S_t e^{-r_f\tau} - K e^{-r_d\tau} \bigg(\tfrac{1}{2} + \tfrac{1}{\pi} \int_{-\infty}^\infty \mathrm{Re} \bigg( e^{-i e^{w} (\ln(K/S)-(r_d-r_f)\tau)} \tfrac{1 - i /e^{w}}{1 + e^{2w}} e^{w} \phi_2(e^{w}) \bigg) \,\D w \bigg)
.\end{split}\end{equation}

The transformed integrand in such form has more even distribution of values, which results in more precise outcome when trapezoidal quadrature is used.


\section{Characteristics of the data used for the tests}

Tests were performed on following time series, which were computed on the basis of mid quotations (i.e. the arithmetic average of bid and offer prices) of instruments from Reuters for the period from 22-07-2010 to 31-08-2015 in the daily frequency:

\begin{itemize}
\item $S$ - EURUSD mid spot rate,
\item $r_f(\tau)$ - foreign logarithmic short interest rates (6 time series in total), for EURUSD pair it is the EUR interest rate, and were obtained by following operations on mid OIS rates (\textit{Overnight Indexed Swap}): $r_f(\tau) = -\tfrac{1}{\tau} \ln((1+\mathrm{OIS}_{\tau} \tau)^{-1})$, the choice of OIS is dictated by the common opinion that it is the best approximation of the risk-free rate, mainly due to relatively low credit risk in comparison with instruments such as deposit or bond (Hull and White (2012)\footnote{~\bibentry{Hull2012}.}),
\item $r_d$ - domestic logarithmic short interest rates (6 time series in total), for EURUSD pair it is the USD interest rate, and were obtained by following operations on mid FX forward rates, FX spot rate and previously computed $r_f(\tau)$ rates: $r_d(\tau) = - \tfrac{1}{\tau}\ln((1+F/S)\exp(-r_f \tau)),$
\item $\sigma(\delta,\tau)$ - implied volatilities for the EURUSD options (30 time series in total). Each series corresponds to a pair of unique parameters, $\{\delta,\tau\}$, where the time to maturity $\tau$ takes values from the set \{1M, 2M, 3M, 6M, 1Y, 2Y\} and the parameter $\delta$ from \{10P, 25P, ATM, 25C, 10C\}. These series were obtained by making the transformation of the formulas from the first chapter on Reuters data for mid quotes of following option strategies: ATM, 25RR, 10RR, 25FLY and 10FLY.
\item $K(\delta,\tau)$ - strike prices that are corresponding to time series of implied volatility $\sigma(\delta,\tau)$, 30 time series in total, and were computed with following formula:

\begin{equation}
K = S e^{(r_d-r_f)\tau} e^{0.5 \sigma_{\delta,\tau}^2 \tau - \mathrm{sgn}(\delta) \sigma_{\delta,\tau} \sqrt{\tau} N^{-1}(|\delta| \exp(r_f\tau))}
,\end{equation}

where: $N^{-1}$ is the inverse cumulative distribution function of the normal distribution and $\delta$ is used in its numerical form, wherein \{10P, 25P, ATM, 25C, 10C\} = \{-0.1, -0.25, 0.5, 0.25, 0.1\}.
\end{itemize}


\section{Test of methods of analytic estimation of model parameters}

\subsection{The design of the study and its objectives}

This section presents a study designed to compare the quality of the fit of option prices from the Heston model using approximate formulas from two methods, namely: Durrleman method and author's Implied Central Moments method. The study used the whole data set with date range of 22-07-2010 to 31-08-2015, which has 1333 working days and all presented time series except historical average time series have this amount of observations, each historical average time series has 1271 observations, i.e. less by 62.

Root mean error-square error has been selected as a method of measuring the effectiveness of model fit. The error is a difference between model implied volatilities and observed implied volatilities:

\begin{equation}
\mathrm{RMSE} = \sqrt{\tfrac{1}{N}\sum^N_{i=1} (\sigma_i^{\mathrm{model}}-\sigma_i^{\mathrm{market}})^2} \\
,\end{equation}

where: $N=30$, because there are five possible values for delta and six maturities.

Since both methods are about estimation of parameters $\rho$ and $\omega$, prior to the test other parameters of the Heston model, i.e. $ \kappa $, $ \theta $, $ \nu_0 $ are numerically calibrated for all days in the data sample according to the following scheme. Firstly, indices of implied variance, which are VIX index equivalents for currency options, are calculated for all dates in the tested range for all maturities. This way a term structure of variance is obtained for each day.

Then, a 3 month moving average of implied variance for 3M maturity is used as historical estimate of parameter $\omega$. It is calculated with following formula:

\begin{equation}
	\omega_{3\mathrm{M}} = \tfrac{1}{\mathrm{VIX}_{1\mathrm{M}}}\sqrt{\mathrm{EWMA}((\Delta \mathrm{VIX}_{1\mathrm{M}}^2)^2,63)}
,\end{equation}

where EWMA stands for the exponentially weighted moving average and is defined same as in the fourth chapter. Similar calculation were done for parameter $ \rho $, which historical estimate with 3 month moving average is equal to:

\begin{equation}
	\rho_{3\mathrm{M}} = \frac{\mathrm{EWMA}((\Delta \ln(S) \times \Delta \mathrm{VIX}_{1\mathrm{M}}^2 ),63)}{\sqrt{\mathrm{EWMA}((\Delta \ln(S))^2,63) \times \mathrm{EWMA}((\Delta \mathrm{VIX}_{1\mathrm{M}}^2)^2,63)}}
.\end{equation}

For the purpose of using historical estimates in numerical calibration methods and for VIX correction, unknown first 62 values of $\omega_{3\mathrm{M}}$ have been set to the value of VIX$_{1\mathrm{M}}$ at given day and unknown first 62 values of $\rho_{3\mathrm{M}}$ have been set to -0.1. This change was done only in aforementioned cases.

After calculating variance term structure, VIX$^2$, it is corrected with methodology presented in the fourth chapter to include small effect of $\rho$ and $\omega$ on it. To compute the modified variance term structure, $\widetilde{\mathrm{VIX}}^2$, historic estimators $\omega_{3\mathrm{M}}$ and $\rho_{3\mathrm{M}}$ have been used. Then, the MSE cost function with differences between implied variances and model theoretical variances as errors is minimized by Nelder-Mead numerical algorithm with a maximum number of iterations set to 8000. This way parameters $\kappa$, $\theta$ and $\nu_0$ can be found numerically. The theoretical value of variance in the Heston model is given by following formula:

\begin{equation}
\sigma(\tau) = \sqrt{\frac{\int_0^\infty \nu(t)\,\D t}{\tau}} = \sqrt{\theta + (\nu_0-\theta)\frac{1-e^{-\kappa T}}{\kappa \tau}}
.\end{equation}

Aforementioned cost function has a following form: $\sum^6_{i=2} (\sigma(\tau_i)- \widetilde{\mathrm{VIX}}(\tau_i))^2$, where as starting points for parameters are taken: $\kappa=2$, $\nu_0 = \widetilde{\mathrm{VIX}}(\tau_{\mathrm{min}})$, $\theta=\widetilde{\mathrm{VIX}}(\tau_{\mathrm{max}})$, i.e. the first and last points of the modelled term structure of the variance. The cost function is calculated on all points of the variance term structure except the first one due to the problems with fitting of the considered function form for the small time to maturity $\tau$.

For the parameter $\kappa$ a starting value of 2 is applied, because this value corresponds to value often presented in the literature on option markets. For example, an important paper on empirical tests of the Heston model and its version with jumps in the returns of the underlying instrument is Bakshi, Cao and Chen (1997)\footnote{~\bibentry{Bakshi1997}.}. In the case of the parameter $ \kappa $ in this paper, its average level for all tested option was 1.15 for the Heston model and 2.03 for the model with the jump-diffusion process. For the sample of the short-term options same averages were respectively 1.62 and 3.93.

In addition, to solve the problem of unexpected and unnatural jumps in the parameters $\nu_0$ and $\theta$ following procedure was used. Firstly, outliers were defined as: $\ln(p_t)- \mathrm{max}(\ln(p_{1}),...,\ln(p_{t-i})) > 0.4 $, where $p$ is parameter $\nu_0$ or $\theta$. In case of an outlier a recalibration was performed for a given day with previous starting value of the outlier parameter being multiplied by 2.

The distributions of estimated parameters is presented in Table \ref{table:5.1}:

\begin{table}[h!]
\centering
\caption{Summary of distributions of calibrated variance term structure parameters}
\label{table:5.1} 
\begin{tabular}{rrrrrrrr}
  \hline
 & Mean & S. D. & Min. & 1st Qu. & Median & 3rd Qu. & Max. \\ 
  \hline
$\nu_0$ & 0.0103 & 0.0065 & 0.0008 & 0.0055 & 0.0082 & 0.0144 & 0.0369 \\ 
  $\theta$ & 0.0174 & 0.0076 & 0.0069 & 0.0112 & 0.0143 & 0.0227 & 0.0406 \\ 
  $\kappa$ & 4.0380 & 5.1578 & 0.0359 & 1.5520 & 2.0680 & 4.7470 & 66.0200 \\ 
  MAE & 0.0111 & 0.0053 & 0.0017 & 0.0065 & 0.0103 & 0.0150 & 0.0292 \\ 
   \hline
\end{tabular}

\end{table}

An alternative to the proposed approach would be to use the Guilliame and Schoutens (2012) approximate formulas for $\nu_0$ and $\theta$ and then estimate $ \kappa $ with numerical minimization algorithm or just use fixed level on the basis of its value from the literature on the Heston model. Knowing parameters $ \kappa $, $ \theta $ and $ \nu_0 $ of the Heston model, the other two parameters are estimated with formulas from one of two possible methods, i.e. Durrleman or ICM.

For comparison of outcomes from approximate formulas with numerical scheme outcomes, all 5 parameters are also calibrated with the Nelder-Mead algorithm using MSE as the cost function with a maximum number of iterations set to 1600. The transformations exp$(x)$ for the parameter $\omega$ and tanh$(x)$ for the parameter $\rho$ were used to have unconstrained real arguments in the cost function. The variance term structure calibration outcomes along with historical estimates of $\rho$ and $\omega$ have been used for the calibration starting points. The calibration was performed in two instances one with  the Feller condition\footnote{~For numerical tests the Feller condition has been implemented in the cost function as the enforced change of the value of the sum of squared errors up to 999 when Feller condition is not fulfilled.} and one without it. Then, values of the RMSE function for all days and for each method were calculated.

Additionally, the Sch\"obel-Zhu model with Nelder-Mead algorithm has been also calibrated to compare the Heston model volatility surface fit capacity. Similar starting point for this calibration has been chosen. Firstly, historical estimates of $\rho$ and $\omega$ were computed, were:

\begin{equation}
	\omega_{3\mathrm{M}} = \sqrt{\mathrm{EWMA}((\Delta \mathrm{VIX}_{1\mathrm{M}})^2,63)}
\end{equation}

and formula for $\rho_{3\mathrm{M}}$ is same as for the Heston model except that it contains VIX instead of VIX$^2$. The Sch\"obel-Zhu volatility term structure parameters were calibrated with the Nelder-Mead calibration with the same setup as in the case of the Heston model and using expected volatility with convexity adjustment, which was calculated from the previously obtained Heston variance term structure parameters and historical estimate of the Heston model parameter $\omega$. The initial value of the parameter $\kappa$ has been set to 0.95 to omit problem of 0 starting value in the Nelder-Mead algorithm.  For the purpose of using historical estimate time series as a starting point in numerical calibration methods, unknown first 62 values of $\omega_{3\mathrm{M}}$ have been set to the value of $\tfrac{1}{2}\mathrm{VIX}_{1\mathrm{M}}$ at given day and unknown first 62 values of $\rho_{3\mathrm{M}}$ have been set to -0.1.
 
Similarly as with the variance term structure, to solve the problem of unexpected and unnatural jumps in calibrated parameters $\nu_0$, $\theta$ or $\omega$ following procedure was used. Firstly, outliers were defined as:
$\ln(p_t)- \mathrm{max}(\ln(p_{1},...,\ln(p_{t-i})) > 0.4 $, where $p$ is parameter $\nu_0$, $\theta$ or $\omega$.  In case of an outlier a recalibration was performed for a given day with previous starting value of the outlier parameter being multiplied by 2. Additionally, the parameter $\kappa$ was multiplied by 100 in the case of the recalibration with the Feller condition for outliers in $\theta$. 

Since ICM method estimates the parameters for a given maturity, in the case of this methods the median values of the parameters $ \rho \omega $ and $ \omega^2 $ are calculated. Then, from these values the parameters $\rho$ and $\omega$ are easily calculated. For the Durrleman method $\widetilde{\mathrm{VIX}}(\tau_{1\mathrm{M}})^2$, which is VIX variance after correction with the method from Chapter 4, was used as a proxy for $\nu_0$ in the formula.

Beside the time series of RMSE, an RMSE of option price errors was calculated for each option from volatility surface grid. This way, Durrleman and ICM methods have been compared in terms of particular volatility surface parts.

In addition to testing the fit, different estimates of the parameter $ \omega $ were summarized in a table and plotted in a figure for comparison purpose. Same analysis was done for the parameter $ \rho $. Besides, for the Heston model which was calibrated without the Feller condition, the condition was tested after the calibration for all days in the sample, to check what fraction of the sample do not fulfill this condition. 

Next, the calibration risk has been tested in order to further investigate both analytic estimation methods. The test involves testing the calibration of 3 parameters, so 2 parameters in model are fixed. The latter parameters are $\rho$ and $\omega$ and are estimated with Durrleman or ICM method. The calibration is done threefold with three different cost functions, then statistics summarizing the differences between the calibrated parameters at different cost functions are computed. The Feller condition is not used in calibration processes.



Methodology of calculation of the calibration risk is similar to the methodology of Detlefsen and H\"ardle (2007) \footnote{~\bibentry{Detlefsen2007}.}. The following definition of calibration risk has been used:

\begin{equation}
	\mathrm{max} (|p^*_{\mathrm{MSE}}-p^*_{\mathrm{MAE}}|,|p^*_{\mathrm{MSE}}-p^*_{\mathrm{MAPE}}|,|p^*_{\mathrm{MAE}}-p^*_{\mathrm{MAPE}}|)
,\end{equation}

where $p^* \in \{\theta^*, \omega^*, \rho^*, \kappa^*, \nu_0^* \}$ is one of the Heston model calibrated parameters, and calibration was made by minimizing one of the following cost functions: MSE, MAE and MAPE. The fitted value was a call option price from the model divided by the market implied sensitivity vega, while the observed value was the market price of the call option divided by the market implied sensitivity vega \footnote{~The market sensitivity vega is understood as the sensitivity of the option price to changes in implied volatility that is calculated from the Black-Scholes formula, i.e. $\nu = Ke^{-r_d*\tau}\sqrt{\tau} \phi(d_2)$, where $ d_2 $ is same as in the Black-Scholes formula.}. For example, the MSE is:

\begin{equation}
\mathrm{MSE} = \sum^N_{i=1} \bigg(\frac{C_i^{\mathrm{model}}-C_i^{\mathrm{market}}}{\nu}\bigg)^2 \\
.\end{equation}

This is the alternative to the calculation of implied volatilities differences. Its main advantage is the speed of computing, which is a significant problem in minimization of the function with five arguments. Big differences in calibrated values of a given parameter are consider as a high calibration risk for that parameter in the used calibration method.

\subsection{The study results}

The results of estimation are presented in Table \ref{table:5.2} and Figures 5.1 and 5.2:

\begin{table}[h!]
\centering
\caption{Summary of distributions of RMSE of option prices for different methods}
\label{table:5.2} 
\begin{tabular}{rcrrrrrr}
  \hline
 & Mean & S. D. & Min. & 1st Qu. & Median & 3rd Qu. & Max. \\ 
  \hline
  Durrleman-Heston & 0.0060*** & 0.0035 & 0.0009 & 0.0029 & 0.0054 & 0.0084 & 0.0213 \\ 
  Hist.-Heston & 0.0091*** & 0.0046 & 0.0019 & 0.0054 & 0.0084 & 0.0120 & 0.0091 \\ 
  ICM-Heston & 0.0049 & 0.0027 & 0.0010 & 0.0030 & 0.0042 & 0.0065 & 0.0174 \\ 
  Calib.-Heston & 0.0022*** & 0.0013 & 0.0004 & 0.0010 & 0.0021 & 0.0031 & 0.0067 \\ 
  Calib.-Heston-Feller & 0.0036*** & 0.0021 & 0.0008 & 0.0019 & 0.0033 & 0.0048 & 0.0137 \\ 
  Calib.-Schöbel-Zhu & 0.0020*** & 0.0011 & 0.0004 & 0.0010 & 0.0019 & 0.0026 & 0.0084 \\ 
   \hline
\multicolumn{8}{l}{\scriptsize{\textit{Note: $^{***}p<0.001$, $^{**}p<0.01$, $^*p<0.05$ are for two-sample t-test for equality with mean of ICM-Heston}}}
\end{tabular}

\end{table}

\begin{figure}[h!]
\centering
\includegraphics[width=150mm]{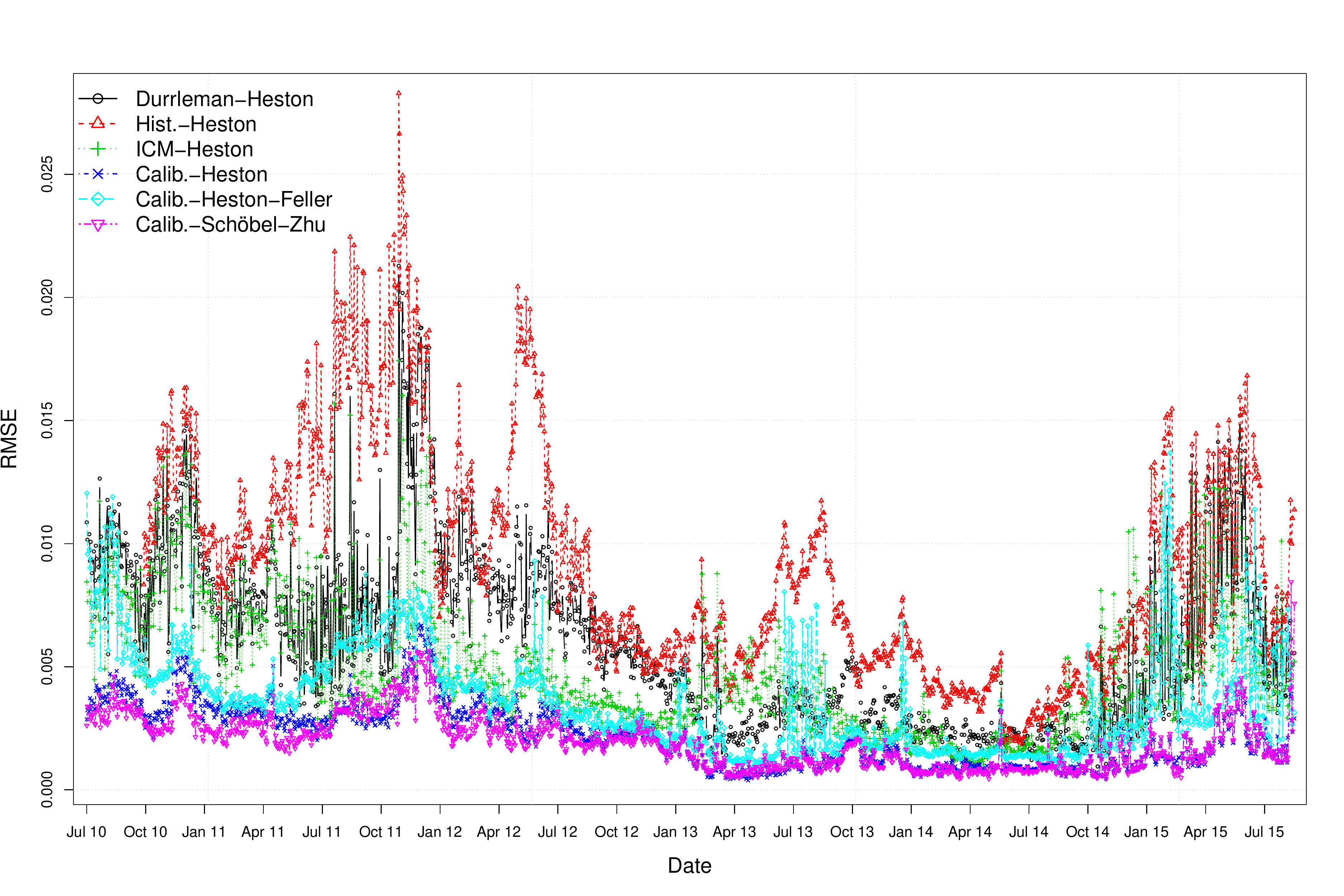}
\caption{The evolution of volatility surface fit errors in considered methods and models}
\label{figure:5.1} 
\end{figure}

Table \ref{table:5.2} and Figure \ref{figure:5.1} show that:
\begin{itemize}
\item The proposed Implied Central Moments method gives the best results among analytic estimation methods in terms of both mean and median value of RMSE, differences to other means are statistically significant
\item There is still some area for the improvement, which can be seen from the difference between errors statistics of the numerical calibration and the ICM method,
\item In the case of the Durrleman (2004) method, there  is relatively higher value of the third quartile and the maximum can be considered as a limitation of the method in terms of the fit to the market data in stressed conditions.
\item The method of historical estimators is the worst of all four methods, which is along with expectations.
\end{itemize}

\begin{figure}[h!]
	\centering     
	\subfigure[In favor of ICM, $\,\mathrm{max(RMSE_{Durr}-RMSE_{ICM},0)}$]{\includegraphics[width=79mm]{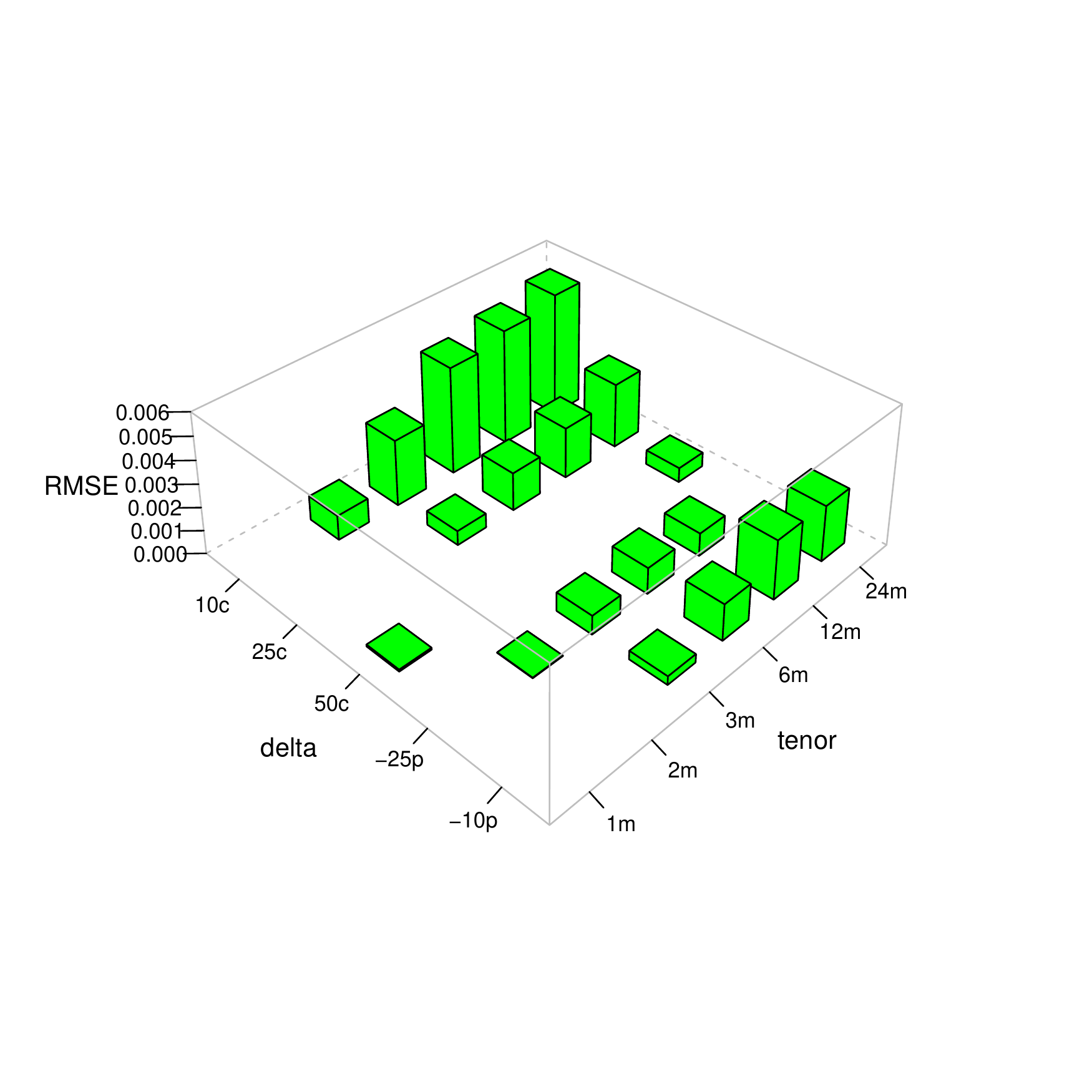}}
	\subfigure[In favor of Durrleman, $\,\mathrm{max(RMSE_{ICM}-RMSE_{Durr},0)}$]{\includegraphics[width=79mm]{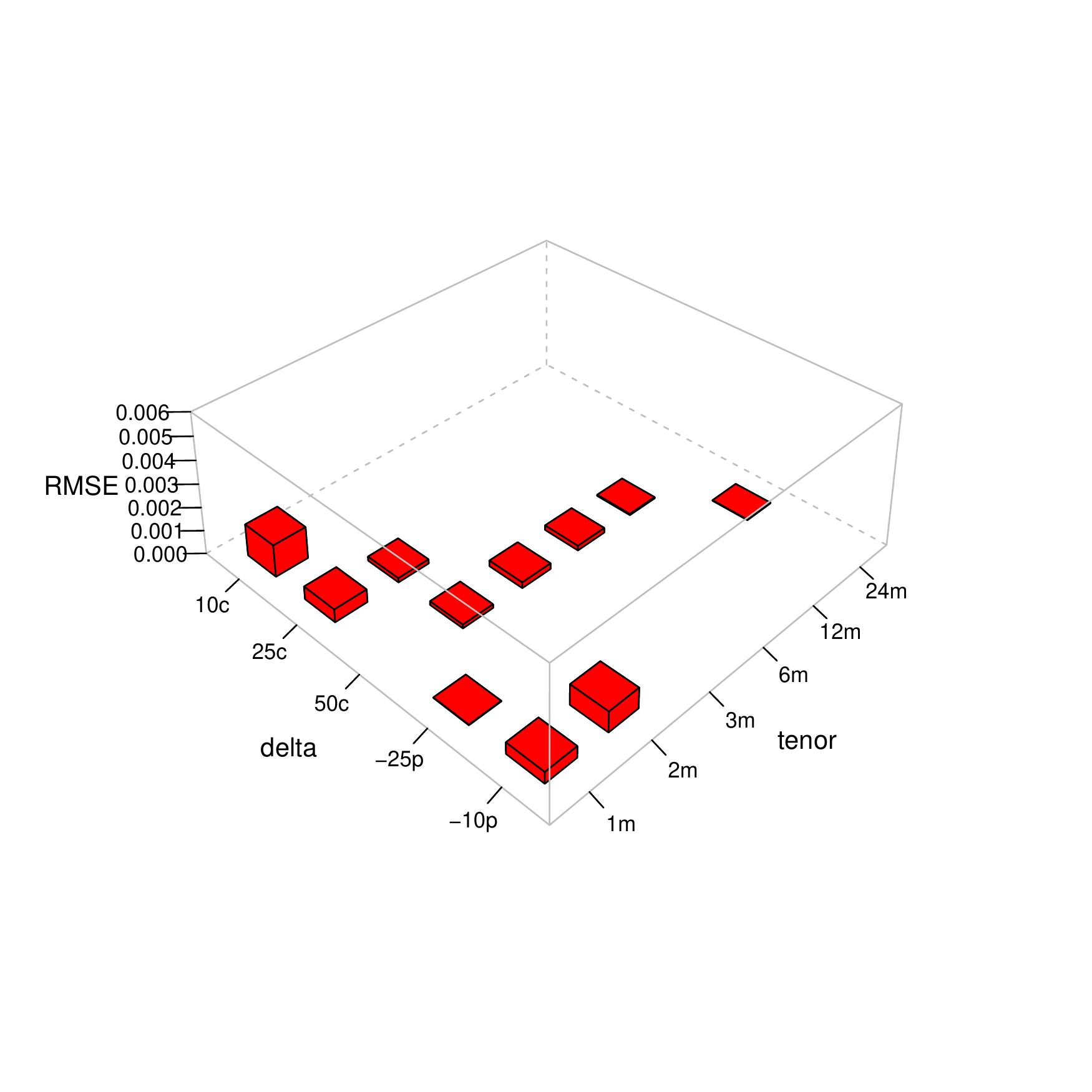}}
	\caption{RMSE differences between ICM and Durrleman method for different options from the volatility grid}
	\label{figure:5.2} 
\end{figure}

It can be seen from Figure \ref{figure:5.2} that the advantage of the ICM method over the Durrleman method is present for the most of options from the volatility surface, with increasing advantage with an increase in the option maturity. The few options, for which the Durrleman method gives better performance, are mostly options with 1M maturity and ATM options. Moreover, in the case of options, for which the ICM method gives a lower fit error, differences between mean errors of both methods are usually clearly higher than in the case of options, for which the Durrleman method gives a lower fit error.


Tests of the Feller condition in the Heston model after the unconditioned calibration show that for 100\% of days in the sample the Feller condition is not satisfied. Assuming that the Heston model is a data generating process for this market and results of the unconditioned Nelder-Mead calibration give a point close to the global minimum, such results could show that the market for EURUSD options is always accounting the possibility of the variance process going to zero. Nevertheless, such conclusion could only be made if the true data generating process would be known and the latter is always uncertain.

\begin{figure}[h!]
\centering
\includegraphics[width=150mm]{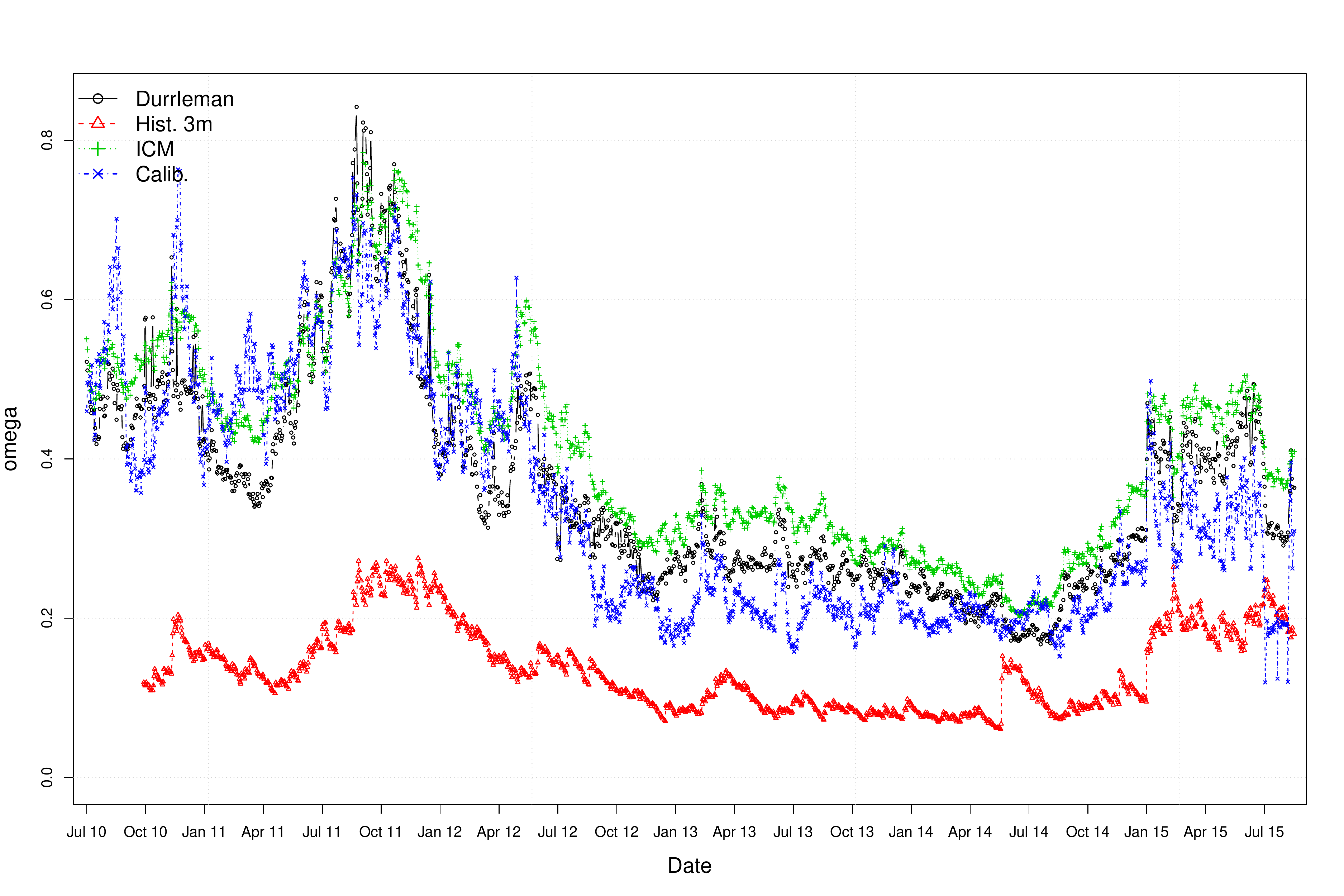}
\caption{The Heston model parameter $ \omega $ from considered methods}
\label{fig:omega}
\label{figure:5.3} 
\end{figure}

\begin{figure}[h!]
\centering
\includegraphics[width=150mm]{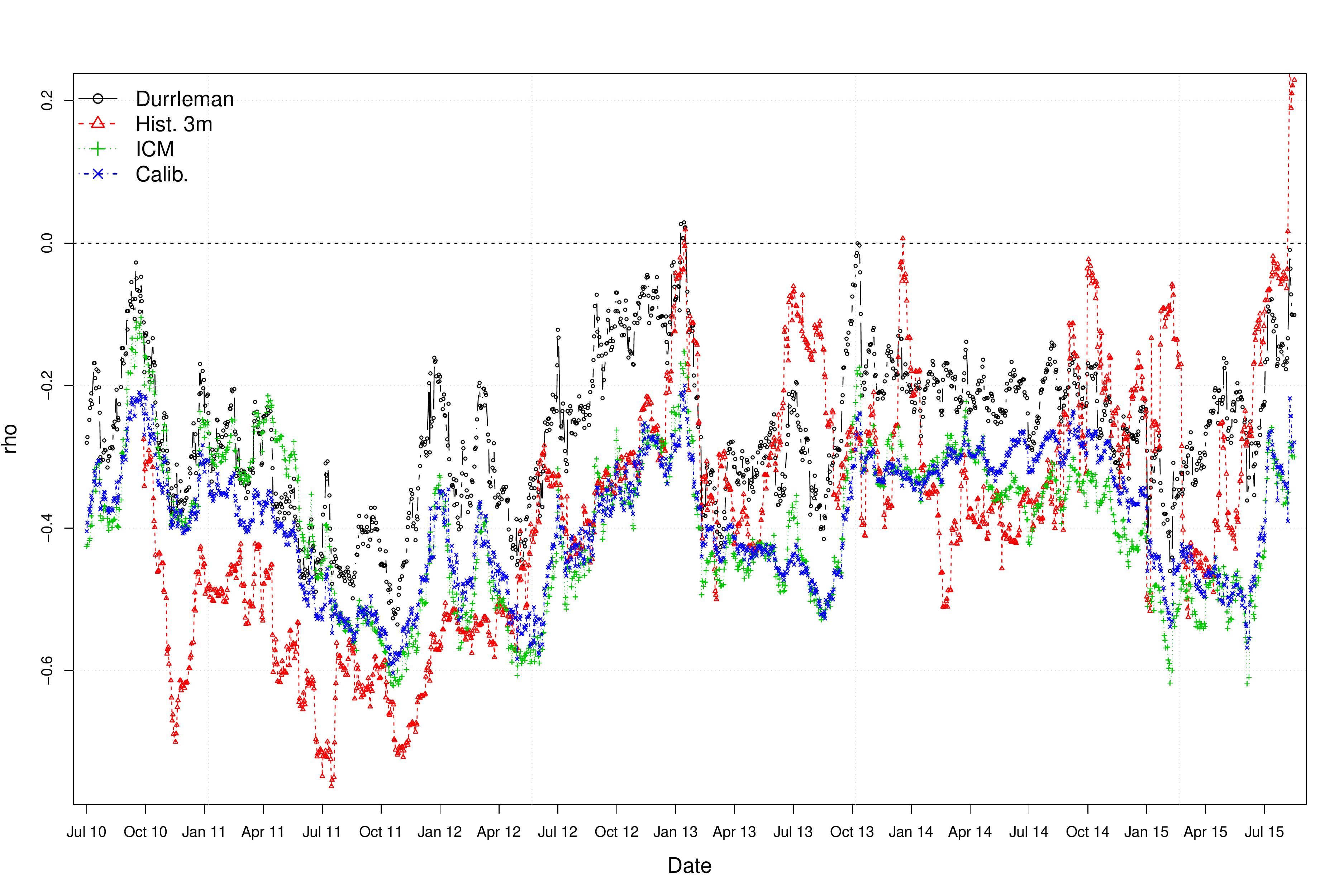}
\caption{The Heston model parameter $ \rho $ from considered methods}
\label{fig:rho} 
\label{figure:5.4} 
\end{figure}

For comparison purpose, Figure \ref{figure:5.3} shows a graphical analysis of parameters $ \omega $, which were calculated by considered methods, including historical average method. Similarly, Figure \ref{figure:5.4} shows a graphical analysis of parameters $ \rho $, which were calculated by considered methods, including historical average method.

As can be seen from Figures \ref{figure:5.3} and \ref{figure:5.4} and Table \ref{table:5.3}, in the case of mean levels of the parameter $ \omega $ Implied Central Moments method gives results which are higher than those from the Durrleman method and can be considered as too high in comparison with the numerical calibration results, mean differences here are statistically significant. On the other hand, in the case of the parameter $ \rho $, the mean and median levels in the Implied Central Moments method are very close to the numerical calibration results, the difference between means in those two methods is not statistically significant. Other conclusion is in the case of the Durrleman (2004) method results, which seem to provide always too low $ \rho $ in absolute terms.

\begin{table}[h!]
\centering
\caption{Summary of distributions of parameters $ \omega $ and $ \rho $ in tested methods}
\label{tab:omegarho} 
\label{table:5.3} 
\begin{tabular}{rcrrrrrc}
  \hline
 & Mean & S. D. & Min. & 1st Qu. & Median & 3rd Qu. & Max. \\ 
  \hline
  \,\,\,\,\,\,\,\,$\omega$ Durr. & 0.363*** & 0.132 & 0.167 & 0.260 & 0.333 & 0.438 & 0.842 \\ 
  \,\,\,\,\,\,\,\,$\omega$ Hist. & 0.136*** & 0.052 & 0.061 & 0.090 & 0.126 & 0.175 & 0.275 \\ 
  \,\,\,\,\,\,\,\,$\omega$ ICM & 0.412 & 0.130 & 0.201 & 0.302 & 0.410 & 0.497 & 0.785 \\ 
  \,\,\,\,\,\,\,\,$\omega$ Calib. & 0.343*** & 0.150 & 0.119 & 0.211 & 0.292 & 0.465 & 0.763 \\ 
  \hline
  \,\,\,\,\,\,\,\,$\rho$ Durr. & -0.262*** & 0.104 & -0.535 & -0.336 & -0.256 & -0.194 & 0.029 \\ 
  \,\,\,\,\,\,\,\,$\rho$ Hist. & -0.376*** & 0.175 & -0.762 & -0.508 & -0.377 & -0.267 & 0.280 \\ 
  \,\,\,\,\,\,\,\,$\rho$ ICM & -0.395 & 0.104 & -0.621 & -0.480 & -0.388 & -0.317 & -0.104 \\ 
  \,\,\,\,\,\,\,\,$\rho$ Calib. & -0.391 & 0.089 & -0.604 & -0.462 & -0.383 & -0.315 & -0.200 \\ 
   \hline
\multicolumn{8}{l}{\scriptsize{\textit{Note: $^{***}p<0.001$, $^{**}p<0.01$, $^*p<0.05$ are for two-sample t-test for equality with given ICM mean}}}
\end{tabular}
\end{table}

\begin{table}[h!]
\centering
\caption{Correlations between considered estimates of parameters $ \omega $ and $ \rho $}
\label{table:5.4} 
\begin{tabular}{r|rrrr|rrrr}
  \hline
 & $\omega$ Durr. & $\omega$ Hist. & $\omega$ ICM & $\omega$ Calib. & $\rho$ Durr. & $\rho$ Hist. & $\rho$ ICM & $\rho$ Calib. \\ 
  \hline
  $\omega$ Durr. & 1.000 & 0.734 & 0.969 & 0.901 & -0.660 & -0.625 & -0.507 & -0.672 \\ 
  $\omega$ Hist. & 0.734 & 1.000 & 0.779 & 0.657 & -0.448 & -0.434 & -0.552 & -0.578 \\ 
  $\omega$ ICM & 0.969 & 0.779 & 1.000 & 0.910 & -0.658 & -0.643 & -0.534 & -0.701 \\ 
  $\omega$ Calib. & 0.901 & 0.657 & 0.910 & 1.000 & -0.696 & -0.750 & -0.399 & -0.578 \\
  \hline 
  $\rho$ Durr. & -0.660 & -0.448 & -0.658 & -0.696 & 1.000 & 0.585 & 0.738 & 0.804 \\ 
  $\rho$ Hist. & -0.625 & -0.434 & -0.643 & -0.750 & 0.585 & 1.000 & 0.302 & 0.452 \\ 
  $\rho$ ICM & -0.507 & -0.552 & -0.534 & -0.399 & 0.738 & 0.302 & 1.000 & 0.895 \\ 
  $\rho$ Calib. & -0.672 & -0.578 & -0.701 & -0.578 & 0.804 & 0.452 & 0.895 & 1.000 \\ 
  \hline
\end{tabular}

\end{table}

In addition, correlations between the parameters $ \omega $ and $ \rho $ from considered methods were verified and are shown in Table \ref{table:5.4}. As shown in the table, both the ICM and the Durrleman (2004) method give estimates of the parameter $ \omega $ which are correlated to similar extent with the parameter $ \omega $ from the numerical calibration. In the case of the parameter $\rho$ considered methods are not equally correlated with the numerical calibration $\rho$ and the difference between their correlations is over 9 percentage points. Reverse logic should be applied when comparing to estimates from historical method. Confirmed facts about historical estimators from the literature, e.g. Canina and Figlewski (1993)\footnote{~\bibentry{Canina1993}.} and Day and Lewis (1993)\footnote{~\bibentry{Day1993}.} for stock options market, Amin and Ng (1997)\footnote{~\bibentry{Amin1997}.} for the interest rate option market, show that the implied variance is a better estimator of the future realized variance than the historical realized variance. Due to this fact, the considered relationship with the historical estimator is rather not desirable.

\begin{table}[h!]
\centering
\caption{Summary of distributions of RMSE and calibration risk of the Heston model parameters}
\label{table:5.5} 
\begin{tabular}{rcrrrrrr}
  \hline
 & Mean & S. D. & Min. & 1st Qu. & Median & 3rd Qu. & Max. \\ 
  \hline
  $\theta$ Durr. & 0.0058 & 0.0050 & 0.0002 & 0.0026 & 0.0046 & 0.0071 & 0.0463 \\ 
  $\theta$ ICM & 0.0024*** & 0.0024 & 0.0000 & 0.0008 & 0.0016 & 0.0032 & 0.0172 \\ 
  \hline
  $\kappa$ Durr. & 3.6850 & 4.5096 & 0.0073 & 1.2550 & 2.4970 & 4.2970 & 38.0200 \\ 
  $\kappa$ ICM & 2.7950*** & 6.8876 & 0.0160 & 0.4585 & 1.0240 & 2.2000 & 88.3700 \\ 
  \hline
  $\nu_0$ Durr. & 0.0013 & 0.0020 & 0.0000 & 0.0002 & 0.0006 & 0.0018 & 0.0175 \\ 
  $\nu_0$ ICM & 0.0012 & 0.0018 & 0.0000 & 0.0004 & 0.0007 & 0.0013 & 0.0180 \\ 
  \hline
  RMSE Durr. & 0.0040 & 0.0018 & 0.0008 & 0.0024 & 0.0040 & 0.0052 & 0.0104 \\ 
  RMSE ICM & 0.0030*** & 0.0011 & 0.0009 & 0.0021 & 0.0030 & 0.0037 & 0.0071 \\ 
   \hline
\multicolumn{8}{l}{\scriptsize{\textit{Note: $^{***}p<0.001$, $^{**}p<0.01$, $^*p<0.05$ are for two-sample t-test for equality with other method mean}}}
\end{tabular}

\end{table}

Table \ref{table:5.5} and Figures \ref{figure:A.1}, \ref{figure:A.2}, \ref{figure:A.3} and \ref{figure:A.4} in Appendix C show the calibration risk of parameters $ \theta $, $ \kappa $ and $ \nu_0 $ and RMSE in both methods. Table \ref{table:5.5} shows that after the numerical calibration of parameters $\theta $, $\kappa$ and $ \nu_0 $ with MSE cost function quartiles and the mean of RMSE errors of the ICM method are lower than analogues ones of the Durrleman method. Similar differences are in calibration risk statistics. The mean level of the parameter $ \theta $ calibration risk in the ICM method is lower by a half in comparison to the mean risk level in the Durrleman method. For the parameter $ \kappa $ the mean risk level is also better for the ICM method, this time by one quarter. The smallest improvements is for the calibration risk of the parameter $\nu_0$, which are on similar levels and the difference between both methods is not statistically significant.
In summary, the Implied Central Moments method allows to apply the partial numerical calibration with parameters $\rho$ and $\omega$ being fixed, which will result in lower calibration risk than in the case of fixing parameters $\rho$ and $\omega$ on the levels from the Durrleman (2004) method.

\section{Tests of stochastic volatility models with two-factor variance}
\label{sec:5.4}

\subsection{The design of the study and its objectives}

This section presents a study which was designed to compare the quality of the model fit to the observed option prices for two-factor variance models. The study used the whole data set with date range of 22-07-2010 to 31-08-2015, which has 1333 working days and all presented time series have this amount of observations. In total there are three models tested in this section:
\begin{enumerate}
\item the Bates two-factor model without the Feller condition,
\item the Bates two-factor model with the Feller condition,
\item the OUOU model.
\end{enumerate}

All models were calibrated twice using two different starting vectors. Main objective was to compare the impact of using different starting vectors on the calibration. Considered starting points for each model are:
\begin{enumerate}
\item Starting point which is computed using Nelder-Mead calibration of 5 parameters of symmetrical version of the model\footnote{~The starting point for the first calibration process is obtained from variance term structure parameters and historical estimates of parameters $\rho$ and $\omega$.}.
\item Starting point which is estimated from Implied Central Moments method coupled together with the Equal Variance Parametrisations method (for unconditioned Bates model) or the Modified Equal Variance Parametrisations methods (for the OUOU model and the Bates model with the Feller condition).
\end{enumerate}

In the first part the model fit in starting points is tested and in the second part the model fit after the Nelder-Mead calibration of 10 parameters is analysed. An additional objective of the study is to compare the quality of the fit to the observed option prices in two models after the calibration: the Bates two-factor variance model and the new OUOU model.

Beside testing methods for two-factor models parameters estimation, it can be considered as further tests of the Implied Central Moments method, because used methods are basing on Implied Central Moments estimates for one-factor model. To make the calibration with the Feller condition ending always with the condition being fulfilled the same mechanism as for one-factor models has been applied.


The RMSE for differences between the model call option prices $ C $ divided by the market sensitivity vega ($ \nu $) to the observed option prices divided by the market parameter vega was taken as the cost function for the calibration algorithm, which was the Nelder-Mead method. The setup is almost the same as in the one-factor model test section. The only exception is the maximum number of iterations, which has been chosen to be 800 due to the longer  and more complicated calibration problem. The square-root of MSE has been chosen as a method for summarizing the efficiency of the fit, same way as in the first part of the chapter:

\begin{equation}
\mathrm{RMSE} = \sqrt{\tfrac{1}{N}\sum^N_{i=1} \bigg(\frac{C_i^{\mathrm{model}}-C_i^{\mathrm{market}}}{\nu}\bigg)^2}
\end{equation}

where $ N = 30 $, because there are five possible values for delta and six for time to maturity. This method is the preferred alternative to computing differences of implied volatilities. Its main advantage is the speed of computing, which is a significant problem in minimizing the function of ten arguments.

For the OUOU model and the Bates two-factor model the effectiveness of the calibration algorithm has been checked in terms of RMSE for each next 10 iterations. Then summary statistics have been computed and means of RMSE for each next 10 iterations have been presented in the plot.

\subsection{The results of the study - the quality of the fit to market prices of options}

\begin{figure}[h!]
\centering
\includegraphics[width=150mm]{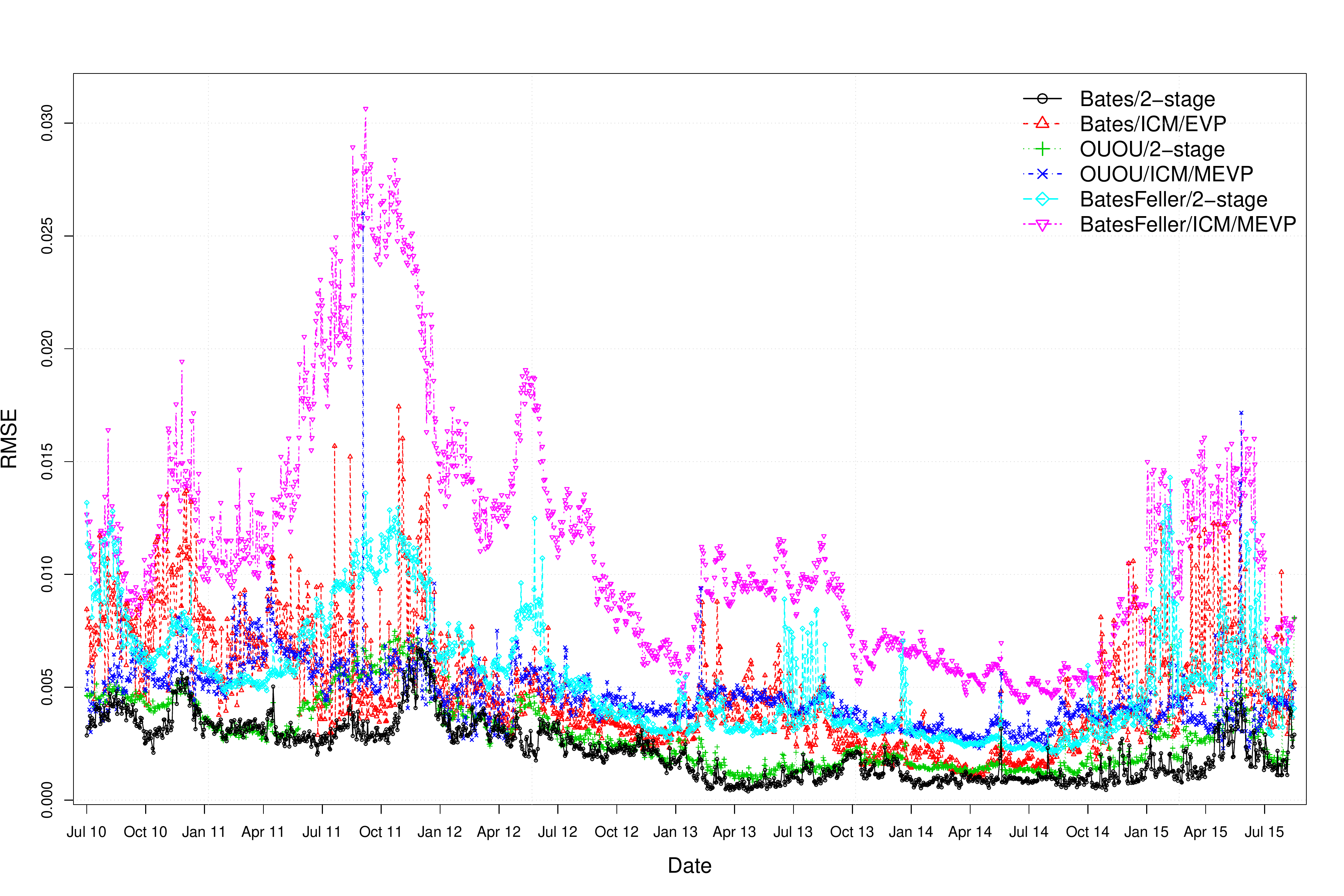}
\caption{The evolution of RMSE for tested starting points}
\label{figure:5.5} 
\end{figure}

The RMSE of the fit to the market data for models using starting points from full calibration of symmetrical version of the model and the Implied Central Moments method with the Equal Variance Parametrisations method are shown in Table \ref{table:5.6} and Figure \ref{figure:5.5}

\begin{table}[h!]
\centering
\caption{Summary of distributions of RMSE of option prices for tested starting points}
\label{table:5.6} 
\begin{tabular}{rcrrrrrl}
  \hline
 & Mean & S. D. & Min. & 1st Qu. & Median & 3rd Qu. & Max. \\ 
  \hline
  Bates/2-stage & 0.0022 & 0.0013 & 0.0004 & 0.0010 & 0.0021 & 0.0031 & 0.0067 \\ 
  Bates/ICM/EVP & 0.0049*** & 0.0027 & 0.0010 & 0.0030 & 0.0042 & 0.0065 & 0.0174 \\ 
  \hline
  OUOU/2-stage & 0.0029 & 0.0015 & 0.0008 & 0.0016 & 0.0027 & 0.0040 & 0.0081 \\ 
  OUOU/ICM/MEVP & 0.0046*** & 0.0015 & 0.0022 & 0.0037 & 0.0044 & 0.0053 & 0.0260 \\ 
  \hline
  BatesFeller/2-stage & 0.0055 & 0.0026 & 0.0020 & 0.0032 & 0.0051 & 0.0070 & 0.0143 \\ 
  BatesFeller/ICM/MEVP & 0.0112*** & 0.0053 & 0.0044 & 0.0070 & 0.0102 & 0.0135 & 0.0306 \\ 
   \hline
\multicolumn{8}{l}{\scriptsize{\textit{Note: $^{***}p<0.001$, $^{**}p<0.01$, $^*p<0.05$ are for two-sample t-test for equality with mean of same model in 2-stage method}}} \\
\multicolumn{8}{l}{\scriptsize{\textit{T-test p-value for means equality within 2/stage results: Bates-OUOU is $<$ 2.2e-16, BatesFeller-OUOU is $<$ 2.2e-16}}} \\
\multicolumn{8}{l}{\scriptsize{\textit{T-test p-value for means equality within ICM/(M)EVP results: Bates-OUOU is 0.00027, BatesFeller-OUOU is $<$ 2.2e-16}}}
\end{tabular}
\end{table}

\begin{table}[h!]
\centering
\caption{Summary of distributions of RMSE of option prices for tested models after the final calibration}
\label{table:5.7} 
\begin{tabular}{rcrrrrrl}
  \hline
 & Mean & S. D. & Min. & 1st Qu. & Median & 3rd Qu. & Max. \\ 
  \hline
  Bates/2-stage & 0.00147 & 0.00076 & 0.00034 & 0.00085 & 0.00132 & 0.00195 & 0.00521 \\ 
  Bates/ICM/EVP & 0.00119*** & 0.00056 & 0.00036 & 0.00074 & 0.00111 & 0.00150 & 0.00537 \\ 
  \hline
  OUOU/2-stage & 0.00159 & 0.00082 & 0.00038 & 0.00098 & 0.00149 & 0.00200 & 0.00787 \\ 
  OUOU/ICM/MEVP & 0.00135*** & 0.00061 & 0.00040 & 0.00083 & 0.00126 & 0.00174 & 0.00458 \\ 
  \hline
  BatesFeller/2-stage & 0.00418 & 0.00214 & 0.00082 & 0.00242 & 0.00358 & 0.00551 & 0.01371 \\ 
  BatesFeller/ICM/MEVP & 0.00228*** & 0.00163 & 0.00040 & 0.00094 & 0.00178 & 0.00319 & 0.01448 \\ 
   \hline
\multicolumn{8}{l}{\scriptsize{\textit{Note: $^{***}p<0.001$, $^{**}p<0.01$, $^*p<0.05$ are for two-sample t-test for equality with mean of same model in 2-stage method}}} \\
\multicolumn{8}{l}{\scriptsize{\textit{T-test p-value for means equality within 2/stage results: Bates-OUOU is 0.00014, BatesFeller-OUOU is $<$ 2.2e-16}}} \\
\multicolumn{8}{l}{\scriptsize{\textit{T-test p-value for means equality within ICM/(M)EVP results: Bates-OUOU is 2.61e-12, BatesFeller-OUOU is $<$ 2.2e-16}}}
\end{tabular}

\end{table}

\begin{figure}[h!]
\centering
\includegraphics[width=150mm]{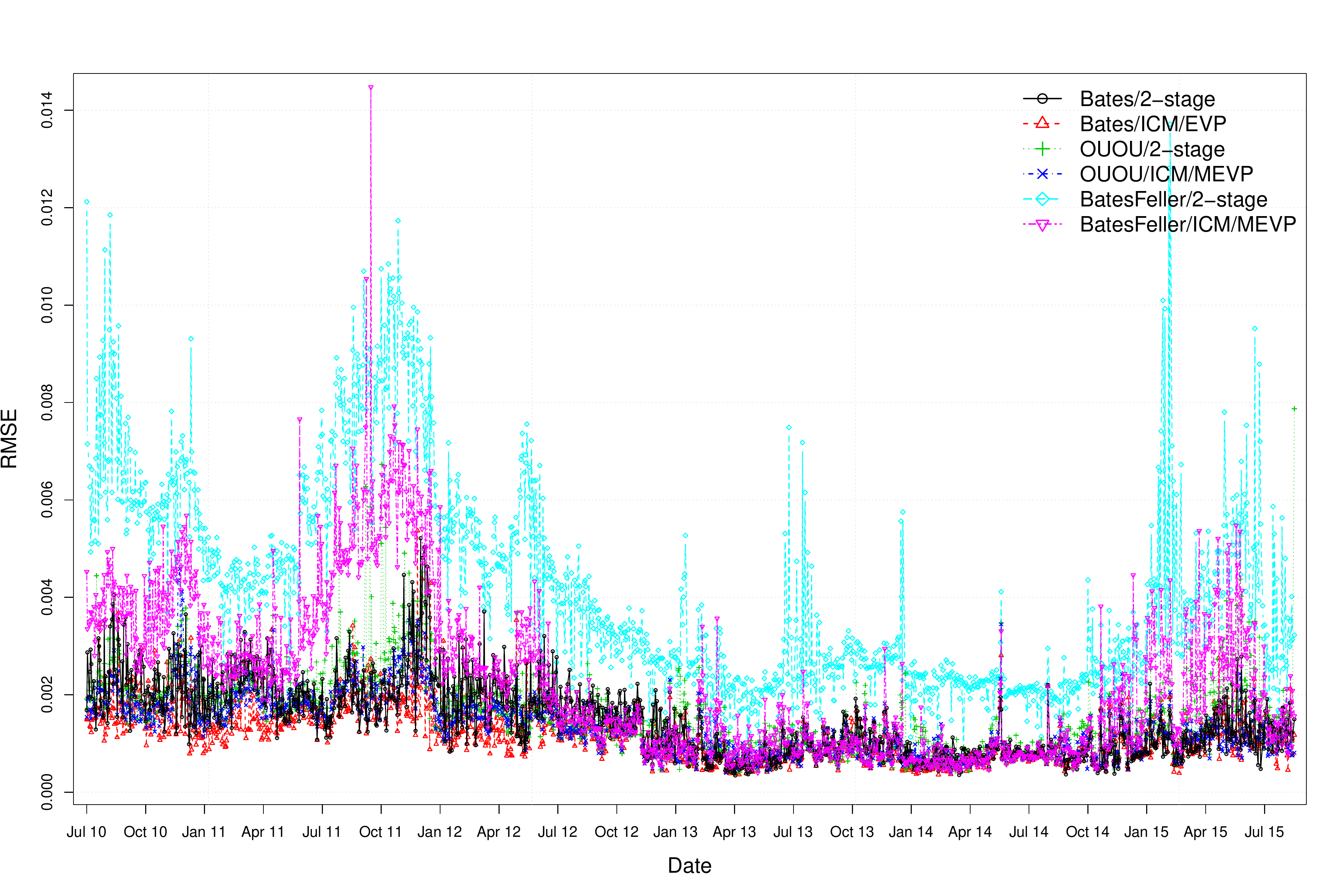}
\caption{The evolution of RMSE for tested models after the final calibration}
\label{figure:5.6} 
\end{figure}

Table \ref{table:5.6} shows that before the final calibration 2-stage methods have lower RMSE for all models, but this costs longer time of execution, because they use numerical calibration in the first stage. Results of the calibrations in terms of RMSE of the fit to the market data using tested starting points are shown in Table \ref{table:5.7} and Figure \ref{figure:5.6}. Table \ref{table:5.7} shows that:
\begin{itemize}
	\item The proposed Implied Central Moments method with Equal Variance Parametrisations method allow a better fit to the market data after the numerical calibration step, differences to means in 2-stage are statistically significant.
	\item All quartiles and the mean of RMSE for the OUOU model are better than for the Bates model with the Feller condition and differences of means are statistically significant.
	\item The results of the OUOU model are worse than the results of the unconditioned Bates model and differences of means are again statistically significant, but the relative distances between those results are small.
\end{itemize}

Besides, in the unconditioned Bates model the overall share of days for which the Feller condition is not true for at least one variance equation is 100\%. This means that although the model can be used as a tool for vanilla option pricing it should not be used for simulations and should not be used for analysis of the EURUSD market dynamics.
Including the fact that in the Heston and the Bates model, when the Feller condition is not satisfied, then they can be still used for simulation although some modification has to be made. The full truncation method is considered to be the best modification\footnote{~\bibentry{Gauthier2009}.}, in which $\nu$ is replaced by $\max(\nu,0)$. Such modification will introduce bias to asset prices, which will be proportional to the ratio $\tfrac{\omega^2}{2 \kappa \theta}$. This is one additional argument for considering the OUOU model results especially good in comparison to the unconditioned Bates model, which is along with expectations presented in the analysis of the OUOU model calibration and the comparison of the symmetrical OUOU and the symmetrical Bates model.

\begin{figure}[h!]
\centering
\includegraphics[width=150mm]{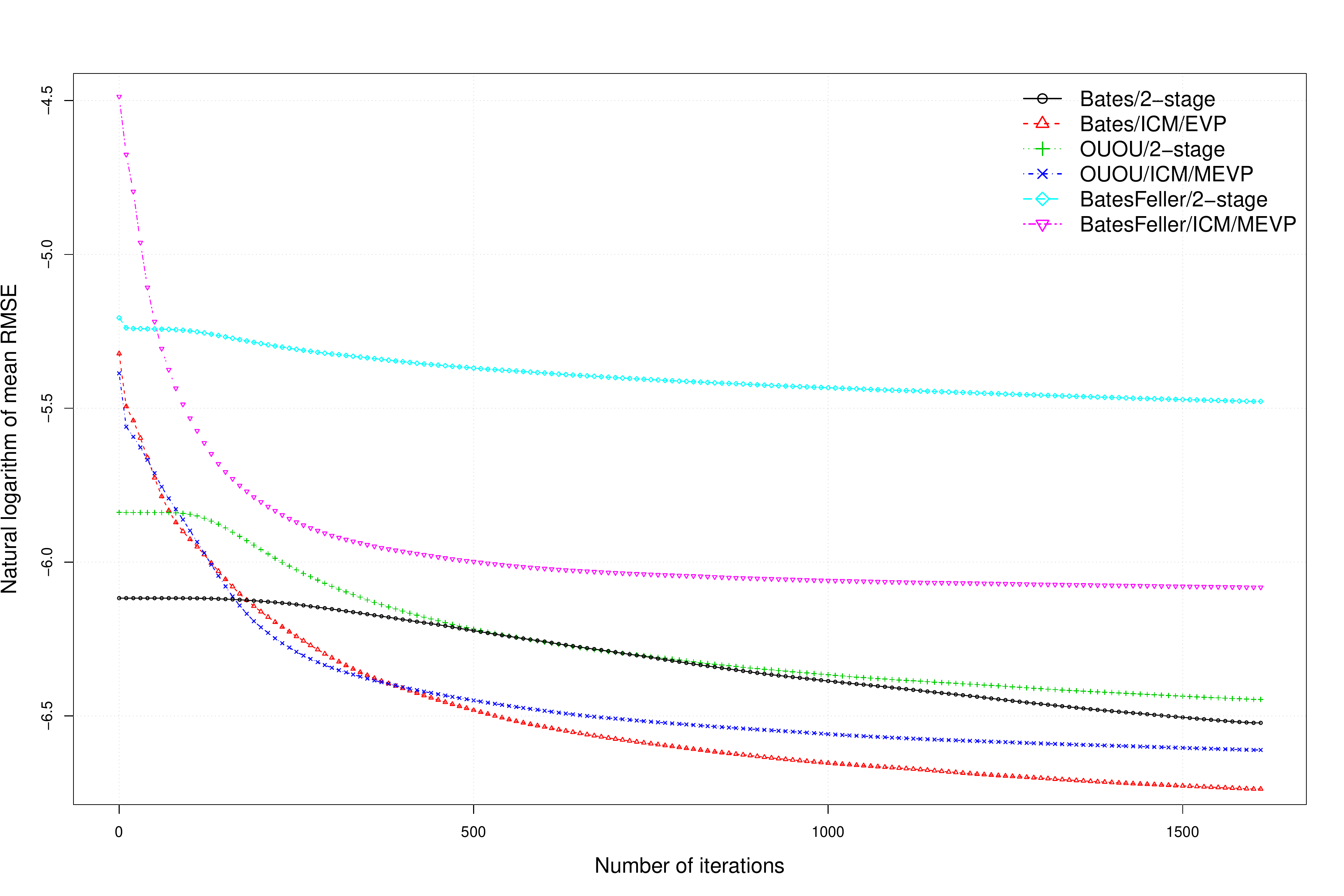}
\caption{Convergence of RMSE for tested starting points}
\label{figure:5.7} 
\end{figure}

Figure \ref{figure:5.7} presents natural logarithms of mean RMSE for each next 10 iterations step during the final calibration process. Logarithms are used due to the better visibility of differences at the calibration end. Figure \ref{figure:5.7} shows that in all cases the two-stage calibration process is starting the final calibration from points, which are closer to the global minimum, but then the convergence to the optimal solution is slower than in the case of Implied Central Moments method with Equal Variance Parametrisations method.




\section{Conclusions}

The research presented in the fifth chapter had two objectives. Firstly, it aimed to test the fit of the Heston model to the market data using approximate formulas for parameters from the fourth chapter, including the author's approach to the problem. This way, the Durrleman (2004) formulas and the author's formulas from Implied Central Moments method were compared and tested. Secondly, the research aimed to test market data fit of the proposed OUOU model against Bates two-factor variance model with the Feller condition, which has the same number of parameters.

Based on the test results the effectiveness and the usefulness of the Implied Central Moments method can be confirmed in both the estimation of the Heston model parameters and in using the estimates as a starting point for the numerical calibration. This confirms the fourth research hypothesis. The presented Implied Central Moments method was more effective than the competing method. The advantage was also present in the calibration risk test, in which all three parameters from the Heston model variance term strucure had less calibration risk. This confirms the fifth research hypothesis.\par

In the second part of the chapter, namely Section \ref{sec:5.4}, the quality of the fit to the market data of stochastic volatility models with two-factor variance or volatility from the second chapter was tested. In this way, for the first time the new original OUOU model has been empirically verified. The Bates two-factor variance model has been selected as a competitive solution, against which the OUOU model has been benchmarked. \par


Firstly, the full calibration using the Implied Central Moments with Equal Variance Parametrisations method starting points gives better results than the two-stage calibration method for the Bates model and the OUOU model, which is the confirmation of the seventh research hypothesis. This way the Implied Central Moments method was also tested in the calibration process of those more complex models.

Secondly, basing on tests results it can be confirmed that the proposed OUOU model allows a better fit to the market data than the Bates two-factor variance model with the Feller condition, which has the same number of parameters. This confirms the eight research hypothesis. The results of fitting of the OUOU model to the market data are similar to findings from the implementation of the third and the fourth additional objectives, which concerned a better fit of the Sch\"obel-Zhu model to the market data than in case of the Heston model. The loss of accuracy between the OUOU model and the unconditioned Bates model is small, which can be considered as a success in terms of finding good model for simulation on the market where the Feller condition is often not fulfilled.\par



\clearpage{\pagestyle{empty}\cleardoublepage}

\chapter*{Research conclusion}
\markboth{\hfill \textit{Research conclusion} \hfill}{}
\fancyhead[RE,LO]{\leftmark}
\addcontentsline{toc}{chapter}{Research conclusion}

The principal objective has been completed, it concerned developing a new method of derivation of approximate formulas for estimation of stochastic volatility diffusion models parameters. Considered models were with exponential-affine characteristic function of price distribution and one-factor or two-factor variance, in particular the Heston model. Considered parameters were corresponding to the volatility of variance and correlation of variance and returns of the underlying instrument.

The new method was supposed to allow deriving an approximate formula for parameters which would generate more appropriate estimates of model parameters, which could be also used as starting points for calibration process by numerical algorithm. This should lead to overall better calibration results as well as reduction of the uncertainty of the calibration with regard to used cost functions, so called calibration risk. Such effects can be obtained by determining the starting point of a closer proximity with respect to the global minimum and thus by increasing the chance of finding the global minimum, or a local minimum, which is at least very close to it by the local minimization algorithm on the objective function. \par

Furthermore, the estimation of parameters with an approximate formula allows to set part of the parameters fixed during the calibration. The reduction of the number of calibrated model parameters is reducing the time and risks of instability between calibrations of parameters for different moments in time. The most popular stochastic volatility diffusion model of exponential-affine class is the Heston model (1993) and therefore most attention in this research was paid to this particular model. However, the aim of this study was to develop and test methods that will work well with other models of this class, such as the Sch\"obel-Zhu (1998) model. Also some part of the research concerns an additional method for using approximate formulas for stochastic volatility diffusion models of exponential-affine class with two-factor variance, such as the Bates two-factor variance model or the new OUOU model. \par

Before the implementation of the principal objective, several additional research problems were considered along with related to them additional objectives. All of them have been completed: \par

\begin{enumerate}
\item The first additional objective was the formulation of the general form for many one-factor variance stochastic volatility models, which is the starting point for further considerations of the research. Implementation of the first additional objective provides an introduction and analysis of the existing literature concerning the valuation of options and stochastic volatility models with a one-factor variance.  In addition, methods of option valuation were presented  beside the implementation of the objective. Among considered methods was the integral of the characteristic function of underlying instrument price distribution, which is the base for the second additional objective with the option pricing problem. \par

\item The second additional objective concerned a derivation of the general equation of replication portfolio dynamics in the prevoisly defined general model and a development of a model which has similarity to the Sch\"obel-Zhu model but has two equations of the dynamics of volatility and derivation of the solution for the characteristic function of the underlying instrument price distribution in this model. As a result, the OUOU model and the analytic formula for the characteristic function of underlying instrument price distribution for this model have been created. The second additional objective is an extension of the theoretical basis, which is important in principal objective. The OUOU model has been used in tests of methods, which were developed during the implementation of the principal objective.\par

Besides, the implementation of the second additional objective is also an extension to the existing literature concerning option pricing in the stochastic volatility diffusion models with exponential-affine characteristic functions of price distribution with one- and two-factor variance. The general form of the equation of the characteristic function segments dynamics for the considered stochastic volatility models enabled the comparison of different models in terms of the possibility of finding an analytic form of the characteristic function. In addition, assumptions, differences and similarities between considered models were discussed during the implementation. \par


\item The third additional objective was to identify statistical factors in the dynamics of implied volatility surface for currency options in both its dimensions, i.e. two dimension separately and together: the option moenyness and the time to maturity. Implementation of the third additional objective is the expansion of the literature concerning the statistical research on the dynamics of the volatility surface. The principal component analysis was mainly used as a tool for this task in the literature. Hence, beside PCA, a Common Factor Analysis was also used in this research.

The first hypothesis was tested during the implementation of the third additional objective. One of the conclusions of the study is the fact that it is harder to explain the variance of the term structure of implied volatility than the variation of volatility smile with the same number of PCA factors. This is the confirmation of the first research hypothesis. This hypothesis refers to previous studies on the implied volatility surface, which were testing both the entire volatility surface in two dimensions and in the dimension of the volatility smile and the term structure of volatility, however, they were lacking a comparison between factors of the volatility smile with factors of the term structure of volatility. Such study is significant in broadening of the modelling of implied volatility in both dimensions, e.g. through the stochastic volatility models development.

The results of the factor analysis suggests that models with four-factors have difficulties with representing of surface movements at the end that is corresponding to the longest tenor. In addition, specific factors are greater at ends of the volatility smile, so four-factor models also have more problems in representing changes in the curvature of the volatility smile. In addition, it can be said that it is worthwhile to increase the concentration of research on more complicated models, i.e. models of two-factor variance, which were considered and developed in the second chapter. It is also the support for the reasonableness of the research on the extended Sch\"obel-Zhu model. \par

\item The fourth additional objective concerned the comparison of the functional form of the variance equation in the Heston and the Sch\"obel-Zhu model. The second hypothesis was tested during implementation of the fourth additional objective. The results of the analysis of autoregressive models made it possible to say that the functional form of the Sch\"obel-Zhu  model is more flexible than the form of the Heston model in explaining of historical data, as its analogous autoregressive model allows to explain the greater part of the variance of implied volatility time series. This confirms the second research hypothesis.

\item The fifth additional objective concerned the comparison of the parametrization of the variance term structure in the Heston and the Sch\"obel-Zhu model in the context of its use in prediction of the direction of implied volatility changes. The third hypothesis was tested during implementation of the fifth additional objective. The discriminant analysis has shown that the parameters of the volatility term structure of the Sch\"obel-Zhu model contain additional and important information for formulation of expectations for probability of future direction of implied volatility changes when they are used in logistic regression models with some other variables from the olatility surface. The Sch\"obel-Zhu parametrization is also more useful in forecasting of direction of volatility changes than the Heston model variance term structure parametrization. This confirms the third research hypothesis.

Both the fourth and fifth additional objectives involved comparisons of the Heston and the Sch\"obel-Zhu model in the context of the use of their calibrated parameters in time series models of the implied volatility and the implied variance. The fourth and the fifth additional objective can be placed in the same mainstream research, in which is the third additional objective. Unlike the third, it have also mathematical elements of option pricing models, however the direct analysis of option pricing is absent. Statistical analyses, which confirm the second and third research hypothesis, also show the advantages of the Sch\"obel-Zhu model and are another argument for meaningfulness of research on the extension of the Sch\"obel-Zhu model.\par

In summary, the implementation of the fourth and the fifth objective allowed the confirmation that stochastic volatility models, especially the five-parameter Sch\"obel-Zhu model, are justified with statistical properties of the implied volatility surface dynamics. \par
\end{enumerate}

The principal objective of the dissertation was completed in the last two chapters, after the implementation of all additional objectives. Its implementation was divided into two parts because it has both the cognitive nature, i.e. introduction of a new theory, and the empirical nature, i.e. model testing: \par

\begin{enumerate}
\item Initially, known calibration methods and the methods of derivation of approximate formulas for estimation of the Heston model parameters were presented. It can be observed that in the subject of approximate formulas for the Heston model parameters, more attention was paid to parameters $ \rho $ and $ \omega $. In case of other three parameters, $ \nu_0 $, $ \theta $, $ \kappa$, which are the parametrisation of variance term structure, their estimation is relatively simpler. They can be found by smaller calibration, which is the calibration of three parameters in the variance term structure or parameters $ \nu_0 $ and $ \theta $ can be estimated with approximate formulas of Guilliame and Schoutens (2010) method.

In the empirical part of the dissertation in all approximate formulas for the parameters $ \rho $ and $ \omega $ parametrization of the variance term structure was used as a given, so the knowledge about values of all five Heston model parameters was used in the analysis of the model performance in the option pricing.\par

After the presentation of known methods of derivation of approximate formulas for estimation of the Heston model parameters in the Section \ref{sec:4.4.3} of the fourth chapter, the development of a new method and corresponding approximate formulas is presented. It uses all of observed option prices, which express fundamental characteristics of the options market. In similar way VIX and SKEW indices are utilizing option prices to show characteristics of the stock index option market. The method has been named Implied Central Moments method because of derivation of approximate formulas for parameters $\rho$ and $\omega$ as a functions of Implied Central Moments of underlying instrument returns. Implied central moments in those equations are known, because they can be calculated from option prices using formulas for expected values of so called power payout portfolios.

The estimation method does not depend on only two option prices, as the method of Rivalle and Gauthier (2009). In this respect, it is similar to the Durrleman (2004) method, but does not contain an estimation of the linear regression of implied volatilities levels versus corresponding moneyness. It is worth mentioning that the Gauthier and Rivalle (2009) method uses only three of all available five points on the volatility smile of currency options, so it is considered to be potentially more risky, as the calculated parameters $ \rho $ and $ \omega $ may not reflect the true degree of the smile convexity, taking into account omitted implied volatilities of 10 - delta options. This is particularly important for the listed options market, where the volatility smile is formed by more than five points. Thus, the only previously known method which allows to determine $ \rho $ and $ \omega $ basing on all of the information from the volatility smile is the Durrleman (2004) method, although it uses only information from the smile with the smallest time to maturity.

Similarly as competing methods, the Implied Central Moments method allows not only to reduce the risk of calibration and to slightly speeds up local minimization algorithms by using a better starting point but alternatively also to significantly shorten the process of model calibration and option pricing by setting $ \rho $ and $ \omega $ on fixed level and calibrating of fewer parameters. The last application may be especially important for option pricing in the market regime of rapidly changing implied volatilities. The implementation of this part of the principal objective was also done for the Sch\"obel-Zhu model and for the Bates two-factor variance model (2000) and the author's OUOU model, through an additional method. \par

\item The second part of the principal objective has been implemented by a number of empirical tests on historical data. Firstly, the quality of the Heston model fit to the market data was tested using Durrleman (2004) method and the Implied Central Moments method. Test results can confirm the effectiveness and usefulness of the Implied Central Moments method in both the estimation of the Heston stochastic volatility model parameters and the partial calibration of that model with the starting point, that has been estimated with the formulas of this method. The estimation of parameters with implied central moments method was more effective than with the competitive method in terms of option pricing errors. This finding has confirmed the fourth research hypothesis. Besides, in the partial model calibration its application to starting points estimation resulted in a smaller calibration risk, which has confirmed the fifth research hypothesis. \par

Then, the quality of the fit to the market data was tested in the case of stochastic volatility models with two-factor variance or volatility. In this way, the Implied Central Moments method has been also tested in a more complex calibration process of the Bates two-factor model and the OUOU model. Using this method with the equal variance parametrzation method proved to be more effective in the final calibration than the two-step calibration method. This is a confirmation of the sixth research hypothesis.\par

In addition, the OUOU model has been empirically verified for the first time. As a competitive solution, against which this model was compared, the Bates two-factor variance model has been selected. Results of the calibration using the starting point that has been estimated with Implied Central Moments and Modified Equal Variance Parametrisations methods, confirm that the proposed OUOU model allows a better fit to the market data with the same number of used parameters as in the case of the Bates model with the Feller condition, which has been calibrated from the starting point from the same method. The results of fitting of the OUOU model to the market data are similar to the findings resulting from the implementation of the third and the fourth additional objectives, which concerned a better fit of the Sch\"obel-Zhu model to the market data than in the case of the Heston model. This confirms the seventh research hypothesis.\par

Besides, in the unconditioned Bates model the overall share of days for which the Feller condition is not fulfilled for at least one variance equation is 100\%. This means that although the model can be used as a tool for vanilla option pricing its application for simulation is problematic and it should not be used for the analysis of the EURUSD market dynamics. The loss of accuracy between the OUOU model and the unconditioned Bates model is small, which can be considered as a success in terms of finding a good model for simulation on the market where the Feller condition is often not fulffiled.\par

Hypotheses sixth and seventh relate to the quality of the application of the newly created method in the Bates model and the new OUOU model, which itself is an expansion of existing theory about stochastic volatility models and was made during implementation of the second additional objective. \par
\end{enumerate}

The principal objective of the research is related with following hypotheses: the fourth, the fifth, the sixth and the seventh. These hypotheses refer directly to the thesis of dissertation, and since all of them have been confirmed, so this is also a confirmation of the thesis of the dissertation. \par

In conclusion, the principal objective of the research in its qualitative as well as empirical part, and all of the additional objectives have been completed, leading also to solve the principal problem of the research and all additional research problems. Also, all hypotheses were verified and confirmed.

The dissertation brings expansion of the quantitative finance scientific theory with a new and more accurate method for deriving of approximate formulas for stochastic volatility diffusion models parameters, whose quality was verified empirically for the Heston model, the Bates two-factor variance model and the OUOU model for options on EURUSD, which is the most traded currency pair. 

On the other hand, in the practical area, the implementation of the methods from the dissertation give banks and other portfolio managers an option to use a new tool. That tool expands the range of the practical methods that are essential for option pricing. It is even more accurate method of estimation of the Heston and the Sch\"obel-Zhu model parameters and also to more general models with two-factor variance. From mentioned models the Heston model is most popular and used often in banking as the base model for pricing of exotic derivatives and simulations of market risk factors for Value at Risk method purposes . The method presented in dissertation is also useful for determining starting points for the calibration of aforementioned models. 

The method established in this study has proven advantages over alternative solutions. Naturally, due to the trade off between the speed of the model calibration and the quality of fit to the market data, option portfolio managers will have to decide if it will help in their daily needs or not. In the latter case they may still use the method for finding starting points for their time-consuming local minimization algorithms. The good starting point can also affect the sequence of the processes performed in a global minimization algorithm.\par

The added value is also present in new findings of Principal Components Analysis applied to the time to maturity and the strike price dimensions of the dynamics of the currency options implied volatility surface and statistical comparisons of the Heston and the Sch\"obel-Zhu model. Furthermore, the theoretical soundness of Implied Central Moments method can be the foundation for other research in this area even in the case when the method results suggest that it is the second preferable method in a given option market. \par

Development of the OUOU model, which is a new stochastic volatility model with a two-factor variance and semi-closed formula for the valuation of options, also brings a new knowledge to the research area and to the financial sector practice. Such model was previously not considered, nor studied in the literature. In addition, the OUOU model has been also examined in relation to a similar model, namely the Bates two-factor variance model, which is sometimes used in banks for the purpose of the market risk factors simulation. This can be particularly useful for pricing and risk assessment of complex derivatives which depend both on the trajectory of the price of the underlying as well as its volatility in the future.\par

Further comparative and statistical analysis of models which are extensions of the Heston and the Sch\"obel-Zhu model with two stochastic factors in volatility or variance equations, as well as research on the Implied Central Moments method in other option markets, e.g. options listed on the stock exchange, or for models from the non-exponential-affine class, can be proposed as a topic for further study in the considered area.


\clearpage{\pagestyle{empty}\cleardoublepage}

\appendix
\appendix
\chapter{Solution to the system of differential equations in the OUOU model}
\markboth{\hfill \textit{Appendix A\quad Solution to the system of differential equations in the OUOU model} \hfill}{}


The considered system of differential equations is given by:

\begin{equation}\begin{split}
& \frac{\partial A_k}{\partial \tau} = \tfrac{1}{N}(r_d - r_f) i u + B_k \kappa_k \hat{\theta_k} + \tfrac{1}{2} \omega_k^2 B_k^2 + 2 C_k ,\\
& \frac{\partial B_k}{\partial \tau} = - B_k b_{j,k} + B_k \rho_k \omega_k i u + 2 B_k C_k \omega_k^2 + 2C_k \kappa_k \hat{\theta_k} ,\\
& \frac{\partial C_k}{\partial \tau} = - 2C_k b_{j,k} + 2 \rho_k \omega_k i u C_k + a_j i u - \tfrac{1}{2} u^2 + 2 \omega_k^2 C_k^2
.\end{split}\end{equation}

with boundary conditions: $A_k(\tau=0)=B_k(\tau=0)=C_k(\tau=0)=0$. Fixing of $k$ does not change the outcome so in further considerations we use simplified notation with $k$ being omitted.

Since the equation for $ C $ has the form of the Riccati equation ($C'=P+Q C+R C^2$), we should make following substitution: $ v = CR $ to get $ v' = E + D v + v ^ 2 $, where $ D = Q $ and $ E = PR $, followed by the substitution $ v = - \tfrac{u'}{u} $ to get $ u''- D u' + E u = 0 $. The latter equation, which is the equation of the second order, has a general solution of the form $K_1 e^{r_1 \tau}+K_2 e^{r_2 \tau}$, where $ r_1 $ and $ r_2 $ are the roots of the equation $ r^2 - r D + E = 0 $. Therefore, the general solution for $ C $ is following:

\begin{equation}
C(\tau) = -\frac{1}{R} \frac{K r_1 e^{r_1 \tau}+r_2 e^{r_2 \tau}}{K e^{r_1 \tau}+e^{r_2 \tau}}
,\end{equation}

where: $r_1 = \tfrac{1}{2}(Q+d)$, $r_2 = \tfrac{1}{2}(Q-d)$, $d=\sqrt{Q^2-4PR}$.

Further, to know the value of the constant $ K $, we should use the knowledge of the particular solution which corresponds to the considered boundary condition, ie. $ C(\tau = 0) = 0 $. It shows that $ K=-\tfrac{r_2}{r_1} $. Thus, the formula for $ C $ is following:

\begin{equation}\begin{split}
C(\tau) & = -\frac{1}{R} \frac{-\tfrac{r_2}{r_1} r_1 e^{r_1 \tau}+r_2 e^{r_2 \tau}}{-\tfrac{r_2}{r_1} e^{r_1 \tau}+e^{r_2 \tau}} = -\frac{1}{R} \frac{r_2(1-e^{r_1 \tau-r_2 \tau})}{1-\tfrac{r_2}{r_1} e^{r_1 \tau-r_2 \tau}} = \frac{-r_2}{R} \frac{1-e^{d \tau}}{1-\tfrac{r_2}{r_1} e^{d \tau}} \\
& = \frac{\beta_j+d_j}{4 \omega^2} \frac{1-e^{d_j \tau}}{1-g_j e^{d_j \tau}}
,\end{split}\end{equation}

where: $\beta_j=2(b_j-\rho\omega iu)$, $d_j=\sqrt{\beta_j^2-4\omega^2(2a_jiu-u^2)}$ and $g_j = \tfrac{\beta_j+d_j}{\beta_j-d_j}$.

Then, using the same logic as with the Heston model element $g_j$ in Duffie, Pan and Singleton (2000), which has been discussed in Chapter 2, after the substitution $G_j = 1/g_j$ we have:

\begin{equation}\begin{split}
C(\tau) = \frac{\beta_j-d_j}{4 \omega^2} \frac{1-e^{-d_j \tau}}{1-G_j e^{-d_j \tau}}
,\end{split}\end{equation}

where: $\beta_j=2(b_j-\rho\omega iu)$, $d_j=\sqrt{\beta_j^2-4\omega^2(2a_jiu-u^2)}$ and $G_j = \tfrac{\beta_j-d_j}{\beta_j+d_j}$.

Knowing the formula for $ C $, the formula for $ B $ will be simplified:

\begin{equation}\begin{split}
& \frac{\partial B}{\partial \tau} + B (b_j - \rho \omega i u - 2 \omega^2 C ) = 2C \kappa \hat{\theta} \\
& \frac{\partial B}{\partial \tau} + B \bigg(\beta_j/2 - \frac{\beta-d_j}{2}\frac{1-e^{-d_j \tau}}{1-G_j e^{-d_j \tau}} \bigg) = \kappa \hat{\theta} \frac{\beta-d_j}{2 \omega^2}\frac{1-e^{-d_j \tau}}{1-G_j e^{-d_j \tau}}
,\end{split}\end{equation}

which is a linear differential equation of a first order \footnote{~Equation of type $ f '(t) + f (t) p (t) = q (t) $ can be solved by computing the integrating factor $r(t)=exp(\int p(t)\, \D t)$, and the solution is of the form $f(t) = \tfrac{1}{r(t)}(\int r(t)q(t) \, \D t + K)$.}, so its solution is following:


\begin{equation}\begin{split}
B(\tau,u) = & \frac{\int e^{\int \beta_j/2 - 2 \omega^2 C \, \D \tau} 2C \kappa \hat{\theta} \, \D \tau + K}{e^{\int \beta_j/2 - 2 \omega^2 C \, \D \tau}} = \frac{\int e^{d_j \tau/2}(1-G_j e^{-d_j \tau}) 2C \kappa \hat{\theta} \, \D \tau + K}{e^{d_j \tau/2}(1-G_j e^{-d_j \tau})} \\
= & \frac{\int e^{d_j \tau/2}(1-G_j e^{-d_j \tau}) \kappa \hat{\theta} \tfrac{\beta-d_j}{2 \omega^2}\frac{1-e^{-d_j \tau}}{1-G_j e^{-d_j \tau}} \, \D \tau + K}{e^{d_j \tau/2}(1-G_j e^{-d_j \tau})} = \frac{\kappa \hat{\theta}}{\omega^2} \frac{\int e^{d_j \tau/2}  \tfrac{\beta-d_j}{2}(1-e^{-d_j \tau}) \, \D \tau + \hat{K}}{e^{d_j \tau/2}(1-G_j e^{-d_j \tau})} \\
= & \frac{\kappa \hat{\theta}}{\omega^2} \frac{\tfrac{\beta-d_j}{d_j} e^{-d_j \tau/2} (1-e^{-d_j \tau}) + \hat{K}}{e^{d_j \tau/2}(1-G_j e^{-d_j \tau})} = \kappa \hat{\theta} \frac{(\beta_j-d_j)(1-e^{-d_j\tau/2})^2}{d_j\omega^2(1-G_j e^{-d_j\tau})}
.\end{split}\end{equation}

Constant $ \hat{K} $ is equal to $ 0 $, because it guarantees the condition $ B (\tau = 0) = 0 $ to be fulfilled.

The formula for $ A $ can be divided into two smaller problems, of which the first is similar to the one from the Heston model:

\begin{equation}\begin{split}
& A(\tau,u) = \tilde{A}(\tau,u) + \hat{A}(\tau,u) ,\\
& \frac{\partial \tilde{A}}{\partial \tau} = \tfrac{1}{N} (r_d - r_f) i u + \omega^2 C ,\\
& \frac{\partial \hat{A}}{\partial \tau} = B \kappa \hat{\theta} + \tfrac{1}{2} \omega^2 B^2
.\end{split}\end{equation}

Knowing the formula for $ C $, the formula for $ \tilde{A} $ can be found after calculating the integral:

\begin{equation}\begin{split}
\tilde{A}(\tau) & = \tfrac{1}{N} (r_d - r_f) i u \tau + \omega^2 \frac{\beta_j+d_j}{4 \omega^2} \int \frac{1-e^{d_j \tau}}{1-\tfrac{\beta_j+d_j}{\beta_j-d_j} e^{d_j \tau}} \, \D \tau \\
& = \tfrac{1}{N} (r_d - r_f) i u \tau + \frac{\beta_j+d_j}{4} \bigg( \tau - \frac{2}{\beta_j+d_j}\ln(\beta_j e^{d_j \tau} - \beta_j + d_j e^{d_j \tau} + d_j) + K \bigg)
.\end{split}\end{equation}

In addition, the restriction $C(\tau=0)=0$ implies that $K = -\ln(2d_j)$, so:

\begin{equation}\begin{split}
\tilde{A}(\tau) & = \tfrac{1}{N} (r_d - r_f) i u \tau + \frac{1}{4} \bigg( (\beta_j+d_j)\tau - 2\ln\frac{-(\beta_j-d_j) + (\beta_j + d_j) e^{d_j \tau}}{\beta_j+d_j-(\beta_j-d_j)} \bigg) \\
& = \tfrac{1}{N} (r_d - r_f) i u \tau + \frac{1}{4} \Bigg( (\beta_j+d_j)\tau - 2\ln\frac{1 - \tfrac{\beta_j+d_j}{\beta_j-d_j} e^{d_j \tau}}{1-\tfrac{\beta_j+d_j}{\beta_j-d_j}} \Bigg)
,\end{split}\end{equation}

which in form with $G_j$ is equal to

\begin{equation}
\tilde{A}(\tau) = \tfrac{1}{N} (r_d - r_f) i u \tau + \frac{1}{4} \Bigg( (\beta_j-d_j)\tau - 2\ln\frac{1 - G_j e^{-d_j \tau}}{1-G_j} \Bigg)
.\end{equation}

The remaining element that has to be calculated is $\frac{\partial \hat{A}(\tau)}{\partial \tau}$ derivative:

\begin{equation}
\frac{\partial \hat{A}(\tau)}{\partial \tau} = B \kappa \hat{\theta} + \tfrac{1}{2} \omega^2 B^2
,\end{equation}

and its solution is equal to:

\begin{equation}\begin{split}
\hat{A}(\tau) = & \int \kappa \hat{\theta} B + \tfrac{1}{2} \omega^2 B^2 \, \D\tau + K\\
= & \int \kappa \hat{\theta} \kappa \hat{\theta} \frac{\beta_j-d_j}{d_j\omega^2} \frac{(1-e^{-d_j\tau/2})^2}{1-G_j e^{-d_j\tau}} + \tfrac{1}{2} \omega^2 \bigg( \kappa \hat{\theta} \frac{\beta_j-d_j}{d_j\omega^2} \frac{(1-e^{-d_j\tau/2})^2}{1-G_j e^{-d_j\tau}} \bigg)^2 \, \D\tau + K\\
= & \frac{\kappa^2 \hat{\theta}^2(\beta_j-d_j)}{2 d_j^2 \omega^2} \int 2 d_j \frac{(1-e^{-d_j\tau/2})^2}{1-G_j e^{-d_j\tau}} + (\beta_j-d_j) \frac{(1-e^{-d_j\tau/2})^4}{(1-G_j e^{-d_j\tau})^2} \, \D\tau + K\\
= & \frac{\kappa^2 \hat{\theta}^2(\beta_j-d_j)}{2 d_j^2 \omega^2} \bigg(\frac{(\beta_j-d_j) (G_j^2 (4 e^{-d_j \tau/2}-1)+G_j (4 e^{-d_j \tau/2}-6)-1)}{-d_j G_j^2 (G_j e^{-d_j \tau}-1)} \\
& \tau (\beta_j + d_j) - \tfrac{1}{G_j^2 d_j}((G_j+1) (\beta_j - d_j - G_j (\beta_j + d_j)) \ln(1-G_j e^{-d_j \tau})) \\
& - \tfrac{1}{G_j^{3/2} d_j} (4 (\beta_j - d_j - G_j (\beta_j + d_j)) \tanh^{-1}( \sqrt{G_j} e^{-d_j \tau/2} )) + \hat{K} \bigg)
.\end{split}\end{equation} 




The last equation after multiplying factors containing $ G_j $, in which $ \beta_j - d_j - G_j (\beta + d_j) = 0 $, simplifies to:

\begin{equation}
\hat{A}(\tau,u) = \frac{\kappa^2 \hat{\theta}^2(\beta_j-d_j)}{2 d_j^2 \omega^2} \bigg( \tau (\beta_j+d_j) + \frac{\beta_j-d_j}{d_j G_j^2} \bigg( \frac{G_j^2 (4 e^{-d_j \tau/2}-1)+G_j (4 e^{-d_j \tau/2}-6)-1}{1-G_j e^{-d_j \tau}} + \bar{K} \bigg) \bigg)
,\end{equation} 

where constant $\bar{K}=-\tfrac{3G_j^2-2G_j-1}{1-G_j} = 3G_j+1 $, because then the condition $\hat{A}(\tau=0)=0$ is fulfilled. Therefore it should be:

\begin{equation}\begin{split}
\hat{A}(\tau,u) = & \frac{\kappa^2 \hat{\theta}^2(\beta_j-d_j)}{2 d_j^2 \omega^2} \bigg( \tau (\beta_j+d_j) + \frac{\beta_j-d_j}{d_j G_j^2} \bigg(\frac{G_j^2 (4 e^{-d_j \tau/2}-1)+G_j (4 e^{-d_j \tau/2}-6)-1}{1-G_j e^{-d_j \tau}}+3G_j+1 \bigg) \bigg) \\
= & \frac{\kappa^2 \hat{\theta}^2(\beta_j-d_j)}{d_j^2 \omega^2} \bigg( \tau \frac{\beta_j+d_j}{2} + \frac{4\beta_j e^{-d_j \tau/2}-(2\beta_j-d_j)e^{-d_j \tau}-2\beta_j-d_j}{d_j(1-G_j e^{-d_j \tau})}\bigg)
.\end{split}\end{equation} 



\clearpage

After completing of all outcomes and switching to full notation, the analytic solution of considered system of equations is:

\begin{equation}\begin{split}
& A_k(\tau,u) = \frac{1}{N} (r_d - r_f) i u \tau + \frac{1}{4}(\beta_{j,k} - d_{j,k}) \tau - \frac{1}{2} \ln \bigg(\frac{G_{j,k}e^{-d_{j,k}\tau}-1}{G_{j,k}-1}\bigg) + \hat{A_k}(\tau,u),\\
& \hat{A}_k(\tau,u) = \frac{\kappa_j^2 \hat{\theta_j}^2(\beta_{j,k}-d_{j,k})}{d_{j,k}^2 \omega_j^2} \bigg( \tau \frac{\beta_{j,k}+d_{j,k}}{2} + \frac{4\beta_{j,k} e^{-d_{j,k} \tau/2}-(2\beta_{j,k}-d_{j,k})e^{-d_{j,k} \tau}-2\beta_{j,k}-d_{j,k}}{d_{j,k}(1-G_{j,k} e^{-d_{j,k} \tau})}\bigg) ,\\
& B_k(\tau,u) = \kappa_k \hat{\theta_k} \frac{(\beta_{j,k}-d_{j,k})(1-e^{-d_{j,k}\tau/2})^2}{d_{j,k}\omega_k^2(1-G_{j,k} e^{-d_{j,k}\tau})} ,\\
& C_k(\tau,u) = \bigg(\frac{\beta_{j,k} - d_{j,k}}{4 \omega_k^2}\bigg) \bigg(\frac{1-e^{-d_{j,k} \tau}}{1-G_{j,k} e^{-d_{j,k} \tau}}\bigg) ,\\
& \beta_{j,k} = 2(b_{j,k} - i \omega_k \rho_k u), \quad d_{j,k} = \sqrt{\beta_{j,k}^2 - 4 \omega_k^2 (2a_j iu-u^2)}, \quad G_{j,k} = \frac{\beta_{j,k}-d_{j,k}}{\beta_{j,k}+d_{j,k}} ,\\
& k=1,2, \quad j=1,2, \quad a_1=\frac{1}{2}, \quad a_2=-\frac{1}{2}, \quad b_1 = \kappa_k + \eta_k - \omega_k \rho_k, \quad b_2 = \kappa_k + \eta_k
.\end{split}\end{equation}

\clearpage

\chapter{Expected value of instantaneous variance in Sch\"obel-Zhu model}
\markboth{\hfill \textit{Appendix B\quad Expected value of instantaneous variance in Sch\"obel-Zhu model} \hfill}{}


The beginning of the considerations is the following modified CIR\footnote{~
For Sch\"obel-Zhu model dynamics of variance, $ \upsilon $, follows this equation with the parametrization: $a=\kappa$, $b=\tfrac{\omega^2}{\kappa}$, $c=2$, $d=2\theta$, $\sigma = 2 \omega$.} equation:
\begin{equation}
\D \upsilon_t = a (b - c \upsilon_t + d \sqrt{\upsilon_t}) \D t + \sigma \sqrt{\upsilon_t} \D W_t
.\end{equation}

In the integrated form it is equal to:
\begin{equation}
\upsilon_t = \upsilon_0 + a \int_0^t (b - c\upsilon_u + d \sqrt{\upsilon_u}) \D u + \sigma \int_0^t \sqrt{\upsilon_u} \D W_u
,\end{equation}

and its expected value is equal to:
\begin{equation}
\mathbb{E}[\upsilon_t] = \upsilon_0 + a \int_0^t (b-c\mathbb{E}[\upsilon_u] + d\mathbb{E}[\sqrt{\upsilon_u}]) \D u
.\end{equation}

After differentiation by $ t $ we have:
\begin{equation}
\tfrac{\partial}{\partial t} \mathbb{E}[\upsilon_t] = ab - ac\mathbb{E}[\upsilon_t] + ad\mathbb{E}[\sqrt{\upsilon_t}]
,\end{equation}

and in addition it is known that:
\begin{equation}
\tfrac{\partial}{\partial t} e^{act} \mathbb{E}[\upsilon_t] = e^{act}(ac\mathbb{E}[\upsilon_t] + \tfrac{\partial}{\partial t} \mathbb{E}[\upsilon_t]) = e^{act}(ab + ad\mathbb{E}[\sqrt{\upsilon_t}])
.\end{equation}

After the integration of the last equation it is:
\begin{equation}
e^{act} \mathbb{E}[\upsilon_t] - \upsilon_0 = ab \int_0^t e^{acu} \D u + ad \int_0^t e^{acu} \mathbb{E}[\sqrt{\upsilon_u}] \D u = \tfrac{b}{c}(e^{act}-1) + ad \int_0^t e^{acu} \mathbb{E}[\sqrt{\upsilon_u}] \D u
,\end{equation}

so

\begin{equation}
\mathbb{E}[\upsilon_t] = \tfrac{b}{c} + e^{-act}(\upsilon_0-\tfrac{b}{c}) + ad e^{-act} \int_0^t e^{acu} \mathbb{E}[\sqrt{\upsilon_u}] \D u
.\end{equation}

Since $ \upsilon_t = \nu_t^2 $, so $\mathbb{E}[\sqrt{\upsilon_t}]$ can be calculated directly from the Sch\"obel-Zhu model, which uses the Ornstein-Uhlenbeck process:
\begin{equation}
\mathbb{E}[\nu_t] = e^{-\kappa t} \nu_0 + \theta(1-e^{-\kappa t})
.\end{equation}

Finally, after entering the values of the parameters, $ a $, $ b $, $ c $ and $ d $, the following result for instantaneous variance is obtained:
\begin{equation}\begin{split}
\mathbb{E}[\upsilon_t] & = \mathbb{E}[\nu_t^2] = \tfrac{\omega^2}{2\kappa} + e^{-2 \kappa t}(\nu_0^2-\tfrac{\omega^2}{2\kappa}) + 2 \theta \kappa e^{-2\kappa t} \int_0^t e^{2\kappa u} (e^{-\kappa u} \nu_0 + \theta(1-e^{-\kappa u})) \D u, \\
& = \tfrac{\omega^2}{2\kappa} + e^{-2 \kappa t}(\nu_0^2-\tfrac{\omega^2}{2\kappa}) + 2 \theta \kappa e^{-2\kappa t} \int_0^t (e^{\kappa u} \nu_0 + \theta (e^{2\kappa u}-e^{\kappa u})) \D u, \\
& = \tfrac{\omega^2}{2\kappa} + e^{-2 \kappa t}(\nu_0^2-\tfrac{\omega^2}{2\kappa}) + \theta e^{-2 \kappa t} (e^{\kappa t} - 1) (\theta (e^{\kappa t} - 1) + 2 \nu_0)
,\end{split}\end{equation}

After integration of the instantaneous variance the total variance yields:

\begin{equation}
\mathbb{E}[\upsilon_T] = \frac{e^{-2 \kappa \tau} (e^{2 \kappa \tau} (\theta^2 (2 \kappa \tau-3)+2 \theta \nu_0+ \tfrac{\omega^2}{2\kappa} (2 \kappa \tau-1)+\nu_0^2)+4 \theta (\theta-\nu_0) e^{\kappa \tau}-(\theta-\nu_0)^2+ \tfrac{\omega^2}{2\kappa} )}{2 \kappa}
.\end{equation}

\clearpage


\chapter{Calibration risk charts}
\markboth{\hfill \textit{Appendix C\quad Calibration risk charts} \hfill}{}

\renewcommand\thefigure{C.\arabic{figure}}   
\setcounter{figure}{0}

\begin{figure}[h!]
\centering
\includegraphics[width=150mm]{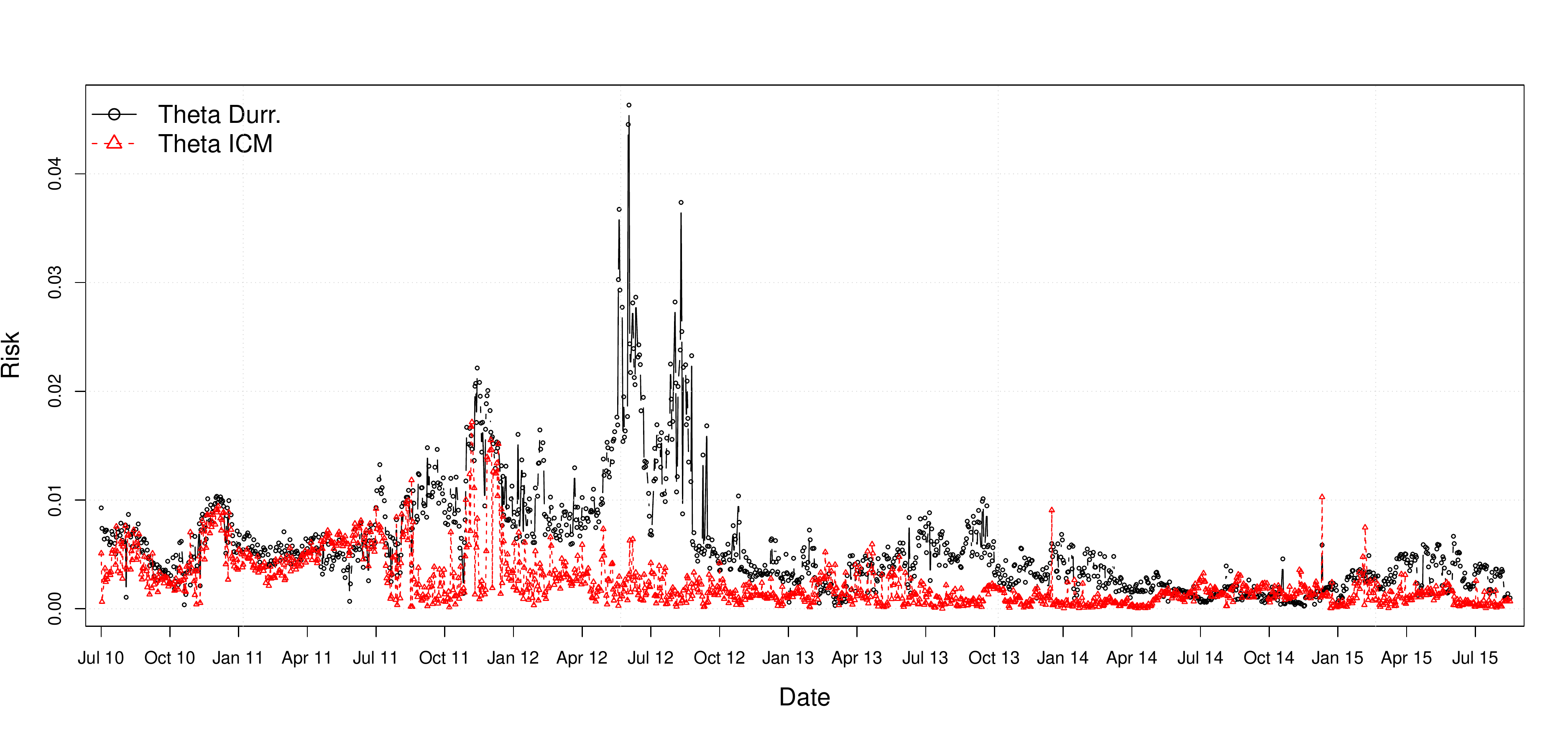}
\caption{Theta calibration risk in considered methods}
\label{figure:A.1} 
\end{figure}

\begin{figure}[h!]
\centering
\includegraphics[width=150mm]{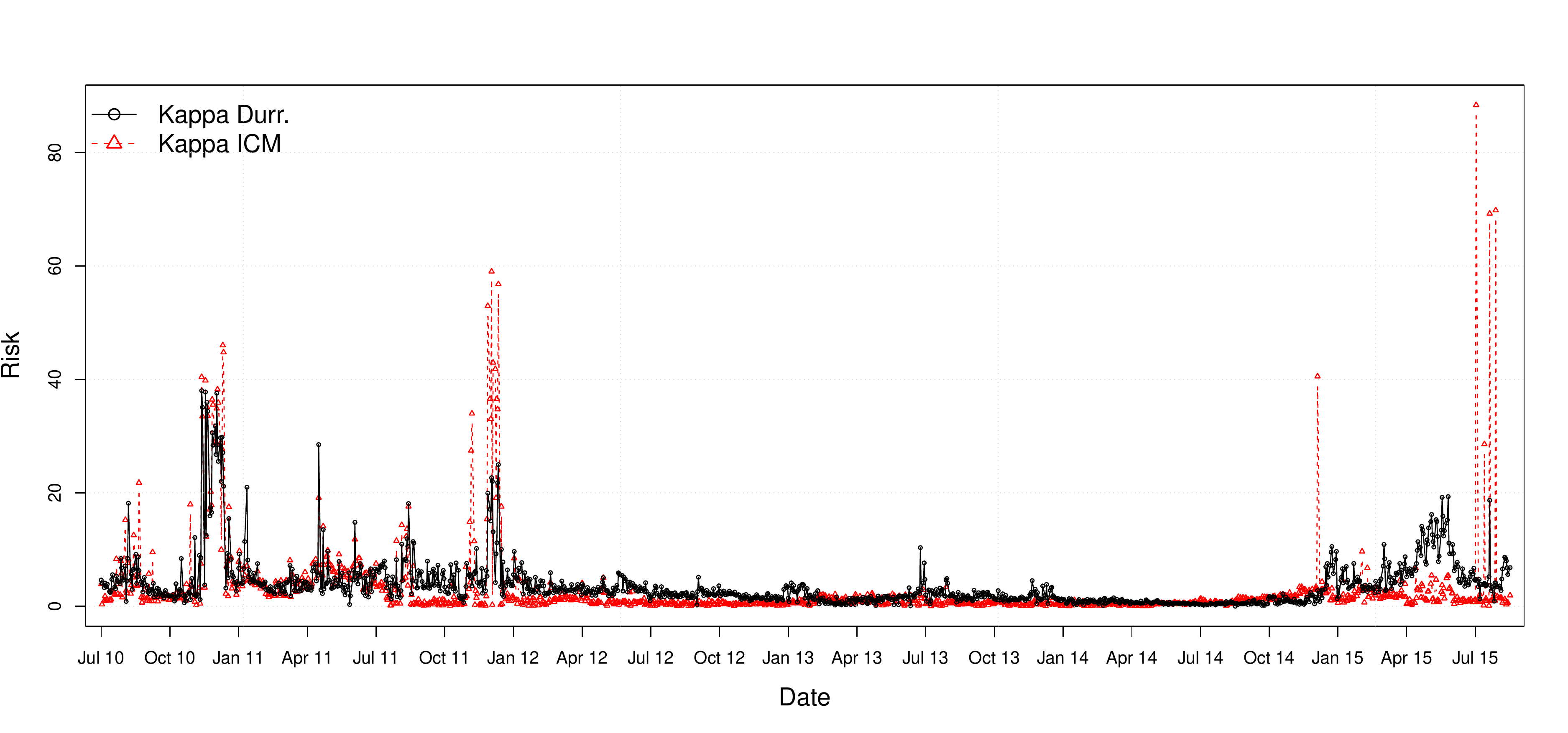}
\caption{Kappa calibration risk in considered methods}
\label{figure:A.2} 
\end{figure}

\begin{figure}[h!]
\centering
\includegraphics[width=150mm]{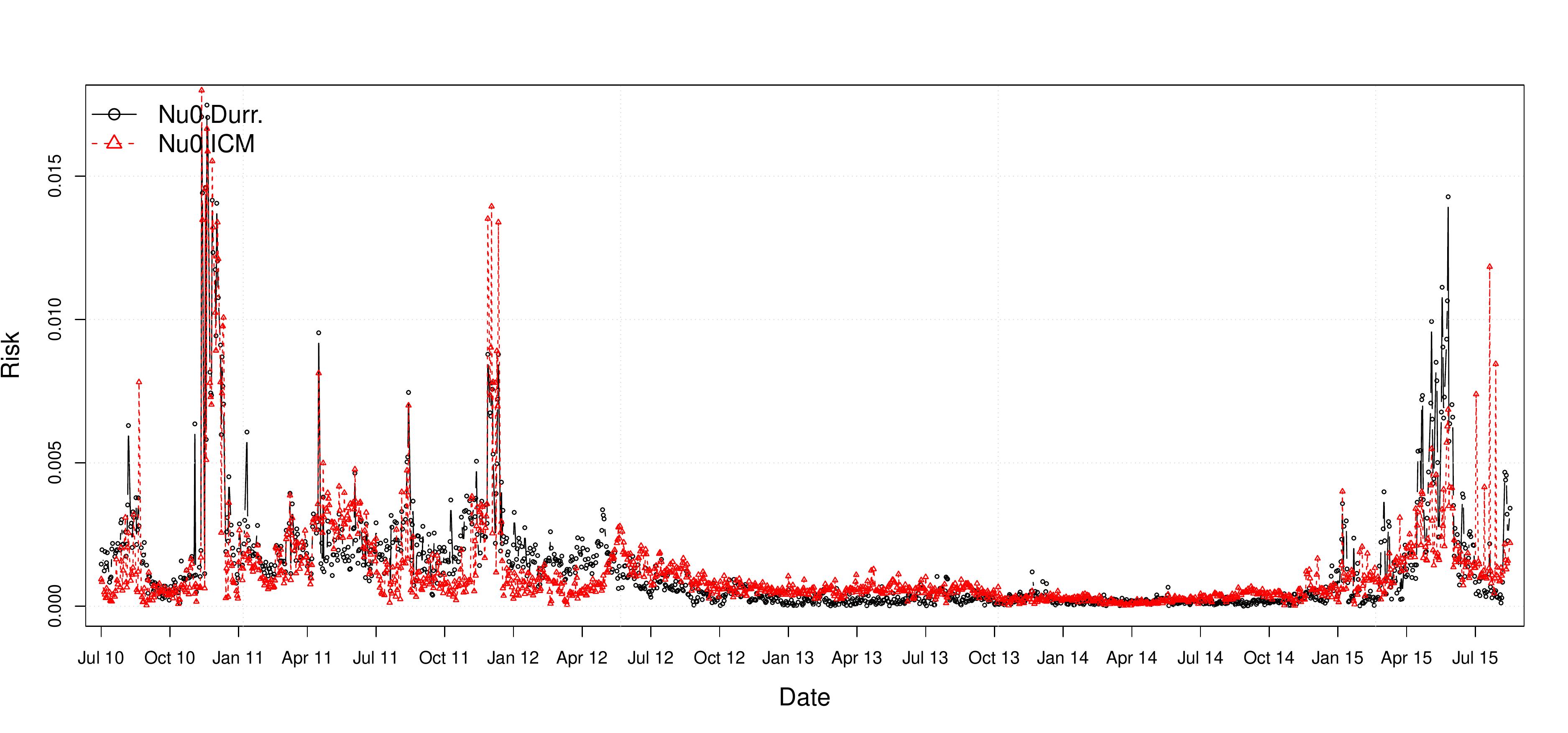}
\caption{Nu0 calibration risk in considered methods}
\label{figure:A.3} 
\end{figure}

\begin{figure}[h!]
\centering
\includegraphics[width=150mm]{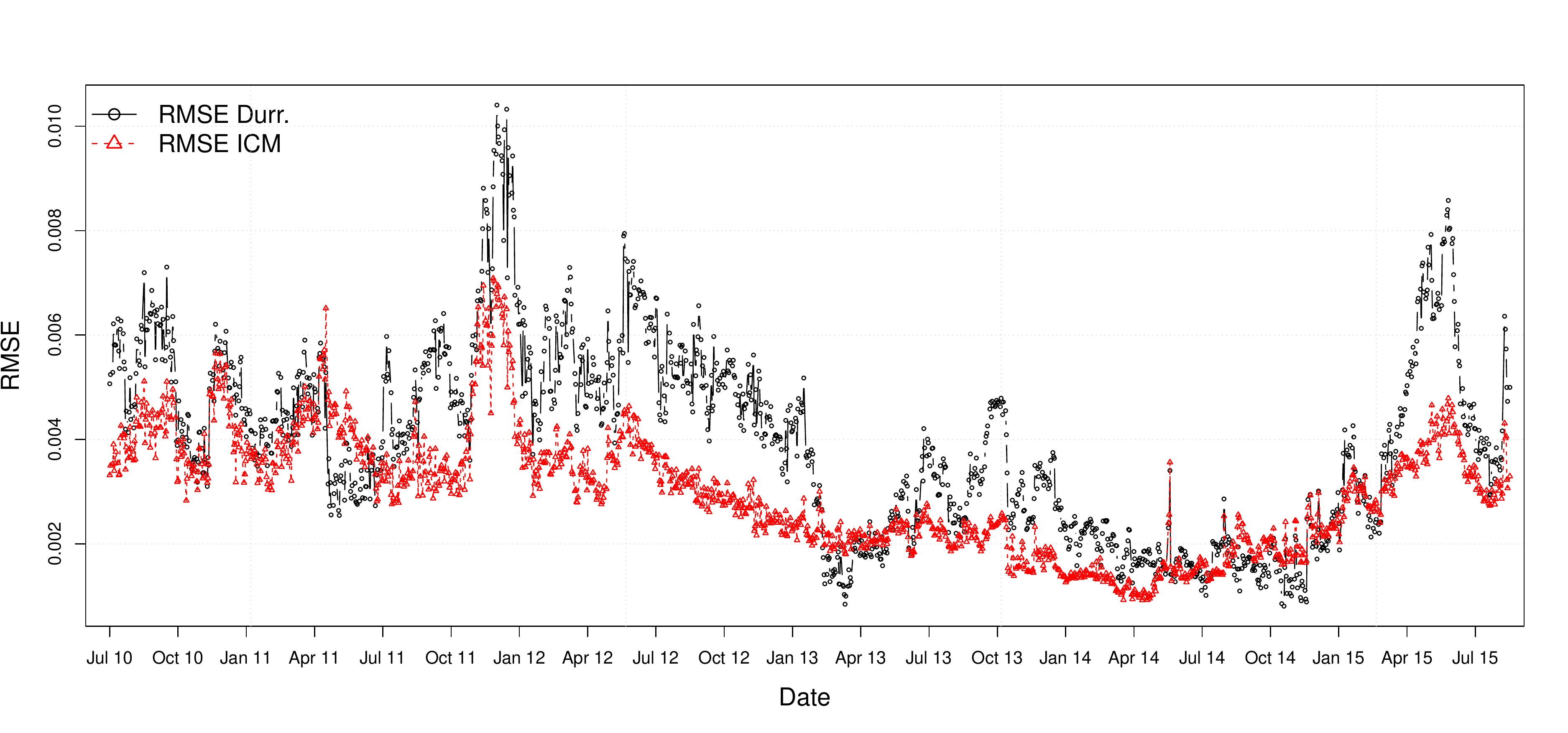}
\caption{RMSE after 3 parameters calibration in considered methods}
\label{figure:A.4} 
\end{figure}
\clearpage{\pagestyle{empty}\cleardoublepage}

\phantomsection

\nocite{*} 

\fancyhead{} 
\fancyhead[C]{\leftmark}
\fancyhead[RO,LE]{\thepage}
\renewcommand{\MakeUppercase}[1]{\textit{#1}}
\renewcommand{\bibname}{References}
\bibliographystyle{plainnat}
\bibliography{dissertation bibliography}

\begin{thebibliography}{134}
\providecommand{\natexlab}[1]{#1}
\providecommand{\url}[1]{\texttt{#1}}
\expandafter\ifx\csname urlstyle\endcsname\relax
  \providecommand{\doi}[1]{doi: #1}\else
  \providecommand{\doi}{doi: \begingroup \urlstyle{rm}\Url}\fi

\bibitem[Ait-Sahalia and Kimmel(2007)]{Ait2007}
Y.~Ait-Sahalia and R.~Kimmel.
\newblock Maximum likelihood estimation of stochastic volatility models.
\newblock \emph{Journal of Financial Economics}, 83\penalty0 (2):\penalty0
  413--452, 2007.

\bibitem[Albrecher et~al.(2007)Albrecher, Mayer, Schoutens, and
  Tistaert]{Albrecher2007}
H.~Albrecher, P.~Mayer, W.~Schoutens, and J.~Tistaert.
\newblock The little {Heston} trap.
\newblock \emph{Wilmott Magazine}, \penalty0 (1):\penalty0 83--92, 2007.

\bibitem[Alexander(2001)]{Alexander2001}
C.~Alexander.
\newblock Principal component analysis of volatility smiles and skews.
\newblock \emph{Risk}, 14:\penalty0 29--32, 2001.

\bibitem[Amin and Ng(1997)]{Amin1997}
K.~Amin and V.~Ng.
\newblock Inferring future volatility from the information in implied
  volatility in {Eurodollar} options: A new approach.
\newblock \emph{Review of Financial Studies}, 10\penalty0 (2):\penalty0
  333--367, 1997.

\bibitem[Andersen(2008)]{Andersen2008}
L.~Andersen.
\newblock Simple and efficient simulation of the {Heston} stochastic volatility
  model.
\newblock \emph{Journal of Computational Finance}, 11\penalty0 (3):\penalty0
  1--42, 2008.

\bibitem[Attari(2004)]{Attari2004}
M.~Attari.
\newblock Option pricing using {Fourier Transforms}: A numerically efficient
  simplification.
\newblock \emph{Working Paper, Charles River Associates, Boston, MA}, 2004.
\newblock \doi{10.2139/ssrn.520042}.

\bibitem[Bachelier(1900)]{Bachelier1900}
L.~Bachelier.
\newblock \emph{Th\'eorie de la Sp\'eculation}.
\newblock Gauthier-Villars, 1900.

\bibitem[Badshah(2010)]{Badshah2010}
I.~Badshah.
\newblock Modeling and forecasting implied volatility. implications for
  trading, pricing, and risk management.
\newblock \emph{Helsinki}, 2010.

\bibitem[Bakshi et~al.(1997)Bakshi, Cao, and Chen]{Bakshi1997}
G.~Bakshi, C.~Cao, and Z.~Chen.
\newblock Empirical performance of alternative option pricing models.
\newblock \emph{The Journal of Finance}, 52\penalty0 (5):\penalty0 2003--2049,
  1997.
\newblock ISSN 1540-6261.
\newblock \doi{10.1111/j.1540-6261.1997.tb02749.x}.
\newblock URL \url{http://dx.doi.org/10.1111/j.1540-6261.1997.tb02749.x}.

\bibitem[Bakshi et~al.(2003)Bakshi, Kapadia, and Madan]{Bakshi2003}
G.~Bakshi, N.~Kapadia, and D.~Madan.
\newblock Stock return characteristics, skew laws, and the differential pricing
  of individual equity options.
\newblock \emph{The Review of Financial Studies}, 16:\penalty0 101--143, 2003.

\bibitem[Bakshi et~al.(2008)Bakshi, Carr, and Wu]{Bakshi2008}
G.~Bakshi, P.~Carr, and L.~Wu.
\newblock Stochastic risk premiums, stochastic skewness in currency options,
  and stochastic discount factors in international economies.
\newblock \emph{Journal of Financial Economics}, 2008.

\bibitem[{Bank for International Settlements}(2013)]{BIS2013}
{Bank for International Settlements}.
\newblock {Triennial Central Bank Survey}. foreign exchange turnover in {April}
  2013: preliminary global results.
\newblock Technical report, September 2013.
\newblock URL \url{http://www.bis.org/publ/rpfx13fx.pdf}.

\bibitem[Barndorff-Nielsen and Shephard(2002)]{Barndorff2002}
O.~E. Barndorff-Nielsen and N.~Shephard.
\newblock Estimating quadratic variation using realized variance.
\newblock \emph{Journal of Applied Econometrics}, 17:\penalty0 457--477, 2002.

\bibitem[Bates(1996)]{Bates1996}
D.~Bates.
\newblock Jumps and stochastic volatility: Exchange rate processes implicit in
  {Deutsche Mark} options.
\newblock \emph{The Review of Financial Studies}, 9\penalty0 (1):\penalty0
  69--107, 1996.

\bibitem[Bates(2000)]{Bates2000}
D.~Bates.
\newblock Post-'87 crash fears in the {S\&P} 500 futures option market.
\newblock \emph{Journal of Econometrics}, 94:\penalty0 181--238, 2000.

\bibitem[Benhamou et~al.(2010)Benhamou, Gobet, and Miri]{Benhamou2010}
E.~Benhamou, E.~Gobet, and M.~Miri.
\newblock Time dependent {Heston} model.
\newblock \emph{SIAM Journal on Financial Mathematics}, 1:\penalty0 289--325,
  2010.

\bibitem[Bin(2007)]{Bin2007}
C.~Bin.
\newblock Calibration of the {Heston} model with application in derivative
  pricing and hedging.
\newblock \emph{Master Thesis, TU Delft}, 2007.
\newblock URL
  \url{http://resolver.tudelft.nl/uuid:25dc8109-4b39-44b8-8722-3fd680c7c4ad}.

\bibitem[Black(1976)]{Black1976}
F.~Black.
\newblock The pricing of commodity contracts.
\newblock \emph{Journal of Financial Economics}, 3:\penalty0 167--179, 1976.

\bibitem[Black and Scholes(1973)]{Black1973}
F.~Black and M.~Scholes.
\newblock The pricing of options and corporate liabilities.
\newblock \emph{Journal of Political Economy}, 81\penalty0 (3):\penalty0
  637--654, 1973.

\bibitem[Bollerslev(1986)]{Bollerslev1986}
T.~Bollerslev.
\newblock {Generalized Autoregressive Conditional Heteroskedasticity}.
\newblock \emph{Journal of Econometrics}, 31\penalty0 (3):\penalty0 307--327,
  1986.

\bibitem[Boness(1964)]{Boness1964}
J.~Boness.
\newblock Elements of a theory of stock-option value.
\newblock \emph{Journal of Political Economy}, 72\penalty0 (2):\penalty0
  163--175, 1964.

\bibitem[Breeden(1979)]{Breeden1979}
D.~Breeden.
\newblock An intertemporal asset pricing model with stochastic consumption and
  investment opportunities.
\newblock \emph{Journal of Financial Economics}, 7:\penalty0 265--296, 1979.

\bibitem[Breeden and Litzenberger(1978)]{Breeden1978}
D.~Breeden and R.~Litzenberger.
\newblock Prices of state-contingent claims implicit in option prices.
\newblock \emph{Journal of Business}, 51\penalty0 (4):\penalty0 621--651, 1978.

\bibitem[Brigo and Mercurio(2002)]{Brigo2002}
D.~Brigo and F.~Mercurio.
\newblock Lognormal-mixture dynamics and calibration to market volatility
  smiles.
\newblock \emph{International Journal of Theoretical and Applied Finance},
  5\penalty0 (4), 2002.

\bibitem[Broadie and Jain(2008)]{Broadie2008}
M.~Broadie and A.~Jain.
\newblock The effect of jumps and discrete sampling on volatility and variance
  swaps.
\newblock \emph{International Journal of Theoretical and Applied Finance},
  11\penalty0 (08):\penalty0 761--797, 2008.
\newblock \doi{10.1142/S0219024908005032}.

\bibitem[Broyden(1970)]{Broyden1970}
C.~G. Broyden.
\newblock The convergence of a class of double-rank minimization algorithms.
\newblock \emph{Journal of the Institute of Mathematics and Its Applications},
  6:\penalty0 76--90, 1970.

\bibitem[Canina and Figlewski(1993)]{Canina1993}
L.~Canina and S.~Figlewski.
\newblock The informational content of implied volatility.
\newblock \emph{The Review of Financial Studies}, 6\penalty0 (3):\penalty0
  659--681, 1993.

\bibitem[Carr and Madan(1999)]{Carr1999}
P.~Carr and D.~Madan.
\newblock Option valuation using the {Fast Fourier Transform}.
\newblock \emph{Journal of Computational Finance}, 2:\penalty0 61--73, 1999.

\bibitem[Carr and Madan(2002)]{Carr2002}
P.~Carr and D.~Madan.
\newblock \emph{Towards a Theory of Volatility Trading}, pages 417--427.
\newblock Risk Publications, 2002.

\bibitem[Carr and Wu(2007)]{Carr2007}
P.~Carr and L.~Wu.
\newblock Stochastic skew in currency options.
\newblock \emph{Journal of Financial Economics}, 86\penalty0 (1):\penalty0
  213--247, 2007.

\bibitem[Castagna and Mercurio(2007)]{Castagna2007}
A.~Castagna and F.~Mercurio.
\newblock {Vanna-Volga} methods applied to {FX} derivatives: from theory to
  market practice.
\newblock \emph{Risk}, \penalty0 (3):\penalty0 39--44, 2007.

\bibitem[CBOE(2010)]{CBOE2010}
CBOE.
\newblock {The CBOE Skew Index - SKEW}.
\newblock 2010.
\newblock URL
  \url{https://www.cboe.com/micro/skew/documents/skewwhitepaperjan2011.pdf}.

\bibitem[CBOE(2014)]{CBOE2014}
CBOE.
\newblock {The CBOE Volatility Index - VIX}.
\newblock 2014.
\newblock URL \url{https://www.cboe.com/micro/vix/vixwhite.pdf}.

\bibitem[Cheng and Scaillet(2002)]{Cheng2002}
P.~Cheng and O.~Scaillet.
\newblock Linear-quadratic jump-diffusion modelling with application to
  stochastic volatility.
\newblock \emph{FAME working paper series}, 67, 2002.

\bibitem[Christoffersen et~al.(2009)Christoffersen, Heston, and
  Jacobs]{Christoffersen2009}
P.~Christoffersen, S.~Heston, and K.~Jacobs.
\newblock The shape and term structure of the index option smirk: Why
  multifactor stochastic volatility models work so well.
\newblock \emph{Management Science}, 55:\penalty0 1914--1932, 2009.

\bibitem[Christoffersen et~al.(2014)Christoffersen, Feunou, Jacobs, and
  Meddahi]{Christoffersen2014}
P.~Christoffersen, B.~Feunou, K.~Jacobs, and N.~Meddahi.
\newblock The economic value of realized volatility: Using high-frequency
  returns for option valuation.
\newblock \emph{Journal of Financial and Quantitative Analysis}, 49\penalty0
  (3):\penalty0 663--697, 2014.

\bibitem[Clark(2011)]{Clark2011}
I.~Clark.
\newblock \emph{{Foreign Exchange} Option Pricing}.
\newblock John Wiley \& Sons, Ltd, 2011.

\bibitem[Cont(2006)]{Cont2006}
R.~Cont.
\newblock Model uncertainty and its impact on the pricing of derivative
  instruments.
\newblock \emph{Mathematical Finance}, 16\penalty0 (3):\penalty0 519--547,
  2006.

\bibitem[Cont and Fonseca(2002)]{Fonseca2002}
R.~Cont and J.~Fonseca.
\newblock Dynamics of implied volatility surfaces.
\newblock \emph{Quantitative Finance}, 2:\penalty0 45--60, 2002.

\bibitem[Cont et~al.(2002)Cont, Fonseca, and Durrleman]{Cont2002}
R.~Cont, J.~Fonseca, and V.~Durrleman.
\newblock Stochastic models of implied volatility surfaces.
\newblock \emph{Economic Notes by Banca Monte dei Paschi di Siena SpA},
  31\penalty0 (2):\penalty0 361--377, 2002.

\bibitem[Corte et~al.(2011)Corte, Sarno, and Tsiakas]{DellaCorte2011}
P.~Della Corte, L.~Sarno, and I.~Tsiakas.
\newblock Spot and forward volatility in foreign exchange.
\newblock \emph{Journal of Financial Economics}, 100\penalty0 (3):\penalty0 496
  -- 513, 2011.
\newblock ISSN 0304-405X.
\newblock \doi{http://dx.doi.org/10.1016/j.jfineco.2011.01.007}.

\bibitem[Cox and Ross(1976)]{Cox1976}
J.~Cox and S.~Ross.
\newblock The valuation of options for alternative stochastic processes.
\newblock \emph{Journal of Financial Economics}, 3:\penalty0 145--166, 1976.

\bibitem[Cox et~al.(1979)Cox, Ross, and Rubinstein]{Cox1979}
J.~Cox, S.~Ross, and M.~Rubinstein.
\newblock Option pricing: A simplified approach.
\newblock \emph{Journal of Financial Economics}, 7\penalty0 (3):\penalty0
  229--263, 1979.

\bibitem[Cox et~al.(1985)Cox, Ingersoll, and Ross]{Cox1985}
J.~Cox, J.~Ingersoll, and S.~Ross.
\newblock A theory of the term structure of interest rates.
\newblock \emph{Econometrica}, 53:\penalty0 385--408, 1985.

\bibitem[Crepey(2004)]{Crepey2004}
S.~Crepey.
\newblock Delta-hedging {Vega} risk.
\newblock \emph{Quantitative Finance}, 4, 2004.

\bibitem[Daglish et~al.(2007)Daglish, Hull, and Suo]{Daglish2007}
T.~Daglish, J.~Hull, and W.~Suo.
\newblock Volatility surfaces: Theory, rules of thumb, and empirical evidence.
\newblock \emph{Quantitative Finance}, 7:\penalty0 507--524, 2007.

\bibitem[Day and Lewis(1993)]{Day1993}
T.~Day and Craig~M. Lewis.
\newblock Forecasting futures market volatility.
\newblock \emph{The Journal of Derivatives}, \penalty0 (4), 1993.

\bibitem[Dennis and Mayhew(2002)]{Dennis2002}
P.~Dennis and S.~Mayhew.
\newblock Risk-neutral skewness: Evidence from stock options.
\newblock \emph{Journal of Financial and Quantitative Analysis}, 37:\penalty0
  471--493, 2002.

\bibitem[Derman(1999)]{Derman1999}
E.~Derman.
\newblock Volatility regimes.
\newblock \emph{Risk}, 12\penalty0 (4):\penalty0 55--59, 1999.

\bibitem[Derman and Kamal(1997)]{Derman1997}
E.~Derman and M.~Kamal.
\newblock The patterns of change in implied index volatilities.
\newblock \emph{Quantitative Strategies Research Notes}, \penalty0 (9), 1997.

\bibitem[Derman and Kani(1993)]{Derman1994}
E.~Derman and I.~Kani.
\newblock Riding on a smile.
\newblock \emph{Risk}, 7:\penalty0 32--39, 1993.

\bibitem[Derman and Kani(1998)]{Derman1998}
E.~Derman and I.~Kani.
\newblock Stochastic implied trees: Arbitrage pricing with stochastic term and
  strike structure of volatility.
\newblock \emph{International Journal of Theoretical and Applied Finance},
  1:\penalty0 61--110, 1998.

\bibitem[Derman et~al.(1996)Derman, Kani, and Zou]{Derman1996}
E.~Derman, I.~Kani, and J.~Zou.
\newblock The local volatility surface: Unlocking the information in index
  options prices.
\newblock \emph{Financial Analysts Journal}, 52\penalty0 (4), 1996.

\bibitem[Detlefsen and H\"ardle(2007)]{Detlefsen2007}
K.~Detlefsen and W.~H\"ardle.
\newblock Calibration risk for exotic options.
\newblock \emph{Journal of Derivatives}, 14\penalty0 (47-63), 2007.

\bibitem[Drost and Nijman(1993)]{Drost1993}
F.~Drost and T.~Nijman.
\newblock Temporal aggregation of {GARCH} processes.
\newblock \emph{Econometrica}, 61\penalty0 (4):\penalty0 909--927, 1993.

\bibitem[Duan(1995)]{Duan1995}
J.~Duan.
\newblock The {GARCH} option pricing model.
\newblock \emph{Mathematical Finance}, 5\penalty0 (1):\penalty0 13--32, 1995.

\bibitem[Duffie et~al.(2000)Duffie, Pan, and Singleton]{Duffie2000}
D.~Duffie, J.~Pan, and K.~Singleton.
\newblock Transform analysis and asset pricing for affine jump-diffusions.
\newblock \emph{Econometrica}, 68:\penalty0 1343--1376, 2000.

\bibitem[Duffie et~al.(2003)Duffie, Filipovi\'c, and Schachermayer]{Duffie2003}
D.~Duffie, D.~Filipovi\'c, and W.~Schachermayer.
\newblock Affine processes and applications in finance.
\newblock \emph{Ann. Appl. Probab.}, 13\penalty0 (3):\penalty0 984--1053, 08
  2003.
\newblock \doi{10.1214/aoap/1060202833}.
\newblock URL \url{http://dx.doi.org/10.1214/aoap/1060202833}.

\bibitem[Dumas et~al.(1998)Dumas, Fleming, and Whaley]{Dumas1998}
B.~Dumas, J.~Fleming, and R.~Whaley.
\newblock Implied volatility functions: Empirical tests.
\newblock \emph{The Journal of Finance}, 53, 1998.

\bibitem[Dupire(1994)]{Dupire1994}
B.~Dupire.
\newblock Pricing with a smile.
\newblock \emph{Risk}, 7:\penalty0 18--20, 1994.

\bibitem[Durrleman(2004)]{Durrleman2004}
V.~Durrleman.
\newblock \emph{From spot to implied volatilities and applications}, pages
  58--74.
\newblock Doctoral Dissertation. Department of Operations Research and
  Financial Engineering. Princeton University, 2004.

\bibitem[Durrleman and Karoui(2008)]{Durrleman2008}
V.~Durrleman and N.~El Karoui.
\newblock Coupling smiles.
\newblock \emph{Quantitative Finance}, 8\penalty0 (6):\penalty0 573--590, 2008.

\bibitem[Engle(1982)]{Engle1982}
R.~Engle.
\newblock {Autoregressive Conditional Heteroscedasticity} with estimates of the
  variance of {United Kingdom} inflation.
\newblock \emph{Econometrica}, 50\penalty0 (4):\penalty0 987--1007, 1982.

\bibitem[Feller(1951)]{Feller1951}
W.~Feller.
\newblock Two singular diffusion problems.
\newblock \emph{Annals of Mathematics}, 54\penalty0 (1):\penalty0 173--182,
  1951.

\bibitem[Fengler(2005)]{Fengler2005}
M.~Fengler.
\newblock Semiparametric modeling of implied volatility.
\newblock \emph{Lecture Notes in Finance, Springer}, 2005.

\bibitem[Fengler et~al.(2003)Fengler, H\"ardle, and Villa]{Fengler2003}
M.~Fengler, W.~H\"ardle, and C.~Villa.
\newblock The dynamics of implied volatilities: A common principle components
  approach.
\newblock \emph{Review of Derivatives Research}, 6:\penalty0 179--202, 2003.

\bibitem[Fengler et~al.(2007)Fengler, H\"ardle, and Mammen]{Fengler2007}
M.~Fengler, W.~H\"ardle, and E.~Mammen.
\newblock A semiparametric factor model for implied volatility surface
  dynamics.
\newblock \emph{Journal of Financial Econometrics}, 5:\penalty0 189--218, 2007.

\bibitem[Fletcher(1970)]{Fletcher1970}
R.~Fletcher.
\newblock A new approach to variable metric algorithms.
\newblock \emph{Computer Journal}, 13\penalty0 (3):\penalty0 317--322, 1970.

\bibitem[Gao and Han(2012)]{Gao2012}
F.~Gao and L.~Han.
\newblock Implementing the {Nelder-Mead} simplex algorithm with adaptive
  parameters.
\newblock \emph{Computational Optimization and Applications}, 51\penalty0
  (1):\penalty0 259--277, 2012.
\newblock ISSN 1573-2894.
\newblock \doi{10.1007/s10589-010-9329-3}.
\newblock URL \url{http://dx.doi.org/10.1007/s10589-010-9329-3}.

\bibitem[Garman and Kohlhagen(1983)]{Garman1983}
M.~Garman and S.~Kohlhagen.
\newblock Foreign currency option values.
\newblock \emph{Journal of International Money and Finance}, 2\penalty0
  (3):\penalty0 231--237, 1983.

\bibitem[Gaspar(2004)]{Gaspar2004}
R.~Gaspar.
\newblock General quadratic term structures of bond, futures and forward
  prices.
\newblock \emph{SSE/EFI Working paper Series in Economics and Finance},
  559\penalty0 (3), 2004.

\bibitem[Gatheral(2004)]{Gatheral2004}
J.~Gatheral.
\newblock A parsimonious arbitrage-free implied volatility parameterization
  with applicationto the valuation of volatility derivatives.
\newblock \emph{Global Derivatives And Risk Management}, 2004.

\bibitem[Gatheral and Kamal(2010)]{Gatheral2010}
J.~Gatheral and M.~Kamal.
\newblock \emph{Implied volatility surface}, pages 926--931.
\newblock John Wiley \& Sons, Ltd, 2010.
\newblock ISBN 9780470061602.

\bibitem[Gauthier and Possama\"i(2011)]{Gauthier2011}
P.~Gauthier and D.~Possama\"i.
\newblock Efficient simulation of the {Double Heston} model.
\newblock \emph{IUP Journal of Computational Mathematics}, 4\penalty0
  (3):\penalty0 23--73, 2011.

\bibitem[Gauthier and Rivaille(2009)]{Gauthier2009}
P.~Gauthier and P.~H. Rivaille.
\newblock Fitting the smile, {Smart Parameters} for {SABR} and {Heston}.
\newblock \emph{Working Paper, Pricing Partners (www.pricingpartners.com)},
  2009.
\newblock \doi{http://dx.doi.org/10.2139/ssrn.1496982}.
\newblock URL \url{https://ssrn.com/abstract=1496982}.

\bibitem[Gil-Pelaez(1951)]{Gil-Pelaez1951}
J.~Gil-Pelaez.
\newblock Note on the inversion theorem.
\newblock \emph{Biometrika}, 38:\penalty0 481--482, 1951.

\bibitem[Gilli and Schumann(2010)]{Gilli2010}
M.~Gilli and E.~Schumann.
\newblock Calibrating the {Heston} model with differential evolution.
  applications of evolutionary computation.
\newblock \emph{Lecture Notes in Computer Science}, pages 242--250, 2010.

\bibitem[Gilli and Schumann(2012)]{Gilli2012}
M.~Gilli and E.~Schumann.
\newblock \emph{Natural Computing in Computational Finance: Volume 4}, chapter
  Calibrating Option Pricing Models with Heuristics, pages 9--37.
\newblock Springer Berlin Heidelberg, Berlin, Heidelberg, 2012.
\newblock ISBN 978-3-642-23336-4.
\newblock \doi{10.1007/978-3-642-23336-4_2}.
\newblock URL \url{http://dx.doi.org/10.1007/978-3-642-23336-4_2}.

\bibitem[Girsanov(1960)]{Girsanov1960}
I.~V. Girsanov.
\newblock On transforming a certain class of stochastic processes by absolutely
  continuous substitution of measures.
\newblock \emph{Theory of Probability and its Applications}, 5\penalty0
  (3):\penalty0 285--301, 1960.

\bibitem[Glosten et~al.(1993)Glosten, Jaganatan, and Runkle]{GJR1993}
L.~Glosten, R.~Jaganatan, and D.~Runkle.
\newblock On the relation between the expected value and the volatility of the
  nominal excess return on stocks.
\newblock \emph{The Journal of Finance}, 48\penalty0 (5):\penalty0 1779--1801,
  1993.

\bibitem[Goldfarb(1970)]{Goldfarb1970}
D.~Goldfarb.
\newblock A family of variable metric updates derived by variational means.
\newblock \emph{Mathematics of Computation}, 24\penalty0 (109):\penalty0
  23--26, 1970.

\bibitem[Goncalves and Guidolin(2006)]{Goncalves2006}
S.~Goncalves and M.~Guidolin.
\newblock Predictable dynamics in the {S\&P} 500 index options volatility
  surface.
\newblock \emph{Journal of Business}, 79:\penalty0 1591--1635, 2006.

\bibitem[Guillaume and Schoutens(2013)]{Guillaume2013}
F.~Guillaume and W.~Schoutens.
\newblock A moment matching market implied calibration.
\newblock \emph{Quantitative Finance}, 13\penalty0 (9):\penalty0 1359--1373,
  2013.
\newblock \doi{10.1080/14697688.2013.794950}.
\newblock URL \url{http://dx.doi.org/10.1080/14697688.2013.794950}.

\bibitem[Guilliame and Schoutens(2010)]{Guilliame2010}
F.~Guilliame and W.~Schoutens.
\newblock Use a reduced {Heston} or reduce the use of {Heston}?
\newblock \emph{Wilmott Journal}, 2\penalty0 (4):\penalty0 171--192, 2010.

\bibitem[Guilliame and Schoutens(2012)]{Guilliame2012}
F.~Guilliame and W.~Schoutens.
\newblock Calibration risk: Illustrating the impact of calibration risk under
  the {Heston} model.
\newblock \emph{Review of Derivative Research}, 15:\penalty0 57--79, 2012.

\bibitem[Hafner(2004)]{Hafner2004}
R.~Hafner.
\newblock Stochastic implied volatility.
\newblock \emph{Lecture Notes in Economics and Mathematical Systems.
  Springer-Verlag Berlin Heidelberg}, 545, 2004.

\bibitem[Hafner and Wallmeier(2001)]{Hafner2001}
R.~Hafner and M.~Wallmeier.
\newblock The dynamics of {DAX} implied volatilities.
\newblock \emph{International Quarterly Journal of Finance}, 1:\penalty0 1--27,
  2001.

\bibitem[Hagan et~al.(2002)Hagan, Kumar, Lesniewski, and Woodward]{Hagan2002}
P.~Hagan, D.~Kumar, A.~Lesniewski, and D.~Woodward.
\newblock Managing smile risk.
\newblock \emph{Wilmott magazine}, \penalty0 (9):\penalty0 84--108, 2002.

\bibitem[Harrisson and Kreps(1979)]{Harrison1979}
M.~Harrisson and D.~Kreps.
\newblock Martingales and multiperiod securities markets.
\newblock \emph{Journal of Economic Theory}, 20:\penalty0 381--408, 1979.

\bibitem[Heston(1993)]{Heston1993}
S.~Heston.
\newblock A closed-form solution for options with stochastic volatility, with
  application to bond and currency options.
\newblock \emph{Review of Financial Studies}, 6:\penalty0 327--343, 1993.

\bibitem[Heston and Nandi(1997)]{Heston1997}
S.~Heston and S.~Nandi.
\newblock A closed form {GARCH} option pricing model.
\newblock \emph{Federal Reserve Bank of Atlanta Working Paper}, 97\penalty0
  (9):\penalty0 1--34, 1997.

\bibitem[Hull and White(1987)]{Hull1987}
J.~Hull and A.~White.
\newblock The pricing of options on assets with stochastic volatilities.
\newblock \emph{The Journal of Finance}, 42:\penalty0 281--300, 1987.

\bibitem[Hull and White(2012)]{Hull2012}
J.~Hull and A.~White.
\newblock {LIBOR vs. OIS}: The derivatives discounting dilemma.
\newblock \emph{Journal of Investment Management}, 11\penalty0 (3):\penalty0
  14--27, 2012.

\bibitem[It\^{o}(1944)]{Ito1944}
K.~It\^{o}.
\newblock Stochastic integral.
\newblock \emph{Proceedings of the Imperial Academy}, 20\penalty0 (8):\penalty0
  519--524, 1944.

\bibitem[Jab\l{}ecki et~al.(2014)Jab\l{}ecki, Kokoszczy\'nski, Sakowski,
  \'Slepaczuk, and W\'ojcik]{Jablecki2014}
J.~Jab\l{}ecki, R.~Kokoszczy\'nski, P.~Sakowski, R.~\'Slepaczuk, and
  P.~W\'ojcik.
\newblock Does historical volatility term structure contain valuable
  information for predicting volatility index futures?
\newblock \emph{Dynamic Econometric Models}, 14:\penalty0 5--28, 2014.

\bibitem[Kac(1949)]{Kac1949}
M.~Kac.
\newblock On distributions of certain {Wiener} functionals.
\newblock \emph{Transactions of the American Mathematical Society}, 65\penalty0
  (1):\penalty0 1--13, 1949.

\bibitem[Karoui et~al.(1998)Karoui, Jeanblanc-Picqu\'e, and
  Shreve]{ElKaroui1998}
N.~El Karoui, M.~Jeanblanc-Picqu\'e, and S.~Shreve.
\newblock Robustness of the {Black and Scholes} formula.
\newblock \emph{Mathematical Finance}, 8\penalty0 (2):\penalty0 93--126, 1998.

\bibitem[Kennedy and Eberhart(1995)]{Kennedy1995}
J.~Kennedy and R.~Eberhart.
\newblock Particle swarm optimization.
\newblock \emph{Proceedings of IEEE International Conference on Neural Networks
  IV}, pages 1942--1948, 1995.

\bibitem[Kilin(2007)]{Kilin2007}
F.~Kilin.
\newblock Accelerating the calibration of stochastic volatility models.
\newblock \emph{Centre for Practical Quantitative Finance Working Paper
  Series}, \penalty0 (6), 2007.

\bibitem[Krylova et~al.(2009)Krylova, Nikkinen, and V\"ah\"amaa]{Krylova2009}
E.~Krylova, J.~Nikkinen, and S.~V\"ah\"amaa.
\newblock Cross-dynamics of volatility term structures implied by foreign
  exchange options.
\newblock \emph{Journal of Economics and Business}, 61\penalty0 (5):\penalty0
  355 -- 375, 2009.
\newblock ISSN 0148-6195.
\newblock \doi{http://dx.doi.org/10.1016/j.jeconbus.2009.01.002}.
\newblock URL
  \url{http://www.sciencedirect.com/science/article/pii/S0148619509000113}.

\bibitem[Lawley and Maxwell(1962)]{Lawley1962}
D.~N. Lawley and A.~E. Maxwell.
\newblock {Factor Analysis} as a statistical method.
\newblock \emph{Journal of the Royal Statistical Society. Series D (The
  Statistician)}, 12\penalty0 (3):\penalty0 209--229, 1962.
\newblock ISSN 00390526, 14679884.
\newblock URL \url{http://www.jstor.org/stable/2986915}.

\bibitem[Lewis(2000)]{Lewis2000}
A.~L. Lewis.
\newblock \emph{Option Valuation under Stochastic Volatility with {Mathematica}
  Code}.
\newblock Finance Press, 2000.

\bibitem[Lipton(2002)]{Lipton2002}
A~Lipton.
\newblock The vol smile problem.
\newblock \emph{Risk Magazine}, 15:\penalty0 61--65, 2002.
\newblock URL \url{http://www.math.ku.dk/~rolf/Lipton_VolSmileProblem.pdf}.

\bibitem[Lord and Kahl(2006)]{LordKahl2006}
R.~Lord and C.~Kahl.
\newblock Why the rotation count algorithm works.
\newblock \emph{Tinbergen Institute Discussion Paper}, 065\penalty0 (2), 2006.

\bibitem[Malz(1997)]{Malz1997}
A.~Malz.
\newblock Estimating the probability distribution of the future exchange rate
  from option prices.
\newblock \emph{The Journal of Derivatives}, 5\penalty0 (2):\penalty0 18--36,
  1997.

\bibitem[Marquardt(1963)]{Marquardt1963}
D.~Marquardt.
\newblock An algorithm for least-squares estimation of nonlinear parameters.
\newblock \emph{Journal of the Society for Industrial and Applied Mathematics},
  11\penalty0 (2):\penalty0 431--441, 1963.

\bibitem[McAleer and Medeiros(2008)]{McAleer2008}
M.~McAleer and M~C. Medeiros.
\newblock Realized volatility: a review.
\newblock \emph{Econometric Reviews}, 27\penalty0 (1-3):\penalty0 10--45, 2008.

\bibitem[Melino and Turnbull(1990)]{Melino1990}
A.~Melino and S.~Turnbull.
\newblock The pricing of foreign currency options with stochastic volatility.
\newblock \emph{Journal of Econometrics}, 45:\penalty0 239--265, 1990.

\bibitem[Melino and Turnbull(1991)]{Melino1991}
A.~Melino and S.~Turnbull.
\newblock The pricing of foreign currency options.
\newblock \emph{Canadian Journal of Economics}, 24:\penalty0 251--281, 1991.

\bibitem[Merton(1973)]{Merton1973}
R.~Merton.
\newblock Theory of rational option pricing.
\newblock \emph{Bell Journal of Economics and Management Science}, 4\penalty0
  (1):\penalty0 141--183, 1973.

\bibitem[Mikhailov and N\"ogel(2003)]{Mikhailov2003}
S.~Mikhailov and U.~N\"ogel.
\newblock {Heston's} stochastic volatility model implementation, calibration
  and some extensions.
\newblock \emph{Wilmott Magazine}, \penalty0 (7):\penalty0 74--94, 2003.

\bibitem[Nelder and Mead(1965)]{NelderMead1965}
J.~Nelder and R.~Mead.
\newblock A simplex method for function minimization.
\newblock \emph{The Computer Journal}, 7\penalty0 (4):\penalty0 308--313, 1965.

\bibitem[Nelson(1990)]{Nelson1990}
D.~Nelson.
\newblock {ARCH} models as diffusion approximations.
\newblock \emph{Journal of Econometrics}, 45:\penalty0 7--38, 1990.

\bibitem[Nikodym(1930)]{Nikodym1930}
O.~Nikodym.
\newblock Sur une g\'en\'eralisation des int\'egrales de {M. J. Radon}.
\newblock \emph{Fundamenta Mathematicae}, 15:\penalty0 131--179, 1930.

\bibitem[Pearson(1901)]{Pearson1901}
K.~Pearson.
\newblock {LIII. On} lines and planes of closest fit to systems of points in
  space.
\newblock \emph{Philosophical Magazine Series 6}, 2\penalty0 (11):\penalty0
  559--572, 1901.
\newblock \doi{10.1080/14786440109462720}.
\newblock URL \url{http://dx.doi.org/10.1080/14786440109462720}.

\bibitem[Pena et~al.(1999)Pena, Rubio, and Serna]{Pena1999}
I.~Pena, G.~Rubio, and G.~Serna.
\newblock Why do we smile? on the determinants of the implied volatility
  function.
\newblock \emph{Journal of Banking and Finance}, 23:\penalty0 1151--1179, 1999.

\bibitem[Powell(1964)]{Powell1964}
M.~Powell.
\newblock An efficient method for finding the minimum of a function of several
  variables without calculating derivatives.
\newblock \emph{The Computer Journal}, 7\penalty0 (2):\penalty0 155--162, 1964.

\bibitem[Prokaj(2013)]{Prokaj2013}
V.~Prokaj.
\newblock The solution of the perturbed {Tanaka} equation is pathwise unique.
\newblock \emph{The Annals of Probability}, 41\penalty0 (3B):\penalty0
  2376--2400, 2013.

\bibitem[Reiswich and Wystup(2012)]{Reiswich2012}
D.~Reiswich and U.~Wystup.
\newblock {FX} volatility smile construction.
\newblock \emph{Wilmott}, 2012\penalty0 (60):\penalty0 58--69, 2012.
\newblock ISSN 1541-8286.
\newblock \doi{10.1002/wilm.10132}.
\newblock URL \url{http://dx.doi.org/10.1002/wilm.10132}.

\bibitem[Sch\"obel and Zhu(1999)]{SchoebelZhu1999}
R.~Sch\"obel and J.~Zhu.
\newblock Stochastic volatility with an {Ornstein-Uhlenbeck} process: An
  extension.
\newblock \emph{European Finance Review}, 3:\penalty0 23--46, 1999.

\bibitem[Schoutens et~al.(2004)Schoutens, Simons, and Tistaert]{Schoutens2004}
W.~Schoutens, E.~Simons, and J.~Tistaert.
\newblock A perfect calibration! {Now} what?
\newblock \emph{Wilmott Magazine}, \penalty0 (3):\penalty0 66--78, 2004.

\bibitem[Scott(1987)]{Scott1987}
L.~Scott.
\newblock Option pricing when the variance changes randomly: Theory,
  estimation, and an application.
\newblock \emph{Journal of Financial and Quantitative Analysis}, 22:\penalty0
  419--438, 1987.

\bibitem[Shano(1970)]{Shano1970}
D.~F. Shano.
\newblock Conditioning of quasi-{Newton} methods for function minimization.
\newblock \emph{Mathematics of Computation}, 24\penalty0 (111):\penalty0
  647--656, 1970.

\bibitem[Shano and Kettler(1970)]{Shano1970b}
D.~F. Shano and P.~C. Kettler.
\newblock Optimal conditioning of quasi-{Newton} methods.
\newblock \emph{Mathematics of Computation}, 24\penalty0 (111):\penalty0
  657--664, 1970.

\bibitem[Shimko(1993)]{Shimko1993}
D.~Shimko.
\newblock Bounds of probability.
\newblock \emph{Risk}, 6\penalty0 (4):\penalty0 34--37, 1993.

\bibitem[Skiadopoulos et~al.(1999)Skiadopoulos, Hodges, and
  Clewlow]{Skiadopoulos1999}
G.~Skiadopoulos, S.~Hodges, and L.~Clewlow.
\newblock The dynamics of the {S\&P} 500 implied volatility surface.
\newblock \emph{Review of Derivatives Research}, 3:\penalty0 263--282, 1999.

\bibitem[Stein and Stein(1991)]{Stein1991}
E.~Stein and J.~Stein.
\newblock Stock price distributions with stochastic volatility: an analytic
  approach.
\newblock \emph{Review of Financial Studies}, 4:\penalty0 727--752, 1991.

\bibitem[Storn and Price(1997)]{StornPrice1997}
R.~Storn and K.~Price.
\newblock Differential evolution - a simple and efficient heuristic for global
  optimization over continuous spaces.
\newblock \emph{Journal of Global Optimization}, 11\penalty0 (4):\penalty0
  341--359, 1997.

\bibitem[Uhlenbeck and Ornstein(1930)]{OU1930}
G.~E. Uhlenbeck and L.~S. Ornstein.
\newblock On the theory of {Brownian Motion}.
\newblock \emph{Physical Review}, 36:\penalty0 823--841, 1930.

\bibitem[Weron and Wystup(2005)]{Weron2005}
R.~Weron and U.~Wystup.
\newblock \emph{{Heston's} Model and the Smile}, pages 161--181.
\newblock Springer Berlin Heidelberg, Berlin, Heidelberg, 2005.
\newblock ISBN 978-3-540-27395-0.
\newblock \doi{10.1007/3-540-27395-6_7}.
\newblock URL \url{http://dx.doi.org/10.1007/3-540-27395-6_7}.

\bibitem[Wiggins(1987)]{Wiggins1987}
J.~Wiggins.
\newblock Option values under stochastic volatilities.
\newblock \emph{Journal of Financial Economics}, 19:\penalty0 351--372, 1987.

\bibitem[{World Bank}(2016)]{WorldBank2016}
{World Bank}.
\newblock {World Development Indicators}.
\newblock 2016.
\newblock URL \url{http://data.worldbank.org/indicator/SP.DYN.LE00.FE.IN}.
\newblock GDP (current US\$) [Data file].

\bibitem[Zhang et~al.(2016)Zhang, Zhen, Sun, and Zhao]{Zhang2016}
J.~E. Zhang, F.~Zhen, X.~Sun, and H.~Zhao.
\newblock The skewness implied in the {Heston} model and its application.
\newblock \emph{Journal of Futures Markets}, pages n/a--n/a, 2016.
\newblock ISSN 1096-9934.
\newblock \doi{10.1002/fut.21801}.
\newblock URL \url{http://dx.doi.org/10.1002/fut.21801}.

\bibitem[Zhao et~al.(2013)Zhao, Zhang, and Chang]{Zhao2013}
H.~Zhao, J.~E. Zhang, and E.~C. Chang.
\newblock The relation between physical and risk-neutral cumulants.
\newblock \emph{International Review of Finance}, 13\penalty0 (3):\penalty0
  345--381, 2013.
\newblock ISSN 1468-2443.
\newblock \doi{10.1111/irfi.12013}.
\newblock URL \url{http://dx.doi.org/10.1111/irfi.12013}.

\end{thebibliography}
\addcontentsline{toc}{chapter}{References}
\markboth{\hfill \textit{References} \hfill}{}
\renewcommand{\MakeUppercase}[1]{\uppercase{#1}}

\clearpage
\phantomsection

\end{document}